\documentclass[10pt,reqno]{amsart}
\usepackage[a4paper, margin=2.5cm,bottom=2.5cm]{geometry}
\usepackage{hyperref} 
\usepackage{amssymb, mathrsfs,bbold}
\usepackage{amsmath, amsthm}
\usepackage{enumerate}
\usepackage{dsfont}
\usepackage{color}
\usepackage{empheq}

\usepackage[utf8]{inputenc}
\usepackage{stmaryrd}
\usepackage{titlesec, wasysym}
\usepackage{appendix}
\usepackage{todonotes, fancyhdr}
\usepackage{chngcntr}
\usepackage{cancel, stmaryrd}
\usepackage{tikz-cd}
\usepackage{comment}

\usepackage{setspace}
\renewcommand{\baselinestretch}{0.99}
\newcommand{\rmd}{\mathrm{d}}

\numberwithin{subsection}{section}
\numberwithin{subsubsection}{subsection}
\numberwithin{equation}{section}

\DeclareRobustCommand{\TitleEquation}[1]{\texorpdfstring{$\boldsymbol{#1}$}{#1}}

\newenvironment{Dem}[1][\unskip]{%
    \begin{list}{\hspace{1.15cm}{\sf \textbf{{Proof #1 --}}}}{   
        \setlength{\topsep}{0pt}%
        \setlength{\leftmargin}{0pt}%
        \setlength{\rightmargin}{0pt}%
        \setlength{\listparindent}{0pt}%
        \setlength{\itemindent}{0pt}%
        \setlength{\parsep}{0pt}%
        \addtolength{\leftmargin}{0pt}  
        \addtolength{\rightmargin}{0pt}%
    } \item }{\hfill $\rhd$\end{list}\smallskip}
    
\newenvironment{Dem*}[1][\unskip]{%
    \begin{list}{\hspace{0cm}{\sf \textbf{{\small Proof #1 --}}}}{%
        \setlength{\topsep}{0pt}%
        \setlength{\leftmargin}{0pt}%
        \setlength{\rightmargin}{0pt}%
        \setlength{\listparindent}{0pt}%
        \setlength{\itemindent}{0pt}%
        \setlength{\parsep}{0pt}%
        \addtolength{\leftmargin}{20pt}%
        \addtolength{\rightmargin}{0pt}%
    } \item }{\hfill $\rhd$\end{list}\smallskip}    
    
\newenvironment{DemLemma}{%
    \begin{list}{\hspace{1.15cm}{\textsf{\textbf{Proof of Lemma \ref{lem:8} --}}}}{%
        \setlength{\topsep}{0pt}%
        \setlength{\leftmargin}{0pt}%
        \setlength{\rightmargin}{0pt}%
        \setlength{\listparindent}{0pt}%
        \setlength{\itemindent}{0pt}%
        \setlength{\parsep}{0pt}%
        \addtolength{\leftmargin}{0pt}%
        \addtolength{\rightmargin}{0pt}%
    } \item }{\hfill{\space $\rhd$}\end{list}\smallskip}

\newenvironment{Rem}[1][\unskip]{%
    \begin{list}{\hspace{1.15cm}{\textsf{\textbf{{\small \textsl{Remark}} --}}}}{%
        \setlength{\topsep}{0pt}%
        \setlength{\leftmargin}{0pt}%
        \setlength{\rightmargin}{0pt}%
        \setlength{\listparindent}{0pt}%
        \setlength{\itemindent}{0pt}%
        \setlength{\parsep}{0pt}%
        \addtolength{\leftmargin}{20pt}%
        \addtolength{\rightmargin}{0pt}%
    } \item }{\hfill \end{list}\esp}

\newenvironment{Examples}[1][\unskip]{%
    \begin{list}{\hspace{1.15cm}{\textsf{\textbf{{\small \textsl{Examples}} }}}}{%
        \setlength{\topsep}{0pt}%
        \setlength{\leftmargin}{0pt}%
        \setlength{\rightmargin}{0pt}%
        \setlength{\listparindent}{0pt}%
        \setlength{\itemindent}{0pt}%
        \setlength{\parsep}{0pt}%
        \addtolength{\leftmargin}{20pt}%
        \addtolength{\rightmargin}{0pt}%
    } \item }{\hfill \end{list}\esp}

\newenvironment{Example}[1][\unskip]{%
    \begin{list}{\hspace{1.15cm}{\textsf{\textbf{{\small \textsl{Example}} }}}}{%
        \setlength{\topsep}{0pt}%
        \setlength{\leftmargin}{0pt}%
        \setlength{\rightmargin}{0pt}%
        \setlength{\listparindent}{0pt}%
        \setlength{\itemindent}{0pt}%
        \setlength{\parsep}{0pt}%
        \addtolength{\leftmargin}{20pt}%
        \addtolength{\rightmargin}{0pt}%
    } \item }{\hfill \end{list}\esp}

\renewcommand\thesection       {\arabic{section}}
\renewcommand\thesubsection    {\thesection{\boldmath $.$}\arabic{subsection}}
\renewcommand\thesubsubsection    {\thesection{\boldmath $.$}\arabic{subsection}{\boldmath $.$}\arabic{subsubsection}}

\titleformat{\section}[block]
{\filcenter\normalfont\sffamily\bfseries\Large}
{{\hspace{-0.7cm}}\thesection \hspace{0.2em} --\vspace{0.3cm}}{0.5em}{}

\titleformat{\subsection}[block]
{\filcenter\normalfont\sffamily\bfseries\large}  						  
{\hspace{-0.7cm}\thesubsection \hspace{0.5em} \vspace{0.3cm}}{.5em}{}  
\titlespacing{\subsection}{-0pc}{1.5ex plus .1ex minus .2ex}{0pc}

\titleformat{\subsubsection}[runin]
{\normalfont\sffamily\bfseries}					  
{\thesubsubsection \vspace{0cm}}{.5em}{}  
\titlespacing{\subsection}{-0pc}{1.5ex plus .1ex minus .2ex}{0pc}

\newtheoremstyle{mystyle}
{3pt}               
{3pt}               
{\it }                      
{}                      
{\sffamily\bfseries}             
{}                      
{0.5em}                 
{#1 #2{\hspace{0.2cm}--\hspace{-0.2cm}}  }

\theoremstyle{mystyle}

\newtheorem{thm}{Theorem}[section]
\newtheorem*{thm*}{Theorem}

\newenvironment{DemThm7}{%
    \begin{list}{\hspace{1.15cm}{\textsf{\textbf{Proof of Theorem~\ref{ThmLpComingDown} --}}}}{%
        \setlength{\topsep}{0pt}%
        \setlength{\leftmargin}{0pt}%
        \setlength{\rightmargin}{0pt}%
        \setlength{\listparindent}{0pt}%
        \setlength{\itemindent}{0pt}%
        \setlength{\parsep}{0pt}%
        \addtolength{\leftmargin}{0pt}%
        \addtolength{\rightmargin}{0pt}%
    } \item }{\hfill{\space $\rhd$}\end{list}\smallskip}

\newtheorem{cor}[thm]{\hspace{-0.15cm}  {Corollary}}
\newtheorem{lem}[thm]{\hspace{-0.14cm}  {Lemma}}
\newtheorem{prop}[thm]{\hspace{-0.13cm} {Proposition}}
\newtheorem{defn}[thm]{ \hspace{-0.32cm} {Definition}}
\newtheorem{rem}[thm]{\hspace{-0.15cm} {Remark}}

\newtheoremstyle{mystyle2}
{3pt}               
{3pt}               
{\it }                      
{}                      
{\sffamily\bfseries}             
{}                      
{0.5em}                 
{\llap{#2 }#1{\hspace{0.2cm}--}}

\theoremstyle{mystyle2}

\newtheorem*{definition*}{Definition}
\newtheorem*{theorem*}{Theorem}
\newtheorem*{Remark*}{Remark}
\newtheorem*{rem*}{\hspace{-0.15cm} {Remark}}
\newtheorem*{lem*}{Lemma}
\newtheorem*{defn*}{Definition}
\newtheorem*{prop*}{Proposition}
\newtheorem*{cor*}{Corollary}

\newcommand{\ssk}{\smallskip}

\renewcommand{\epsilon}{\varepsilon}
\newcommand{\eps}{\epsilon}

\makeatletter
\newcommand*{\bigcdot}{}
\DeclareRobustCommand*{\bigcdot}{%
  \mathbin{\mathpalette\bigcdot@{}}%
}
\newcommand*{\bigcdot@scalefactor}{.5}
\newcommand*{\bigcdot@widthfactor}{1.15}
\newcommand*{\bigcdot@}[2]{%
  \sbox0{$#1\vcenter{}$}
  \sbox2{$#1\cdot\m@th$}%
  \hbox to \bigcdot@widthfactor\wd2{%
    \hfil
    \raise\ht0\hbox{%
      \scalebox{\bigcdot@scalefactor}{%
        \lower\ht0\hbox{$#1\bullet\m@th$}%
      }%
    }%
    \hfil
  }%
}
\makeatother

\newcommand{\dens}{\mathrm{dens}}

\newcommand\bbE{\mathbb{E}}

\newcommand\bbN{\mathbb{N}}
\newcommand\bbR{\mathbb{R}}
\newcommand\bbP{\mathbb{P}}

\newcommand{\mcA}{\mathcal{A}}
\newcommand{\mcB}{\mathcal{B}} 
\newcommand{\mcC}{\mathcal{C}} 
\newcommand{\mcD}{\mathcal{D}}
\newcommand{\mcE}{\mathcal{E}}
\newcommand{\mcF}{\mathcal{F}}
\newcommand{\mcG}{\mathcal{G}}

\newcommand{\mcL}{\mathcal{L}}
\newcommand{\mcM}{\mathcal{M}}

\newcommand\mcO{\mathcal{O}}
\newcommand{\esp}{\textcolor{white}{a}}
\newcommand{\mcP}{\mathcal{P}}
\newcommand{\mcQ}{\mathcal{Q}}
\newcommand\mcS{\mathcal{S}}

\newcommand{\mcR}{\mathcal{R}}
\newcommand\mcT{\mathcal T}
\newcommand\mcU{\mathcal U}

\newcommand\mcX{\mathcal X}
\newcommand\mcY{\mathcal Y}

\makeatletter
\newcommand*{\defeq}{\overset{\text{\tiny{$\mathrm{def}$}}}{=}}
\makeatother

\makeatletter
\newcommand*{\eqdef}{=\mathrel{\rlap{\raisebox{0.3ex}{$\m@th\cdot$}}%
					                     \raisebox{-0.3ex}{$\m@th\cdot$}}%
                     }
\makeatother

\newcommand{\Com}{\mathrm{Com}}
\usepackage{tikz-cd}
\usetikzlibrary{calc}
\usetikzlibrary{shapes.misc}
\usetikzlibrary{shapes.symbols}
\usetikzlibrary{snakes}
\usetikzlibrary{decorations}
\usetikzlibrary{decorations.markings}

\makeatletter
\pgfdeclareshape{crosscircle}
{
  \inheritsavedanchors[from=circle] 
  \inheritanchorborder[from=circle]
  \inheritanchor[from=circle]{north}
  \inheritanchor[from=circle]{north west}
  \inheritanchor[from=circle]{north east}
  \inheritanchor[from=circle]{center}
  \inheritanchor[from=circle]{west}
  \inheritanchor[from=circle]{east}
  \inheritanchor[from=circle]{mid}
  \inheritanchor[from=circle]{mid west}
  \inheritanchor[from=circle]{mid east}
  \inheritanchor[from=circle]{base}
  \inheritanchor[from=circle]{base west}
  \inheritanchor[from=circle]{base east}
  \inheritanchor[from=circle]{south}
  \inheritanchor[from=circle]{south west}
  \inheritanchor[from=circle]{south east}
  \inheritbackgroundpath[from=circle]
  \foregroundpath{
    \centerpoint%
    \pgf@xc=\pgf@x%
    \pgf@yc=\pgf@y%
    \pgfutil@tempdima=\radius%
    \pgfmathsetlength{\pgf@xb}{\pgfkeysvalueof{/pgf/outer xsep}}%
    \pgfmathsetlength{\pgf@yb}{\pgfkeysvalueof{/pgf/outer ysep}}%
    \ifdim\pgf@xb<\pgf@yb%
      \advance\pgfutil@tempdima by-\pgf@yb%
    \else%
      \advance\pgfutil@tempdima by-\pgf@xb%
    \fi%
    \pgfpathmoveto{\pgfpointadd{\pgfqpoint{\pgf@xc}{\pgf@yc}}{\pgfqpoint{-0.707107\pgfutil@tempdima}{-0.707107\pgfutil@tempdima}}}
    \pgfpathlineto{\pgfpointadd{\pgfqpoint{\pgf@xc}{\pgf@yc}}{\pgfqpoint{0.707107\pgfutil@tempdima}{0.707107\pgfutil@tempdima}}}
    \pgfpathmoveto{\pgfpointadd{\pgfqpoint{\pgf@xc}{\pgf@yc}}{\pgfqpoint{-0.707107\pgfutil@tempdima}{0.707107\pgfutil@tempdima}}}
    \pgfpathlineto{\pgfpointadd{\pgfqpoint{\pgf@xc}{\pgf@yc}}{\pgfqpoint{0.707107\pgfutil@tempdima}{-0.707107\pgfutil@tempdima}}}
  }
}
\makeatother

\colorlet{symbols}{black}    
\colorlet{testcolor}{green!60!black}
\colorlet{supcolor}{red!60!black}

\tikzset{
	root/.style={circle, fill=testcolor!70, draw=testcolor, inner sep=1pt, minimum size=0.5mm},
	dot/.style={circle, draw=black, fill=black, inner sep=0pt, minimum size=0.2mm},
	noise/.style={circle, draw=black, fill=white, inner sep=0pt, minimum size=1mm},
	noiseblue/.style={circle, fill=blue!40, draw=blue, inner sep=0pt, minimum size=1mm},
	noisegray/.style={circle, fill=gray!40, draw=gray, inner sep=0pt, minimum size=1mm},
	blackdot/.style={circle, draw=black, fill=black, inner sep=0pt, minimum size=1.2mm},
	K/.style= {semithick, shorten >=0pt,shorten <=0pt,-},
	DK/.style={thick, densely dotted, shorten >=0pt,shorten <=0pt},   
	}

\newcommand{\X}{\raisebox{-.2ex}{\includegraphics[scale=1]{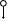}}}
\newcommand{\Xtwo}{\raisebox{0ex}{\includegraphics[scale=1]{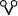}}} 

\newcommand{\IXtwo}{\raisebox{-.2ex}{\includegraphics[scale=1]{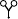}}}
\newcommand{\IXthree}{\raisebox{-.3ex}{\includegraphics[scale=1]{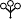}}}

\newcommand{\XtwoIXtwo}{\raisebox{-.2ex}{\includegraphics[scale=1]{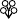}}}

\newcommand{\XtwoIXthree}{\raisebox{-.2ex}{\includegraphics[scale=1]{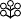}}}
\newcommand{\XIXthree}{\raisebox{-.2ex}{\includegraphics[scale=1]{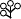}}}

\begin{document}
\thispagestyle{fancy}
\begin{center}
{\LARGE\sffamily{$\Phi^4_3$ measures on compact Riemannian $3$-manifolds   \vspace{0.5cm}}}
\end{center}

\begin{center}
{\sf I. BAILLEUL, N. V. DANG, L. FERDINAND and T.D. T\^O}
\end{center}

\vspace{1cm}

\begin{center}
\begin{minipage}{0.8\textwidth}
\renewcommand\baselinestretch{0.7} \scriptsize \textbf{\textsf{\noindent Abstract.}} We construct the $\Phi^4_3$ measure on an arbitrary 3-dimensional compact Riemannian manifold without boundary as an invariant probability measure of a singular stochastic partial differential equation. Proving the nontriviality and the covariance under Riemannian isometries of that measure gives a non-perturbative, non-topological interacting Euclidean quantum field measure on curved spaces in dimension 3. To control analytically several Feynman diagrams appearing in the construction of a number of random fields, we introduce a novel approach of renormalisation using microlocal and harmonic analysis. This allows to obtain a renormalized equation which involves some universal constants independent of the manifold. In a companion paper \cite{BDFTCompanion}, we develop all the tools from paradifferential and microlocal analysis that we use to build in our manifold setting a number of analytic and probabilistic objects.
\end{minipage}
\end{center}

\setcounter{tocdepth}{2}

\makeatletter
\def\@dotsep{1.5}
\renewcommand{\l@section}  {\@dottedtocline{1}{0em}{1em}}
\renewcommand{\l@subsection}  {\@dottedtocline{2}{2em}{2em}}
\makeatother

{\small 
\tableofcontents
}

\section{Introduction}
\label{SectionIntro}

In the setting of the discrete $d$-dimensional torus $\Lambda_d =\left(\frac{\mathbb{Z}}{N\mathbb{Z}}\right)^d $, for $d \geqslant 2$, a field is represented by a real-valued function $\sigma$ on $\Lambda_d$. Write $i\sim j$ when two points $i$ and $j$ are neighbours in $\Lambda_d$. One can assign an energy
\begin{equation} \label{EqEnergy}
S(\sigma) = \frac{1}{2} \sum_{i\sim j} \vert\sigma_i-\sigma_j\vert^2 + \frac{1}{2}\sum_{i}\vert\sigma_i\vert^2
\end{equation}
to any field $\sigma$ and define a Gibbs probability measure $\nu_{\Lambda_d}$ proportional to 
$$
e^{-S(\sigma)}\prod_{i\in\Lambda_d} \rmd\sigma_i.
$$
A random variable with values in the space $\mcE_d=\bbR^{\Lambda_d}$ of fields, with law $\nu_{\Lambda_d}$, is called a discrete Gaussian free field. The continuum analogue of this random variable is the Gaussian free field on the torus $\mathbb{T}^d$, characterized by the fact that it is a random centered Gaussian field $\zeta$ with covariance $(1-\Delta)^{-1}$ on the torus $\mathbb{T}^d$. This means that for any smooth real-valued test function $f\in C^\infty(\mathbb{T}^d)$ the random variable $\zeta(f)$ is Gaussian with zero mean and covariance $\langle f, (1-\Delta)^{-1}f\rangle_{L^2(\mathbb{T}^d)}$. One can construct $\zeta$ as a random distribution that is almost surely of Besov-H\"older regularity $-\frac{d-2}{2}-\epsilon$, for all $\epsilon>0$. Note the dependence of the regularity exponent on the dimension. Back to the discrete setting, we denote by $\Delta_{\mcE_d}$ the canonical Laplace operator on the field space $\mcE_d\simeq \mathbb{R}^{\Lambda_d}$ and by ${\bf 1}$ the constant function on $\mcE_d$ equal to $1$. A dynamical picture of the Gibbs measure $\nu_{\Lambda_d}$ can be obtained from the genuine identity
$$
\nabla_{\mcE_d}({\bf 1}_{\mcE_d}) = 0
$$
by rewriting it under the form
$$
\nabla_{\mcE_d}\big(e^{-S}\nabla_{\mcE_d}e^S\big)(e^{-S}) = 0.
$$
The density $e^{-S}$ appears here as an element of the kernel of the dual of the conjugated operator 
$$
\big\{\nabla_{\mcE_d}\big(e^{-S}\nabla_{\mcE_d}e^S\big)\big\}^* = \Delta_{\mcE_d} - (\nabla_{\mcE_d} S)\cdot\nabla_{\mcE_d}.
$$
This allows a construction of $\nu_{\Lambda_d}$ as the invariant measure of the Markov process on $\mcE_d$ with generator $\Delta_{\mcE_d} - (\nabla_{\mcE_d} S)\cdot\nabla_{\mcE_d}$ -- provided this Markov process has indeed a unique invariant probability measure. The diffusion associated with the operator $\Delta_{\mcE_d} - (\nabla_{\mcE_d} S)\cdot\nabla_{\mcE_d}$ is the solution of the stochastic differential equation in $\mcE_d$
\begin{equation*}
\rmd z_t = \sqrt{2}\rmd w_t - \nabla_{\mcE_d} S(z_t)\rmd t,
\end{equation*}
for a Brownian motion $w$ in $\mcE_d$.

\ssk

The energy $S$ in \eqref{EqEnergy} contains a kinetic term $\sum_{i\sim j}$ and a potential term $\sum_i$. One can add to the potential term an additional bit of the form 
$$
Q(\sigma) = \sum_i Q(\sigma_i).
$$
for a real-valued function $Q$, typically a real polynomial bounded from below for the so--called Ginzburg--Landau models. The corresponding Gibbs probability measure $\mu_{\Lambda_d}$ can then be seen as a perturbation of the discrete Gaussian free field probability measure $\nu_{\Lambda_d}$
$$
\mu_{\Lambda_d}(\rmd\sigma) \sim e^{-Q(\sigma)}\nu_{\Lambda_d}(\rmd\sigma).
$$
The dynamics on $\mcE_d$ associated with this measure is given as above by the stochastic differential equation
\begin{equation} \label{EqSDE}
\rmd z_t = \sqrt{2} \rmd w_t - \nabla_{\mcE_d} (S+Q)(z_t)\rmd t.
\end{equation}
There is a problem for taking the formal continuum limit of these measures as the Gaussian free field measure is supported on a set of distributions of negative regularity, so an expression like $\int Q(\sigma)$ does not make sense for a nonlinear function $Q$ of a distribution $\sigma$. It is the aim of the Euclidean quantum field theory of scalar fields to make sense of and construct such measures for some particular examples of potentials $Q$. The $\Phi^4_3$ measure corresponds to $Q(a)=a^4/4$ in a $3$-dimensional setting.

\ssk

Quantum field theory was developed as a theory describing interactions of the elementary particles. It is arguably one of the most successful physical theory of the $20$th century and has led to remarkable physical predictions with unprecedented numerical accuracy. However in spite of its success in theoretical physics a complete mathematical formulation and understanding of quantum field theory is still work in progress. A major difficulty in the subject comes from the divergences inherent to the formulation of the theory. In quantum field theory the perturbative calculation of any physical process involves a summation over an infinite number of virtual multi-particle states which is generically divergent, hence produces infinities. The divergences of perturbation theory in quantum field theory are directly linked to its short distance structure which is highly non-trivial because its description involves the infinity of multi-particle states. These divergences must be carefully subtracted in some organized way compatible with physical requirements such as locality, causality, unitarity. The methods developped to deal with these infinities were called {\sl renormalization}. Constructive quantum field theory provides one of the mathematically  rigorous approaches to quantum field theory. It was developped in the 70s with seminal contributions of Albeverio, Brydges, Feldman, Fr\"ohlich, Gallavotti, Gawedzki, Glimm, Guerra, Jaffe, Kupiainen, Nelson, Rivasseau, Seiler, Sénéor, Spencer, Simon, Symanzik and Wightman to name but a few -- see e.g. Glimm \& Jaffe's book \cite{GlimmJaffe} and the references inside for an account of the early achievements in this domain. One of the first successes of constructive quantum field theory was the construction of the so called $P(\phi)_2$-model on $\mathbb{R}^2$ and of the $\phi^4_3$ theory on $\mathbb{R}^3$ by Glimm \& Jaffe \cite{Glimm68,GJ73} in the 70s. The recent breakthroughs by Hairer \cite{Hairer} and Gubinelli, Perkowski \& Imkeller \cite{GIP} allowed several authors to recover the results of Glimm \& Jaffe following the stochastic quantization program of Parisi \& Wu \cite{ParisiWu}, using only PDE and probabilistic techniques without the intricate combinatorial methods from the constructive school. However we would like to emphasize that the powerful cluster expansion techniques developed by the constructive school allowed, for both $P(\phi)_2,\phi^4_3$ theories, to show the Borel summability of the partition function in the couplings, to control the mass gap (exponential decay of the correlations), to study phase transitions (when there is no longer exponential decay of correlations), to investigate fine properties of the spectrum and scattering properties of these theories and finally, one of the deepest results of the constructive school was the stability proofs of Yang-Mills in $3$ and $4$ dimensions due to Balaban~\cite{Balaban3d,Balaban4d}. Many of these results are currently out of reach of the stochastic methods: We refer to the book of Rivasseau \cite{Rivasseau} for more information on the topic. The dynamical construction of the $\Phi^4_3$ measure was first done in finite volume on the flat $3$-dimensional torus by Mourrat \& Weber \cite{MourratWeber}, Hairer \& Mattingly \cite{HairerMattingly}, Hairer \& Sch\"onbauer \cite{HairerSchonbauer}, Albeverio \& Kusuoka \cite{AlbeverioKusuoka1}, then in the infinite volume $3$-dimensional Euclidean space by different authors -- Albeverio \& Kusuoka \cite{AlbeverioKusuoka2}, Moinat \& Weber \cite{MoinatWeber}, Gubinelli \& Hofmanov\'a \cite{GubinelliHofmanova1, GubinelliHofmanova2}, Barashkov \& Gubinelli \cite{BarashkovGubinelli, BarashkovGubinelliGirsanov},  and Duch \cite{Duch21,Duch22}, using different methods. A crucial integrability property of the $\phi^4_3$ measure was proved in Hairer \& Steele's work \cite{HairerSteele}. Gubinelli's lecture notes \cite{LN1Gubinelli, LN2Gubinelli} provide a remarkable source of inspiration and information on the stochastic quantization approach to the construction of the $\Phi^4_3$ measure.
      
Yet most results in constructive quantum field theory are proved in the geometric settings of either $\mathbb{R}^3$ or the flat torus $\mathbb{T}^3$. On the other hand quantum field theory on curved spacetimes has been studied since the 70s. Recent breakthroughs by Brunetti \& Fredenhagen \cite{BF00}, Hollands \& Wald \cite{HW1,HW2} and Rejzner \cite{Rejzner} on Lorentzian manifolds (see \cite{Dangthese} for a detailed mathematical exposition of part of this approach) and Kopper \& M\"uller \cite{KM07} in a Riemannian setting, lead to complete proofs of {\sl perturbative renormalization} on (pseudo)-Riemannian manifolds of dimension less than or equal to $4$. We also mention the approach of Costello \cite{Cos09} that is designed to work on Riemannian manifolds, possibly with a boundary after Albert's work \cite{AlbertCostello}. However all these results are perturbative and construct quantum field theory objects as formal power series. They do not provide any probability measures, Hilbert spaces or operators in a straightforward way.
Constructive quantum fields on manifolds have been earlier addressed only on {\it compact surfaces} in~\cite{Pickrell} by Pickrell and \cite{Dimock} by Dimock for the $P(\varphi)_2$ theories, in \cite{Levy} by L\'evy for the $2d$ Yang-Mills theory and in \cite{GuillarmouKupiainenRhodesVargas} by Guillarmou, Kupiainen, Rhodes \& Vargas for the Liouville field theory on Riemann surfaces. In Lorentz signature, we would also like to mention the work~\cite{BarataJaekelMund} of Barata, Jaekel \& Mund who managed to define the $P(\varphi)_2$ theory on $2$-dimensional de Sitter space, extending previous work~\cite{FigariHoeghKrohnNappi} of Figari, Høegh-Krohn \& Nappi. From the PDE side, the constructions of Gibbs measures on Riemannian surfaces that we are aware of, come from~\cite{BurqThomannTzvetkov} by Burq, Thomann \& Tzvetkov for dynamical $P(\Phi)_2$ and from~\cite{OhRobertTzvetkovWang} by Oh, Robert, Tzvetkov \& Wang for the dynamical Liouville model.

\bigskip

{\subsection{Stochastic quantization and $\Phi^4_3$ measures}}
\label{SectionIntroStochQuant}

Let $(M,g)$ stand for a closed $3$-dimensional Riemannian manifold. This work is dedicated to constructing the $\Phi^4_3$ measure over $M$, formally the ill-defined functional integral measure 
$$
\frac{1}{Z}{\exp\Big({-\frac{1}{2}\int_M(\vert\nabla u\vert^2+u^2) -\frac{1}{4}\int_M {u^4}  } \Big)
\rmd u},
$$ 
as an invariant probability measure of the dynamics
\begin{equation} \label{EqSPDE}
(\partial_t+P)u =  - u^3+\sqrt{2} \xi\,,
\end{equation}
where $\xi$ stands for a spacetime white noise and 
\[
P\defeq (1-\Delta_g).
\] 
The noise $\xi$ plays in a continuum setting the role of the Brownian motion $w$ in \eqref{EqSDE} in a discrete setting while the terms $P u$ and $u^3$ come from the gradient terms of the energy $S$ and the quartic potential $Q$ respectively. The construction of the $\Phi^4_3$ measure as the hopefully unique invariant probability measure of this dynamics was first put forward by Parisi \& Wu in a famous work of the early 80s; it comes under the name of {\sl stochastic quantization}. Note that we define the $\Phi^4_3$ measure as an invariant measure from Equation \eqref{EqSPDE}. This is a priori not equivalent to obtaining the measure as a scaling limit of lattice models, a highly non-trivial issue on manifolds since there is no canonical way of discretising quantum field theories on manifolds. We do not try to relate here our construction of the $\Phi^4_3$ measure with any such limiting procedure.

Equation \eqref{EqSPDE} involves the fundamental problem of considering a nonlinear function of a distribution. Spacetime white noise on a $3$-dimensional Riemannian manifold has indeed almost surely a parabolic~\footnote{Meaning the time variable has weight $2$ whereas space variables have weight $1$.} Besov-H\"older regularity $-5/2-\epsilon$, for all $\epsilon>0$, so one does not expect from a possible solution $u$ to Equation \eqref{EqSPDE} that it has parabolic regularity better than $-1/2-\epsilon$, as a consequence of Schauder estimate. The term $u^3$ in \eqref{EqSPDE} is thus ill-defined. This kind of problem in a stochastic partial differential equation (PDE) is characteristic of the class of singular stochastic PDEs, a field that was opened around $2014$ by the groundbreaking works of M. Hairer on regularity structures \cite{Hairer} and Gubinelli, Imkeller \& Perkowski \cite{GIP} on paracontrolled calculus. The tools of regularity structures and paracontrolled calculus were used to run the stochastic quantization approach for constructing of the $\Phi^4_3$ measure over a $3$-dimensional torus and Euclidean space in a series of works. Local well-posedness of Equation \eqref{EqSPDE} was proved first by Hairer in \cite{Hairer} -- see also Catellier \& Chouk's work \cite{CatellierChouk} for a proof of that result with the tools of paracontrolled calculus. Mourrat \& Weber proved in \cite{MourratWeber} an a priori estimate that gives the long time existence (and well-posedness) of the solution to \eqref{EqSPDE} and the existence of an invariant probability measure. The uniqueness of this invariant probability measure comes from the works of Hairer \& Mattingly \cite{HairerMattingly} on the strong Markov property of transition semigroups associated to singular stochastic PDEs, and Hairer \& Sch\"onbauer \cite{HairerSchonbauer} on the support of the laws of solutions to singular stochastic PDEs. See Hairer \& Steele's work \cite{HairerSteele} for more references.

\ssk

None of the previous works are readily available in a manifold setting, and so far the only works on singular stochastic PDEs in a manifold setting can be divided in two groups:
\begin{itemize}
	\item[--] The works \cite{BB1, BB2, BB3} of Bailleul \& Bernicot, and the works \cite{DahlqvistDiehlDriver}, \cite{Mouzard} and \cite{BDM} of Dahlqvist, Diehl \& Driver, Mouzard and Bailleul, Dang \& Mouzard on the Anderson operator on a $2$-dimensional Riemannian manifold. We refer to~\cite{Mouzardthese} for a lucid exposition of some of the above results.   \vspace{0.1cm}

	\item[--] The works \cite{BurqThomannTzvetkov} by
Burq, Thomann \& Tzvetkov for dynamical $P(\Phi)_2$ and from~\cite{OhRobertTzvetkovWang} by
Oh, Robert, Tzvetkov \& Wang for the dynamical Liouville model, both works deal with interacting QFT measures 
on surfaces.
\end{itemize}

We would also like to mention the work of Hairer \& Singh \cite{HairerSingh} which develops a generalisation of the original Theory of Regularity Structures which is able to treat SPDEs on manifolds with values in vector bundles in full generality. Their work gives \emph{only short time existence} for the SPDE which overlaps with our own result. However, to build the QFT measure requires the long time existence and the coming down from infinity property for the solutions of the SPDE, which are not covered by~\cite{HairerSingh}.

\ssk

The aim of the present work, together with \cite{Phi43Uniqueness}, is to develop all the tools needed to run the analysis in a $3$ dimensional closed Riemannian manifold. On the purely analytical side
\begin{itemize}
	\item[--] We follow Jagannath \& Perkowski's simple approach \cite{JagannathPerkowski} of Equation \eqref{EqSPDE} to prove that this equation is locally well-posed. Their formulation of the problem avoids the use of regularity structures or paracontrolled calculus.  We freely use several tools of pseudodifferential and paradifferential calculus on manifolds which are developed in detail in~\cite{BDFTCompanion}.   \vspace{0.1cm}

	\item[--] We give a simple and short proof of an $L^p$ ``{\sl coming down from infinity}'' property satisfied by the solution to Equation \eqref{EqSPDE} using energy methods. The longtime existence of a unique solution to Equation \eqref{EqSPDE}, {and the existence of an invariant measure for the associated Markovian dynamics}, follow as a consequence.   \vspace{0.1cm}
	
	\item[--] As usual in the study of singular stochastic PDEs we need to feed the analytic machinery with a number of random distributions whose formal definitions involve some ill-defined products and whose actual definitions involve some probabilistic constructions based on regularization and renormalization. Our approach to the renormalization problem relies on microlocal analysis. It is a far reaching generalization of the Epstein-Glaser point of view where we benefit from the many improvements contained in \cite{PopineauStora, BF00, HW1, HW2}. We reduce the problem of renormalization to an extension problem for distributions on a configuration space defined outside all the diagonals, for which we develop a general machinery. We believe this construction to be of independent interest.    \vspace{0.1cm}
    
    \item[--] To control analytically the Feynman amplitudes appearing in the stochastic bounds we feed our renormalization machine with several microlocal estimates of distributional kernels which are done separately in~\cite{BDFTCompanion}. 
\end{itemize}

\ssk

We note that there is also a new approach to stochastic PDE's relying on the Epstein-Glaser renormalization in the works \cite{DDRZ} by  Dappiaggi, Drago, Rinaldi \& Zambotti and \cite{BDR} by Bonicelli, Dappiaggi \& Rinaldi. However it seems that these authors work only at a perturbative level whereas our results are non-perturbative.

\ssk

One remarkable feature of our approach is that we are able to renormalize Equation \eqref{EqSPDE} using some universal counterterms -- they do not depend on the metric on $M$; {Section \ref{SectionLocallyCovariant} gives some comments on this point}. In the following statement we take a positive constant $r\in(0,1/8)$ and let $\xi_r\defeq e^{-rP}(\xi)$ stand for a {\it space regularization} of the spacetime white noise $\xi$ by the heat operator, so $\xi_r$ is still white in time. Set
\begin{equation} \label{EqDefnArBr} \begin{split}
a_r &\defeq \frac{r^{-1/2}}{4\sqrt{2} \, \pi^{3/2}}\,,   \\
b_r &\defeq \frac{\vert\hspace{-0.03cm}\log r\vert}{16\pi^2}.
\end{split} \end{equation}

\ssk

\begin{thm} \label{ThmSolutionTheory}
Pick $\phi\in \mcC^{-1/2-\epsilon}(M)$. The equation
\begin{equation} \label{EqRenormalizedSPDE}
(\partial_t - \Delta + 1)u_r = \sqrt{2}\xi_r - u_r^3 + 3(a_r - b_r)u_r
\end{equation}
with initial condition $\phi$ has a unique solution over $[0,\infty)\times M$ in some appropriate function space. For any $0<T<\infty$ this random variable converges in probability in $C\big([0,T], \mcC^{-1/2-\epsilon}(M)\big)$ as $r>0$ goes to $0$ to a limit $u$. 
\end{thm}

\ssk

The function $u$ is what we define as the solution to Equation \eqref{EqSPDE} and it turns out to be a Markov process.  The a priori estimate encoded in the coming down from infinity property provides a compactness statement from which the existence of an invariant probability measure for the Markovian dynamics follows. 

\ssk

\begin{thm} \label{ThmMain2}
The dynamics of $u$ is Markovian and its associated semigroup on $\mcC^{-1/2-\epsilon}(M)$ has an invariant non-Gaussian probability measure.
\end{thm}

\ssk

A $\Phi^4_3$ measure over $M$ is such an invariant measure. We would like to point out some properties of such an invariant measures which we establish:
\begin{itemize}
	\item[--] The constants $a_r, b_r$ in \eqref{EqRenormalizedSPDE} are universal, in the sense that they do not depend on the Riemannian metric on $M$. While they depend however on the regularization scheme we use, here the heat regularization in space, our choice of regularization and counterterms is {\it fully covariant} with respect to the Riemannian structure, that is to say respects both covariance and locality. One will find in Section \ref{SectionLocallyCovariant} some comments on this feature of our construction. In the case where the manifold has some isometries, the invariant measures are invariant in law under isometries, see Theorem~\ref{prop:3} below. This proof relies on the companion work \cite{Phi43Uniqueness}.   \vspace{0.1cm}

	\item[--] We prove in Section~\ref{SubsectionNonTriviality} that this invariant measure is non-Gaussian.   \vspace{0.1cm}

	\item[--] In Section~\ref{SectionCouplingFunction}, we extend our analysis to the case where the coupling $\lambda$ in front of the cubic term of the dynamic is a smooth function. Indeed, since there is no translation invariance, there is no reason to select constant couplings, so the coupling constants naturally become \textbf{position dependent}. 
The counterterms needed to renormalize the SPDE are \textbf{now local functionals of the coupling functions}, see Equation \ref{EqSPDELambda}. 
\end{itemize}

\ssk

The next theorem, proved in Section~\ref{sec:423}, emphasizes the functorial behaviour of our construction; its proof relies on the uniqueness properties of the invariant measure established in \cite{Phi43Uniqueness}.

\ssk

\begin{thm}   \label{prop:3}
Let $ \chi: (M_1,g_1)\rightarrow (M_2,g_2) $ be a smooth diffeomorphism of smooth closed $3$ manifolds satisfying $\chi^*g_2=g_1$. For $i\in\{1,2\}$, denote by $\mu_i$ the $\Phi^4_3$ measure on $M_i$, that is to say the unique invariant measure of the limiting dynamics of~\ref{EqRenormalizedSPDE}. The pull-back $\chi^*:\mathcal{C}^{-\frac{1}{2}-\kappa}(M_2)\rightarrow \mathcal{C}^{-\frac{1}{2}-\kappa}(M_1)$ induces some push-forward of measures $T_\chi: \mathrm{Proba}( \mathcal{C}^{-\frac{1}{2}-\kappa}(M_2) ) \rightarrow \mathrm{Proba}( \mathcal{C}^{-\frac{1}{2}-\kappa}(M_1) )  $. Then, we have 
\begin{align*}
    \big(T_{\chi}\big)_\#\mu_2=\mu_1\,.
\end{align*}
\end{thm}

\ssk

We choose in this work to use the heat regularization. This has several obvious advantages, first the mollifier commutes with the heat operator so we can use the full power of the functional calculus of the heat kernel. Secondly, the heat kernel expansion allows us to calculate explicitly the counterterms for the dynamics to be well--defined. 

\medskip

$\bullet$ \textit{What is missing to get a full construction of the $\Phi^4_3$ quantum field theory and interesting problems raised by our work.} Here are a number of points that are left unexplored in our work. 
\begin{enumerate}
    \item A proof that our invariant measure can be also obtained as \textbf{weak limits of
    interacting Gibbs measures}, each Gibbs measure being invariant under some regularized SPDE where we \textbf{mollified the non linearity} instead of the white noise. This would allow us to really qualify our measure as a $\Phi^4_3$ quantum field theory measure. To establish such claim, one should establish that the limiting dynamics can be approximated by a sequence of equations with regularised non-linearities, that admit as invariant measure a Gibbs measure. It was proved by the first author in \cite{Phi43Uniqueness} that the Markovian dynamics of Theorem \ref{ThmMain2} on $\mathcal{C}^{-\frac{1}{2}-\kappa}(M), \kappa>0$, endowed with the $\Phi^4_3$ measure, is \textbf{uniquely ergodic}. Combining this with the previous point, this would conclude that the $\Phi^4_3$ measure we construct can be explicitly described as \textbf{the limit of a sequence of renormalized Gibbs measures}.    \vspace{0.1cm}

    \item We already proved that our measure $\mu_{\Phi^4_3}$ has finite moments, it would be desirable to prove it has \textbf{exponential moments} in the spirit of the work of Hairer \& Steele. This exponential moment condition is required for the Osterwalder-Schrader axioms to hold. But this would require much more work.    \vspace{0.1cm}

    \item We should prove that our measure is reflection positive, this is highly non obvious since most regulators on manifolds break the locality and the mollified $\Phi^4_3$ field would loose the spatial Markov property.    \vspace{0.1cm}

    \item More general than the reflection positivity, we should try to prove some spatial Markov property for our measure in the full manifold case. An important preliminary case was already done in \cite{BarashkovGunaratnam} for cylindrical domains for flat torus geometries.    \vspace{0.1cm}
   
   \item Related to the previous questions, is it possible to prove analogues of the Segal axioms for the $\Phi^4_3$ measure in the spirit of~\cite{Lin24}? A very important first step in the flat case and certain cylindrical amplitudes is covered by \cite{BarashkovGunaratnam}. Probably combing the tecniques therein with those we develop in the present work might lead to a construction of a $\Phi^4_3$ theory on $3$-manifolds with boundary fully compatible with Segal's axioms.    \vspace{0.1cm}
   
   \item Devise a version of the Osterwalder--Schrader axioms which is adapted to certain curved Riemannian manifolds which are reflection positive (this notion already appeared in the work of Jaffe \& Ritter \cite{JaffeRitter}). Then a Wick rotation of the $\Phi^4_3$ measure when $M=\Sigma\times \mathbb{R}$ with the obvious cylindrical metric (this is a noncompact manifold) and when  $M=\mathbb{S}^3$ (in which case the Wick rotated theory is expected to live on the de three dimensional de Sitter space $\mathrm d\mathbb{S}^3$) would yield a genuine relativistic Quantum Field Theory on some Lorentzian space--time.
 \end{enumerate}

\bigskip

{\subsection{Organization of the document}   }
\label{SectionOrganization}
The study of Equation \eqref{EqSPDE} on an arbitrary fixed time interval $[0,T]$ is the object of Section \ref{SectionWellPosedness}. We follow  ~\cite{JagannathPerkowski} which yields a robust formulation of Equation \eqref{EqSPDE} which avoids the use of regularity structures or paracontrolled calculus. The local in time well-posedness of \eqref{EqSPDE} is proved in Section \ref{SubsectionLongTimeWellPosedness}. We get the long time existence from the $r$-uniform $L^p$ `{\sl coming down from infinity}' property satisfied by $u_r$, proved in Section \ref{SubsectionComingDown}. The results of Section \ref{SectionWellPosedness} show that $u_r$ depends continuously on a finite family $\widehat\xi_r$ of multilinear functionals of $\xi_r$. {Anticipating over the subsequent development, we take for granted in Section \ref{SectionInvariantMeasure} the convergence of $\widehat\xi_r$ and use the $r$-uniform $L^p$ coming down property from Section \ref{SubsectionComingDown} to get the existence of an invariant probability measure for this dynamics. The non-triviality of this invariant probability measure is proved in Section \ref{SubsectionNonTriviality} using a classical reasoning.
We discuss in Section \ref{SectionLocallyCovariant} the locally covariant character of our renormalization procedure.}

{Section \ref{SectionMotivation} serves as a road-map and a motivation for the development of some systematic tools that we use for the study of the convergence of $\widehat{\xi}_r$. The results proved in the subsequent sections are of broader interest than their sole application to the $\Phi^4_3$ dynamics. An abstract version of part of} the functional setting needed to prove the convergence { of $\widehat\xi_r$} in some appropriate space as $r$ goes to $0$ is detailed in Section \ref{SectionExtension}. A crucial role is played here by a set of distributions with given wavefront sets and a certain scaling property with respect to some submanifolds. The notion of scaling field is introduced in Section \ref{SubsectionScalingFields} and the preceding set of distributions is introduced in Section \ref{SubsectionFunctionSpacesScaling}. We prove our main workhorse in Section \ref{SubsectionCanonicalExtension}, Theorem \ref{ThmCanonicalExtension}. It provides a numerical criterion for a distribution defined outside a submanifold, with some wavefront set bound and some scaling property with respect to that submanifold, to have a possibly unique extension to the whole manifold. {Section \ref{SectionKolmogorovFeynman} makes clear the link between the convergence problem for $\widehat{\xi}_r$ and the study of the convergence of some Feynman graphs, which we attack from the Epstein-Glaser extension problem point of view. The material of Section \ref{SectionExtension} is introduced for that purpose.} We draw some consequences of the general statements of Section \ref{SectionExtension} for the particular case of a configuration space in Section \ref{SubsectionConfigurationSpace}. The convergence of $\widehat\xi_r$ in an appropriate space is the object of Section \ref{SectionStochasticBounds}.

Last, Section \ref{SectionCouplingFunction} is dedicated to the construction of a $\Phi^4_3$ measure corresponding to a space-dependent coupling constant. 
Detailed proofs of a number of basic tools from microlocal and harmonic analysis are given in our companion paper \cite{BDFTCompanion}. They are mainly put to work in our analysis of the convergence of the enhanced noise $\widehat\xi_r$ in Section \ref{SectionStochasticBounds}.

\medskip

{\noindent {\textbf{\textsf{Reading guide --}}} We invite a reader who is primarily interested in the $\Phi^4_3$ dynamics and not so much in the geometric setting to read Section \ref{SectionWellPosedness} and  Section \ref{SectionInvariantMeasure}. Some of the results that we prove here are new, even in the setting of the $3$-dimensional torus. The results of Section \ref{SectionExtension} on the extension problem for distributions are of independent interest and can be read with no previous knowledge on the matter. 
Except from some analytic results on some kernels that we borrow from our companion work \cite{BDFTCompanion}, we tried to give some details which partly explains} the length of the present work. 
\bigskip

\noindent {\textbf{\textsf{Acknowledgements --}}} We would like to thank C. Bellingeri, C. Brouder, N. Burq, C. Dappiaggi, P. Duch, C. Gérard, C. Guillarmou, F. Hélein, K. L\^e, D. Manchon, P.T. Nam, S. Nonnenmacher, V. Rivasseau, G. Rivière, N. Tzvetkov, F. Vignes-Tourneret for interesting questions, remarks, comments on the present work when we were in some preliminary stage and also simply for expressing some interest and motivating us to pursue. Special thanks are due to Y. Bonthonneau, C. Brouder, J. Derezinski, M. Hairer and M. Wrochna for their very useful comments on a preliminary version of our work. Special thanks to A. Mouzard who found an important mistake in a previous version and for his strong support in the revision process.
Special thanks to  Rongchan Zhu and Xiangchan Zhu who motivated us with the revision.
We also would like to thank M. Gubinelli, Y. Dabrowski, M. Lewin who suggested us that certain techniques from microlocal analysis used for QFT on curved spaces could be merged with constructive or SPDE methods in order to prove the present results. We would like to thank the referees for helpful comments
and suggestions. N.V.D. acknowledges the support of the Institut Universitaire de France. The authors would like to thank the ANR grant SMOOTH "ANR-22-CE40-0017", QFG "ANR-20-CE40-0018" and MARGE "ANR-CE40-0011-01" for support.

\bigskip

\noindent \textbf{\textsf{Notations --}} {\it We gather here a number of notations that are used throughout the text.}   \vspace{0.1cm}

-- \textit{Given $0<T<\infty$ and a Banach space $B$ we write $C_TB$ for $C([0,T],B)$. The parabolic Besov-H\"older spaces on space time are denoted by $\mcC^\gamma((a,b)\times M)$, the Besov spaces on the manifold $M$ are denoted by $B^a_{p,q}(M)$, these spaces being defined in Definition \ref{def:besov}.   \vspace{0.1cm}}

-- \textit{The cotangent space to $M$ is denoted by $T^*M$, the conormal to a submanifold $E$ of $M$ is denoted by $N^*(E)$, and we denote by $\mathrm{dv}_g$ the volume form on $M$. Throughout this paper we are given a finite cover of $M$ with some open charts $M=\cup_{i\in I}U_i$, where $I$ is a finite set. By localising on these charts, we define in Appendix~\ref{SectionLPProjectors}, using the flat Laplacian, some Littlewood-Paley blocks $(P^i_j,\widetilde P^i_j)^{i\in I}_{j\geqslant-1}$ on $M$ -- see \eqref{eq:defLPblocks}, along with some paraproduct and resonant operators $(\prec_i,\succ_i,\odot_i)_{i\in I}$ such that for any smooth functions $a,b\in\mcD(M)$ one has   }

\begin{align*}
ab &= \sum_{i\in I} \Big(a\prec_i b + a\odot_i b + a\succ_i b \Big)   \\
     &= \sum_{i\in I}\bigg(\sum_{ j\leqslant k-1}(P^i_ja)(\widetilde{P}^i_kb) + \sum_{ |j- k|\leqslant1} (P^i_ja)(\widetilde{P}^i_kb) + \sum_{ j\geqslant k+1} (P^i_ja)(\widetilde{P}^i_kb) \bigg).
\end{align*}

\bigskip

\section{Long time well-posedness and a priori estimate}
\label{SectionWellPosedness}

We prove the existence of a unique solution to Equation \eqref{EqSPDE} over any fixed time interval $[0,T]$, for an arbitrary initial condition in $\mcC^{-1/2-\epsilon}(M)$, for $\epsilon>0$ small enough. We adopt here the robust approach of~\cite{JagannathPerkowski}. They use a clever change of variable to reformulate the equation as a non-singular parabolic partial differential Equation \eqref{EqJPFormulation} with random coefficients of regularity no worse than $-1/2-\epsilon$. This allows to solve the equation locally in time by a fixed point argument set in a classical functional space without resorting to regularity structures or paracontrolled calculus. Section \ref{SubsectionComingDown} is dedicated to proving an $L^p$ estimate on the solution to Equation \eqref{EqJPFormulation} that is independent of the initial condition. This plays a crucial role in proving the existence of an invariant probability measure for \eqref{EqSPDE} by an argument using compactness.

\ssk

\subsection{Local in time well-posedness}
\label{SubsectionLongTimeWellPosedness}

This section is dedicated to proving the local well-posedness of a solution to Equation \eqref{EqSPDE}, uniformly over $r>0$.

\ssk

Some distributions in the list \eqref{EqEnhancedNoise} below involve an operator $\odot$, called {\it resonant operator}, that we introduce in Appendix {\sf \ref{SectionLPProjectors}}; its precise definition here does not matter other than the fact that it is well-defined and continuous from $B^{\alpha_1}_{p_1,q_1}(M)\times B^{\alpha_2}_{p_2,q_2}(M)$ into some Besov space if and only if $\alpha_1+\alpha_2>0$, in which case it takes values in $B^{\alpha_1+\alpha_2}_{p,q}(M)$, for some integrability exponents $p,q$ whose precise value does not matter here. For  $\Lambda_1\in B^{\alpha_1}_{p_1,q_1}(M)$ and $\Lambda_2\in B^{\alpha_2}_{p_2,q_2}(M)$, the product $\Lambda_1\Lambda_2$ is well-defined if and only if $\Lambda_1\odot\Lambda_2$ is well-defined.

\ssk

\subsubsection{The enhanced noise.} We regularize $\xi$ in space (only) using the heat kernel and set 
$$
\xi_r \defeq e^{-rP}(\xi)\,.
$$
$\xi_r$ is a white noise in time with values in a space of regular functions. The fact that $\xi$ appears in an additive form in \eqref{EqSPDE} does not make it necessary to regularize it in time. Regularising $\xi$ only in space makes clear the Markovian character of the renormalized Equation \eqref{EqRenormalizedSPDE}. Denote by $\mcL^{-1}$, respectively $\underline{\mcL}^{-1}$, the resolvent operator of $\partial_t+P$ with null initial condition at time $0$, respectively at time $-\infty$. Explicitly, $\mcL^{-1}f(t,\cdot)=\int_0^t e^{-(t-s)P}f(s,\cdot)\rmd s $ and $\underline{\mcL}^{-1}f(t)=\int_{-\infty}^t e^{-(t-s)P}f(s,\cdot)\rmd s $ for any function $f$ on $\mathbb{R}\times M$ with at most polynomial growth in time. 
The operator $\underline{\mcL}^{-1}$ provides stationary solutions. Recall from \eqref{EqDefnArBr} the definitions of the constants $a_r$ and $b_r$ and set
$$
\begin{tikzpicture}[scale=0.3,baseline=0cm]
\node at (0,0)  [dot] (1) {};
\node at (0,0.8)  [noise] (2) {};
\draw[K] (1) to (2);
\end{tikzpicture}_r \defeq \sqrt{2} \underline{\mcL}^{-1}(\xi_r)\,,
$$
and 
$$
\begin{tikzpicture}[scale=0.3,baseline=0cm]
\node at (0,0) [dot] (0) {};
\node at (0.3,0.6)  [noise] (noise1) {};
\node at (-0.3,0.6)  [noise] (noise2) {};
\draw[K] (0) to (noise1);
\draw[K] (0) to (noise2);
\end{tikzpicture}_r \defeq (\begin{tikzpicture}[scale=0.3,baseline=0cm]
\node at (0,0)  [dot] (1) {};
\node at (0,0.8)  [noise] (2) {};
\draw[K] (1) to (2);
\end{tikzpicture}_r)^2 - a_r, \qquad \begin{tikzpicture}[scale=0.3,baseline=0cm]
\node at (0,0) [dot] (0) {};
\node at (0,0.4) [dot] (1) {};
\node at (-0.3,0.8)  [noise] (noise1) {};
\node at (0.3,0.8)  [noise] (noise2) {};
\draw[K] (0) to (1);
\draw[K] (1) to (noise1);
\draw[K] (1) to (noise2);
\end{tikzpicture}_r \defeq \underline{\mcL}^{-1} (\begin{tikzpicture}[scale=0.3,baseline=0cm]
\node at (0,0) [dot] (0) {};
\node at (0.3,0.6)  [noise] (noise1) {};
\node at (-0.3,0.6)  [noise] (noise2) {};
\draw[K] (0) to (noise1);
\draw[K] (0) to (noise2);
\end{tikzpicture}_r)\,,\quad
\begin{tikzpicture}[scale=0.3,baseline=0cm]
\node at (0,0) [dot] (0) {};
\node at (-0.4,0.5)  [noise] (noise1) {};
\node at (0,0.7)  [noise] (noise2) {};
\node at (0.4,0.5)  [noise] (noise3) {};
\draw[K] (0) to (noise1);
\draw[K] (0) to (noise2);
\draw[K] (0) to (noise3);
\end{tikzpicture}_r \defeq (\begin{tikzpicture}[scale=0.3,baseline=0cm]
\node at (0,0)  [dot] (1) {};
\node at (0,0.8)  [noise] (2) {};
\draw[K] (1) to (2);
\end{tikzpicture}_r)^3 - 3a_r \begin{tikzpicture}[scale=0.3,baseline=0cm]
\node at (0,0)  [dot] (1) {};
\node at (0,0.8)  [noise] (2) {};
\draw[K] (1) to (2);
\end{tikzpicture}_r, \qquad
\IXthree_r \defeq \underline{\mcL}^{-1} (\begin{tikzpicture}[scale=0.3,baseline=0cm]
\node at (0,0) [dot] (0) {};
\node at (-0.4,0.5)  [noise] (noise1) {};
\node at (0,0.7)  [noise] (noise2) {};
\node at (0.4,0.5)  [noise] (noise3) {};
\draw[K] (0) to (noise1);
\draw[K] (0) to (noise2);
\draw[K] (0) to (noise3);
\end{tikzpicture}_r)\,,
$$
alongside

\begin{align*}
    \XIXthree_r\defeq \IXthree_r\odot\X_r\,,\quad \XtwoIXtwo\defeq \IXtwo_r\odot\Xtwo_r-\frac{b_r}{3}\,,\quad\text{and}\quad\XtwoIXthree_r\defeq\IXthree_r\odot\Xtwo_r-b_r\X_r\,.
\end{align*}
Furthermore, we set
\begin{equation} \label{EqEnhancedNoise}
\widehat{\xi}_r \defeq \Big(\xi_r, \Xtwo_r, \; 
\IXthree_r, \;
\XIXthree_r\,,\; 
\XtwoIXtwo_r\,, \; 
\big\vert\nabla \IXtwo_r \big\vert^2 - \frac{b_r}{3}, \;  
\XtwoIXthree_r \,\Big)\, ,
\end{equation}
where $\nabla$ denotes the gradient with respect to the Riemannian metric $g$, that is
\[
\int_M \langle X, \nabla f\rangle \mathrm{dv}_g\defeq -\int_M ({\rm div} X)f \mathrm{dv}_g
\] 
for any $f\in \mathcal{D}'(M)$ and vector field $X$ on $M$.

One has $\xi_r\in \mcC^{-5/2-\epsilon}([0,T]\times M)$ and the restriction to any time interval $[0,T]$ of the other components of $\widehat\xi_r$ is seen as an element of the product space
\begin{equation} \label{EqSpaceEnhancedNoise}
{C_T\mcC^{-1-2\epsilon}(M) \times C_T\mcC^{1/2-3\epsilon}(M) \times C_T \mcC^{-4\epsilon}(M)^3 \times C_T \mcC^{-1/2-5\epsilon}(M) }.
\end{equation}
The enhancement $\widehat\xi_r$ can be seen as a placeholder for a number of products that are not well-defined in the zero regularization limit. We will see in Section \ref{SectionStochasticBounds} that $\widehat{\xi}_r$ converges in all the $L^p(\Omega)$ spaces, $1\leqslant p<\infty$, as $r>0$ goes to $0$, to a limit that does not depend on the mollification used to define $\xi_r$ from $\xi$. Using the operator $\underline{\mcL}^{-1}$ rather than the operator $\mcL^{-1}$ in the definitions of $\IXtwo_r$ and $\IXthree_r$ builds some random distributions that are stationary in time. This property will be useful in Section \ref{SectionInvariantMeasure} to get a compactness statement on the family of laws of the solutions to \eqref{EqRenormalizedSPDE}.

\ssk

\subsubsection{Jagannath \& Perkowski's formulation of Equation \eqref{EqRenormalizedSPDE}.} Set 
$$
v_{r,\textrm{ref}} \defeq 3 \, \underline{\mcL}^{-1}\Big(e^{3\begin{tikzpicture}[scale=0.3,baseline=0cm]
\node at (0,0) [dot] (0) {};
\node at (0,0.4) [dot] (1) {};
\node at (-0.3,0.8)  [noise] (noise1) {};
\node at (0.3,0.8)  [noise] (noise2) {};
\draw[K] (0) to (1);
\draw[K] (1) to (noise1);
\draw[K] (1) to (noise2);
\end{tikzpicture}_r} \Big\{ \begin{tikzpicture}[scale=0.3,baseline=0cm]
\node at (0,0) [dot] (0) {};
\node at (0,0.5) [dot] (1) {};
\node at (-0.4,1)  [noise] (noise1) {};
\node at (0,1.2)  [noise] (noise2) {};
\node at (0.4,1)  [noise] (noise3) {};
\draw[K] (0) to (1);
\draw[K] (1) to (noise1);
\draw[K] (1) to (noise2);
\draw[K] (1) to (noise3);
\end{tikzpicture}_r  \, \begin{tikzpicture}[scale=0.3,baseline=0cm]
\node at (0,0) [dot] (0) {};
\node at (0.3,0.6)  [noise] (noise1) {};
\node at (-0.3,0.6)  [noise] (noise2) {};
\draw[K] (0) to (noise1);
\draw[K] (0) to (noise2);
\end{tikzpicture}_r - b_r\big( \begin{tikzpicture}[scale=0.3,baseline=0cm]
\node at (0,0)  [dot] (1) {};
\node at (0,0.8)  [noise] (2) {};
\draw[K] (1) to (2);
\end{tikzpicture}_r + \begin{tikzpicture}[scale=0.3,baseline=0cm]
\node at (0,0) [dot] (0) {};
\node at (0,0.5) [dot] (1) {};
\node at (-0.4,1)  [noise] (noise1) {};
\node at (0,1.2)  [noise] (noise2) {};
\node at (0.4,1)  [noise] (noise3) {};
\draw[K] (0) to (1);
\draw[K] (1) to (noise1);
\draw[K] (1) to (noise2);
\draw[K] (1) to (noise3);
\end{tikzpicture}_r\,\big) \Big\}\Big).
$$
This is an element of $C_T\mcC^{1-\epsilon}(M)$. {We take as our starting point Jagannath \& Perkowski's reformulation of the renormalized form \eqref{EqRenormalizedSPDE} of Equation \eqref{EqSPDE}. (It was introduced in Equation (2.4) of \cite{JagannathPerkowski}.) It states} that $u_r$ is a solution to \eqref{EqRenormalizedSPDE} if and only if
\begin{equation} \label{EqFromUtoV}
v_r \defeq e^{3\begin{tikzpicture}[scale=0.3,baseline=0cm]
\node at (0,0) [dot] (0) {};
\node at (0,0.4) [dot] (1) {};
\node at (-0.3,0.8)  [noise] (noise1) {};
\node at (0.3,0.8)  [noise] (noise2) {};
\draw[K] (0) to (1);
\draw[K] (1) to (noise1);
\draw[K] (1) to (noise2);
\end{tikzpicture}_r} \Big(u_r - \begin{tikzpicture}[scale=0.3,baseline=0cm]
\node at (0,0)  [dot] (1) {};
\node at (0,0.8)  [noise] (2) {};
\draw[K] (1) to (2);
\end{tikzpicture}_r + \begin{tikzpicture}[scale=0.3,baseline=0cm]
\node at (0,0) [dot] (0) {};
\node at (0,0.5) [dot] (1) {};
\node at (-0.4,1)  [noise] (noise1) {};
\node at (0,1.2)  [noise] (noise2) {};
\node at (0.4,1)  [noise] (noise3) {};
\draw[K] (0) to (1);
\draw[K] (1) to (noise1);
\draw[K] (1) to (noise2);
\draw[K] (1) to (noise3);
\end{tikzpicture}_r\Big) - v_{r,\textrm{ref}}
\end{equation}
is a solution of a particular equation of the form
\begin{equation} \label{EqJPFormulation}
(\partial_t+P)v_r = -6\nabla \begin{tikzpicture}[scale=0.3,baseline=0cm]
\node at (0,0) [dot] (0) {};
\node at (0,0.4) [dot] (1) {};
\node at (-0.3,0.8)  [noise] (noise1) {};
\node at (0.3,0.8)  [noise] (noise2) {};
\draw[K] (0) to (1);
\draw[K] (1) to (noise1);
\draw[K] (1) to (noise2);
\end{tikzpicture}_r \cdot \nabla v _r- e^{-6 \begin{tikzpicture}[scale=0.3,baseline=0cm]
\node at (0,0) [dot] (0) {};
\node at (0,0.4) [dot] (1) {};
\node at (-0.3,0.8)  [noise] (noise1) {};
\node at (0.3,0.8)  [noise] (noise2) {};
\draw[K] (0) to (1);
\draw[K] (1) to (noise1);
\draw[K] (1) to (noise2);
\end{tikzpicture}_r} v_r^3 + Z_{2,r} v_r^2 +  Z_{1,r} v_r + Z_{0,r}\,, 
\end{equation}
where $Z_{2,r}, Z_{1,r}, Z_{0,r}$ are elements of $C_T\mcC^{-1/2-\eta}(M)$, for all $\eta>0$, that depend continously on $\widehat{\xi}_r$. 
(We deduce the regularity properties of the $Z_i$ from the fact that $\mcL^{-1}$ sends continuously $\mcC^\gamma([0,T]\times M)$ into $C_T\mcC^{\gamma+2}(M)$ when $-2<\gamma<0$). 

\smallskip

We now solve Equation \eqref{EqJPFormulation} with an arbitrary initial condition in $\mcC^{-1/2-\epsilon}(M)$ -- \cite{JagannathPerkowski} only considered the case of an initial condition that differs from $\begin{tikzpicture}[scale=0.3,baseline=0cm] \node at (0,0)  [dot] (1) {}; \node at (0,0.8)  [noise] (2) {}; \draw[K] (1) to (2); \end{tikzpicture}_r(0)$ by an element of $\mcC^{3/2-\epsilon}(M)$. For that purpose, and for {some} exponents $\alpha>0,\beta\in\bbR_{\geqslant0}$, we introduce the spaces $\llparenthesis \alpha,\beta\rrparenthesis$ made up of all functions $v\in C((0,T],C^\beta(M))$ such that
$$
t^\alpha \Vert u(t)\Vert_{L^\infty} \underset{t\downarrow 0}{\longrightarrow} 0
$$
and
$$
\Vert v\Vert_{\llparenthesis \alpha,\beta\rrparenthesis} \defeq  \max\left\{ \underset{0<t\leqslant T}{\sup}\,t^\alpha\Vert v(t)\Vert_{C^\beta},\sup_{0\leqslant t\neq s\leqslant T} \frac{\| t^\alpha v(t)-s^\alpha v(s)  \|_{L^\infty}}{|s-t|^{\beta/2}} \right\} < \infty.
$$
(The use of such weighted spaces is suggested in \cite{JagannathPerkowski}; we use here the same spaces as in Section 6 of Gubinelli \& Perkowski's work \cite{GubinelliPerkowskiKPZReloaded}.) The free propagation map
$$
(\mcF a)(t) \defeq e^{-tP}a
$$
sends for instance $C^\beta(M)$ into $\llparenthesis \alpha,\beta+2\alpha\rrparenthesis$, for all $\beta\in\{\bbR_{\geqslant0}\backslash\bbN\}$ and $\alpha\in (0,1)$, and one has for all $0 \leqslant \delta<\min(\beta,2\alpha)$
\begin{equation} \label{EqWeightedContinuityResolvent}
\Vert \mcL^{-1}(f)\Vert_{\llparenthesis\alpha-\delta/2,\beta-\delta\rrparenthesis} \lesssim \underset{0<t\leqslant T}{\sup}\,t^\alpha\Vert f\Vert_{C^{\beta-2}}.
\end{equation}
This inequality allows to trade some explosion rate against some regularity. We also have for the same range of exponents and all $f\in \llparenthesis\alpha,\beta\rrparenthesis$

\begin{equation} \label{EqWeightedContinuityGain}
\Vert f\Vert_{\llparenthesis\alpha-\delta/2,\beta-\delta\rrparenthesis} \lesssim \Vert f\Vert_{\llparenthesis\alpha,\beta\rrparenthesis}.
\end{equation}
These statements correspond in our setting to Lemma 6.6 and Lemma 6.8 in Gubinelli \& Perkowski's work \cite{GubinelliPerkowskiKPZReloaded} -- a proof is given in our companion work~\cite[Lemma 2.3]{BDFTCompanion}. Note that since the different components of the enhanced noise are stationary they do not take value $0$ at time $0$. The initial condition for $v_r$ is thus different from the initial condition for $u_r$. We keep the notation $\phi$ for the initial condition for $u_r$ and write $\phi'$ for the initial condition for $v_r$.
We will repeatedly use the estimate
\begin{equation}\label{eq_product_dist}
\| fg\|_{C^{\alpha\wedge \beta} }\leqslant \|f\|_{C^\alpha}\|g\|_{C^\beta}\,,
\end{equation}
if $\alpha+\beta>0$ which follows immediately from Proposition~\ref{PropContinuityParaproductResonance}.

\ssk

 \begin{prop} \label{PropLocalWellPosedness}
 Pick $\epsilon' = 4\epsilon$ and set $\alpha_0 \defeq 3/4+(\epsilon+\epsilon')/2$. For any $\phi'\in \mcC^{-1/2-\epsilon}(M)$ there exists a positive time $T^*$ such that for all $0<T<T^*$ Equation \eqref{EqJPFormulation} has a unique solution
$$
v_r \in C_T\mcC^{-1/2-\epsilon}(M) \cap \llparenthesis \alpha_0, 1+\epsilon'\rrparenthesis
$$
with initial condition $\phi'$. This solution depends continuously on $\widehat{\xi}_r$ and  $\phi'\in \mcC^{-1/2-\epsilon}(M)$, and for any small positive $\lambda$ these exist  $T_\lambda \in (\lambda, T^*)$ such that $  u\in C\big([\lambda, T_\lambda], \mcC^{3/2-4\epsilon}(M)\big)$. 
\end{prop}

\ssk

\begin{Dem}
First, remark that  $\lim_{t\downarrow 0}t^{\alpha_0}\mcF(\phi') =0 $, since $\|\mcF(\phi') \|_{L^\infty}\leqslant t^{-1/4-\epsilon/2} \| \phi'\|_{-1/2-\epsilon}$, so 
\[
\mcF(\phi')  \in C_T\mcC^{-1/2-\epsilon}(M) \cap \llparenthesis \alpha_0, 1+\epsilon'\rrparenthesis.
\] 
We use a standard Picard iteration argument for the map
\begin{eqnarray}
F(v) \defeq \mcF(\phi') + \mcL^{-1}\Big( -6\nabla \begin{tikzpicture}[scale=0.3,baseline=0cm]
\node at (0,0) [dot] (0) {};
\node at (0,0.4) [dot] (1) {};
\node at (-0.3,0.8)  [noise] (noise1) {};
\node at (0.3,0.8)  [noise] (noise2) {};
\draw[K] (0) to (1);
\draw[K] (1) to (noise1);
\draw[K] (1) to (noise2);
\end{tikzpicture}_r \cdot \nabla v - e^{-6 \begin{tikzpicture}[scale=0.3,baseline=0cm]
\node at (0,0) [dot] (0) {};
\node at (0,0.4) [dot] (1) {};
\node at (-0.3,0.8)  [noise] (noise1) {};
\node at (0.3,0.8)  [noise] (noise2) {};
\draw[K] (0) to (1);
\draw[K] (1) to (noise1);
\draw[K] (1) to (noise2);
\end{tikzpicture}_r} v^3 + Z_{2,r} v^2 + Z_{1,r} v + Z_{0,r} \Big).
\end{eqnarray}
Denote $B_R$ the ball of radius $R=4\|\phi'\|_ {\mathcal{C}^{-1/2-\epsilon}}$ in $C_T\mcC^{-1/2-\epsilon}(M)\cap \llparenthesis \alpha_0, 1+\epsilon'\rrparenthesis$. Let $v_1, v_2\in B_R$.

Our first goal is to get a bound of the form
\begin{equation*} \begin{split}
\| F(v_2) - F(v_1) \|_{\llparenthesis\alpha_0,1+\epsilon'\rrparenthesis} &\lesssim_{\widehat{\xi}_r} T^{\epsilon/2} (R+R^2) \| v_2-v_1\|_{\llparenthesis\alpha_0,1+\epsilon'\rrparenthesis} .
\end{split} \end{equation*}
meaning $F$ is a contraction for the $\llparenthesis\alpha_0,1+\epsilon'\rrparenthesis$ norm by choosing $T$ small enough. We have
\begin{equation*} \begin{split}
\| F(v_2) - F(v_1) \|_{\llparenthesis\alpha_0,1+\epsilon'\rrparenthesis} &\leqslant \big\| \mcL^{-1}\big(6\nabla \begin{tikzpicture}[scale=0.3,baseline=0cm]
\node at (0,0) [dot] (0) {};
\node at (0,0.4) [dot] (1) {};
\node at (-0.3,0.8)  [noise] (noise1) {};
\node at (0.3,0.8)  [noise] (noise2) {};
\draw[K] (0) to (1);
\draw[K] (1) to (noise1);
\draw[K] (1) to (noise2);
\end{tikzpicture}_r \cdot \nabla (v_2 - v_1)\big)\big\|_{\llparenthesis\alpha_0,1+\epsilon'\rrparenthesis} + \big\| \mcL^{-1}\big( e^{-6 \begin{tikzpicture}[scale=0.3,baseline=0cm]
\node at (0,0) [dot] (0) {};
\node at (0,0.4) [dot] (1) {};
\node at (-0.3,0.8)  [noise] (noise1) {};
\node at (0.3,0.8)  [noise] (noise2) {};
\draw[K] (0) to (1);
\draw[K] (1) to (noise1);
\draw[K] (1) to (noise2);
\end{tikzpicture}_r} (v_2^3 - v_1^3)\big) \big\|_{\llparenthesis\alpha_0,1+\epsilon'\rrparenthesis}   \\
&\quad+ \big\| \mcL^{-1}\big(Z_{2,r} (v_2^2 - v_1^2 \big) \big\|_{\llparenthesis\alpha_0,1+\epsilon'\rrparenthesis} + \big\| \mcL^{-1}\big(Z_{1,r}(v_2-v_1) \big) \big\|_{\llparenthesis\alpha_0,1 +\epsilon'\rrparenthesis}.
\end{split} \end{equation*}
Since $\nabla \begin{tikzpicture}[scale=0.3,baseline=0cm]
\node at (0,0) [dot] (0) {};
\node at (0,0.4) [dot] (1) {};
\node at (-0.3,0.8)  [noise] (noise1) {};
\node at (0.3,0.8)  [noise] (noise2) {};
\draw[K] (0) to (1);
\draw[K] (1) to (noise1);
\draw[K] (1) to (noise2);
\end{tikzpicture}_r \in C_T\mcC^{-\eta}(M)$ for all $\eta>0$ by the definition for Besov norm of vector fields \eqref{eq:def1Besovnorm_vector}, we first use the estimate \eqref{EqWeightedContinuityResolvent} with $\delta= 0$ 
and \eqref{eq_product_dist}
to get
\begin{equation*} \begin{split}
\big\| \mcL^{-1}\big(\nabla \begin{tikzpicture}[scale=0.3,baseline=0cm]
\node at (0,0) [dot] (0) {};
\node at (0,0.4) [dot] (1) {};
\node at (-0.3,0.8)  [noise] (noise1) {};
\node at (0.3,0.8)  [noise] (noise2) {};
\draw[K] (0) to (1);
\draw[K] (1) to (noise1);
\draw[K] (1) to (noise2);
\end{tikzpicture}_r \cdot \nabla (v_2-v_1) \big)\big\|_{\llparenthesis\alpha_0,1+\epsilon'\rrparenthesis} & \lesssim  \sup_{0<t\leqslant T} t^{\alpha_0 }  \| \big(\nabla \begin{tikzpicture}[scale=0.3,baseline=0cm]
\node at (0,0) [dot] (0) {};
\node at (0,0.4) [dot] (1) {};
\node at (-0.3,0.8)  [noise] (noise1) {};
\node at (0.3,0.8)  [noise] (noise2) {};
\draw[K] (0) to (1);
\draw[K] (1) to (noise1);
\draw[K] (1) to (noise2);
\end{tikzpicture}_r \cdot \nabla (v_2-v_1) \big)\big\|_{C^{-\eta}}   \\
&\lesssim_{\widehat{\xi}_r}  \sup_{0<t\leqslant T} t^{\alpha_0 } \| \nabla (v_2-v_1)\|_{C^{2 \eta}}   \\
& \lesssim T^{\alpha_0 - \alpha_1 } \| v_2-v_1\|_{(|\alpha_1, 1+2\eta|)},   \quad (\alpha_1=3/4+(\epsilon+2\eta)/2) \\ 
& \lesssim T^{\epsilon'/2-\eta} \|v_2-v_1\|_{\llparenthesis\alpha_1,1+2\eta\rrparenthesis},  \quad (\eta=\epsilon'/4)\\
& \lesssim T^{\epsilon'/4} \|v_2-v_1\|_{\llparenthesis\alpha_0,1+\epsilon'
\rrparenthesis},  \quad (\text{by \eqref{EqWeightedContinuityGain} with }\, \delta= \epsilon'-2\eta).
\end{split} \end{equation*}
Now using again \eqref{EqWeightedContinuityResolvent},  \eqref{EqWeightedContinuityGain}, \eqref{eq_product_dist} and the fact that $\exp\big(-6 \begin{tikzpicture}[scale=0.3,baseline=0cm]
\node at (0,0) [dot] (0) {};
\node at (0,0.4) [dot] (1) {};
\node at (-0.3,0.8)  [noise] (noise1) {};
\node at (0.3,0.8)  [noise] (noise2) {};
\draw[K] (0) to (1);
\draw[K] (1) to (noise1);
\draw[K] (1) to (noise2);
\end{tikzpicture}_r\big) \in \mcC^{1-\eta}(M)$ for all $\eta>0$, we have, for $\epsilon' = 4\epsilon$
\begin{equation*} \begin{split}
&\big\|\mcL^{-1}\big( \exp(-6 \begin{tikzpicture}[scale=0.3,baseline=0cm]
\node at (0,0) [dot] (0) {};
\node at (0,0.4) [dot] (1) {};
\node at (-0.3,0.8)  [noise] (noise1) {};
\node at (0.3,0.8)  [noise] (noise2) {};
\draw[K] (0) to (1);
\draw[K] (1) to (noise1);
\draw[K] (1) to (noise2);
\end{tikzpicture}_r) (v_1^3-v_2^3)\big) \big\|_{\llparenthesis\alpha_0,1+\epsilon'\rrparenthesis}\lesssim_{\widehat{\xi}_r}  \sup_{0<t\leqslant T} t^{\alpha_0 } \big\|(v_2-v_1)(v_2^2+v_1^2+v_2v_1)(t)\big\|_{C^{\eta}}. \\
&\lesssim T^{\alpha_0-3\alpha_0'} \| v_2-v_1 \|_{\llparenthesis\alpha_0',\eta\rrparenthesis} \big(\| v_2\|_{\llparenthesis\alpha'_0,\eta\rrparenthesis}^2 + \|v_1\|_{\llparenthesis\alpha'_0,\eta\rrparenthesis}^2 + \| v_2\|_{\llparenthesis\alpha'_0,\eta\rrparenthesis} \|v_1\|_{\llparenthesis\alpha'_0,\eta\rrparenthesis} \big),  \quad (\alpha_0' = 1/4 + (\epsilon+\eta)/2)  \\
&\lesssim T^{(\epsilon'-2\epsilon-3\eta)/2} \|v_2-v_1\|_{\llparenthesis\alpha'_0,\eta\rrparenthesis} \big(\| v_2\|_{\llparenthesis\alpha'_0,\eta\rrparenthesis}^2 + \|v_1\|_{\llparenthesis\alpha'_0,\eta\rrparenthesis}^2 + \|v_2\|_{\llparenthesis\alpha'_0,\eta\rrparenthesis} \|v_1\|_{\llparenthesis\alpha'_0,\eta\rrparenthesis}\big)   \\
&\lesssim T^{\epsilon/2} \|v_2-v_1\|_{\llparenthesis\alpha_0,1+\epsilon'\rrparenthesis} \big(\|v_2\|_{\llparenthesis\alpha_0,1+\epsilon'\rrparenthesis} + \|v_1\|_{\llparenthesis\alpha_0,1+\epsilon'\rrparenthesis}^2 + \|v_2\|_{\llparenthesis\alpha_0,1+\epsilon'\rrparenthesis} \|v_1\|_{\llparenthesis\alpha_0,1+\epsilon'\rrparenthesis} \big),
\end{split} \end{equation*}
choosing $\eta= \epsilon/3$ and using \eqref{EqWeightedContinuityGain} in the last inequality. Next, with the same argument we have for $\epsilon'=4\epsilon$, and setting $\alpha_0' = \frac{1}{2}+ (\epsilon+\eta)/2,  \alpha_0'' = \frac{1}{4}+ (\epsilon+\eta)/2$ in the third inequality

\begin{equation*} \begin{split}
\big\| \mcL^{-1}\big(Z_{2,r} (v_2^2-v_1^2)\big)\big\|_{\llparenthesis\alpha_0,1+\epsilon'\rrparenthesis} &\lesssim_{\widehat{\xi}_r}  \sup_{0<t\leqslant T}{t^{\alpha_0}} \big\| (v_2^2-v_1^2)(t) \big\|_{C^{1/2+\eta}}  \\
&\lesssim  \sup_{0<t\leqslant T}{t^{\alpha_0}}\| v_2(t) - v_1(t)\|_{C^{1/2+\eta}}\big(\|v_2(t)\|_{\eta}+\|v_1(t)\|_{\eta}\big)  \\
& \leqslant T^{\alpha_0-\alpha_0'-\alpha_0''} \| v_2-v_1 \|_{\llparenthesis\alpha_0',1/2+\eta\rrparenthesis} \big(\|v_2\|_{\llparenthesis\alpha_0'', \eta\rrparenthesis} + \|v_1\|_{\llparenthesis\alpha_0'', \eta\rrparenthesis} \big)  \\
\end{split} \end{equation*}
\begin{equation*} \begin{split}
&\leqslant T^{(\epsilon' -2\epsilon-2\eta)/2}  \| v_2-v_1 \|_{\llparenthesis\alpha_0,1+\epsilon'\rrparenthesis} \big(\|v_2\|_{\llparenthesis\alpha_0,1+\epsilon'\rrparenthesis} + \|v_1\|_{\llparenthesis\alpha_0,1+\epsilon'\rrparenthesis}\big)  \\
&\leqslant T^{\epsilon/2}  \| v_2-v_1\|_{\llparenthesis\alpha_0,1+\epsilon'\rrparenthesis} \big(\|v_2\|_{\llparenthesis\alpha_0,1+\epsilon'\rrparenthesis} + \|v_1\|_{\llparenthesis\alpha_0,1+\epsilon'\rrparenthesis}\big), 
\end{split} \end{equation*}
using \eqref{EqWeightedContinuityGain} in the fourth inequality and choosing $\eta= \epsilon/2$ in the last inequality. Similarly, we get 
$$
\big\| \mcL^{-1}\big(Z_{1,r} (v_2-v_1)\big)\big\|_{\llparenthesis\alpha_0,1+\epsilon'\rrparenthesis} \lesssim_{\widehat{\xi}_r} T^{\epsilon/2}  \|v_2-v_1\|_{\llparenthesis\alpha_0,1+\epsilon'\rrparenthesis}. 
$$
Therefore 
\begin{equation*} \begin{split}
\| F(v_2) - F(v_1) \|_{\llparenthesis\alpha_0,1+\epsilon'\rrparenthesis} &\lesssim_{\widehat{\xi}_r} T^{\epsilon/2} (R+R^2) \| v_2-v_1\|_{\llparenthesis\alpha_0,1+\epsilon'\rrparenthesis} .
\end{split} \end{equation*}
The rest  is  to estimate $F(v_1)-F(v_2) $ in $C_T\mcC^{-1/2-\epsilon}$. Since $v_1, v_2\in B_R$, for any $s\in [0,T]$,  $i=1,2$ and $0\leq\delta<1+\epsilon'$, it follows from \eqref{EqWeightedContinuityGain} that $\|v_i\|_{(|\alpha_0-\delta/2,1+\epsilon'-\delta|)}\leqslant R$, hence we have
\begin{equation}\label{eq_short_time_1}
\|v_i(s)\|_{C^{-1/2-\epsilon}} \leqslant R \, \text{ and } \|v_i\|_{C^{1+\epsilon' -\delta}}\leqslant R s^{-(\alpha_0-\delta/2)}. 
\end{equation}  
Recall the following estimate which is used many times below: if $\|u(s)\|_{C^{\beta}}\lesssim s^{-\gamma},$ for $\gamma<1$, then 
\begin{equation}\label{eq:ineq_int}
 \|\mcL^{-1}u\|_{C^{\beta}}\lesssim \int_0^Ts^{-\gamma}\lesssim T^{1-\gamma}. 
\end{equation}
Using \eqref{eq_short_time_1} and  the fact that $\nabla \begin{tikzpicture}[scale=0.3,baseline=0cm]
\node at (0,0) [dot] (0) {};
\node at (0,0.4) [dot] (1) {};
\node at (-0.3,0.8)  [noise] (noise1) {};
\node at (0.3,0.8)  [noise] (noise2) {};
\draw[K] (0) to (1);
\draw[K] (1) to (noise1);
\draw[K] (1) to (noise2);
\end{tikzpicture}_r \in C_T\mcC^{-\epsilon'/4}(M)$,   we have
\begin{equation*} \begin{split}
\big\|  \nabla \begin{tikzpicture}[scale=0.3,baseline=0cm]
\node at (0,0) [dot] (0) {};
\node at (0,0.4) [dot] (1) {};
\node at (-0.3,0.8)  [noise] (noise1) {};
\node at (0.3,0.8)  [noise] (noise2) {};
\draw[K] (0) to (1);
\draw[K] (1) to (noise1);
\draw[K] (1) to (noise2);
\end{tikzpicture}_r \cdot \nabla (v_2-v_1)(s) \big\|_{C^{-1/2-\epsilon}} \lesssim_{\widehat{\xi}_r} \big\| \nabla (v_2-v_1) (s)\big\|_{C^{\epsilon'}} \lesssim  \|v_2-v_1\|_{C^{1+\epsilon'}} \lesssim s^{-\alpha_0} \|v_2-v_1\|_{(|\alpha_0, 1+\epsilon'|)},
\end{split} \end{equation*}
hence by \eqref{eq:ineq_int}
\begin{equation*} \begin{split}
 \big\| \mcL^{-1}\big( \nabla \begin{tikzpicture}[scale=0.3,baseline=0cm]
\node at (0,0) [dot] (0) {};
\node at (0,0.4) [dot] (1) {};
\node at (-0.3,0.8)  [noise] (noise1) {};
\node at (0.3,0.8)  [noise] (noise2) {};
\draw[K] (0) to (1);
\draw[K] (1) to (noise1);
\draw[K] (1) to (noise2);
\end{tikzpicture}_r \cdot \nabla (v_2-v_1) \big) \big\|_{C^{-1/2-\epsilon}} &\lesssim_{\widehat{\xi}_r} T ^{1-\alpha_0} \|v_2-v_1\|_{(|\alpha_0, 1+\epsilon'|)} = T ^{1/4 -\epsilon/2-\epsilon'/2} \|v_2-v_1\|_{(|\alpha_0, 1+\epsilon'|)}.
\end{split} \end{equation*}
Again, by \eqref{eq_short_time_1} and the fact that $ \begin{tikzpicture}[scale=0.3,baseline=0cm]
\node at (0,0) [dot] (0) {};
\node at (0,0.4) [dot] (1) {};
\node at (-0.3,0.8)  [noise] (noise1) {};
\node at (0.3,0.8)  [noise] (noise2) {};
\draw[K] (0) to (1);
\draw[K] (1) to (noise1);
\draw[K] (1) to (noise2);
\end{tikzpicture}_r \in C_T\mcC^{1-\eta}(M)$, and $ \|u v\|_{C^{\beta}} \leqslant \|u\|_{L^\infty}\|v\|_{C^\beta}+ \|v\|_{L^\infty}\|u\|_{C^\beta}$, for $\beta\in (0,1)$,
\begin{align*} 
\big\| \big( \exp(-6 \begin{tikzpicture}[scale=0.3,baseline=0cm]
\node at (0,0) [dot] (0) {};
\node at (0,0.4) [dot] (1) {};
\node at (-0.3,0.8)  [noise] (noise1) {};
\node at (0.3,0.8)  [noise] (noise2) {};
\draw[K] (0) to (1);
\draw[K] (1) to (noise1);
\draw[K] (1) to (noise2);
\end{tikzpicture}_r) (v_1^3-v_2^3)\big)(s) \big\|_{-1/2-\epsilon}  &\lesssim_{\widehat{\xi}_r} \big\|(v_2-v_1)(v_2^2+v_1^2+v_2v_1)(t) \big\|_{C^{-1/2-\epsilon}} \\
& \lesssim \quad \|v_2-v_1\|_{C^{-1/2-\epsilon}} \|v_2^2+v_1^2+v_2v_1\|_{C^{1/2+2\epsilon}}\\
&\lesssim \quad   \|v_2-v_1\|_{C^{-1/2-\epsilon}} \sum_{i, j=1}^2 \|v_i\|_{C^{\epsilon}}\|v_j\|_{C^{1/2+2\epsilon}}   \\
&\lesssim \quad  \|v_2-v_1\|_{C^{-1/2-\epsilon}}  s^{-3/4-5\epsilon/2 } R^2\,,
\end{align*}
where we use \eqref{eq_short_time_1}  for both $\|v_i\|_{C^{\epsilon}}$ and $\|v_j\|_{C^{1/2+2\epsilon}}$ in the last inequality. Hence by \eqref{eq:ineq_int} we have
\begin{equation*} \begin{split}
&\big\| \mcL^{-1}\big(  \big( \exp(-6 \begin{tikzpicture}[scale=0.3,baseline=0cm]
\node at (0,0) [dot] (0) {};
\node at (0,0.4) [dot] (1) {};
\node at (-0.3,0.8)  [noise] (noise1) {};
\node at (0.3,0.8)  [noise] (noise2) {};
\draw[K] (0) to (1);
\draw[K] (1) to (noise1);
\draw[K] (1) to (noise2);
\end{tikzpicture}_r) (v_1^3-v_2^3)\big)\big) \big\|_{-1/2-\epsilon} \lesssim_{\widehat{\xi}_r}  T^{1/4-5\epsilon/2} R^2 \|v_2-v_1\|_{C^{-1/2-\epsilon}} .
\end{split} \end{equation*}
Now, with the same argument and \eqref{EqWeightedContinuityGain}, for $ \eta>\epsilon$, we have 
\begin{equation*} \begin{split}
\big\| \big(Z_{2,r} (v_2^2-v_1^2)\big)(s)\big\|_{C^{-1/2-\epsilon}} &\lesssim_{\widehat{\xi}_r}  \big\| (v_2^2-v_1^2)(s) \big\|_{C^{1/2+\eta}}  \\
&\lesssim (\|v_1 \|_{C^\eta} + \|v_2\|_{C^\eta} )\|v_1-v_2\|_{1/2+\eta} \\
&\lesssim  s^{-(\alpha_0-(1+\epsilon'-\eta)/2)}  R s^{-1/2-\epsilon/2-\eta/2}  \|v_1-v_2\|_{\llparenthesis\frac{1}{2}+\frac{\epsilon+\eta}{2},\frac{1}{2}+\eta\rrparenthesis} \\
& \lesssim s^{-(3/4+\epsilon+\eta)} R\|v_1-v_2\|_{\llparenthesis \alpha_0, 1+\epsilon' \rrparenthesis },
\end{split} \end{equation*}
hence, choosing $\eta= 2\epsilon$ yields
\begin{equation*} \begin{split}
\big\| \mcL^{-1}\big( Z_{2,r} (v_2^2-v_1^2)\big)\big\|_{-1/2-\epsilon}
\lesssim_{\widehat{\xi}_r} T^{1/4-3\epsilon}R\|v_1-v_2\|_{\llparenthesis\alpha_0, 1+\epsilon'\rrparenthesis }.
\end{split} \end{equation*}
Similarly, we get 
$$
\big\| \mcL^{-1}\big(Z_{1,r} (v_2-v_1)\big)\big\|_{C^{-1/2-\epsilon}} \lesssim_{\widehat{\xi}_r} T^{1/2 -3\epsilon/2}\|v_1-v_2\|_{\llparenthesis\alpha_0, 1+\epsilon'\rrparenthesis }.
$$
Therefore
\begin{equation*} \begin{split}
\| F(v_2) - F(v_1) \|_{C^{-1/2-\epsilon}}  &\lesssim_{\widehat{\xi}_r} T^{1/4-5\epsilon/2} R^2 \|v_2-v_1\|_{C^{-1/2-\epsilon}}   \\
&\quad \quad
+ \big(T ^{1/4 -\epsilon/2-\epsilon'/2} + T^{1/4-3\epsilon}R + T^{1/2 -3\epsilon/2}\big) \|v_1-v_2\|_{\llparenthesis \alpha_0, 1+\epsilon'\rrparenthesis }.  
\end{split} \end{equation*}
Combining the estimates above we infer that for $T>0$ sufficiently small, depending on $\|\phi'\|_ {\mathcal{C}^{-1/2-\epsilon}}$, $\widehat{\xi}_r$, the map $F$ is a contraction on the ball of radius $4\|\phi'\|_ {\mathcal{C}^{-1/2-\epsilon}}$ in $C_T\mcC^{-1/2-\epsilon}(M)\cap \llparenthesis \alpha_0, 1+\epsilon'\rrparenthesis$. The unique fixed point is our solution on $[0,T]$. Taking the supremum of all such $T$ gives the maximal existence time $T^*$. Once we know that $v$ takes values in $\llparenthesis \alpha_0, 1+\epsilon'\rrparenthesis$ we can restart the fixed point procedure from a positive time, with an initial condition that is now of H\"older regularity $(1+\epsilon')$. It is elementary to adapt the preceding estimates to see that now the solution will take values in $\mcC^{3/2-4\epsilon}(M)$.   \vspace{0.15cm}

For the continuous dependence on $\widehat \xi_r$ and the initial data, we define 
$$
F(\widehat \xi_r , \phi, v) \defeq e^{-tP}(\phi) + \mcL^{-1}\Big( -6\nabla \begin{tikzpicture}[scale=0.3,baseline=0cm]
\node at (0,0) [dot] (0) {};
\node at (0,0.4) [dot] (1) {};
\node at (-0.3,0.8)  [noise] (noise1) {};
\node at (0.3,0.8)  [noise] (noise2) {};
\draw[K] (0) to (1);
\draw[K] (1) to (noise1);
\draw[K] (1) to (noise2);
\end{tikzpicture}_r  (\xi_r)\cdot \nabla v - e^{-6 \begin{tikzpicture}[scale=0.3,baseline=0cm]
\node at (0,0) [dot] (0) {};
\node at (0,0.4) [dot] (1) {};
\node at (-0.3,0.8)  [noise] (noise1) {};
\node at (0.3,0.8)  [noise] (noise2) {};
\draw[K] (0) to (1);
\draw[K] (1) to (noise1);
\draw[K] (1) to (noise2);
\end{tikzpicture}_r(\xi_r) } v^3 + Z_{2,r}( \xi_r) v^2 + Z_{1,r}( \xi_r) v + Z_{0,r} (\xi_r)\Big).
$$
Let $K>0$ be a uniform constant satisfying
\begin{equation}
\|e^{-tP} \phi\|_{C^{-1/2-\epsilon}}\leqslant K \|\phi\|_{C^{-1/2-\epsilon}}   \, \text{  and  }  \|e^{-tP} \phi\|_{\llparenthesis\alpha_0, 1+\epsilon'\rrparenthesis}\leqslant K \|\phi\|_{C^{-1/2-\epsilon}}.   
\end{equation}
Take the ball $B_R$ in $\mcC^{-1/2-\epsilon}(M)$.
 Since $F$  depends  linearly on $\widehat{\xi}_r$ and $ \exp(-6 \begin{tikzpicture}[scale=0.3,baseline=0cm]
\node at (0,0) [dot] (0) {};
\node at (0,0.4) [dot] (1) {};
\node at (-0.3,0.8)  [noise] (noise1) {};
\node at (0.3,0.8)  [noise] (noise2) {};
\draw[K] (0) to (1);
\draw[K] (1) to (noise1);
\draw[K] (1) to (noise2);
\end{tikzpicture}_r)$, by the same arguments  above,   for any $\phi\in B_R$ and  we can choose  $T=T(R, \widehat \xi_r,\widehat \xi'_r)$  small enough such that   $C(T)<1/2$ and
 $$ 
 \|F(\widehat \xi_{r},\phi, v_1)-F(\widehat{\xi}'_r,\phi, v_2 )\|_{C_T\mcC^{1/2-\epsilon}}  \leqslant  C(T) \big(\|v_1-v_2\|_{C_T\mcC^{1/2-\epsilon}}+\|v_1-v_2\|_{\llparenthesis\alpha_0, 1+\epsilon'\rrparenthesis} + \|\widehat \xi_r -\widehat\xi_r'\|\big)  
$$
and
$$   
\big\|F(\widehat \xi_{r},\phi, v_1)-F(\widehat{\xi}'_r,\phi, v_2 )\big\|_{\llparenthesis \alpha_0, 1+\epsilon\rrparenthesis}  \leqslant C(T) \big(\|v_1-v_2\|_{\llparenthesis\alpha_0, 1+\epsilon'\rrparenthesis}+   \|\widehat \xi_r -\widehat\xi_r'\|\big). 
$$
Now  for $\phi_1, \phi_2\in B_R$ we have

\begin{equation*} \begin{split}
&\|v_r(\widehat{\xi}_{r} ,\phi_1)-v_r(\widehat \xi_{r}' , \phi_2)\|_{C_T\mcC^{1/2-\epsilon}} =  \big\|F(\widehat\xi_{r},\phi_1,v_r(\phi_1)) - F(\widehat \xi_{r}',\phi_2,v_r(\phi_2))\big\|_{C_T\mcC^{1/2-\epsilon}}   \\
&\leqslant\big \|F(\widehat\xi_{r},\phi_1, v_{r}(\phi_1))-F(\widehat\xi_{r},\phi_2,v_{r}(\phi_1)) \big\|_{C_T\mcC^{1/2-\epsilon}}   +  \big\|F(\widehat\xi_r, \phi_2, v_{r}(\phi_1))-F(\widehat \xi_r',\phi_2,v_{r}(\phi_2)) \big\|_{C_T\mcC^{1/2-\epsilon}}  \\
&\leqslant  K\|\phi_1-\phi_2\|_{C^{-1/2-\epsilon}}  \hspace{-0.05cm}+ \hspace{-0.05cm} C(T) \Big(\|v_r(\phi_1)-v_r(\phi_2)\|_{C_T\mcC^{1/2-\epsilon}}  \hspace{-0.05cm} + \|v_r(\phi_1)-v_r(\phi_2)\|_{\llparenthesis\alpha_0, 1+\epsilon'\rrparenthesis}  \hspace{-0.05cm}+ \|\widehat \xi_r -\widehat\xi_r'\|\Big). 
\end{split} \end{equation*}
Similarly we have
\begin{equation*} \begin{split}
\|v_r(\phi_1)-v_r(\phi_2)\|_{\llparenthesis\alpha_0, 1+\epsilon'\rrparenthesis} &= \big\|F(\widehat\xi_{r},\phi_1,v_r(\phi_1)) - F(\widehat\xi_{r}',\phi_2,v_r(\phi_2)) \big\|_{\llparenthesis\alpha_0, 1+\epsilon'\rrparenthesis}   \\
&\leqslant  K\|\phi_1-\phi_2\|_{C^{-1/2-\epsilon}} +   C(T)\big( \|v_r(\phi_1)-v_r(\phi_2)\|_{\llparenthesis\alpha_0, 1+\epsilon'\rrparenthesis} +\|\widehat \xi_r -\widehat\xi_r'\|\big),
\end{split} \end{equation*}
so
\begin{equation*} \begin{split}
\|v_r(\phi_1)-v_r(\phi_2)\|_{C_T\mcC^{1/2-\epsilon}} + \|v_r(\phi_1)&-v_r(\phi_2)\|_{\llparenthesis\alpha_0, 1+\epsilon'\rrparenthesis}   \\
&\leqslant  2K\|\phi_1-\phi_2\|_{C_T\mcC^{1/2-\epsilon}}  + C(T) \|v_r(\phi_1)-v_r(\phi_2)\|_{C_T\mcC^{-1/2-\epsilon}}  \\ 
&\qquad+   2C(T) \|v_r(\phi_1)-v_r(\phi_2)\|_{\llparenthesis \alpha_0, 1+\epsilon'\rrparenthesis} + 2\|\widehat \xi_r -\widehat\xi_r'\|, 
\end{split} \end{equation*}
and we read on the estimate
\begin{equation*} \begin{split}
\|v_r(\phi_1)-v_r(\phi_2)\|_{C_T\mcC^{1/2-\epsilon}}  &+ \|v_r(\phi_1)-v_r(\phi_2)\|_{\llparenthesis\alpha_0, 1+\epsilon'\rrparenthesis}   \\
&\leqslant  \frac{1 }{1-2C(T)} \left(2K \|\phi_1-\phi_2\|_{C^{-1/2-\epsilon}} + 2C(T)\|\widehat \xi_r -\widehat\xi_r'\|\right)
\end{split} \end{equation*}
the continuous dependence of the solution on $\widehat \xi$ and  the initial  data.
\end{Dem}

\ssk

We see from the proof that $T^*$ depends only on $\widehat\xi_r$ and $\phi'\in \mcC^{-1/2-\epsilon}(M)$. The following additional piece of information will be useful when proving the coming down from infinity property by energy methods in the next section. 

\ssk

\begin{lem} Let $0<t_0<t_1$.  For  $\beta=3/2-\epsilon$ and  any $\kappa\leqslant \beta/2$, then  $t\mapsto v_r(t,x)$ is  $\kappa-$ H\"older continuous as a function from $[t_0,t_1]$ to $L^\infty$. 
\end{lem}

\ssk

\begin{Dem} By the change of variable $t\mapsto t-t_0$, we can assume $t_0= 0, t_1=T>0$ and $v_r\in C_T\mcC^{3/2-\epsilon}(M)$. We now show the Hölder regularity of $v_r$ at time $0$, the adaptation to arbitrary times is straightforward. We have
$$
v_r(t,\cdot) = e^{-tP} v_r(0) + \int_{0}^t e^{-(t-s)P} \hspace{-0.1cm} \left(-6 \nabla \begin{tikzpicture}[scale=0.3,baseline=0cm] \node at (0,0) [dot] (0) {}; \node at (0,0.4) [dot] (1) {}; \node at (-0.3,0.8)  [noise] (noise1) {}; \node at (0.3,0.8)  [noise] (noise2) {}; \draw[K] (0) to (1); \draw[K] (1) to (noise1); \draw[K] (1) to (noise2); \end{tikzpicture}_r \cdot \nabla v_r - e^{-6 \begin{tikzpicture}[scale=0.3,baseline=0cm] \node at (0,0) [dot] (0) {}; \node at (0,0.4) [dot] (1) {}; \node at (-0.3,0.8)  [noise] (noise1) {}; \node at (0.3,0.8)  [noise] (noise2) {}; \draw[K] (0) to (1); \draw[K] (1) to (noise1); \draw[K] (1) to (noise2); \end{tikzpicture}_r} v_r^3 + Z_{2,r} v_r^2 + Z_{1,r} v_r + Z_{0,r} \right)\hspace{-0.1cm}(s)\rmd s.
$$
We first remark that 
\begin{eqnarray}
| v_r(t,\cdot) - v_r(0) | \leqslant | e^{-tP}v_r(0) - v_r(0)| +\left|   \int_{0}^t e^{-(t-s)P} (f_r(s))\rmd s\right|,
\end{eqnarray}
where $f_r = -6 \nabla \begin{tikzpicture}[scale=0.3,baseline=0cm] \node at (0,0) [dot] (0) {}; \node at (0,0.4) [dot] (1) {}; \node at (-0.3,0.8)  [noise] (noise1) {}; \node at (0.3,0.8)  [noise] (noise2) {}; \draw[K] (0) to (1); \draw[K] (1) to (noise1); \draw[K] (1) to (noise2); \end{tikzpicture}_r \cdot \nabla v_r - e^{-6\begin{tikzpicture}[scale=0.3,baseline=0cm] \node at (0,0) [dot] (0) {}; \node at (0,0.4) [dot] (1) {}; \node at (-0.3,0.8)  [noise] (noise1) {}; \node at (0.3,0.8)  [noise] (noise2) {}; \draw[K] (0) to (1); \draw[K] (1) to (noise1); \draw[K] (1) to (noise2); \end{tikzpicture}_r} v_r^3 + Z_{2,r} v_r^2 + Z_{1,r}v_r + Z_{0,r}$. It follows from the time regularity of the heat flow that $\|(1-e^{-tP}) h \|_{L^\infty} \lesssim t^{\frac{\beta}{2}}\|h\|_{C^{\beta}} $ for $0 \leqslant \beta\leqslant 2 $, hence 
\begin{equation}
 \| e^{-tP}v_r(0) - v_r(0) \|_{L^\infty} \lesssim t^{\beta/2} \| v_r(0)\|_{C^\beta}. 
\end{equation}
Since $v_r \in C_T \mcC^{\beta}$, with $\beta=3/2-\epsilon$,  $ \nabla \begin{tikzpicture}[scale=0.3,baseline=0cm] \node at (0,0) [dot] (0) {}; \node at (0,0.4) [dot] (1) {}; \node at (-0.3,0.8)  [noise] (noise1) {}; \node at (0.3,0.8)  [noise] (noise2) {}; \draw[K] (0) to (1); \draw[K] (1) to (noise1); \draw[K] (1) to (noise2); \end{tikzpicture}_r \in C_T\mcC^{-\epsilon}$, $e^{-6\begin{tikzpicture}[scale=0.3,baseline=0cm] \node at (0,0) [dot] (0) {}; \node at (0,0.4) [dot] (1) {}; \node at (-0.3,0.8)  [noise] (noise1) {}; \node at (0.3,0.8)  [noise] (noise2) {}; \draw[K] (0) to (1); \draw[K] (1) to (noise1); \draw[K] (1) to (noise2); \end{tikzpicture}_r}\in C_T\mcC^{1-\epsilon}$, $ {\bf Z}_r = \{Z_{0,r}, Z_{1,r}, Z_{2,r}\} \subset C_T\mcC^{-\alpha}$ with  $\alpha= 1/2+\epsilon'$, using $  \| gh\|_{C^{\alpha'\wedge\beta'}} \lesssim  \|g\|_{C^{\alpha'}}\| h\|_{C^{\beta'}}$  for $\alpha'+\beta'>0$ we have
$$
 \| f_r(s, \cdot)\|_{C^{-\alpha}}\lesssim_{{\bf Z}_r} \|v_r\|_{C_T\mcC^{\beta}} \lesssim 1.
 $$
Then the estimate $\|e^{-tP} h\|_{C^{\eta+\gamma}}\lesssim t^{-\gamma/2}\|h\|_{C^\eta} $, for $\gamma\geqslant 0$ implies
$$
 \left\| \int_0^t e^{-(t-s)P} (f_r(s))\rmd s\right\|_{C^{\epsilon'}} \rmd s \leqslant  \int_0^t (t-s)^{-(\alpha-\epsilon')/2}\|f_r(s)\|_{C^{-\alpha}}\rmd s \lesssim t^{1 - \alpha/2 -\epsilon'/2}.
$$
By choosing $\epsilon'\geqslant  \epsilon/2$  we have $\kappa\defeq 1-\alpha/2 -\epsilon'/2 \leqslant \beta/2$, therefore $u$  is $\kappa$-H\"older from $[0,T]$ to $L^\infty$.
\end{Dem}

\ssk

As in Proposition 6.8 of Mourrat \& Weber's work \cite{MourratWeberPlane}, it follows from this property that the function 
\[
t\in (0,T] \mapsto \Vert v_r(t)\Vert_{L^p}^p 
\]
satisfies the equation
\begin{equation} \label{EqIntegralLpNormIdentity} \begin{split}
    \frac{1}{p} \, \big(\Vert v_r(t)\Vert^p_{L^p}  -  \Vert v_r(s)\Vert^p_{L^p} \big) &= \int_s^t \big(v_r^{p-1},\Delta v_r\big) - \int_s^t \Big(\int_M \hspace{-0.05cm} e^{-6\begin{tikzpicture}[scale=0.3,baseline=0cm] \node at (0,0) [dot] (0) {}; \node at (0,0.4) [dot] (1) {}; \node at (-0.3,0.8)  [noise] (noise1) {}; \node at (0.3,0.8)  [noise] (noise2) {}; \draw[K] (0) to (1); \draw[K] (1) to (noise1); \draw[K] (1) to (noise2); \end{tikzpicture}_r(s_1)} v_r^{p+2}(s_1)\Big)\rmd s_1   \\
    	&\quad+ \int_s^t \Big( \frac{1}{p}\,(-6 \nabla \begin{tikzpicture}[scale=0.3,baseline=0cm] \node at (0,0) [dot] (0) {}; \node at (0,0.4) [dot] (1) {}; \node at (-0.3,0.8)  [noise] (noise1) {}; \node at (0.3,0.8)  [noise] (noise2) {}; \draw[K] (0) to (1); \draw[K] (1) to (noise1); \draw[K] (1) to (noise2); \end{tikzpicture}_r , \nabla v_r^p) + (Z_2,v_r^{p+1}) + (Z_1,v_r^p) + (Z_0,v_r^{p-1})\Big).
\end{split} \end{equation}

\bigskip

\subsection{Long time existence and coming down from infinity}
\label{SubsectionComingDown}

We show in this section that the superlinear attractive drift $- \exp\big(-6 \begin{tikzpicture}[scale=0.3,baseline=0cm]
\node at (0,0) [dot] (0) {};
\node at (0,0.4) [dot] (1) {};
\node at (-0.3,0.8)  [noise] (noise1) {};
\node at (0.3,0.8)  [noise] (noise2) {};
\draw[K] (0) to (1);
\draw[K] (1) to (noise1);
\draw[K] (1) to (noise2);
\end{tikzpicture}_r\big) \, v_r^3$ in Equation \eqref{EqJPFormulation} entails an a priori bound on the $L^p(M)$ norm of the solution away from the initial time that is independent of the initial condition. This bound entails the long time existence of the solution $v_r$ to \eqref{EqJPFormulation} and is the key to proving the existence of an invariant probability measure for the dynamics \eqref{EqSPDE} via a compactness argument. This point will be developed in Section \ref{SectionInvariantMeasure}.

\medskip

We rewrite Equation \eqref{EqJPFormulation} in the form
\begin{equation} \label{EqJPFormulationSynthetic}
\big(\partial_t+P + B_r\nabla\big) v_r = - A_r v_r^3 + Z_{2,r} v_r^2 + Z_{1,r} v_r + Z_{0,r},
\end{equation}
with
$$
B_r \defeq 6\nabla \begin{tikzpicture}[scale=0.3,baseline=0cm] \node at (0,0) [dot] (0) {}; \node at (0,0.4) [dot] (1) {}; \node at (-0.3,0.8)  [noise] (noise1) {}; \node at (0.3,0.8)  [noise] (noise2) {}; \draw[K] (0) to (1); \draw[K] (1) to (noise1); \draw[K] (1) to (noise2); \end{tikzpicture}_r \in C_T\mcC^{-\eta}(M), \quad A_r \defeq e^{-6\begin{tikzpicture}[scale=0.3,baseline=0cm] \node at (0,0) [dot] (0) {}; \node at (0,0.4) [dot] (1) {}; \node at (-0.3,0.8)  [noise] (noise1) {}; \node at (0.3,0.8)  [noise] (noise2) {}; \draw[K] (0) to (1); \draw[K] (1) to (noise1); \draw[K] (1) to (noise2); \end{tikzpicture}_r} \in C_T \mcC^{1-\eta}(M),
$$ 
and $Z_{i,r}\in C_T\mcC^{-1/2-\eta}(M)$, for all $\eta>0$.

\ssk

\begin{thm} \label{ThmLpComingDown}
The solution $v_r(t) \in \mcC^{3/2-\epsilon}(M)$ exists for all times $t>0$. Pick an even integer $p\geqslant 8$. There is a random variable $C\big(p,\widehat\xi_{r\vert[0,t]}\big)$ that depends only on the restriction to the interval $[0,t]$ of $\widehat\xi_r$ such that one has
\begin{equation} \label{EqUpperBoundComingDown}
\Vert v_r(t)\Vert_{L^p(M)} \leqslant  C\big(p,\widehat\xi_{r\vert[0,t]}\big) \max\left\{\frac{1}{\sqrt{t}}, 1 \right\}
\end{equation}
for all $t > 0$, independently of the initial condition $\phi' \in \mcC^{-1/2-\epsilon}(M)$.
\end{thm}

\ssk

The upper bound in \eqref{EqUpperBoundComingDown} is in particular independent of the initial condition in \eqref{EqJPFormulationSynthetic}; this phenomenon is called {\sl coming down from infinity}. We note for later use that keeping track of the implicit constants in the computations below gives {as a possible choice of $C\big(p,\widehat\xi_{r\vert[0,t]}\big)$ for $0<t\leqslant2$}
\begin{equation} \label{EqEstimateCabZ}
{C\big(p,\widehat\xi_{r\vert[0,t]}\big) =} \big(1+\Vert\widehat\xi_{ r\vert[0,2]}\Vert\big)^\gamma \Big(\exp\big( \gamma' \Vert \begin{tikzpicture}[scale=0.3,baseline=0cm] \node at (0,0) [dot] (0) {}; \node at (0,0.4) [dot] (1) {}; \node at (-0.3,0.8)  [noise] (noise1) {}; \node at (0.3,0.8)  [noise] (noise2) {}; \draw[K] (0) to (1); \draw[K] (1) to (noise1); \draw[K] (1) to (noise2); \end{tikzpicture}_r\Vert_{L^\infty([0,2]\times M)} \big) + 1\Big)
\end{equation}
for some positive constants $\gamma=\gamma(p), \gamma'=\gamma'(p)$, up to a multiplicative constant. We denoted here by $\Vert\widehat\xi_{r\vert[0,2]}\Vert$ the norm of $\widehat\xi_r$ seen as an element of the product space where $\widehat\xi_r$ takes its values. We use a priori energy estimates to prove Theorem \ref{ThmLpComingDown}, following the strategy initiated by Mourrat \& Weber in their proof of a similar result in \cite{MourratWeber}, Theorem 7.1 therein. Gubinelli \& Hofmanov\'a also used energy estimates in their work \cite{GubinelliHofmanova1} on the $\Phi^4_3$ measure on $\bbR^3$. See also the proof of Proposition 3.7 in the work \cite{TsatsoulisWeber} of Tsatsoulis \& Weber for an implementation of that strategy in the $2$-dimensional torus.

{Jagannath \& Perkowski used a different strategy in \cite{JagannathPerkowski} to prove the long time existence of solutions to \eqref{EqJPFormulation}, based on the maximum principle. Their approach works verbatim here as well and leads to global in time well-posedness. This is not sufficient for our purpose, though. The coming down from infinity estimate \eqref{EqUpperBoundComingDown} will entail, by a well-known reasoning, a tightness result on the laws of $v_r,u_r$, and $v,u$ that will lead in Section \ref{SectionInvariantMeasure} to the existence of an invariant measure for the dynamical $\Phi^4_3$ equation. Such a result cannot be obtained as a consequence of a global in time well-posedness result.}
 
\medskip

In the remainder of this section, we use the shorthand notation
$$
B^\gamma_p(M) \defeq B^\gamma_{p,\infty}(M)
$$
for any $\gamma\in\bbR$ and $1\leqslant p\leqslant \infty$. 

The main intermediate step toward proving Theorem~\ref{ThmLpComingDown} is the following result.
\begin{lem}  \label{lem:8}
Set 
$$
F_r(t) \defeq \| v_r(t)\|^{p}_{L^{p+2}}+ \|v_r(t)\|^{\frac{p}{3}}_{B^{1+2\epsilon}_{\frac{p+2}{3}}}.
$$
For every $0<T_0\leqslant s<t\leqslant T<T^*\wedge 1$, it holds
\begin{equation} \label{EqMainEstimateComingDown}
\Vert v_r(t)\Vert_{L^p}^p + \int_s^t F_r(s_1)^{\frac{p+2}{p}}\rmd s_1 \lesssim_{\widehat\xi_r} 1 + F_r(s).
\end{equation} 
\end{lem}
We postpone the proof of Lemma~\ref{lem:8} to the end of the section, and directly turn to the proof of Theorem~\ref{ThmLpComingDown}.
\begin{DemThm7}
    We start by noticing that the inequality \eqref{EqMainEstimateComingDown} specialises to 
\begin{equation} 
 \int_s^t F_r(s_1)^{\frac{p+2}{p}}\rmd s_1 \lesssim_{\widehat\xi_r} 1 + F_r(s)
\end{equation} 
for all $0<T_0\leqslant s<t\leqslant T$.

It then follows from a modified version of Mourrat \& Weber's comparison test recalled in Proposition \ref{PropComparisonTest} that  there is an integer $N\geq1$ and sequence of times  $T_0=t_0<t_1<t_2<\dots<t_N=T$ such that for all $n\in\{0,\dots,N-1\}$,
$$
F_r(t_n) \lesssim_{\widehat\xi_r} 1 + {t_{n+1}}^{-\frac{p}{2}},
$$
for an implicit constant that does not depend on $T_0,T$. Pick $t\in[T_0,T]$. There exists $n\in\{0,\cdots,N-1\} $ such that $t\in[t_n,t_{n+1}]$. Moreover, by \eqref{EqMainEstimateComingDown} with $s=t_n$, we have
$$ 
\Vert v_r(t)\Vert^p_{L^p}\lesssim_{\widehat\xi_r}1+F_r(t_n)\lesssim_{\widehat\xi_r}1+ {t_{n+1}}^{-\frac{p}{2}}
\lesssim_{\widehat\xi_r}1+ {t}^{-\frac{p}{2}}\,.
$$

This bound holds for $T_0$ arbitrarily small and $T=1$, and can be repeated on $[1,2]$, etc, so that the uniform estimate \eqref{EqUpperBoundComingDown} follows. Recall $p>6$ so the space $L^p(M)$ is continuously embedded into the space $\mcC^{-1/2-\epsilon}(M)$. Given that $T^*$ depends only on the restriction to $[0,T^*]$ of $\widehat\xi_r$ and the initial condition $\phi'\in \mcC^{-1/2-\epsilon}(M)$ the uniform estimate \eqref{EqUpperBoundComingDown} and the continuous injection of $L^p(M)$ into $\mcC^{-1/2-\epsilon}(M)$ imply we can extend the solution through $T^*\wedge 1$, hence $T^*>1$. Then  we can repeat the same argument on the interval $[1, 2]$ and so on to get the long time existence of $v_r$.
\end{DemThm7}

On a technical level, our proof of Lemma~\ref{lem:8} will only use the fractional Leibniz rule from Proposition \ref{PropLeibniz} and the elementary interpolation result from Proposition \ref{PropInterpolation}. Last, recall Young inequality that gives the existence for any positive $\delta$ of a constant $C_\delta$ such that one has 
$$
ab \leqslant \delta a^{p'} + \delta^{-\frac{q'}{p'}} b^{q'},   \vspace{0.1cm}
$$
for all positive $a,b$ and exponent $1<p'<\infty$ with conjugate exponent $q'$.

The proof of Lemma~\ref{lem:8} requires two intermediate results stated as lemmas. \vspace{0.1cm}


\begin{lem} \label{LemMainIngredientComingDown}
For every $0< s<t\leqslant T$, we have 
 \begin{align} \label{eq:lemma1}
     \Vert v_r(t)\Vert_{L^p}^p + \int_s^t\Vert v_r(s_1)\Vert^{p+2}_{L^{p+2}}\,\rmd s_1 \lesssim_{\widehat\xi_r} 1 + \Vert v_r(s)\Vert_{L^{p+2}}^p +\int_s^t\Vert v_r(s_1)\Vert^{\frac{p+2}{3}}_{B_{\frac{p+2}{3}}^{1+\epsilon}}\,\rmd s_1.
 \end{align}
\end{lem}
\begin{Dem}
Pairing the equation with $v^{p-1}$ with respect to the $L^2$ scalar product yields identity \eqref{EqIntegralLpNormIdentity}. As $A$ is positive and bounded below and $(v^{p-1},-\Delta v)$ is positive, since $p$ is an even integer greater than $4$, we obtain
\begin{equation} \label{EqIntermediateLpEstimate}
\Vert v_r(t)\Vert^p_{L^p} +\int_s^t \Vert v_r\Vert^{p+2}_{L^{p+2}} \lesssim_{\widehat\xi_r} \Vert v_r(s)\Vert^p_{L^p} + \int_s^t (B_r,\nabla v_r^p) + (Z_{2,r},v_r^{p+1}) + (Z_{1,r},v_r^p) + (Z_{0,r},v_r^{p-1}).
\end{equation}
where the implicit constant is $p\Big(\exp\big( 6 \Vert \begin{tikzpicture}[scale=0.3,baseline=0cm] \node at (0,0) [dot] (0) {}; \node at (0,0.4) [dot] (1) {}; \node at (-0.3,0.8)  [noise] (noise1) {}; \node at (0.3,0.8)  [noise] (noise2) {}; \draw[K] (0) to (1); \draw[K] (1) to (noise1); \draw[K] (1) to (noise2); \end{tikzpicture}_r\Vert_{L^\infty([0,2]\times M)} \big)+1\Big)$.

We bound the different terms in the right hand side of \eqref{EqIntermediateLpEstimate}. Recall that since $B_r$ is an element of $B^{-\epsilon'}_{\infty,\infty}(M)$ for all $\epsilon'>0$ it is an element of $B_{1,\infty}^{-\epsilon}(M)$. By the fractional Leibniz rule from Proposition \ref{PropLeibniz} and Young inequality we have for $\big|(B_r,\nabla v_r^p)\big|$, up to a $\widehat\xi_r$-dependent multiplicative constant, the upper bound
$$
\Vert \nabla v_r^p\Vert_{B^\epsilon_1}\lesssim\Vert  v_r^p\Vert_{B^{1+\epsilon}_1}\lesssim\Vert v_r^{p-1}\Vert_{L^{\frac{p+2}{p-1}}}\Vert v_r\Vert_{B^{1+\epsilon}_{\frac{p+2}{3}}}\lesssim\Vert v_r\Vert_{L^{p+2}}^{p-1}\Vert v_r\Vert_{B^{1+\epsilon}_{\frac{p+2}{3}}} \lesssim \delta \Vert v_r\Vert^{p+2}_{L^{p+2}} +  \delta^{-\frac{p-1}{3}}\Vert v_r\Vert_{B^{1+\epsilon}_{\frac{p+2}{3}}}^{\frac{p+2}{3}},
$$
where $\delta>0$ is arbitrarily small.
For the other terms, we have first 
\begin{align*}
    \big|(Z_{2,r},v_r^{p+1})\big| &\lesssim_{\widehat\xi_r}\Vert v_r^{p+1}\Vert_{B^{\frac{1+\epsilon}{2}}_1}\lesssim\Vert v^p\Vert_{L^{\frac{p+2}{p}}}\Vert v_r\Vert_{B^{\frac{1+\epsilon}{2}}_{\frac{p+2}{2}}}\lesssim\Vert v_r\Vert^p_{L^{p+2}}\Vert v_r\Vert_{B^{\frac{1+\epsilon}{2}}_{\frac{p+2}{2}}}\,.
\end{align*}
Here we interpolate the last term to obtain
\begin{align*}
    \Vert v_r\Vert_{B^{\frac{1+\epsilon}{2}}_{\frac{p+2}{2}}}\lesssim\Vert v_r\Vert_{L^{p+2}}^{\frac12}\Vert v_r\Vert^{\frac12}_{B_{\frac{p+2}{3}}^{1+\epsilon}}\,,
\end{align*}
and we deduce that 
\begin{align*}
    \big|(Z_{2,r} , v_r^{p+1})\big| &\lesssim_{\widehat\xi_r}
    \Vert v_r\Vert^{p+\frac{1}{2}}_{L^{p+2}}\Vert v_r\Vert_{B^{1+\epsilon}_{\frac{p+2}{3}}}^{\frac{1}{2}}
    \leqslant \delta \Vert v_r\Vert^{p+2}_{L^{p+2}} + C_\delta\Vert v_r\Vert_{B^{1+\epsilon}_{\frac{p+2}{3}}}^{\frac{p+2}{3}},
\end{align*}
using Young inequality in the second inequality,
here $C_\delta= \delta^{-\frac{2p+1}{6}}$.

 Similar estimates hold for the $Z_{1,r}$ and $Z_{0,r}$ terms. We have
\begin{align*}
    \big|(Z_{1,r} , v_r^{p})\big| &\lesssim_{\widehat\xi_r} \Vert v_r^{p}\Vert_{B^{1+\epsilon}_1} \lesssim \Vert v_r^{p-1}\Vert_{L^{\frac{p+2}{p-1}}}\Vert v_r\Vert_{B^{1+\epsilon}_{\frac{p+2}{3}}} \lesssim \Vert v_r\Vert^{p-1}_{L^{p+2}}\Vert v_r\Vert_{B^{1+\epsilon}_{\frac{p+2}{3}}}\\
    &\lesssim \delta \Vert v_r\Vert^{p+2}_{L^{p+2}} + \delta ^{-\frac{p-1}{3}}\Vert v_r\Vert_{B^{1+\epsilon}_{\frac{p+2}{3}}}^{\frac{p+2}{3}}.
\end{align*}
and 
\begin{align*}
    \big|(Z_{0,r} , v_r^{p-1})\big| &\lesssim_ {\widehat\xi_r} \Vert v_r^{p-1}\Vert_{B^{1+\epsilon}_1} \lesssim \Vert v_r^{p-2}\Vert_{L^{\frac{p+2}{p-2}}}\Vert v_r\Vert_{B^{1+\epsilon}_{\frac{p+2}{4}}} \lesssim  \Vert v_r\Vert^{p-2}_{L^{p+2}}\Vert v_r\Vert_{B^{1+\epsilon}_{\frac{p+2}{3}}}    \\
    &\lesssim \delta \Vert v_r\Vert^{\frac{(p+2)(p-2)}{p-1}}_{L^{p+2}} + \delta ^{-\frac{p-1}{3}}\Vert v_r\Vert_{B^{1+\epsilon}_{\frac{p+2}{3}}}^{\frac{p+2}{3}} \lesssim 1+ \delta^{\frac{p-1}{p-2}} \Vert v_r\Vert^{p+2}_{L^{p+2}} + \delta ^{-\frac{p-1}{3}}\Vert v_r\Vert_{B^{1+\epsilon}_{\frac{p+2}{3}}}^{\frac{p+2}{3}}.
\end{align*}

One can then absorb the $\delta$ terms of these upper bounds in the corresponding $L^{p+2}$ term in the left hand side of \eqref{EqIntermediateLpEstimate} to get the result by integrating in time on the interval $(s,t)$.
Since we choose $\delta\lesssim  (1+\Vert\widehat\xi_r\Vert)^{-1} $,  then $\, C_\delta, \delta^{-\frac{p-1}{3}}\gtrsim (1+\Vert\widehat\xi_r\Vert)^\gamma$ for some $\gamma>0$ depending on $p$. Combining with the implicit constant in \eqref{EqIntermediateLpEstimate} we obtain that the implicit constant in \eqref{eq:lemma1} is of form  $$
(1+\Vert\widehat\xi_r\Vert)^\gamma \Big(\exp\big( \gamma' \Vert \begin{tikzpicture}[scale=0.3,baseline=0cm] \node at (0,0) [dot] (0) {}; \node at (0,0.4) [dot] (1) {}; \node at (-0.3,0.8)  [noise] (noise1) {}; \node at (0.3,0.8)  [noise] (noise2) {}; \draw[K] (0) to (1); \draw[K] (1) to (noise1); \draw[K] (1) to (noise2); \end{tikzpicture}_r\Vert_{L^\infty([0,2]\times M)} \big) + 1\Big)$$  for some $\gamma,\gamma'>0$ depending on $p$. The implicit constants in the next steps will be obtained in the same way.
\end{Dem}

\vspace{-2pt}

\begin{lem} \label{LemSecondIngredientComingDown}
For $0\leqslant s<t<T<T^*\wedge 1$ we have
\begin{equation} \label{eq:corollary1} \begin{split}
\int_s^t \Vert v_r(s_1)\Vert^{\frac{p+2}{3}}_{B^{1+2\epsilon}_{\frac{p+2}{3}}} \, \rmd s_1  \lesssim_{\widehat\xi_r} 1 + F_r(s) + \int_s^t \Vert v_r(s_1)\Vert^{p+2}_{L^{p+2}} \, \rmd s_1.
\end{split} \end{equation}
\end{lem}
\begin{Dem}
We proceed in two steps.   \vspace{0.15cm}

{\it Step 1.} We first prove that one has

 \begin{equation} \label{eq:theorem1} \begin{split}
     \Vert v_r(t)\Vert_{B^{1+2\epsilon}_{\frac{p+2}{3}}} \lesssim_{\widehat\xi_r}
      \Vert e^{-(t-s)P} v_r(s)\Vert_{B^{1+2\epsilon}_{\frac{p+2}{3}}} + \bigg(
     \int_s^t \Vert v_r(s_1)\Vert^{p+2}_{L^{p+2}} \, \rmd s_1\bigg)^{\frac{3}{p+2}} + \bigg( \int_s^t\Vert v_r(s_1)\Vert^{\frac{p+2}{3}}_{B^{1+\epsilon}_{\frac{p+2}{3}}}\,\rmd s_1\bigg)^{\frac{3}{p+2}}\,.
 \end{split} \end{equation}
We look at each term in the expression for $v_r(t) - e^{-(t-s)P} v_r(s) $
\begin{align*}
 \int_s^t e^{-(t-s_1)P}\Big(
    -  A_r(s_1) v_r(s_1)^3 + B_r(s_1)\nabla v_r(s_1) + Z_{2,r}(s_1) v_r(s_1)^2 + Z_{1,r}(s_1) v_r(s_1) + Z_{0,r}(s_1)\Big)
    \rmd s_1.
\end{align*}
One has
\begin{align*}
    \Big\Vert \int_s^t e^{-(t-s_1)P}\big(A_r(s_1)v_r(s_1)^3\big)\,\rmd s_1\Big\Vert_{B^{1+2\epsilon}_{\frac{p+2}{3}}} &\lesssim \int_s^t \big\Vert e^{-(t-s_1)P}\big(A_r(s_1)v_r(s_1)^3\big) \big\Vert_{B^{1+2\epsilon}_{\frac{p+2}{3}}} \, \rmd s_1   \\
     &\lesssim\int_s^t(t-s_1)^{-\frac{1+2\epsilon}{2}} \big\Vert A_r(s_1)v_r(s_1)^3\big\Vert_{L^{\frac{p+2}{3}}} \, \rmd s_1   \\    
     &\lesssim_{\widehat\xi_r}  \int_s^t(t-s_1)^{-\frac{1+2\epsilon}{2}}\Vert v_r(s_1)\Vert^3_{L^{p+2}} \, \rmd s_1   \\
     &\lesssim\bigg( \int_s^t\Vert v_r(s_1)\Vert_{L^{p+2}}^{p+2}\bigg)^{\frac{3}{p+2}},
\end{align*}
where we used Hölder inequality, the integrability in time of $(t-s_1)^{-\frac{(1+2\epsilon)(p+2)}{2(p-1)}}$ and the fact that $s\leqslant s_1 \leqslant t<T<1$. Similarly, we have
\begin{equation*} \begin{split}
     \Big\Vert \int_s^t e^{-(t-s_1)P}\big( B_r(s_1)\nabla v_r(s_1)\big)\rmd s_1 \Big\Vert_{B^{1+2\epsilon}_{\frac{p+2}{3}}} &\lesssim \int_s^t \Big\Vert e^{-(t-s_1)P}\big( B_r(s_1) \nabla v_r(s_1)\big) \Big\Vert_{B^{1+2\epsilon}_{\frac{p+2}{3}}} \, \rmd s_1   \\
     &\lesssim \int_s^t(t-s_1)^{-\frac{1+3\epsilon}{2}}\Vert B(s_1)\nabla v_r(s_1)\Vert_{B^{-\epsilon}_{\frac{p+2}{3}}} \, \rmd s_1   \\
     &\lesssim_{\widehat\xi_r} \int_s^t(t-s_1)^{-\frac{1+3\epsilon}{2}}\Vert v_r(s_1)\Vert_{B^{1+\epsilon}_{\frac{p+2}{3}}} \, \rmd s_1 \lesssim \bigg( \int_s^t\Vert v_r(s_1)\Vert^{\frac{p+2}{3}}_{B^{1+\epsilon}_{\frac{p+2}{3}}}\bigg)^{\frac{3}{p+2}}.
\end{split} \end{equation*}
Next we have
\begin{align*}
     \left\Vert \int_s^t e^{-(t-s_1)P} \big(Z_{2,r}(s_1) \, v_r(s_1)^2\big)\rmd s_1 \right\Vert_{B^{1+2\epsilon}_{\frac{p+2}{3}}} &\lesssim
     \int_s^t  \Big\Vert e^{-(t-s_1)P} \big(Z_{2,r}(s_1) \, v_r(s_1)^2\big) \Big\Vert_{B^{1+2\epsilon}_{\frac{p+2}{3}}} \, \rmd s_1   \\
     &\lesssim\int_s^t(t-s_1)^{-\frac{1+2\epsilon+\frac{1+\epsilon}{2}}{2}} \big\Vert Z_{2,r}(s_1) \, v_r^2(s_1) \big\Vert_{B^{-\frac{1+\epsilon}{2}}_{\frac{p+2}{3}}} \, \rmd s_1.
    \end{align*}
Using the interpolation result from Proposition \ref{PropInterpolation} and Young inequality we have

\begin{align*} 
\big\Vert Z_{2,r} v_r^2\big\Vert_{B^{-\frac{1+\epsilon}{2}}_{\frac{p+2}{3}}}&\lesssim_{\widehat\xi_r}
\Vert v_r^2\Vert_{B^{\frac{1+\epsilon}{2}}_{\frac{p+2}{3}}}\lesssim  \Vert v_r\Vert_{L^{p+2}} \Vert v_r\Vert_{B^{\frac{1+\epsilon}{2}}_{\frac{p+2}{2}}}\lesssim  \Vert v_r\Vert^{\frac32}_{L^{p+2}} \Vert v_r\Vert_{B^{1+\epsilon}_{\frac{p+2}{3}}}^{\frac12} \lesssim \Vert v_r\Vert^{p+2}_{L^{p+2}}+ \Vert v_r\Vert_{B^{1+\epsilon}_{\frac{p+2}{3}}}^{\frac{p+2}{3}}\,,
\end{align*}    
and the desired estimate follows as in the previous terms. One proceeds in exactly the same way to prove similar estimates on the $Z_{1,r}$ and $Z_{0,r}$ terms. We leave the details to the interested reader.   \vspace{0.15cm}

{\it Step 2.} We first rewrite  \eqref{eq:theorem1} at $t=s_1$ and rise this inequality to the power $\frac{p+2}{3}$. It yields the upper bound
\begin{align*}
     \Vert v_r(s_1)\Vert^{\frac{p+2}{3}}_{B^{1+2\epsilon}_{\frac{p+2}{3}}} 
&\lesssim_{\widehat\xi_r}
      \Vert e^{-(s_1-s)P} v_r(s)\Vert^{\frac{p+2}{3}}_{B^{1+2\epsilon}_{\frac{p+2}{3}}} + 
     \int_s^{t} \Vert v_r\Vert^{p+2}_{L^{p+2}} +\int_s^{t}\Vert v_r\Vert^{\frac{p+2}{3}}_{B^{1+\epsilon}_{\frac{p+2}{3}}}\,,    
\end{align*}
which holds for $t\geqslant s_1$. 
Integrating on $s_1\in[s,t]$, we obtain
\begin{equation} \label{EqAlmostThere} \begin{split}
\bigg( \int_s^t \Vert v_r(s_1)\Vert^{\frac{p+2}{3}}_{B^{1+2\epsilon}_{\frac{p+2}{3}}}\,&\rmd s_1 \bigg)^{\frac{3}{p+2}} \lesssim_{\widehat\xi_r} \bigg(\int_s^t \big\Vert e^{-(s_1-s)P} v_r(s) \big\Vert^{\frac{p+2}{3}}_{B^{1+2\epsilon}_{\frac{p+2}{3}}} \, \rmd s_1 \bigg)^{\frac{3}{p+2}}   \\
&\quad+ (t-s)^{\frac{3}{p+2}}\bigg[\bigg( \int_s^t \Vert v_r(s_1)\Vert^{p+2}_{L^{p+2}} \, \rmd s_1\bigg)^{\frac{3}{p+2}} + \bigg( \int_s^t\Vert v_r(s_1)\Vert^{\frac{p+2}{3}}_{B^{1+\epsilon}_{\frac{p+2}{3}}} \, \rmd s_1\bigg)^{\frac{3}{p+2}} \bigg],
\end{split} \end{equation}
as the function $[\,\dots]$ after $(t-s)^{\frac{3}{p+2}}$ in the right hand side is increasing. We bound the first term in \eqref{EqAlmostThere} by $\Vert v_r(s)\Vert^{\frac{p+2}{3}}_{B^{(1+2\epsilon)(1-\frac{3}{p+2})}_{\frac{p+2}{3}}}$ using the fact that the linear continuous map 
$$
e^{-(s_1-s)P} : B^{(1+2\epsilon)(1-\frac{3}{p+2})}_{\frac{p+2}{3}}(M) \rightarrow B^{1+2\epsilon}_{\frac{p+2}{3}}(M)
$$
has a norm bounded above by $(s_1-s)^{-\frac{3(1+2\epsilon)}{2(p+2)}}\leqslant (s_1-s)^{-1/2}$, a quantity that is integrable over the interval $(s,t)$. The interpolation estimate
\begin{align*}
\Vert v_r(s)\Vert^{\frac{p+2}{3}}_{B^{(1+2\epsilon)(1-\frac{3}{p+2})}_{\frac{p+2}{3}}} \lesssim\Vert v_r(s)\Vert_{L^{\frac{p+2}{3}}}\Vert v(s)\Vert^{\frac{p+2}{3}-1}_{B^{1+2\epsilon}_{\frac{p+2}{3}}} \lesssim \Vert v_r(s)\Vert^p_{L^{\frac{p+2}{3}}} + \Vert v_r(s)\Vert^{\frac{p}{3}}_{B^{1+2\epsilon}_{\frac{p+2}{3}}} = F_r(s),
\end{align*}
gives $F_r(s)$ as a final upper bound for this term. Now, with $v_r$ evaluated at time $s_1$ and $\theta_\epsilon=\frac{1+\epsilon}{1+2\epsilon}$, we have
\begin{align*}
\Vert v_r\Vert_{B^{1+\epsilon}_{\frac{p+2}{3}}}  \lesssim
\Vert v_r\Vert_{L^{\frac{p+2}{3}}}^{1-\theta_\epsilon} \Vert v_r\Vert_{B^{\gamma}_{\frac{p+2}{3}}}^{\theta_\epsilon}\lesssim  
\Vert v_r\Vert_{B^{1+2\epsilon}_{\frac{p+2}{3}}}^{\frac{3\theta_\epsilon}{2+\theta_\epsilon}} + \Vert v_r\Vert_{L^{p+2}}^{3} \leqslant 
\delta \Vert v_r\Vert_{B^{1+2\epsilon}_{\frac{p+2}{3}}} + C_\delta\big(1+ \Vert v_r\Vert_{L^{p+2}}^{3}\big),
\end{align*} 
for some $C_\delta>0$ and  $\delta$  small  enough so that the term related to $\delta \Vert v_r\Vert_{B^{1+2\epsilon}_{\frac{p+2}{3}}}$ can be absorbed by the left hand side of \eqref{EqAlmostThere}. This gives inequality \eqref{eq:corollary1}. 
\end{Dem}

With Lemmas~\ref{LemMainIngredientComingDown} and \ref{LemSecondIngredientComingDown} in hand, we are ready to give the proof of Lemma~\ref{lem:8}, which will conclude the demonstration of Theorem~\ref{ThmLpComingDown}.

\ssk

\begin{DemLemma}
The claim stems from the fact that we have a $B^{1+\epsilon}_{\frac{p+2}{3}}$ norm involved in \eqref{eq:lemma1} while in \eqref{eq:corollary1} we estimate a stronger $B^{1+2\epsilon}_{\frac{p+2}{3}}$ norm.

We start by using the interpolation estimate
\begin{align*}
    \Vert v_r\Vert_{B_{\frac{p+2}{3}}^{1+\epsilon}}\lesssim \Vert v_r\Vert^{1-\theta_\epsilon}_{L^{\frac{p+2}{3}}}\Vert v_r\Vert^{\theta_\epsilon}_{B_{\frac{p+2}{3}}^{1+2\epsilon}} &\leqslant 
    \delta\Vert v_r\Vert^3_{L^{\frac{p+2}{3}}} + C_\delta \Vert v_r\Vert_{B_{\frac{p+2}{3}}^{1+2\epsilon}}^{\sigma_\epsilon} \leqslant \delta\Vert v_r\Vert^3_{L^{p+2}} + C_\delta \Vert v_r\Vert_{B_{\frac{p+2}{3}}^{1+2\epsilon}}^{\sigma_\epsilon}\,,
\end{align*}
with 
$$
\theta_\epsilon \defeq \frac{1+\epsilon}{1+2\epsilon}<1, \quad \sigma_\epsilon \defeq \frac{3\theta_\epsilon}{2+\theta_\epsilon}<1.
$$ 
Using Young's inequality once more, uniform in small $\delta$ and $\eta$ we thus have
\begin{align*}
     \Vert v_r\Vert^{\frac{p+2}{3}}_{B_{\frac{p+2}{3}}^{1+\epsilon}} \lesssim \delta\Vert v_r\Vert^{p+2}_{L^{p+2}} + C_\delta \Vert v_r\Vert_{B_{\frac{p+2}{3}}^{1+2\epsilon}}^{\sigma_\epsilon {\frac{p+2}{3}} }\lesssim
\delta\Vert v_r\Vert^{p+2}_{L^{p+2}} 
+
     \eta \Vert v_r\Vert^{\frac{p+2}{3}}_{B_{\frac{p+2}{3}}^{1+2\epsilon}} + C_{\delta,\eta}
     \,.
\end{align*}
We feed this estimate inside the RHS of \eqref{eq:lemma1}. Taking $\delta$ small enough the contribution of the small factor involving the $L^{p+2}$ norm of $v_r$ can be absorbed in the corresponding term in the left hand side of \eqref{eq:lemma1}. Therefore, \eqref{eq:lemma1} implies
\begin{align} \label{EqAlmostAlmost}
\Vert v_r(t)\Vert_{L^p}^p + \int_s^t\Vert v_r(s_1)\Vert^{p+2}_{L^{p+2}} \, \rmd s_1 &\lesssim 1+ \Vert v_r(s)\Vert_{L^{p+2}}^p + \eta\int_s^t\Vert v_r(s_1)\Vert^{\frac{p+2}{3}}_{B_{\frac{p+2}{3}}^{1+2\epsilon}} \, \rmd s_1\,,\\
&\lesssim 
1+ F_r(s) + \eta\int_s^t\Vert v_r(s_1)\Vert^{p+2}_{L^{p+2}}\rmd s_1\,.
\nonumber
\end{align}
To go from the first to the second line we used that $\Vert v_r(s)\Vert_{L^{p+2}}^p\leqslant F_r(s)$ and controlled the ${B_{\frac{p+2}{3}}^{1+2\epsilon}}$ of $v_r(s_1)$ using \eqref{eq:corollary1}. By choosing $\eta$ small enough we can absorb the $L^{p+2}$ term coming from \eqref{eq:corollary1} in the left hand side of \eqref{EqAlmostAlmost}, and we obtain
\begin{align} \label{EqAlmostAlmostAlmost}
\Vert v_r(t)\Vert_{L^p}^p + \int_s^t\Vert v_r(s_1)\Vert^{p+2}_{L^{p+2}} \, \rmd s_1 
&\lesssim 
1+ F_r(s) \,.
\end{align}
On the other hand, by definition of $F_r$, we have
\begin{align*}
    \Vert v_r(t)\Vert_{L^p}^p + \int_s^t F_r(s_1)^{\frac{p+2}{p}}\rmd s_1 &\lesssim  \Vert v_r(t)\Vert_{L^p}^p + \int_s^t\Vert v_r(s_1)\Vert^{p+2}_{L^{p+2}} \, \rmd s_1+\int_s^t\Vert v_r(s_1)\Vert^{\frac{p+2}{3}}_{B_{\frac{p+2}{3}}^{1+2\epsilon}} \, \rmd s_1\\
    &\lesssim 1+F_r(s)+\int_s^t\Vert v_r(s_1)\Vert^{p+2}_{L^{p+2}} \, \rmd s_1 \lesssim1+F_r(s)\,.
\end{align*}
To go from the first to the second inequality, we used \eqref{EqAlmostAlmostAlmost} to control the first two terms of the RHS, and \eqref{eq:corollary1} to control the integral of the ${B_{\frac{p+2}{3}}^{1+2\epsilon}}$ norm. Using \eqref{eq:corollary1} creates a new integral of the $L^{p+2}$ norm, which again can be controlled using \eqref{EqAlmostAlmostAlmost}. This concludes the proof.\end{DemLemma}

\bigskip

\section{Invariant measure}
\label{SectionInvariantMeasure}

{We will prove below that the enhanced regularized noise $\widehat{\xi}_r$ converges in $L^p(\bbP)$ to some limit enhanced noise $\widehat{\xi}$.} We prove in Section \ref{SubsectionMarkov} that the dynamics generated by Equation \eqref{EqSPDE} is Markovian and that its semigroup has the Feller property. The existence of an invariant measure is obtained from a {classical} compactness argument building on the $L^p$ coming down from infinity property of Theorem \ref{ThmLpComingDown}. We prove in Section \ref{SubsectionNonTriviality} that the invariant probability measure is non-Gaussian.

\medskip

\subsection{A Markovian dynamics}
\label{SubsectionMarkov}

Denote by $\mcF_t$ the usual augmentation of the $\sigma$-algebra generated by random variables $\xi(f)$, for functions $f\in L^2(\mathcal M)$ that are null in $[t,\infty)\times M$. Since $\xi_r$ is white in time the dynamics
$$
(\partial_t+P) u_r =  - u_r^3 + 3(a_r-b_r)u_r +\sqrt{2} \xi_r
$$
generates a $(\mcF_t)_{t\geqslant 0}$-Markov process. For looking at the restriction to a finite time interval $[0,T]$ of this process it is convenient to extend any function on $[0,T]$ into a function on $[0,+\infty)$ that is constant on $[T,+\infty)$. For any $t\in\bbR$ and any $(s,x)\in\mathcal M$ set 
$$
\tau_t(s,x) \defeq (s-t,x).
$$
Denote by $\theta_s : \Omega\rightarrow\Omega, \, s\geqslant 0$ a family of measurable shifts on $(\Omega,\mcF)$ such that one has
$$
(\xi\circ\theta_s , f) = (\xi, f\circ\tau_s)
$$
for all $s$ and all $L^2$ test functions $f$. The Markov property for 
$$
u_r : \Omega\times [0,T]\times \mcC^{-1/2-\epsilon}(M)\rightarrow \mcC^{-1/2-\epsilon}(M)
$$ 
reads
\begin{equation} \label{EqMarkov}
\bbE\big[F\big(u_r(s+\cdot , \phi)\big){\bf 1}_E\big] = \bbE\big[F\big(u_r\circ\theta_s(\cdot , u_r(s,\phi))\big){\bf 1}_E\big],
\end{equation}
for any bounded measurable cylindrical functional $F$ on $C\big((0,T], \mcC^{-1/2-\epsilon}(M)\big)$ and all events $E\in\mcF_s$, with $0\leqslant s\leqslant T$ arbitrary. We need the following quantitative stability result to pass to the zero $r$ limit in \eqref{EqMarkov}.

\ssk

\begin{lem} \label{LemLocalLipschitzPhi}
Fix some positive times $t_1<\cdots<t_k$. There exists two positive constants $\gamma, \gamma'$ such that the restriction of the functions
$$
\phi\in \mcC^{-1/2-\epsilon}(M) \mapsto u_r(t_i,\phi) \in \mcC^{-1/2-\epsilon}(M), \quad(1\leqslant i\leqslant k)
$$
to any centered ball of $\mcC^{-1/2-\epsilon}(M)$ with radius $R>0$ is Lipschitz continuous, with Lipschitz constant bounded above by an explicit function of $R$ and $\widehat\xi_r$.
\end{lem}

\ssk

\begin{Dem}
This result is obtained from the exact same statement for the functions $v_r(t_i,\cdot)$. The relation
$$
u_r(t,\phi_1) - u_r(t,\phi_2) =  e^{-3 \begin{tikzpicture}[scale=0.3,baseline=0cm] \node at (0,0) [dot] (0) {}; \node at (0,0.4) [dot] (1) {}; \node at (-0.3,0.8)  [noise] (noise1) {}; \node at (0.3,0.8)  [noise] (noise2) {}; \draw[K] (0) to (1); \draw[K] (1) to (noise1); \draw[K] (1) to (noise2); \end{tikzpicture}_r(t)} \big(v_r(t,\phi'_1) - v_r(t,\phi'_2)\big),
$$
with 
\begin{equation*}
\phi_i = \begin{tikzpicture}[scale=0.3,baseline=0cm]
\node at (0,0)  [dot] (1) {};
\node at (0,0.8)  [noise] (2) {};
\draw[K] (1) to (2);
\end{tikzpicture}_r(0) - \begin{tikzpicture}[scale=0.3,baseline=0cm]
\node at (0,0) [dot] (0) {};
\node at (0,0.5) [dot] (1) {};
\node at (-0.4,1)  [noise] (noise1) {};
\node at (0,1.2)  [noise] (noise2) {};
\node at (0.4,1)  [noise] (noise3) {};
\draw[K] (0) to (1);
\draw[K] (1) to (noise1);
\draw[K] (1) to (noise2);
\draw[K] (1) to (noise3);
\end{tikzpicture}_r(0)
+ e^{-3\begin{tikzpicture}[scale=0.3,baseline=0cm]
\node at (0,0) [dot] (0) {};
\node at (0,0.4) [dot] (1) {};
\node at (-0.3,0.8)  [noise] (noise1) {};
\node at (0.3,0.8)  [noise] (noise2) {};
\draw[K] (0) to (1);
\draw[K] (1) to (noise1);
\draw[K] (1) to (noise2);
\end{tikzpicture}_r(0)} \big(\phi'_i + v_{\textrm{ref}}(0)\big)
\end{equation*}
allows to transport the locally Lipschitz character of $v_r$ to $u_r$. It suffices to prove the statement with $k=1$ and $t_1=1$; we prove in that case that 
$\phi \in \mcC^{-1/2-\epsilon}(M) \mapsto v_r(1,\phi) \in \mcC^{-1/2-\epsilon}(M)$ is locally Lipschitz. Define 
\begin{eqnarray}
F(\phi, v) \defeq e^{-tP}(\phi) + \mcL^{-1}\Big( -6\nabla \begin{tikzpicture}[scale=0.3,baseline=0cm]
\node at (0,0) [dot] (0) {};
\node at (0,0.4) [dot] (1) {};
\node at (-0.3,0.8)  [noise] (noise1) {};
\node at (0.3,0.8)  [noise] (noise2) {};
\draw[K] (0) to (1);
\draw[K] (1) to (noise1);
\draw[K] (1) to (noise2);
\end{tikzpicture}_r \cdot \nabla v - e^{-6 \begin{tikzpicture}[scale=0.3,baseline=0cm]
\node at (0,0) [dot] (0) {};
\node at (0,0.4) [dot] (1) {};
\node at (-0.3,0.8)  [noise] (noise1) {};
\node at (0.3,0.8)  [noise] (noise2) {};
\draw[K] (0) to (1);
\draw[K] (1) to (noise1);
\draw[K] (1) to (noise2);
\end{tikzpicture}_r} v^3 + Z_{2,r} v^2 + Z_{1,r} v + Z_{0,r} \Big).
\end{eqnarray}
Let $K>0$ be a uniform constant satisfying
\begin{equation}
\|e^{-tP} \phi\|_{C^{-1/2-\epsilon}}\leqslant K \|\phi\|_{C^{-1/2-\epsilon}}   \, \text{  and  }  \|e^{-tP} \phi\|_{(|\alpha_0, 1+\epsilon'|)}\leqslant K \|\phi\|_{C^{-1/2-\epsilon}}.   
\end{equation}
Take the ball $B_R$ in $\mcC^{-1/2-\epsilon}(M)$. It follows from the proof of Proposition \ref{PropLocalWellPosedness} that for any $\phi\in B_R$, there exists $T= T(\widehat \xi_r{}_{|[0,2]}, R)$ and a constant $C(T)<1/2$ only depending on $T$  such that  
$$ 
\|F(\phi, v_1)-F(\phi, v_2 )\|_{C_T\mcC^{-1/2-\epsilon}}  \leqslant C(T) \big(\|v_1-v_2\|_{C_T\mcC^{-1/2-\epsilon}}+\|v_1-v_2\|_{\llparenthesis \alpha_0, 1+\epsilon'\rrparenthesis}\big)
$$
and
$$  
\|F(\phi, v_1)-F(\phi, v_2 )\|_{\llparenthesis\alpha_0, 1+\epsilon\rrparenthesis} \leqslant C(T)\|v_1-v_2\|_{\llparenthesis\alpha_0, 1+\epsilon'\rrparenthesis}). 
$$
Now by the same argument in the proof of   Proposition \ref{PropLocalWellPosedness}, we infer that  for $\phi_1, \phi_2\in B_R$,
\begin{equation*} \begin{split}
&\|v_r(\cdot,\phi_1)-v_r(\cdot,\phi_2)\|_{C_T\mcC^{-1/2-\epsilon}}   \\
&\leqslant  K\|\phi_1-\phi_2\|_{C^{-1/2-\epsilon}} +   C(T) \big(\|v_r(\cdot,\phi_1)-v_r(\cdot,\phi_2)\|_{C_T\mcC^{-1/2-\epsilon}} + \|v_r(\cdot,\phi_1)-v_r(\cdot,\phi_2)\|_{\llparenthesis\alpha_0, 1+\epsilon'\rrparenthesis}\big)
\end{split} \end{equation*}
and
$$
\|v_r(\cdot,\phi_1)-v_r(\cdot,\phi_2)\|_{\llparenthesis\alpha_0, 1+\epsilon'\rrparenthesis} \leqslant  K\|\phi_1-\phi_2\|_{C^{-1/2-\epsilon}} +   C(T) \|v_r(\cdot,\phi_1)-v_r(\cdot,\phi_2)\|_{\llparenthesis\alpha_0, 1+\epsilon'\rrparenthesis}. 
$$
This implies that
\begin{equation*} \begin{split}
\|v_r(\cdot,\phi_1)-v_r(\cdot,\phi_2)\|_{C_T\mcC^{-1/2-\epsilon}} &+ \|v_r(\cdot,\phi_1)-v_r(\cdot,\phi_2)\|_{\llparenthesis\alpha_0, 1+\epsilon'\rrparenthesis}   \\
&\leqslant  2K\|\phi_1-\phi_2\|_{C_T\mcC^{-1/2-\epsilon}} + C(T) \|v_r(\cdot,\phi_1)-v_r(\cdot,\phi_2)\|_{C_T\mcC^{-1/2-\epsilon}}   \\
&\quad+ 2C(T) \|v_r(\cdot,\phi_1)-v_r(\cdot,\phi_2)\|_{\llparenthesis\alpha_0, 1+\epsilon'\rrparenthesis},
\end{split} \end{equation*}
hence
$$
\|v_r(\cdot,\phi_1)-v_r(\cdot,\phi_2)\|_{C_T\mcC^{-1/2-\epsilon}} \leqslant  \frac{2K }{1-C}\|\phi_1-\phi_2\|_{C^{-1/2-\epsilon}}.  
$$
By the $L^p$ a priori estimate of Theorem \ref{ThmLpComingDown}, if  $T<1$ then 
$$
\|v_r(t, \phi)\|_{C^{-1/2-\epsilon}}\leqslant R'\defeq \frac{C(\widehat \xi_r{}_{|[0,2]})}{T^{1/2}}, \quad (T/2\leqslant t\leqslant 1).
$$
Then as above we can get a short time $T' =T'(\widehat \xi_r{}_{|[0,2]}, R')$ such that the map $F(\cdot,\cdot)$ is contracting with a constant $C(T')<1/2$ for any initial condition in $B_{R'}\subset \mcC^{-1/2-\epsilon}(M)$. Since  $v_r(t)\in B_{R'}$ for all $t\in [T/2, 1]$ we can divide the interval $[T/2, 1]$ into subintervals $[t_j,t_j+T']$  and repeat our process above to get 
$$
\big\|v_r(t,\phi_1)-v_r(t,\phi_2)\big\|_{C^{-1/2-\epsilon}} \leqslant  \frac{2K }{1-C(T')}\big\|v_r(t_j,\phi_1) - v_r(t_j,\phi_2)\big\|_{C^{-1/2-\epsilon}},
$$
on each small interval $[t_j,t_j+T']$. Combining all yields that $\phi\mapsto v_r(1,\phi)$ is locally Lipschitz.
\end{Dem}

\ssk

Given the convergence of $\widehat{\xi}_r$ to some limit $\widehat{\xi}$, the results of Section \ref{SectionWellPosedness} give the existence of a limit dynamics $u$ for the solutions $u_r$ to \eqref{EqRenormalizedSPDE}. We denote by $(\mcF_t)_{t\geq 0}$ the $\bbP$-complete filtration generated by the spacetime white noise $\xi$ and set for any bounded measurable function $F$ on $\mcC^{-1/2-\epsilon}(M)$ and $\phi\in\mcC^{-1/2-\epsilon}(M)$
\[
\mcP_t(F)(\phi) = \bbE\big[F(u(t,\phi))\big].
\]

\ssk

\begin{prop}\label{prop:semigroup}
The dynamics of $u$ is Markovian and its associated semigroup $(\mcP_t)_{t\geqslant 0}$ on {the space} $\mcC^{-1/2-\epsilon}(M)$ has the Feller property.
\end{prop}

\ssk

\begin{Dem}
We note two facts as a preliminary remark: (1) For any $\phi\in\mcC^{-1/2-\epsilon}(M)$ and any $s\geq 0$ the random variable $u\circ\theta_s(t,\phi)$ is independent of $\mcF_s$; (2) One has $\bbP$-almost surely the flow property $u_\circ\theta_s(\cdot,u(s,\phi))=u(\cdot,\phi)$. The first point is a direct consequence of the $\bbP$-independence of $\mcF_s$ and the $\sigma$-algebra generated by the $\xi(g)$ with $g\in\mcD([0,\infty)\times M)$ with support in $[s,\infty)$. For the second point, given any $\eta>0$ it follows from Lemma \ref{LemLocalLipschitzPhi} there is an $R(\eta)>0$ such that outside an event of probability $\eta$ the random variables $(u_r\circ\theta_s(\cdot,\cdot))_{0<r\leqslant 1}$ are ($r$-uniformly) uniformly  continuous in their second argument on the centered ball of $\mcC^{-1/2-\epsilon}(M)$ of radius $R(\eta)$ and $u_r(s,\phi)$ is converging to $u(s,\phi)$ with $\vert u(s,\phi)\vert_{C^{-1/2-\epsilon}}\leqslant R(\eta)$. The process $u_r\circ\theta_s(\cdot , u_r(s,\phi))$ is thus converging in probability to $u\circ\theta_s(\cdot , u(s,\phi))$, hence $\bbP$-almost surely along a subsequence $(r_k)$. One then gets the \eqref{EqMarkov} for a cylindrical function $F$ that is continuous by passing to the zero $r$ limit in \eqref{EqMarkov} along that subsequence $r_k$, using dominated convergence. A standard approximation argument extends this identity to all bounded measurable cylindrical functions $F$. This kind of proof is classical, with archetype the proof of Theorem 9.14 in da Prato \& Zabczyk's book \cite{daPratoZabczyk}.   \vspace{0.1cm}

The Feller property of the semigroup $(\mcP_t)_{t\geqslant 0}$, that is the fact that it sends the space of continuous functions on $\mcC^{-1/2-\epsilon}(M)$ into itself, is a direct consequence of the pathwise continuous dependence of the solution $u$ to $\eqref{EqSPDE}$ with respect to the initial condition $\phi$ and dominated convergence in the expression $(\mcP_t f)(\phi) = \bbE\big[f(u(t,\phi))\big]$. This concludes the proof.
\end{Dem}

\ssk

\begin{prop}\label{prop_existence_inv_measure}
The semigroup $(\mcP_t)_{t\geqslant 0}$ has an invariant probability measure.
\end{prop}

\ssk

\begin{Dem}
Recall we turned Equation \eqref{EqRenormalizedSPDE} on $u_r$ into Equation \eqref{EqJPFormulation} on $v_r$, with abstract form \eqref{EqJPFormulationSynthetic}. Coming back to
\begin{equation} \label{EqUFromV}
u = \begin{tikzpicture}[scale=0.3,baseline=0cm]
\node at (0,0)  [dot] (1) {};
\node at (0,0.8)  [noise] (2) {};
\draw[K] (1) to (2);
\end{tikzpicture} - \begin{tikzpicture}[scale=0.3,baseline=0cm]
\node at (0,0) [dot] (0) {};
\node at (0,0.5) [dot] (1) {};
\node at (-0.4,1)  [noise] (noise1) {};
\node at (0,1.2)  [noise] (noise2) {};
\node at (0.4,1)  [noise] (noise3) {};
\draw[K] (0) to (1);
\draw[K] (1) to (noise1);
\draw[K] (1) to (noise2);
\draw[K] (1) to (noise3);
\end{tikzpicture}
+ e^{-3\begin{tikzpicture}[scale=0.3,baseline=0cm]
\node at (0,0) [dot] (0) {};
\node at (0,0.4) [dot] (1) {};
\node at (-0.3,0.8)  [noise] (noise1) {};
\node at (0.3,0.8)  [noise] (noise2) {};
\draw[K] (0) to (1);
\draw[K] (1) to (noise1);
\draw[K] (1) to (noise2);
\end{tikzpicture}} (v + v_{\textrm{ref}}),
\end{equation}
seen as an element of $C_T\mcC^{-1/2-\epsilon}(M)$, one can write for any fixed time 
\begin{equation*} \begin{split}
\big\{\Vert u(t)&\Vert_{C^{-1/2-\epsilon}} > 3m\big\}   \\
&\subset \Big\{\Vert \begin{tikzpicture}[scale=0.3,baseline=0cm] \node at (0,0)  [dot] (1) {}; \node at (0,0.8)  [noise] (2) {}; \draw[K] (1) to (2); \end{tikzpicture}(t)\Vert_{C^{-1/2-\epsilon}} > m\Big\} \cup \Big\{\Vert\begin{tikzpicture}[scale=0.3,baseline=0cm] \node at (0,0) [dot] (0) {}; \node at (0,0.5) [dot] (1) {}; \node at (-0.4,1)  [noise] (noise1) {}; \node at (0,1.2)  [noise] (noise2) {}; \node at (0.4,1)  [noise] (noise3) {}; \draw[K] (0) to (1); \draw[K] (1) to (noise1); \draw[K] (1) to (noise2); \draw[K] (1) to (noise3); \end{tikzpicture}(t)\Vert_{C^{-1/2-\epsilon}} > m\Big\} \cup \Big\{\big\Vert e^{-3\begin{tikzpicture}[scale=0.3,baseline=0cm] \node at (0,0) [dot] (0) {}; \node at (0,0.4) [dot] (1) {}; \node at (-0.3,0.8)  [noise] (noise1) {}; \node at (0.3,0.8)  [noise] (noise2) {}; \draw[K] (0) to (1); \draw[K] (1) to (noise1); \draw[K] (1) to (noise2); \end{tikzpicture}} (v + v_{\textrm{ref}}) \big\Vert_{C^{-1/2-\epsilon}} > m\Big\},
\end{split} \end{equation*}
with 
$$
\bbP\big(\Vert \begin{tikzpicture}[scale=0.3,baseline=0cm] \node at (0,0)  [dot] (1) {}; \node at (0,0.8)  [noise] (2) {}; \draw[K] (1) to (2); \end{tikzpicture}(t)\Vert_{C^{-1/2-\epsilon}} > m\big) + \bbP\big(\Vert\begin{tikzpicture}[scale=0.3,baseline=0cm] \node at (0,0) [dot] (0) {}; \node at (0,0.5) [dot] (1) {}; \node at (-0.4,1)  [noise] (noise1) {}; \node at (0,1.2)  [noise] (noise2) {}; \node at (0.4,1)  [noise] (noise3) {}; \draw[K] (0) to (1); \draw[K] (1) to (noise1); \draw[K] (1) to (noise2); \draw[K] (1) to (noise3); \end{tikzpicture}(t)\Vert_{C^{-1/2-\epsilon}} > m\big) = o_m(1)
$$
uniformly in $t\geqslant 0$ by stationarity. We also have

\begin{equation*} \begin{split}
\bbP\Big(\big\Vert e^{-3\begin{tikzpicture}[scale=0.3,baseline=0cm] \node at (0,0) [dot] (0) {}; \node at (0,0.4) [dot] (1) {}; \node at (-0.3,0.8)  [noise] (noise1) {}; \node at (0.3,0.8)  [noise] (noise2) {}; \draw[K] (0) to (1); \draw[K] (1) to (noise1); \draw[K] (1) to (noise2); \end{tikzpicture}(t)} (v + v_{\textrm{ref}})(t) &\big\Vert_{C^{-1/2-\epsilon}} > m\Big)   \\
&\leqslant \bbP\Big(\big\Vert e^{-3\begin{tikzpicture}[scale=0.3,baseline=0cm] \node at (0,0) [dot] (0) {}; \node at (0,0.4) [dot] (1) {}; \node at (-0.3,0.8)  [noise] (noise1) {}; \node at (0.3,0.8)  [noise] (noise2) {}; \draw[K] (0) to (1); \draw[K] (1) to (noise1); \draw[K] (1) to (noise2); \end{tikzpicture}(t)} \big\Vert_{C^{-1/2-\epsilon}} \geqslant c\Big) + \bbP\Big(\Vert (v + v_{\textrm{ref}})(t)\Vert_{C^{-1/2-\epsilon}} > \frac{m}{c}\Big)   \\
\end{split} \end{equation*}
\begin{equation*} \begin{split}
&\leqslant o_c(1) + \bbP\Big(\Vert v(t) \Vert_{C^{-1/2-\epsilon}} > \frac{m}{2c}\Big) + \bbP\Big(\Vert v_{\textrm{ref}}(t) \Vert_{C^{-1/2-\epsilon}} > \frac{m}{2c}\Big)   \\
&\leqslant o_c(1) +o_{m/c}(1).
\end{split} \end{equation*}
The $o_m(1)$ function does not depend on $t$ by stationarity. In the last step we used the $\phi$-independent the estimate \eqref{EqEstimateCabZ} quantifying the upper bound \eqref{EqUpperBoundComingDown} in the coming down from infinity property together with the stationarity of $v_{\textrm{ref}}$. This gives the $t$-uniform and $\phi$-independent estimate
\begin{equation} \label{EqBoundUForInvariantProbability}
\bbP\big(\Vert u(t)\Vert_{C^{-1/2-\epsilon}} > 3m\big) = o_m(1).
\end{equation}
We have been cautious to construct an enhanced noise whose law is stationary in time. This property together with the independence of the estimate \eqref{EqBoundUForInvariantProbability} with respect to the initial condition allows then to propagate \eqref{EqBoundUForInvariantProbability} uniformly in time by restarting fictively the dynamics every integer time while keeping an upper bound $o_m(1)$ that does not depend on the interval considered. The family of laws $\mathscr{L}(u_r(t,\phi))$ of $u_r(t,\phi)$ is thus tight in $\mcC^{-1/2-2\epsilon}(M)$, independently of the regularization parameter $r\in[0,1]$ and the initial condition $\phi\in \mcC^{-1/2-\epsilon}(M)$, uniformly in $t\geqslant 1$. It follows that, for any initial condition $\phi\in \mcC^{-1/2-2\epsilon}(M)$, introducing the probability measure $\nu_{\mathrm{Bir},T}$ on $\mcC^{-1/2-\epsilon}(M)$ obtained by taking the average of the laws of the solution to the dynamic at different times, called Birkhoff average, that is to say setting
\begin{align}
    \label{eq:Bir}
\nu_{\mathrm{Bir},T}\eqdef\frac{1}{T-1}\int_1^T\delta_{\mathscr{L}(u(t,\phi))}\,\rmd t\,,  \qquad (T\geqslant 2)\,,
\end{align}
the sequence of Birkhoff averages $(\nu_{\mathrm{Bir},T})_T$ has a weak limit along a subsequence of times $T$ tending to infinity. The Feller property of the semigroup generated by \eqref{EqSPDE} ensures that this weak limit is an invariant probability measure of the dynamics.
\end{Dem}

\ssk

The first author proved in \cite[Thm 1.1 and Thm 2.1 of Section 2]{Phi43Uniqueness} the following result.

\ssk

\begin{thm}\label{thm:uniquenessPhi43measure}
The semigroup $(\mcP_t)_{t\geqslant0}$ acting on $\mathcal{C}^{-\frac{1}{2}-\varepsilon}(M)$ has a unique invariant probability measure.
\end{thm}

\ssk

We will thus freely talk in the sequel of {\it the} $\Phi^4_3$ measure as this unique invariant probability measure.

\bigskip

\subsection{Non-triviality of the \TitleEquation{\Phi^4_3} measure}
\label{SubsectionNonTriviality}


\subsubsection{Integrability of $u$.}  The representation proposed by Jagannath \& Perkowski 
$$
u_r = \begin{tikzpicture}[scale=0.3,baseline=0cm] \node at (0,0)  [dot] (1) {}; \node at (0,0.8)  [noise] (2) {}; \draw[K] (1) to (2); \end{tikzpicture}_r -  \begin{tikzpicture}[scale=0.3,baseline=0cm] \node at (0,0) [dot] (0) {}; \node at (0,0.5) [dot] (1) {}; \node at (-0.4,1)  [noise] (noise1) {}; \node at (0,1.2)  [noise] (noise2) {}; \node at (0.4,1)  [noise] (noise3) {}; \draw[K] (0) to (1); \draw[K] (1) to (noise1); \draw[K] (1) to (noise2); \draw[K] (1) to (noise3); \end{tikzpicture}_r +  e^{-3\begin{tikzpicture}[scale=0.3,baseline=0cm] \node at (0,0) [dot] (0) {}; \node at (0,0.4) [dot] (1) {}; \node at (-0.3,0.8)  [noise] (noise1) {}; \node at (0.3,0.8)  [noise] (noise2) {}; \draw[K] (0) to (1); \draw[K] (1) to (noise1); \draw[K] (1) to (noise2); \end{tikzpicture}_r}(v_r + v_{r,\textrm{ref}})
$$
involves a subtle point when it comes to taking the expectation of some quantities. This is related to the fact that the random variable $\begin{tikzpicture}[scale=0.3,baseline=0cm] \node at (0,0) [dot] (0) {}; \node at (0.3,0.6)  [noise] (noise1) {}; \node at (-0.3,0.6)  [noise] (noise2) {}; \draw[K] (0) to (noise1); \draw[K] (0) to (noise2); \end{tikzpicture}$ being a quadratic polynomial of a Gaussian noise the random variable $\exp(-3\begin{tikzpicture}[scale=0.3,baseline=0cm] \node at (0,0) [dot] (0) {}; \node at (0,0.4) [dot] (1) {}; \node at (-0.3,0.8)  [noise] (noise1) {}; \node at (0.3,0.8)  [noise] (noise2) {}; \draw[K] (0) to (1); \draw[K] (1) to (noise1); \draw[K] (1) to (noise2); \end{tikzpicture})$ may not be integrable. A priori it is not immediately clear how to infer the integrability of  $u_r(t)$ from its description in terms of $v_r(t)$. To circumvent this problem we follow Jagannath \& Perkowski's suggestion to trade $\exp(-3\begin{tikzpicture}[scale=0.3,baseline=0cm] \node at (0,0) [dot] (0) {}; \node at (0,0.4) [dot] (1) {}; \node at (-0.3,0.8)  [noise] (noise1) {}; \node at (0.3,0.8)  [noise] (noise2) {}; \draw[K] (0) to (1); \draw[K] (1) to (noise1); \draw[K] (1) to (noise2); \end{tikzpicture})$ for $\exp(-3P_{\geqslant N}\begin{tikzpicture}[scale=0.3,baseline=0cm] \node at (0,0) [dot] (0) {}; \node at (0,0.4) [dot] (1) {}; \node at (-0.3,0.8)  [noise] (noise1) {}; \node at (0.3,0.8)  [noise] (noise2) {}; \draw[K] (0) to (1); \draw[K] (1) to (noise1); \draw[K] (1) to (noise2); \end{tikzpicture})$ in their change of unknown \eqref{EqFromUtoV}, where for $P_{\geqslant N}$ is defined for any $N\geq 1$ as 
$$
P_{\geqslant N} = \sum_{i\in I}\sum_{\vert k\vert \geqslant N} P^i_k\,.
$$
Note that $P_{\geqslant N}$ removes a number of initial terms of a Littlewood-Paley expansion. We choose $N$ as a \textit{random integer} such that 
$$
\Vert  P_{\geqslant N} \begin{tikzpicture}[scale=0.3,baseline=0cm] \node at (0,0) [dot] (0) {}; \node at (0,0.4) [dot] (1) {}; \node at (-0.3,0.8)  [noise] (noise1) {}; \node at (0.3,0.8)  [noise] (noise2) {}; \draw[K] (0) to (1); \draw[K] (1) to (noise1); \draw[K] (1) to (noise2); \end{tikzpicture} \Vert_{C_T\mcC^{1-\eta}} \leqslant 1.
$$
 We define a random  element of $C_T C^\infty(M)$ setting
$$
 f_N \defeq \big(1 - P_{\geqslant N}\big) (\begin{tikzpicture}[scale=0.3,baseline=0cm] \node at (0,0) [dot] (0) {}; \node at (0,0.4) [dot] (1) {}; \node at (-0.3,0.8)  [noise] (noise1) {}; \node at (0.3,0.8)  [noise] (noise2) {}; \draw[K] (0) to (1); \draw[K] (1) to (noise1); \draw[K] (1) to (noise2); \end{tikzpicture} )\equiv\sum_{i\in I}\sum_{\vert k\vert \leqslant N-1} P^i_k( \begin{tikzpicture}[scale=0.3,baseline=0cm] \node at (0,0) [dot] (0) {}; \node at (0,0.4) [dot] (1) {}; \node at (-0.3,0.8)  [noise] (noise1) {}; \node at (0.3,0.8)  [noise] (noise2) {}; \draw[K] (0) to (1); \draw[K] (1) to (noise1); \draw[K] (1) to (noise2); \end{tikzpicture} ).
$$
 As $P^i_kP^j_\ell = 0$ for $\vert k-\ell\vert$ greater than a fixed constant we have
$$
\Vert f_N\Vert_{C_T\mcC^{1-2\epsilon}} \lesssim \Vert \begin{tikzpicture}[scale=0.3,baseline=0cm] \node at (0,0) [dot] (0) {}; \node at (0,0.4) [dot] (1) {}; \node at (-0.3,0.8)  [noise] (noise1) {}; \node at (0.3,0.8)  [noise] (noise2) {}; \draw[K] (0) to (1); \draw[K] (1) to (noise1); \draw[K] (1) to (noise2); \end{tikzpicture}\Vert_{C_T\mcC^{1-2\epsilon}}
$$
{\it uniformly in our definition of the random integer} $N$. 
This change of unknown adds a term $f_Nv$ into the equation for $v$, which only changes $Z_1$ for  $Z_1+f_N$, that is still an element of $C_T\mcC^{-1/2-\epsilon}(M)$ and is a polynomial of the noise, so it has finite moments of any order. We deduce from the estimates \eqref{EqUpperBoundComingDown} and \eqref{EqEstimateCabZ} quantifying of the coming down, with $\exp(-\begin{tikzpicture}[scale=0.3,baseline=0cm] \node at (0,0) [dot] (0) {}; \node at (0,0.4) [dot] (1) {}; \node at (-0.3,0.8)  [noise] (noise1) {}; \node at (0.3,0.8)  [noise] (noise2) {}; \draw[K] (0) to (1); \draw[K] (1) to (noise1); \draw[K] (1) to (noise2); \end{tikzpicture})$ now replaced by $\exp(-P_{\geqslant N}\begin{tikzpicture}[scale=0.3,baseline=0cm] \node at (0,0) [dot] (0) {}; \node at (0,0.4) [dot] (1) {}; \node at (-0.3,0.8)  [noise] (noise1) {}; \node at (0.3,0.8)  [noise] (noise2) {}; \draw[K] (0) to (1); \draw[K] (1) to (noise1); \draw[K] (1) to (noise2); \end{tikzpicture})$, and from the formula \eqref{EqFromUtoV} relating $u$ and $v$, that 
\begin{equation} \label{EqRegularityDecomposition}
u(1) =  \begin{tikzpicture}[scale=0.3,baseline=0cm]
\node at (0,0)  [dot] (1) {};
\node at (0,0.8)  [noise] (2) {};
\draw[K] (1) to (2);
\end{tikzpicture}(1) + \begin{tikzpicture}[scale=0.3,baseline=0cm]
\node at (0,0) [dot] (0) {};
\node at (0,0.5) [dot] (1) {};
\node at (-0.4,1)  [noise] (noise1) {};
\node at (0,1.2)  [noise] (noise2) {};
\node at (0.4,1)  [noise] (noise3) {};
\draw[K] (0) to (1);
\draw[K] (1) to (noise1);
\draw[K] (1) to (noise2);
\draw[K] (1) to (noise3);
\end{tikzpicture} (1)
+ (\star)
\end{equation}
for an element $(\star)\in \mcC^{1-\eta}(M)$ whose norm belongs to all the $L^q(\Omega)$ spaces, $1\leqslant q <\infty$, uniformly with respect to the initial condition $\phi$ of $u$. We now see clearly that $u(1)\in \mcC^{-1/2-\epsilon}(M)$ belongs to all the $L^q(\Omega)$ spaces, $1\leqslant q <\infty$, {\it uniformly with respect to the initial condition $\phi$}. We assume in the remainder of this section that we work with this version of Jagannath \& Perkowski's equation.   \vspace{0.15cm}

\subsubsection{Non-triviality of the $\Phi^4_3$ measures.} The mechanics of the proof that the $\Phi^4_3$ measure is non-Gaussian is well-known. We write it here for completeness and follow for that purpose the lecture notes \cite{LN2Gubinelli} of Gubinelli -- Section 6.4 therein, after Gubinelli \& Hofmanov\'a's work \cite{GubinelliHofmanova2}. Assume $\phi$ is random, with law the invariant measure of the dynamics, so $u(1)$ itself has the same law. Consider the heat regularisation $e^{r \Delta}u(1)$ of our solution $u$ at time $1$. In this subsection, for simplicity, we shall assume that we used true Wick ordering for the renormalisation which simplifies the discussion and allows to use true orthogonality properties of the Wiener chaos decomposition. Our argument is of semiclassical nature, we will use the small $r$ asymptotic behaviour of heat kernels to justify nontriviality -- so $r$ somehow plays the role of a semiclassical parameter. If the $\Phi^4_3$ measure were Gaussian the random variable $e^{-rP}(u(1))$ would also be Gaussian uniformly when $r>0$ goes to $0$. So its truncated four point function

\begin{equation*} \begin{split}
C_4^r &= C_4\Big(e^{-rP}(u(1)), e^{-rP}(u(1)), e^{-rP}(u(1)), e^{-rP}(u(1))\Big)   \\
&\defeq \bbE\big[\big\{e^{-rP}(u(1))\big\}^4\big] - 3\,\bbE\big[\big\{e^{-rP}(u(1))\big\}^2\big]^2
\end{split} \end{equation*}
would be identically null uniformly in $r\in(0,1]$. The sufficient integrability of the different elements of the decomposition
$$
u(1) = \begin{tikzpicture}[scale=0.3,baseline=0cm]
\node at (0,0)  [dot] (1) {};
\node at (0,0.8)  [noise] (2) {};
\draw[K] (1) to (2);
\end{tikzpicture}(1) - \begin{tikzpicture}[scale=0.3,baseline=0cm]
\node at (0,0) [dot] (0) {};
\node at (0,0.5) [dot] (1) {};
\node at (-0.4,1)  [noise] (noise1) {};
\node at (0,1.2)  [noise] (noise2) {};
\node at (0.4,1)  [noise] (noise3) {};
\draw[K] (0) to (1);
\draw[K] (1) to (noise1);
\draw[K] (1) to (noise2);
\draw[K] (1) to (noise3);
\end{tikzpicture}(1)
+ e^{-3P_{\geqslant n}\begin{tikzpicture}[scale=0.3,baseline=0cm]
\node at (0,0) [dot] (0) {};
\node at (0,0.4) [dot] (1) {};
\node at (-0.3,0.8)  [noise] (noise1) {};
\node at (0.3,0.8)  [noise] (noise2) {};
\draw[K] (0) to (1);
\draw[K] (1) to (noise1);
\draw[K] (1) to (noise2);
\end{tikzpicture}(1)} \big(v(1) + v_{\textrm{ref}}(1)\big),
$$
allows to plug it inside the formula for the fourth order cumulant and use Wick's Theorem to get
\begin{equation*} \begin{split}
C_4&\Big(e^{-rP}(u(1)), e^{-rP}(u(1)), e^{-rP}(u(1)) ,e^{-rP}(u(1))\Big) \\
&= 24\int_{-\infty}^{t} \Big(G_r^{(3)}(t-s) \, e^{-(t-s+r)P}\Big)(x,x) \rmd s + 216 \int_{(-\infty,t]^2} \int_{y_1,y_2\in U^2} G^{(2)}_0(s_1-s_2,y_1,y_2)   \\
&\quad\times e^{-(r+t-s_1)P}(y_1,x)e^{-(r+t-s_2)P}(y_2,x)G^{(1)}_r(t-s_1,y_1,x)G^{(1)}_r(t-s_2,y_2,x) \, \rmd y_{12}\rmd s_{12}
\end{split} \end{equation*}
\begin{equation*} \begin{split}
&\quad+ \mathbb{E}\left[ 
e^{-rP}
\begin{tikzpicture}[scale=0.3,baseline=0cm]
\node at (0,0)  [dot] (1) {};
\node at (0,0.8)  [noise] (2) {};
\draw[K] (1) to (2);
\end{tikzpicture}(1) Q\big(e^{-rP}\begin{tikzpicture}[scale=0.3,baseline=0cm]
\node at (0,0) [dot] (0) {};
\node at (0,0.5) [dot] (1) {};
\node at (-0.4,1)  [noise] (noise1) {};
\node at (0,1.2)  [noise] (noise2) {};
\node at (0.4,1)  [noise] (noise3) {};
\draw[K] (0) to (1);
\draw[K] (1) to (noise1);
\draw[K] (1) to (noise2);
\draw[K] (1) to (noise3);
\end{tikzpicture}(1), e^{-rP} e^{-3P_{\geqslant n}\begin{tikzpicture}[scale=0.3,baseline=0cm]
\node at (0,0) [dot] (0) {};
\node at (0,0.4) [dot] (1) {};
\node at (-0.3,0.8)  [noise] (noise1) {};
\node at (0.3,0.8)  [noise] (noise2) {};
\draw[K] (0) to (1);
\draw[K] (1) to (noise1);
\draw[K] (1) to (noise2);
\end{tikzpicture}(1)} \big(v(1) + v_{\textrm{ref}}(1)\big)\big)\right]
\end{split} \end{equation*}
where $Q$ is some polynomial function in its two arguments and, for $1\leqslant i\leqslant3$,
\[
G_r^{(i)}(t-s)(x,y=\Big(\frac{e^{-(2r+\vert t-s\vert)P}}{P}(x,y)\Big)^i.
\]
We have many cancellations in the above expression since Gaussian cumulants only retain connected Feynman graphs and we also use orthogonality of homogeneous Wiener chaoses of different degrees. The remainder has the corresponding decay
$$
\mathbb{E}\left[ e^{-rP} \begin{tikzpicture}[scale=0.3,baseline=0cm]
\node at (0,0)  [dot] (1) {};
\node at (0,0.8)  [noise] (2) {};
\draw[K] (1) to (2);
\end{tikzpicture}(1) \, Q\big(e^{-rP}\begin{tikzpicture}[scale=0.3,baseline=0cm]
\node at (0,0) [dot] (0) {};
\node at (0,0.5) [dot] (1) {};
\node at (-0.4,1)  [noise] (noise1) {};
\node at (0,1.2)  [noise] (noise2) {};
\node at (0.4,1)  [noise] (noise3) {};
\draw[K] (0) to (1);
\draw[K] (1) to (noise1);
\draw[K] (1) to (noise2);
\draw[K] (1) to (noise3);
\end{tikzpicture}(1), e^{-rP}e^{-3P_{\geqslant n}\begin{tikzpicture}[scale=0.3,baseline=0cm]
\node at (0,0) [dot] (0) {};
\node at (0,0.4) [dot] (1) {};
\node at (-0.3,0.8)  [noise] (noise1) {};
\node at (0.3,0.8)  [noise] (noise2) {};
\draw[K] (0) to (1);
\draw[K] (1) to (noise1);
\draw[K] (1) to (noise2);
\end{tikzpicture}(1)} \big(v(1) + v_{\textrm{ref}}(1)\big)\big)\right] = \mathcal{O}(r^{-\frac{1}{4}})
$$ 
since we just need to recall that the remainder only involves the terms, 
$$
e^{-rP}\begin{tikzpicture}[scale=0.3,baseline=0cm]
\node at (0,0)  [dot] (1) {};
\node at (0,0.8)  [noise] (2) {};
\draw[K] (1) to (2);
\end{tikzpicture}(1) = \mathcal{O}(r^{-\frac{1}{4}}), \quad e^{-rP}\begin{tikzpicture}[scale=0.3,baseline=0cm]
\node at (0,0) [dot] (0) {};
\node at (0,0.5) [dot] (1) {};
\node at (-0.4,1)  [noise] (noise1) {};
\node at (0,1.2)  [noise] (noise2) {};
\node at (0.4,1)  [noise] (noise3) {};
\draw[K] (0) to (1);
\draw[K] (1) to (noise1);
\draw[K] (1) to (noise2);
\draw[K] (1) to (noise3);
\end{tikzpicture}(1) = \mathcal{O}(1), \quad e^{-rP}e^{-3P_{\geqslant n}\begin{tikzpicture}[scale=0.3,baseline=0cm]
\node at (0,0) [dot] (0) {};
\node at (0,0.4) [dot] (1) {};
\node at (-0.3,0.8)  [noise] (noise1) {};
\node at (0.3,0.8)  [noise] (noise2) {};
\draw[K] (0) to (1);
\draw[K] (1) to (noise1);
\draw[K] (1) to (noise2);
\end{tikzpicture}(1)} \big(v(1) + v_{\textrm{ref}}(1)\big) = \mathcal{O}(1)
$$ 
since they are H\"older regular in $\mcC^{\frac{1}{2}-}$  and $\mcC^{1-}$ respectively. We used the fact that we can probe the space H\"older regularity by testing against heat kernels
\[
\sup_{\varepsilon\in (0,1]} \varepsilon^{-\frac{s}{2}}\Vert e^{-\varepsilon P}u\Vert_{L^\infty(M)} \lesssim\Vert u \Vert_{C^s(M)} 
\]
and also we made an implicit use of Besov embeddings $ \Vert \bigcdot\Vert_{C^{s-\frac{d}{p}-\delta}}  \lesssim \Vert \bigcdot\Vert_{B^s_{p,p}}$, for all $\delta>0$, together with hypercontractive estimates which allows us to consider expectations of H\"older norms.  \vspace{0.1cm}


Let us study in detail the asymptotics of the first term on the right hand side of the equation for $C_4^r$ which has a Feynman integral interpretation. For every $x\in U$, choose some cut-off function ${\ell}\in C^\infty_c(U)$ which equals $1$ near $x$ that we use to localize the asymptotics as in the calculation of counterterms, then we can extract the small $r$ leading asymptotics as
\begin{equation*} \begin{split}
\int_{-\infty}^t \big(G_r^{(3)}(t-s) &\, e^{-(t-s+r)P}\Big)(x,x)\rmd s   \\
&\simeq \int_0^\infty \int_{y\in U}{\ell}(y) \left(\int_{[a+r,+\infty)^3} \prod_{i=1}^3 e^{-s_iP}(x,y)\rmd s_i\right)e^{-(a+r)P}(y,x)\rmd a.
\end{split} \end{equation*}
We compute the integral with respect to $y$ first; this reads
\begin{equation*} \begin{split}
\int &\left(\prod_{i=1}^3 K^0(s_i,x,y)\right) K^0(a+r,x,y) {\ell}(y) \rmd y   \\
&\simeq \frac{e^{-(s_1+s_2+s_3+a+r)}(4\pi)^{-6}}{(s_1s_2s_3(a+r))^{\frac{3}{2}}} \int_U e^{-\frac{\vert x-y\vert_{g(x)}}{4} (s_1^{-1}+s_2^{-1}+s_3^{-1}+(a+r)^{-1}) } {\ell}(y)\det(g)^{\frac{1}{2}}_y\rmd y   \\
&\simeq \frac{(4\pi)^{-\frac{9}{2}} e^{-(s_1+s_2+s_3+a+r)} }{ (s_1s_2s_3(a+r))^{\frac{3}{2}}(s_1^{-1}+s_2^{-1}+s_3^{-1}+(a+r)^{-1})^{\frac{3}{2}}   }   \\
&\hspace{3cm}+ \mathcal{O}\left((s_1s_2s_3(a+r))^{-\frac{3}{2}}(s_1^{-1}+s_2^{-1}+s_3^{-1}+(a+r)^{-1})^{-\frac{5}{2}} \right)
\end{split} \end{equation*}
where we only keep the leading terms in the heat asymptotic expansion and use a stationary phase estimate. It is possible, as we did for the counterterms, to show that the integral with respect to $a,s_1,s_2,s_3$ of the $\mathcal{O}(\cdots)$ term gives subleading asymptotics compared to the leading term. We are reduced after a change of variables to the asymptotics of the following integral

$$
\int_0^1  (a+r)^{-\frac{3}{2}} \left(  \int_{[1,+\infty)^3} \big(a_2a_3 + a_1a_3 + a_1a_2+a_1a_2a_3\big)^{-\frac{3}{2}} \, \rmd a_{123} \right)  \rmd a \simeq c r^{-\frac{1}{2}}
$$
for some non-null constant $c$. The next term in the formula for $C_4^r$ is

\begin{equation*} \begin{split}
\int_{(-\infty,t]^2}\int_{y_1,y_2\in U^2} \hspace{-0.3cm} G^{(2)}_0(s_1-s_2,y_1,y_2) &\,e^{-(r+t-s_1)P}(y_1,x) \, e^{-(r+t-s_2)P}(y_2,x)   \\
&\times G^{(1)}_r(t-s_1,y_1,x) \, G^{(1)}_r(t-s_2,y_2,x) \, \rmd y_1\rmd y_2\rmd s_{12}
\end{split} \end{equation*}
which is bounded by a constant multiple of the integral over $(-\infty,t]^2\times U^2$ of 
$$
\left(\sqrt{\vert s_1-s_2\vert}+\vert y_1-y_2\vert\right)^{-2}\left(\sqrt{\vert r+t-s_1\vert}+\vert y_1-x\vert\right)^{-4}\left(\sqrt{\vert r+t-s_2\vert}+\vert y_2-x\vert\right)^{-4}.
$$
Making first the change of variables $s_i\mapsto r^2(s_i-t)+s_i$, $y_i\mapsto r(y_i-x)+x $ and then using polar coordinates gives $\mathcal{O}(\vert\hspace{-0.05cm}\log r\vert)$ as an upper bound for that integral, therefore the cumulant $C_4^r$ blows-up like $24cr^{-\frac{1}{2}}$ when $r>0$ goes to $0$. This shows that $C_4^r$ does not vanish asymptotically and that the $\Phi^4_3$ measure is non-Gaussian.

\ssk

For a $\Phi^4_3$ measure $\nu$ obtained as above as a weak limit of the Birkhoff averages \eqref{eq:Bir}, the covariance property under Riemannian isometries is clear from its construction and the fact that the renormalisation constants $a_r, b_r$ do not depend on which Riemannian metric is used:
given a field $\phi$ on $(M,g)$ whose law is a $\Phi^4_3$ measure, let $f:M^\prime\mapsto M$ be a smooth diffeomorphism, then the pulled--back field $f^*\phi$ on $(M^\prime,f^*g)$ will have the law of a $\Phi^4_3$ measure of the stochastic PDE \eqref{EqSPDE} for the metric $f^*g$. 
 Such measure gives for the first time a non-perturbative, non-topological interacting quantum field theory on $3$-dimensional curved Riemannian spaces. We prove in \cite{Phi43Uniqueness} that the semigroup on $\mcC^{-1/2-\epsilon}(M)$ generated by the dynamics \eqref{EqSPDE} has a unique invariant probability measure $\mu$. This uniqueness result yields a stronger notion of covariance.

\ssk

{More properties of the $\Phi^4_3$ measure $\mu$ on $\mcC^{-1/2-\epsilon}(M)$ are known in the case where $M$ is the $3$-dimensional torus. It was for instance proved by Hairer \& Steele \cite{HairerSteele} that for any test function $\psi$ the random variable $\exp(c\,\omega(\psi)), \, \omega\in \mcC^{-1/2-\epsilon}(M)$, belongs to $L^1(\mu)$, for $c>0$ small enough. We do not prove this kind of interesting property in this work.

}

\bigskip

\subsection{Locally covariant renormalization}
\label{SectionLocallyCovariant}

Let us discuss  our choice of counterterms. We renormalise the dynamical $\Phi^4_3$ model using metric-independent counterterms given in Equation \eqref{EqDefnArBr}. It turns out that the philosophy we adopted in order to obtain these counterterms was to set up a \textbf{locally covariant renormalisation} whose clean mathematical formulation follows the seminal work of Gilkey on the local index theory \cite{Gilkey, ABP}. We formalize this notion in the following definition:

\ssk

    \begin{defn} \label{def:localcounterterms}
    A local counterterm $c[g](x)$ at a point $x$ of our manifold  is \textbf{\textit{locally covariant}} if it only depends smoothly on finite jets of the metric $g$ at $x$ in the following precise sense, in any coordinate system \cite{ABPerrata}, \cite[p.161]{Gilkey}
     \begin{equation}\label{eq:countertermform}
      c[g](x)=\sum_{\alpha} c_\alpha(g(x))m_\alpha 
      \end{equation}
      where the sum is finite, $c_\alpha$ is smooth in its arguments and $m_\alpha$ stands for a monomial in partial derivatives of the metric $g$
    and the counterterm must transform in a natural way: for any diffeomorphism $\varphi:\widetilde{M} \mapsto M$ the counterterm
    on $\widetilde{M}$ is obtained by pulling back the counterterm on $M$ and obeys the consistency equation \cite[eq 2.3 p.~282]{ABP}
    $$ 
    c[\varphi^*g](x) = \varphi^*\left(c[g](x) \right).
    $$
    \end{defn}
    
\ssk    
    
The key point is that our counterterms trivially satisfy the above definition since they do not depend on the Riemannian metric! Note that the diagonal value $\mathbf{G}(x,x)$ of the Green function $\mathbf{G}$ of the Laplacian, though covariant, is not locally covariant. In fact, the prescription of local covariance restricts the form of the counterterms considerably since it should have the form given by Equation \eqref{eq:countertermform}. Moreover the derivatives of the metric $g$ which are allowed to appear in the expansion \eqref{eq:countertermform} is also constrained by the dimension of the coupling constant as usual in renormalization theory (the notion of dimension in quantum field theory is closely related to the weights in the algebra of invariant polynomials of the metric introduced by Gilkey in~\cite[Lemma 2.4.1 p.~162]{Gilkey}).

So for each counterterm appearing in the Lagrangian, there is only a finite dimensional space of choices. We do not give more details on these aspects and refer the reader to the very comprehensive work of Khavkine \& Moretti \cite{KM06} where all Wick powers of the free field on globally hyperbolic spacetimes are defined in a locally covariant way and renormalization ambiguities in the definition of the Wick polynomials are carefully classified in detail~\cite[Def 3.5 p.~602 and Theorem 3.1 p.~604]{KM06}. Let us stress that the definitions in~\cite{KM06} extend to the Euclidean setting in a rather straightforward way. Such an expansion of the counterterms stems from the near diagonal, short time asymptotic expansion of the heat kernel on closed manifolds. In fact, it is an important result of Gilkey~\cite[Lemma 2.4.2 p.~163]{Gilkey} that the coefficients appearing in the heat kernel expansion are local polynomial invariants in the derivatives of the metric~\cite[Definition p.~161]{Gilkey}. In the case of the dynamical $\Phi^4_3$ model, for the two counterterms only $c_0$ would be divergent with the regularisation parameter, so we can take them to be constant. 
    
\ssk
    
\subsubsection{Mass shift covariance.}

We would like to give some more motivation for using a locally covariant renormalisation. To simplify the notation, we discuss the case of the $\Phi^4_2$ measure $\nu^{k,t,\ell}$, which is obtained as weak limit of the sequence
    \begin{align*}
        \nu^{k,\ell,t}_r(\rmd\phi)\propto \exp\big(-k\int :\phi_r^4:_t-\ell\int :\phi_r^2:_t\big)\mu^t(\rmd\phi)\,,
          \end{align*}
         when $r\downarrow 0$ where
        \begin{align*}
:\phi_r^4:_t(x)\eqdef\phi_r^4(x)-6C_{t,r}(x,x)\phi_r^2+3C^2_{t,r}(x,x)\,,\quad :\phi_r^2:_t(x)\eqdef\phi_r^2(x)-6C_{t,r}(x,x),
    \end{align*}
$r>0$ denotes a regularisation scale, $\phi_r$ denotes the regularised field at scale $r$, and $\mu^t$ stands for the Gaussian field of covariance $(t-\Delta)^{-1}$, whose regularised kernel is denoted by $C_{t,r}$. One question one could ask about the $\Phi^4_2$ measures is whether the unit mass measure $\nu^{k,0,1}$ is equal to another $\Phi^4_2$ measure $\nu^{k,\ell_t,t}$, for some function $\ell_t$.
Said differently, can we fine tune the coupling constants $k,\ell$ in the interacting part of our Lagrangian $\big(-k\int :\phi_r^4:_t-\ell\int :\phi_r^2:_t\big)$ in order to absorb some simple shift $1-\Delta\mapsto t+1-\Delta$ of the massive Laplacian defining our GFF so that we get the same quantum field theory measure? This is the question of the \textnormal{mass-shift covariance of the $\Phi^4_2$ measure}. This question is natural even in the case of the free field, and it turns out that in dimension lower or equal to three, the answer is yes because the massive Laplace inverse $(1-\Delta)^{-1}$ is Hilbert--Schmidt, and by the results of~\cite[section 9]{GlimmJaffe} one has
 \begin{align*}
     \mu^1(\rmd\phi) \propto \exp\Big((t-1)\int :\phi^2:_t\Big)\mu^t(\rmd\phi)\,.
 \end{align*}
The above formula confirms the intuition that putting the mass in the Gaussian or in the density should be equivalent and that the \textsl{shift in the coupling in front of $:\phi^2:$ should be the same as a shift of the squared mass}. In other words, for $k=0$, the $\Phi^4_2$ measure is mass-shift invariant, that is to say that as expected it holds $\nu^{0,0,1}=\nu^{0,1-t,t}$. However, turning on the coupling constant $k$ the interaction changes things. In the translation invariant case of the torus $\mathbb{T}^2$ endowed with a flat metric, because $C_{t,r}(x,x)$ does not depend on $x$, one would obtain
\begin{align*}
    \nu^{k,0,1} = \nu^{k,\ell_t,t}\quad\text{with}\quad \ell_t=6\lim_{r\downarrow0}(C_{t,r}-C_{1,r})(x,x)+1-t.
\end{align*}
Therefore, on the torus, mass-shift invariance is broken, but translation-invariance helps restore mass-shift covariance, with a complicated mass function $\ell_t$. On the other hand, in the general case of a manifold with a generic Riemannian metric $k$, the isometry group contains only the identity element, hence the difference term $\lim_{r\downarrow0}(C_{t,r}-C_{1,r})(x,x)$ does depend on $x$, and mass-shift covariance is broken. However, if instead of performing the usual Wick renormalisation, one works with a locally covariant renormalisation, $C_{t,r}$ and $C_{1,r}$ are replaced by the exact same divergent counterterm, and not only the $\Phi^4_2$ measures are mass-shift covariant, but also we have the natural identity $\nu^{k,0,1}=\nu^{k,-t,t} $ for all $t\geqslant 0$ hence the measures are even mass-shift invariant as in the non interacting $k=0$ case! To conclude, locally covariant renormalisation is necessary to preserve the symmetries of the $\Phi^4$ measures on manifolds and, even in the case of the torus, where Wick renormalisation and locally covariant renormalisation differ by a finite number, locally covariant renormalisation is more respectful of symmetries.

\subsubsection{Compatibility with Segal cutting and gluing.}

Another even stronger argument in favour of locally covariant renormalization, coming from physics, is to make renormalization compatible with cutting and gluing, i.e. make renormalization compatible with the Riemannian version of the Segal axioms in Euclidean quantum field theory \cite{KandelMnevWernli, Lin24, Pickrell}. As above, let us illustrate our claim by some simple example related to some quadratic perturbation of the GFF. Assume we are given a closed compact Riemannian surface $(M,g)$ obtained by gluing two Riemannian surfaces $M_L,M_R$ with boundaries along some common boundary curve $\Sigma$, $M=M_L\cup_\Sigma M_R$. By the spatial Markov property, the GFF $\Phi$ on $M$ decomposes as an independent sum: $ \Phi=\Phi_{L,D}+\Phi_{R,D} + H\Phi_\Sigma $ where $\Phi_{R,D}$ (respectively  $\Phi_{L,D} $) is a Dirichlet GFF on $M_R$ (resp $M_L$) and $H\Phi_\Sigma$ is the harmonic extension of the traced field on the hypersurface $\Sigma$. It is well-known that the field $H\Phi_\Sigma$ is Gaussian with covariance the inverse $DN^{-1}$ of  the Dirichlet to Neumann operator; we denote by $\mu_\Sigma(\rmd H\phi_\Sigma)$ the corresponding boundary measure. Using functional integral notations, for each piece $M_{R/L}$, consider
\begin{align*}
\Psi_{L/R}(\phi_\Sigma) := \int_{\Phi_{L/R,D}} \exp\left( -S(\Phi_{L/R,D}+H\Phi_\Sigma)  \right)\rmd\phi_{L/R,D}    
\end{align*}
where $S(\Phi_{L/R,D}+H\Phi_\Sigma)=\int_M\vert \nabla \phi_{L/R}\vert^2 +m^2(\Phi_{L/R,D}+H\Phi_\Sigma)^2+g(\Phi_{L/R,D}+H\Phi_\Sigma)^4   $. One has to think of $\Psi_{L/R}$  as $L^2$ vectors for the measure $\mu(\rmd\phi_\Sigma)$ on boundary fields, in fact states in the quantum physics terminology. In fact, it is proved in~\cite{Lin24} that $\Psi_{L/R}$ is an actual $L^2$ function. Then in \cite[Section 5.1 p.~1885]{KandelMnevWernli}, it is proved that if one renormalizes the $\Phi^4_2$ theory with the usual Wick renormalization, the theory obtained does not satisfy Segal gluing
\begin{align*}
  \int_{\Phi} \exp\left( -S(\Phi)  \right)\rmd\phi      \neq \int_{\Phi_\Sigma}  \Psi_{L}(\phi_\Sigma)\Psi_{R}(\phi_\Sigma)\rmd\mu(\Phi_\Sigma) \,, 
\end{align*}
where the integral on the right hand side  is taken over the boundary fields.

\subsubsection{Proof of Theorem~\ref{prop:3}.}
\label{sec:423}
We conclude this section with the proof of Theorem \ref{prop:3} from the introduction.
Start from the PDE
\begin{equation} 
(\partial_t - \Delta_{g_2} + 1)u_r = \sqrt{2}\xi_{r,g_2} - u_r^3 + 3(a_r - b_r)u_r
\end{equation}
on the Riemannian manifold $(M_2,g_2)$ where $\Delta_{g_2}$ means the Laplace--Beltrami operator for the metric $g_2$ with initial data $u_r(0,.)=0$. By functoriality of all the involved objects $\chi^*\left(\Delta_{g_2}u_r \right)=\Delta_{\chi^*g_2} \chi^*u_r$, $\chi^*e^{r\Delta_{g_2}}=e^{r\Delta_{\chi^*g_2}}\chi^*$ and $\chi^*\xi_r$ has the law of the regularized white noise on $M_1$ for the metric $g_1=\chi^*g_2$, 
$\chi^*u_r$ will be the unique solution (this is by Cauchy well--posedness) of the PDE
\begin{equation} 
(\partial_t - \Delta_{g_1} + 1) \chi^*u_r = \sqrt{2}\chi^*\xi_r - (\chi^*u_r)^3 + 3(a_r - b_r)\chi^*u_r
\end{equation}
still with initial condition $\chi^*u_r(0,.)=0$
where this time all the dynamics occurs on $M_1,g_1=\chi^*g_2$.

The Birkhoff average $\frac{1}{T-1}\int_1^T \delta_{\chi^*u_r(t,\cdot)} \rmd t $ describes the pushforward of the Birkhoff average $\frac{1}{T-1}\int_1^T \delta_{u_r(t,\cdot)} \rmd t $
under the map $T_\chi$ by definition: $ \frac{1}{T-1}\int_1^T \delta_{\chi^*u_r(t,\cdot)} \rmd t=\left( T_\chi\right)_\sharp \left(\frac{1}{T-1}\int_1^T \delta_{u_r(t,\cdot)} \rmd t \right)  $. By unique ergodicity of the limiting Markovian dynamics on $\mathcal{C}^{-\frac{1}{2}-\kappa}(M)$, $\kappa>0$, letting $r\downarrow 0$ and then $T\uparrow \infty $, both Birkhoff averages converge to the respective limiting $\Phi^4_3$ measures $\mu_1$ on $M_1$ and $\mu_2$ on $M_2$, this implies that
$\left( T_\chi\right)_\sharp \mu_2=\mu_1 $
by continuity of the pushforward acting on Borel measures on $\mathcal{C}^{-\frac{1}{2}-\kappa}(M)$ for all $\kappa>0$ which concludes the proof. 

\section{Enhanced noise, Feynman diagrams and extension of distributions}
\label{SectionMotivation}

\subsection{A road map for the construction of the enhanced noise \texorpdfstring{$\boldsymbol{\widehat{\xi}}$}{\widehat{\xi}}}

The remainder of this work is dedicated to the development of some tools of independent interest that we use in Section \ref{SectionStochasticBounds} to give a complete proof of the convergence of the enhanced, regularized, renormalized noise $\widehat{\xi}_r$ in any $L^q(\bbP)$ space ($1\leqslant q<\infty$) to some limit in the product space 
\[
\mcC^{-5/2-\epsilon}([0,T]\times M)\times C_T\mcC^{-1-2\epsilon} \times C_T\mcC^{1/2-3\epsilon} \times \big(C_T\mcC^{-4\epsilon}\big)^3 \times C_T\mcC^{-1/2-5\epsilon},
\]
where we use the generic notation $C_T\mcC^\gamma$ for $C\big([0,T],C^\gamma(M)\big)$ for any $\gamma\in\bbR$. The method that we use for this purpose seems more important than its application to the dynamical $\Phi^4_3$ equation. We import from quantum field theory the Epstein-Glaser point of view that renormalization is fundamentally a question of extension of some distributions. The convergence of $\widehat{\xi}_r$ to some limit can indeed be quantified by the convergence of some expectations involving $\widehat{\xi}_{r_1}-\widehat{\xi}_{r_2}$, via some Kolmogorov type convergence theorem. Some natural upper bounds on these expectations are given by some multiple integrals that are structured by some Feynman graphs, some of whose kernels are regularized. The non-integrated functions that appear in these multiple integrals are usually called {\sl Feynman amplitudes}. They are typically given by some $(r_1,r_2)$-dependent functions of all the integration variables that are smooth outside of all the diagonals, where two or more integration variables coincide, for all positive values of the parameters. However we typically loose control on the regularity of these functions near the diagonals as $r_1$ or $r_2$ or both approach $0$. The existence, in some appropriate distribution space over the full space of integration variables, of some amplitude defined in the extended space $(r_1,r_2)\in [0,+\infty)^2$ of parameters, including $0$, implies that $(\widehat{\xi}_r)_{0<r\leqslant1}$ is a Cauchy family in $L^q(\bbP)$, so it has a limit as $r>0$ goes to $0$.   

\ssk

 The (regularized) kernels typically involved in quantum field theory, and the (regularized) kernels involved in the construction of $\widehat\xi$, have a negative power behavior when (at least) two of their arguments come close. In the example of a regularized kernel $K_r$ with two arguments, this means that $K_r(z_1,s_2)\propto \vert r + z_1-z_2\vert^{-\gamma}$, for some exponent $\gamma>0$, or that $K_r(z_1,s_2) - c\vert r + z_1-z_2\vert^{-\gamma}$ has a continuous extension up to $r=0$, for some constant $c$. This scaling type behavior will play a pivotal role in the sequel. In this simple example, there is a cheap way of extending the distribution $\vert z_1-z_2\vert^{-\gamma}$ on $\{(z_1,z_2)\in \bbR^d\,;\,z_1\neq z_2\}$ into a distribution on $(\bbR^d)^2$. If $0<\gamma<d$ the function $K(z_1,z_2)=\vert z_1-z_2\vert^{-\gamma}$ is actually in $L^1_{\textrm{loc}}(\bbR^d\times\bbR^d)$ and its has a natural extension as a distribution defined on the full configuration space. If $d<\gamma$, we can follow Hadamard and use Taylor expansion on the diagonal to define its extension by the formula
\begin{align*}
K(\varphi) &= \int_{ \vert z_1-z_2\vert \leqslant 1  } K(z_1,z_2) \Big(\varphi(z_1,z_2) - \sum_{0\leqslant\vert k\vert<\lfloor\gamma-d\rfloor} \frac{1}{k!} \, (\partial^k_{z_2}\varphi)(z_1,z_1) (z_2-z_1)^k \Big)\,\rmd z_1\rmd z_2\\
&\quad+\int_{ \vert z_1-z_2\vert \geqslant 1  } K(z_1,z_2) \varphi(z_1,z_2) \,\rmd z_1\rmd z_2
\end{align*}
for any test function $\varphi$. We obtain this extension by adding to $K$ the distribution 
\[
-\sum_{0\leqslant\vert k\vert<\lfloor\gamma-d\rfloor} \frac{(-1)^k}{k!} \partial^k_{z_2}\delta_0(z_1-z_2)\,\int_{ \vert y-z_1\vert \leqslant 1 } (y-z_1)^k K(z_1,y) \rmd y\,,
\] 
that has support on the diagonal $\{z_1=z_2\}$. This is exactly the mechanics that will be involved in the general setting of Section \ref{SectionExtension}, in particular in Theorem \ref{ThmCanonicalExtension} and Theorem \ref{Thm:extensiontwosteps}. These statements deal with the general task of extending a distribution that is defined on the complement of a submanifold into a distribution defined on the whole space. This is the core of the extension problem involved in the study of Feynman graphs. The particular features involved in that case will be studied in Section \ref{SectionInductionGraphs}.

\ssk

We use a geometric language to develop the extension machinery. The inverse power behavior of a kernel will be encoded in the notion of scaling introduced in Section \ref{SectionExtension}, where we will introduce some spaces of distributions with some scaling behavior near some submanifold. In the example of some Feynman amplitude this submanifold would be the union of all the diagonals in the configuration space. The extension procedure will then be done in the spirit of the example above. 

It will be useful in our analysis to have a fine description of the distributions we work with. This will come under the form of a control of their scaling properties and their wavefront sets. We will recall in Section \ref{SubsectionFunctionSpacesScaling} what we need about wavefront sets. The relevance of this information to the extension problem for Feynman amplitudes should be clear from a well-known theorem of H\"ormander stating that the product of two distributions is a well defined and continuous map on a class of distributions whose wavefront sets satisfy some particular ``transversality condition'' recalled in Section \ref{SubsectionHormander}.

The configuration space of a Feynman amplitude, together with all its diagonals, has some particular feature compared to the general setting of a manifold and a submanifold. The union of all the diagonals comes with a stratification that will result in the following strategy for the extension problem. Denote by $\textbf{d}_i$ the open submanifold of configurations where exactly $i$ of the $n$ integration variables coincide, for $2\leqslant i\leqslant n$. The extension from the complement $U$ of all the diagonals to the whole space will be done inductively by first extending a distribution to $U\cup \textbf{d}_2$ then $U\cup \textbf{d}_2\cup\textbf{d}_3$, and so on. This type of construction will be the object of Section \ref{SectionInductionGraphs}.

We will eventually put these tools in action in Section \ref{SectionStochasticBounds} on the concrete example of the study of the convergence of $\widehat{\xi}_r$ to some limit enhanced noise $\widehat\xi$.

\ssk

Before embarking on this journey, we would like to illustrate the extension procedure on the examples of two Feynman amplitudes from the perturbative Euclidean $\Phi^4_3$ theory. They will involve some notions of scaling, wavefront set, conormal bundle that will be defined later. Their detailed understanding is not mandatory to understand the mechanics involved here. We invite the reader to come back to these examples after reading Section \ref{SectionExtension}.

\medskip

\subsection{{A motivating example}}
\label{ss:motivatingexample}

We will denote by $\mathbf{G}$ the massive Green function, in other words, $\mathbf{G}$ is the Schwartz kernel of the inverse massive Laplacian $(1-\Delta)^{-1}$. Concretely, $\mathbf{G}(x_1-x_2)= (2\pi)^{-3} \int_{\mathbb{R}^3}  e^{i\xi\cdot(x_1-x_2)}\left( \vert \xi\vert^2+1\right)^{-1}\rmd\xi $. {Denote by $\Phi$ a Gaussian free field on $\bbR^3$ with covariance $(1-\Delta)^{-1}$.} Consider a test function ${f}\in C^\infty_c(\mathbb{R}^3)$ which localizes on some compact region $D$ of $\mathbb{R}^3$ and consider the amplitudes
\begin{align*}
{}&\mathbb{E}\left[\Phi(x_1)\Phi(x_2) :\Phi^4:(f):\Phi^4:(f) \right]-C\mathbb{E}\left[\Phi(x_1)\Phi(x_2)\right] \mathbb{E}\left[  :\Phi^4:(f):\Phi^4:(f) \right]\\
&= C_1 \int_{D^2}\mathbf{G}(x_1,y_1)  \mathbf{G}^3(y_1,y_2) {f}(y_1){f}(y_2) \mathbf{G}(y_2,x_2)\rmd y_1 \rmd y_2 \,,   
\end{align*}
and 
\begin{align*}
{}&\mathbb{E}\left[ :\Phi^4:({f}):\Phi^4:({f}):\Phi^4:({f})   \right] \\
&= C_2 \int_{D^3}  \mathbf{G}^2(y_1,y_2) \mathbf{G}^2(y_2,y_3) \mathbf{G}^2(y_3,y_1) {f}(y_1){f}(y_2){f}(y_3) \rmd y_1\rmd y_2\rmd y_3\,,
\end{align*}
where $C,C_1,C_2$ are some combinatorial factors. The Feynman graph corresponding to the first amplitude is the so-called sunset graph.

\ssk

The first Feynman amplitude appears in the perturbative expansion of the two point function
$$
\mathbb{E}\left[ \Phi(x_1)\Phi(x_2) \exp\left(-c\int_{\mathbb{R}^3}  :\Phi^4(x):{f}(x)\rmd x \right) \right]
$$
in formal power series in $c$ whereas the second one appears in the perturbative expansion of the partition function
$$
\mathbb{E}\left[  \exp\left(-c\int_{\mathbb{R}^3}  :\Phi^4(x):{f}(x)\rmd x \right) \right]
$$
and is sometimes called a vacuum diagram. For each of these amplitudes, we expressed them also in terms of expectations of the Gaussian Free Field.

\ssk

Let us show how to make sense of the two amplitudes. The first idea is to measure singularities of some distribution by scaling. For any distribution $T$ on $\mathbb{R}^3\times \mathbb{R}^3$, we denote by $e^{-\ell\rho *} T:=T(x_1,e^{-\ell}(x_2-x_1)+x_1) $ for $\ell\in \mathbb{R}_{\geqslant 0}$ the scaled distribution $T$ along the diagonal. Our choice of notation $e^{-\ell\rho}$ will be justified later but just note that if $\rho$ denotes the scaling field $\rho:=(x_2-x_1)\cdot\partial_{x_2}$, then $e^{-\ell\rho}:\mathbb{R}^3\times \mathbb{R}^3\mapsto \mathbb{R}^3\times \mathbb{R}^3$ means the flow of $\rho$ which is a one parameter family of diffeomorphisms. The Green kernel $\mathbf{G}(x_1-x_2)$ is singular along the diagonal $\{x_1=x_2\}\subset \mathbb{R}^3_{x_1}\times \mathbb{R}^3_{x_2}$ and from the usual bound
$$ 
\vert \mathbf{G}(x_1-x_2)\vert \lesssim \min(1 , \vert x_1-x_2 \vert)^{-1}, 
$$ 
we immediately deduce that the family
$ 
\left(e^{-\ell }e^{-\ell\rho*}\mathbf{G}\right)_{\ell\geqslant 0}     
$ 
forms a bounded family of distributions in $\mathcal{S}^\prime(\mathbb{R}^3)$ whose wave front set is uniformly bounded in the conormal bundle  $N^*\left( \{x_1=x_2\}\right)$ of the diagonal $\{x_1=x_2\}$. The uniform bound on the wave front set just translates the fact that the family $ 
\left(e^{-\ell }e^{-\ell\rho*}\mathbf{G}\right)_{\ell\geqslant 0}     
$ of distributional kernels is \textbf{translation invariant} for all $\ell\geqslant 0$.
Now let us deal with the first amplitude, $\mathbf{G}^3(y_1,y_2)$ which is well-defined and  smooth outside the diagonal $\{y_1=y_2\}$, weakly homogeneous of degree $-3$ with respect to  $\rho$ where $\rho$ is the scaling field of $\{y_1=y_2\}$. To make sense of $\int \mathbf{G}(x_1,y_1)  \mathbf{G}^3(y_1,y_2)  \mathbf{G}(y_2,x_2)f(y_1)f(y_2)\rmd y_1 \rmd y_2   $, we need to extend the product $ \mathbf{G}^3$ which is defined on $\mathbb{R}^3\times \mathbb{R}^3\setminus \{y_1=y_2\}$ over the whole configuration space $\mathbb{R}^3\times\mathbb{R}^3$.
The key observation is that minus the weak degree of $\mathbf{G}^3$ is equal to $3$, which is also the codimension of the diagonal $\{y_1=y_2\}$. Therefore, by our general extension Theorem \ref{ThmCanonicalExtension}, we will have to renormalize in order to extend the 
product $\mathbf{G}^3$ as a distribution.
To define such an extension, we first mollify each Green function as $\mathbf{G}_\varepsilon$, then we prove that there exists a divergent counterterm $c_\varepsilon\in\bbR$ such that $\lim_{\varepsilon\downarrow0}\mathbf{G}^3_\varepsilon(y_1,y_2) - c_\varepsilon\,\delta_{\{0\}}^{\mathbb{R}^3}(y_1-y_2)$
extends as a distribution on $\mathbb{R}^3\times \mathbb{R}^3$. 
Once found the corresponding counterterm $c_\varepsilon$, we can give a probabilistic interpretation 
of the subtraction we just performed, defining the renormalised expectation as
\begin{align*}
&\mathbb{E}\left[\Phi(x_1)\Phi(x_2) :\Phi^4_\varepsilon:({f}):\Phi^4_\varepsilon:({f}) -c_\varepsilon  \Phi(x_1)\Phi(x_2)  :\Phi^2({f}^2): \right]-C\mathbb{E}\left[\Phi(x_1)\Phi(x_2)\right] \mathbb{E}\left[  :\Phi^4_\varepsilon:(f):\Phi^4_\varepsilon:(f) \right]
\\
&=C_1 \int \mathbf{G}(x_1,y_1)  \mathbf{G}^3_\varepsilon
(y_1,y_2) {f}(y_1){f}(y_2) \mathbf{G}(y_2,x_2)\rmd y_1 \rmd y_2 -C_1c_\varepsilon \int \mathbf{G}(x_1,y) {f}^2(y) \mathbf{G}(y,x_2)\rmd y   \\
&=C_1 \int\mathbf{G}(x_1,y_1)  \mathbf{G}^3_\varepsilon
(y_1,y_2) {f}(y_1){f}(y_2) \mathbf{G}(y_2,x_2)\rmd y_1 \rmd y_2  \\
&\quad-C_1c_\varepsilon\int\mathbf{G}(x_1,y_1)  \delta^{\mathbb{R}^3}_{\{0\}}(y_1-y_2) {f}(y_1){f}(y_2) \mathbf{G}(y_2,x_2)\rmd y_1 \rmd y_2\,.
\end{align*}
Therefore, the counterterm from the extension procedure can be identified with some mass counterterm in the Lagrangian, more precisely when $\varepsilon\downarrow0$ the following quantity $$
\mathbb{E}\left[\Phi(x_1)\Phi(x_2) \exp\left(-c\int_{\mathbb{R}^3}  :\Phi^4_\varepsilon(x):{f}(x)\rmd x + c^2\,c_\varepsilon\int_{\mathbb{R}^3}  :\Phi^2(x):{f}^2(x)\rmd x \right) \right] 
$$ 
has a well-defined limit.

\ssk

The treatment of the second amplitude is more subtle and the extension procedure is \textbf{inductive}, it relies on a two steps procedure. 
\begin{enumerate}
	\item We first initialize the induction 
by proving each $\mathbf{G}^2(y_i-y_j)$ is smooth on $\mathbb{R}^3\times \mathbb{R}^3\setminus \{y_i=y_j\} $. Therefore the triple product
$$ \mathbf{G}^2(y_1,y_2) \mathbf{G}^2(y_2,y_3) \mathbf{G}^2(y_3,y_1)  $$
is well-defined and $C^\infty$ on $\mathbb{R}^3\times \mathbb{R}^3\times \mathbb{R}^3\setminus \big( \{y_1=y_2\}\cup\{y_2=y_3\}\cup\{y_1=y_3\} \big)$, namely the region of configuration space where pairs of points are \textbf{not allowed to collide}.   \vspace{0.1cm}
	
	\item Second observation: $\mathbf{G}^2(y_i-y_j)$ is weakly homogeneous of degree $-2$ with respect to  the scaling $(y_i,y_j) \mapsto (y_i,e^{-t}(y_j-y_i)+y_i)$ colliding the two points $y_i,y_j$. The corresponding infinitesimal generator of the flow is the linear vector field which reads
$\rho=(y_j-y_i)\cdot\partial_{y_j}$.
Since minus the weak homogeneity of $\mathbf{G}^2$ $=2<\text{codimension}(\{y_i=y_j\})=3$, we know by Theorem~\ref{ThmCanonicalExtension} that $\mathbf{G}^2$ extends \textbf{canonically} as a $L^1_{loc}$ function hence as distribution on $\mathbb{R}^3\times \mathbb{R}^3$. So there is no need to subtract any counterterm. The wave front sets of each term $ \mathbf{G}^2(y_1,y_2), \mathbf{G}^2(y_2,y_3)$, and $ \mathbf{G}^2(y_3,y_1)  $ is now contained in the conormal bundles $N^*(\{y_1=y_2\}\subset (\mathbb{R}^3)^3)$, $N^*(\{y_2=y_3\}\subset (\mathbb{R}^3)^3)$, and $N^*(\{y_1=y_3\}\subset (\mathbb{R}^3)^3)$ respectively. By transversality of the wave front sets, it is thus immediate that the product of distributions $ \mathbf{G}^2(y_1-y_2) \mathbf{G}^2(y_2-y_3) \mathbf{G}^2(y_3-y_1)  $ is well-defined on $(\mathbb{R}^3)^3\setminus \{y_1=y_2=y_3\}$ where $ \{y_1=y_2=y_3\}$ is the deepest diagonal in $(\mathbb{R}^3)^3$. Intuitively, we are allowing pairs of points to collide but not the three points to collide.   \vspace{0.1cm}

	\item We come to the third step. Now consider the new scaling 
$$ (y_1,y_2,y_3)\mapsto (y_1,e^{-t}(y_2-y_1)+y_1,e^{-t}(y_3-y_1)+y_1) $$ which collapses everything on the deepest diagonal $ \{y_1=y_2=y_3\}$. With respect to this new scaling, the distributional product $  \mathbf{G}^2(y_1-y_2) \mathbf{G}^2(y_2-y_3) \mathbf{G}^2(y_3-y_1) $   is weakly homogeneous of degree $-6$ which happens to be equal to $-\text{codim}(\{y_1=y_2=y_3\})  $. The equality of minus the homogeneity degree and the codimension of the deep diagonal tells us we will need to subtract some singular counterterm to make sense of the distributional extension of the Feynman amplitude. Again, 
after mollifying the Green function $\mathbf{G}$, there exists a divergent counterterm $c_\varepsilon$ such that 
$$  \mathbf{G}^2_\varepsilon(y_1-y_2) \mathbf{G}^2_\varepsilon(y_2-y_3) \mathbf{G}^2_{\varepsilon}(y_3-y_1)-c_\varepsilon\delta_{0}^{\mathbb{R}^3}(y_1-y_2)\delta_{0}^{\mathbb{R}^3}(y_2-y_3) $$
has a limit in $\mathcal{S}^\prime((\mathbb{R}^{3})^3)$ when $\varepsilon\downarrow 0$ which extends the product $  \mathbf{G}^2(y_1-y_2) \mathbf{G}^2(y_2-y_3) \mathbf{G}^2(y_3-y_1) $ from  $\mathbb{R}^3\times \mathbb{R}^3\times \mathbb{R}^3\setminus \left( \{y_1=y_2\}\cup\{y_2=y_3\}\cup\{y_1=y_3\}\right)$ to the whole $(\mathbb{R}^3)^3$.   \vspace{0.1cm}
\end{enumerate}

As for the above example, this means that 
$ 
\mathbb{E}\big[ :\Phi^4_\varepsilon:({f}):\Phi^4_\varepsilon:({f}):\Phi^4_\varepsilon:({f})-c_{2,\varepsilon}\int_{\mathbb{R}^3} {f}^3 \big]  
$ 
has a well-defined limit when $\varepsilon\downarrow 0$ hence the action functional with counterterms reads
$$ 
 - c:\Phi^4_\varepsilon:({f}): + c^2 \, c_{1,\varepsilon} :\Phi^2:({f}^2)  + c^3 \, c_{2,\varepsilon} \int_{\mathbb{R}^3}{f}^3
$$
and, modulo powers of $c$ of degree higher than $4$, the limit as $\varepsilon\downarrow 0$ of all Feynman graphs exists.

\medskip

\subsection{The difficulties when working on curved spaces.}

We comment here on the difficulties we need to deal with when we work on curved spaces. The main problem is contrary to the torus case, the manifold setting lacks translation invariance. For example this prevents us from using stationarity of the white noise. This affects the Kolmogorov estimates and even the Kolmogorov criterion has a tedious statement. Global Fourier modes no longer exist and even in local charts the Fourier decomposition no longer commutes with the Laplace operator and the heat operator. On the other hand, using the Laplace eigenfunctions is not easily compatible with products of functions. Concretely when we expand the product of eigenfunctions $e_{\lambda_1}e_{\lambda_2}=\sum_{\lambda\in \sigma(\Delta)} C_{\lambda_1\lambda_2}^\lambda e_\lambda $ in terms of the basis $(e_\lambda)_{\lambda\in \sigma(\Delta)}$ of Laplace eigenfunctions, we have rather poor control on the coefficients $C_{\lambda_1\lambda_2}^\lambda$ when $\lambda\rightarrow +\infty$ on generic manifolds contrary to the torus case. Therefore we cannot treat the usual Feynman amplitudes in terms of simple lattice sums with constraints as in the work \cite{MWX} of  Mourrat, Weber \& Xu and instead, we are forced to work in position space. Microlocal analysis is therefore a perfectly adapted tool when we want to work in position space while keeping informations about frequencies.  However even in position space, since our diagrams contain the Schwartz kernel of the resonant product which is defined in local charts in terms of Littlewood-Paley blocks which uses Fourier techniques, this mixture of Fourier versus position space argument makes our arguments quite complicated (this difficulty does not appear in Hairer \& Singh's work because in the regularity structure approach, one does not need resonant products and therefore they do not appear in the stochastic estimates).  Another difficulty is the fact that functions having their Fourier support in some fixed ball or corona is not a diffeomorphism invariant concept and is not stable by cut-off. This explains why we have to be very careful in the second part and use several layers of cut-off together with several layers of commutators arguments. Finally, our techniques allowing to deal with variable coefficient metrics also allow us to handle non constant coupling functions.

\medskip

\section{Scaling fields, regularity and microlocal extension}
\label{SectionExtension}

We state in this section an extension result, Theorem \ref{ThmCanonicalExtension}, that provides some conditions under which a distribution defined outside a submanifold of an ambiant manifold can be extended to the whole manifold. The quantification of this extension result involves the notion of scaling field that is introduced in Section \ref{SubsectionScalingFields}. Such vector fields are also known as Euler vector fields. Some function spaces associated with a given scaling field are introduced in Section \ref{SubsectionFunctionSpacesScaling}, they allow to measure the singularity of distributions when we scale along certain submanifolds of some given ambient space. They generalise the weakly homogeneous distributions introduced by Meyer in~\cite{Meyer}. An extension Theorem~\ref{ThmCanonicalExtension} is proven in Section~\ref{SubsectionCanonicalExtension}, given a distribution whose blow up is moderate along a given submanifold, it allows to extend canonically the distribution to the whole ambient space in the spirit of the definition of the principal value. This statement is put to work in the particular setting of a configuration space in Section~\ref{SubsectionConfigurationSpace} to give a useful extension result for a class of Feynman amplitudes -- see Theorem~\ref{ThmBlackBoxFeynmanGraphs}. Moreover in Theorem~\ref{ThmBlackBoxFeynmanGraphs}, we will need to control the blow up of our Feynman amplitudes in two steps, first when all points collapse on the deepest diagonal, then when all points collapse to a single given point -- this difficulty comes from the absence of translation invariance. This requires a variant of the extension Theorem~\ref{ThmCanonicalExtension} stated in Theorem~\ref{Thm:extensiontwosteps}, in which we scale with respect to a given submanifold first, and then with respect to  a given point, in order to control the blow up of our distributions with respect to  both scalings.

\medskip

We work in Sections~\ref{SubsectionScalingFields} to \ref{SubsectionCanonicalExtension} in the setting of a smooth manifold $\mcX$ where a smooth submanifold $\mcY\subset\mcX$ is given.

\medskip

\subsection{Scaling fields}
\label{SubsectionScalingFields}

\begin{defn*}
Let $\mathfrak{I}_{\mcY}$ be the ideal of smooth real valued functions on $\mcX$ that vanish on $\mcY$. Set for $k\geqslant 1$:
$$
\mathfrak{I}^k_\mcY \defeq \Big\{f_1 \dots f_k  \,;\, (f_1,\dots,f_k)\in\mathfrak{I}_{\mcY}\times \dots \times \mathfrak{I}_{\mcY} \Big\}.
$$
A vector field $\rho$ defined on a neighbourhood of $\mcY$ is called an isotropic \textbf{\textsf{scaling field for $\mcY\subset\mcX$}} if for all $f\in\mathfrak{I}_{\mcY}$
$$
 \rho f-f\in\mathfrak{I}_{\mcY}^2.
$$
This type of vector field is also called an {\sl Euler vector field} in the literature.
\end{defn*}

\ssk

{
    Let us pause a moment to discuss the definition of the scaling fields. In fact they are generalisation of the classical notion of Euler vector fields as reviewed in our first example. 

\ssk
    
\begin{Example}-- 
    The first naive example is the origin $0\in \mathbb{R}^n$. 
    Scaling with respect to the origin is the one parameter group of diffeomorphism 
    $x\in \mathbb{R}^n \mapsto e^tx\in \mathbb{R}^n $ whose infinitesimal generator is the 
    vector field $\rho_{\mathrm{dil}}=\sum_{i=1}^n x^i\partial_{x^i}$.
    {The} homogeneous polynomials are eigenfunctions of the 
    differential operator $\rho_{\mathrm{dil}}$ with eigenvalue equal to the degree of the polynomial:
    \begin{align*}
       \rho_{\mathrm{dil}}\left((x^1)^{j_1}\dots (x^n)^{j_n} \right)=\left(\sum_{i=1}^n j_i\right)\left((x^1)^{j_1}\dots (x^n)^{j_n} \right). 
    \end{align*}
    But the differential operator $\rho_{\mathrm{dil}}$ can be seen to have the following property:
    the ideal $\mathfrak{I}_{\{0\}} $ is the closed ideal of smooth functions vanishing at the origin and is finitely generated as $C^\infty$ module by $(x^i)_{i=1}^n$, concretely it means that every $f\in \mathfrak{I}_{\{0\}}$ 
    reads $f=\sum_{i=1}^n x^ih_i$ where $h_i\in C^\infty$. So a direct calculation using
    $\rho_{\mathrm{dil}}(x^i) = x^i $ implies that $\rho_{\mathrm{dil}}f=\rho_{\mathrm{dil}} \left( \sum_{i=1}^n x^ih_i\right)=\left( \sum_{i=1}^n x^ih_i\right)+ \sum_{i,j=1}^n x^ix^j \left( \partial_{x^j} h_i\right)$ where the second sum is seen to vanish at order $2$ at $0\in \mathbb{R}^n$ therefore it belongs to the ideal squared $\mathfrak{I}_{\{0\}}^2$. So for every $f\in \mathfrak{I}_{\{0\}}$, we have the identity 
    $ \rho_{\mathrm{dil}}f-f\in \mathfrak{I}_{\{0\}}^2$. In fact the curious reader can generalize the above fact by checking that for every {$d\in \bbN$ and} $f\in \mathfrak{I}_{\{0\}}^d$ we have $ \rho_{\mathrm{dil}}f-df\in \mathfrak{I}_{\{0\}}^{d+1}$ which means that the differential operator $\rho_{\mathrm{dil}}-d$ kills the {part of $f\in \mathfrak{I}_{\{0\}}^d$ that is} homogeneous of degree $d$.
\end{Example}

\ssk

Our next example of scaling field is especially relevant for quantum field theory applications since it involves scaling on diagonals of configuration space. 

\ssk

\begin{Example}-- 
The second paradigmatic example is that of a diagonal of a vector space. For concreteness, take $\mathcal X=\mathbb{R}^2$ and $\mathcal Y=\{(x,x):x\in\mathbb{R}\}$. Scaling with respect to $\mathcal{Y}$ amounts to sending $(x,y)$ onto $(x,x+\lambda(y-x))$, which is generated by $\rho_{\mathrm{dil}}:=(y-x)\partial_y$. On the other hand, the ideal $\mathfrak{I}_{\mcY}$ is generated by monomials $(y-x)$, from which we read that for every $f\in\mathfrak{I}_{\{0\}}$, $f$ has the form $f(x,y)=(y-x)h(x,y) $ for $h\in C^\infty(\mathbb{R}^2)$ {the function} $\rho_{\mathrm{dil}}f-f$ {vanishes} at order $2$ along the diagonal by a calculation similar as in the previous example. Therefore, scaling fields are chosen to be small perturbations of $\rho_{\mathrm{dil}}$, in the sense that $\rho$ is a scaling field provided $\rho-\rho_{\mathrm{dil}}:\mathfrak{I}_{\mcY}\rightarrow\mathfrak{I}_{\mcY}^2$.    
\end{Example}

\ssk

We next give another example of scaling field with parabolic homogeneity.  

\ssk

\begin{Example} -- 
We provide another elementary example, which plays an important role in the sequel. We are interested in space-time kernels $K(t,x,y)$, and we want to probe {their} scaling properties (typically their blow-up) on the space diagonal $\{x=y\}$ and on the null-time hyperplane $\{t=0\}$. In the flat case $(t,x,y)\in \bbR^3$ and with a parabolic scaling giving a weight two to the first component, the scaling on  $\{x=y\}\cap\{t=0\}$ is thus generated by
$$
\rho_{\mathrm{dil}} = (y-x)\partial_y+2t\partial_t\,.
$$
This vector field over $\bbR^3$ will be our model scaling field in a parabolic setting. In the manifold setting, scaling fields can thus be thought of as small perturbations of this reference case.
\end{Example}
}
\ssk

\ssk
 Denote by $n$ the dimension of $\mcX$ and by $d$ the dimension of $\mcY$. If $\rho$ is a scaling field for $\mcY\subset\mcX$ there exists a neighbourhood of $\mcY$ that is stable by the backward semiflow $(e^{-s\rho})_{s\geqslant 0}$ of $\rho$ and every point $y\in\mcY$ has a neighbourhood $U_y$ in $\mcX$ on which some coordinates 
$$
h=(h_1,\dots, h_n) : U \rightarrow\bbR^n
$$
are defined and such that 
$$
U_y \cap \mcY = h^{-1}(\bbR^d)
$$
with $\bbR^d\subset\bbR^n$, and 
$$
\rho = \sum_{i=d+1}^n h^i\partial_{h^i}.
$$
A proof of existence of a stable neighbourhood can be found in Lemma 2.4 of \cite{DangWrochna} and the normal form theorem can be found in Proposition 2.5 of \cite{DangWrochna} -- see also Lemma 2.1 in \cite{Meinrenkensurvey}. 

The example of the configuration space of $\ell$ points in $\bbR^k$ will be particularly relevant for us. The scaling field $\rho$ whose flow reads
$$
e^{-t\rho}(x_1,\dots,x_\ell) = \Big(x_1,e^{-t}(x_2-x_1)+x_1, \dots, e^{-t}(x_\ell-x_1)+x_1\Big)
$$
will move all points towards the deepest diagonal $\big\{(x_1,\dots,x_1)\,;\,x_1\in \bbR^k\big\}$ and its dynamics is tangent to all the larger diagonals, that is to say those submanifolds indexed by a non-empty strict subset $I$ of $[\ell]$ and defined by $\big\{(x_1,\dots,x_\ell):x_i   =x_{\min I} \,\forall i\in I\big\}$.  In the sequel we will only work with product (sub)manifolds of the form
\begin{equation} \label{EqModelParabolicSituation} \begin{split}
\mcX &= \bbR^p\times X,   \\
\mcY &= \big(\{{ 0_{p-q}}\}\times\bbR^q\big)\times Y
\end{split} \end{equation}
with $(\{{ 0_{p-q}}\}\times\bbR^q)\subset\bbR^p$ and $Y\subset X$ {two smooth manifolds}, and some non-isotropic scaling fields, of the form
\begin{equation}\label{eq:scalingfieldsparabolic}
\rho = \sum_{j=1}^{p-q} 2t_j\partial_{t_j} + \rho_Y
\end{equation}
for the canonical coordinates $(t_j)_{1\leqslant j\leqslant p}$ on $\bbR^p$ and a scaling field $\rho_Y$ for $Y\subset X$. This is an example of a \textsf{\textbf{weighted vector field on a weighted manifold}}. 

We give a formal definition in the case where our submanifold $\mcY$ is the transverse intersection $\mcY_1\cap \mcY_2$ where $\mcY_1 = (\{ 0_{p-q}\}\times\bbR^q)\times X$ and $\mcY_2=\left(\mathbb{R}^p\times Y\right)$. Then one has a description of the ideal $\mathfrak{I}_{\mathcal{Y}}$ as the product $\mathfrak{I}_{\mcY_1}\mathfrak{I}_{\mcY_2} $ so the ideal $\mathfrak{I}_{\mcY}$ has a bifiltration
$$
\mathfrak{I}_{\mcY_1}^m \mathfrak{I}_{\mcY_2}^n \subset \dots \subset \mathfrak{I}_{\mcY_1} \mathfrak{I}_{\mcY_2} = \mathfrak{I}_{\mcY}. 
$$

\ssk

\begin{Example}[Linear model case]\label{Examplebifiltration}
 In the model case where $Y=\bbR^{d_1}$ and $X=\bbR^{d_1+d_2}$, with $x=(y,z)\in \bbR^{d_1}\times \bbR^{d_2}$, and we use the linear coordinates $(s,t)\in \bbR^{p-q}\times\bbR^q$ on the factor $\bbR^p$ in $\mcX$. The submanifold $\mcY$ (here it is a vector subspace) has equation $\mcY=\{s=0, z=0\}$. Thus a function in $\mathfrak{I}_{\mathcal{Y}}$ has the form $f(x) =\sum_{i=1}^{p-q} s_i g_i(s,t,y,z)+\sum_{j=1}^{d_2} z_jh_j(s,t,y,z) $ for some smooth functions  $g_i,h_j\in C^\infty(\mathbb{R}^p\times \mathbb{R}^{d_1+d_2} )$, and an element of $\mathfrak{I}_{\mcY_1}^m \mathfrak{I}_{\mcY_2}^n$ has the form $f(x) =\sum_{I,J, \vert I\vert=m, \vert J\vert=n} s^I  z^Jh_{I,J}(s,t,y,z)$ where $I,J$ are multi--indices and $h_{I,J}\in C^\infty(\mathbb{R}^p\times \mathbb{R}^{d_1+d_2} )$.  
\end{Example}

\ssk

Now assume we want to put a weight $2$ to powers of $\mathfrak{I}_{\mcY_1}$ and weight $1$ to powers of $\mathfrak{I}_{\mcY_2}$. We want to give an intrinsic characterization for vector fields of the form given by Equation \eqref{eq:scalingfieldsparabolic}.

\ssk

\begin{defn*}
A vector field $\rho$ defined on a neighbourhood of $\mcY$ is called a parabolic \textbf{\textsf{scaling field for $\mcY\subset\mcX$}} if for all $(m,n)$ and for all $f\in \mathfrak{I}_{\mcY_1}^m \mathfrak{I}_{\mcY_2}^n $ one has
\begin{eqnarray*}
\rho f-(2m+n)f\in \Big(\mathfrak{I}_{Y_1}^{m+1} \mathfrak{I}_{Y_2}^n + \mathfrak{I}_{Y_1}^m \mathfrak{I}_{Y_2}^{n+1}\Big).
\end{eqnarray*}
\end{defn*}

\ssk

\begin{Example}
In the linear model case from Example \ref{Examplebifiltration}, and keeping the same notations, a natural choice for $\rho$ would be
\begin{align*}
\rho:=2\sum_{i=1}^{p-q} s_i\partial_{s_i}+\sum_{j=1}^{d_2} z_j\partial_{z_j}.    
\end{align*}
We can check that for any $f\in \mathfrak{I}_{\mcY_1}^m \mathfrak{I}_{\mcY_2}^n$, the function $\rho f$ is equal to

\begin{align*}
\rho \bigg(\sum_{I,J, \vert I\vert=m, \vert J\vert=n} &s^I  z^Jh_{I,J}(s,t,y,z) \bigg)   \\
&=\sum_{I,J, \vert I\vert=m, \vert J\vert=n} (\rho s^I)  z^Jh_{I,J}(s,t,y,z)+s^I (\rho z^J)h_{I,J}(s,t,y,z)+\underbrace{s^Iz^J(\rho h_{I,J})(s,t,y,z)}   \\
&=\sum_{I,J, \vert I\vert=m, \vert J\vert=n} (2\vert I\vert+\vert J\vert)  s^I  z^Jh_{I,J}(s,t,y,z)+\underbrace{s^Iz^J(\rho h_{I,J})(s,t,y,z)}  \\
&= (2m+n)f \; \text{mod} \Big(\mathfrak{I}_{Y_1}^{m+1} \mathfrak{I}_{Y_2}^n + \mathfrak{I}_{Y_1}^m \mathfrak{I}_{Y_2}^{n+1}\Big) 
\end{align*}
where the term underbraced belongs to the ideal $\Big(\mathfrak{I}_{Y_1}^{m+1} \mathfrak{I}_{Y_2}^n + \mathfrak{I}_{Y_1}^m \mathfrak{I}_{Y_2}^{n+1}\Big)$.
\end{Example}

In the sequel we will simply call `scaling fields' some parabolic scaling fields as we will only work with such fields.
The \textbf{\textsf{weighted co-dimension}} of $\mcY$ is defined here as
$$
\textrm{codim}_w(\mcY\subset\mcX) \defeq 2(p-q) + \textrm{dim}(X) - \textrm{dim}(Y).
$$

\medskip

\subsection{Function spaces associated with scaling fields}
\label{SubsectionFunctionSpacesScaling}
{
We recall the notion of wave front set of distributions on $\mcX$.
}

\subsubsection{{Recollection on the notion of wave front set.}}

{
{The $\text{singular support}$  of a distribution $T\in \mathcal{D}^\prime(\mcX)$ is the closed subset of $\mcX$ where $T$ is not locally smooth.} The wave front set 
of $T$ denoted by $WF(T)$ generalizes the notion of singular support of the distribution $T$; it measures singularities of the distribution $T$ in the \emph{phase space} (or cotangent space) instead of position space {only}. More precisely, the wave front set of $T$ will be a certain subset of the cotangent space $T^*\mcX$ of $\mcX$ which tells us the location and codirection of singularities.
We also let $U\subset \mcX$ be an open set and $\Gamma$ a closed {\it conic set} in $T^*U\setminus{\bf 0}$, where 
\[
{{\bf 0}\defeq \big\{(x;0)\in T^*U; x\in U\big\} \subset T^*U}
\] 
denotes the \emph{zero section} of $T^*U$. By conic subset of $T^*U\setminus{\bf 0}$, we mean that $(x,\xi)\in\Gamma$ implies that $(x,\lambda \xi)\in\Gamma$ for every $\lambda>0$. We start by recalling the definition of the wave front set for distributions on $\mathbb{R}^n$. The manifold case follows immediately using charts and using the fact that the wave front set transforms naturally under local diffeomorphisms.

\ssk

\begin{defn}\label{def:wavefrontsetRn}
Let $U\subset \mathbb{R}^n$ be an open set. For every distribution $T\in \mcD^\prime(U)$ we define the wave front set $WF(T)$ to be a closed conic set $\Gamma\subset T^*U\setminus {\bf 0}$
defined as the \textbf{complement of the set} of $(x_0;\xi_0)\in T^*\mathbb{R}^n, \xi_0 \neq 0$ such that 
there exists ${\varphi}\in C^\infty_c(U)$, ${\varphi}(x_0)\neq 0$,  a closed conic neighbourhood $V\subset \mathbb{R}^{n*}$ of $\xi_0$ 
such that 
\begin{align*}
\forall N\in \mathbb{N}, \exists C_{N,V}>0, \forall \xi\in V, \quad  \big\vert \widehat{T{\varphi}}(\xi) \big\vert \leqslant C_{N,V} \big(1+\vert \xi\vert\big)^{-N},
\end{align*}
where $ \widehat{T{\varphi}}$ is the Fourier transform of the compactly supported distribution $T{\varphi}\in \mathcal{D}^\prime(U)$.
\end{defn}

\ssk

To extend the above notion of wave front set to the manifold case,
we must localize using smooth test functions and local charts. Given a distribution $T\in \mathcal{D}^\prime(U)$ for $U\subset \mcX$ an open subset of some smooth manifold $\mcX$, 
then $(x_0;\xi_0)\notin WF(T)$ if there exists some  chart $\kappa:U\mapsto \mathbb{R}^n$, some test function ${\varphi}\in C^\infty_c(U)$, ${\varphi}(x_0)\neq 0$ and some closed conic neighbourhood $V\subset \mathbb{R}^{n*}$ of $ \left( ^t \rmd\kappa_{x_0} \right)^{-1}\left( \xi_0 \right)$ 
such that  
\begin{align*}
\forall N\in \mathbb{N}, \exists C_{N,V}>0, \forall \eta\in V, \quad \big\vert \widehat{\kappa_*\left( T{\varphi} \right)}(\eta) \big\vert \leqslant C_{N,V} \big(1+\vert \eta\vert\big)^{-N}.
\end{align*}
We do not prove it here, but it is a classical fact from microlocal analysis that the notion of wave front is intrinsic and does not depend on the various choices we made above. Now the goal of the next definition is to endow the space of distributions whose wave front set lies in a given closed conic set $\Gamma$ of a topology. The definition takes the test functions $\varphi$ and test cones $V$ from Definition \ref{def:wavefrontsetRn} to define {some} continuous seminorms for this topology.  
}

\begin{defn}
{For a closed conic set $\Gamma\subset T^*U$} we denote by $\mcD'_\Gamma(U)$ the space of distributions on $U$ whose wave front set is contained in $\Gamma$.
This is a locally convex topological vector space endowed with a natural topology called \emph{normal} topology. The seminorms defining this topology are
\begin{itemize}
    \item given a chart $\kappa:\Omega\subset U\mapsto \mathbb{R}^{\dim(\mcX)}$, an integer $N\in\bbN$, ${\varphi}\in C^\infty_c(\kappa(\Omega))$, and a cone $V\subset \mathbb{R}^{n*}$ such that
$$
\text{supp}({\varphi})\times V\cap \kappa_{*}\Gamma=\emptyset, \qquad \text{ where } \qquad  \kappa_{*}\Gamma = \Big\{ \big(\kappa(x) \,;\, (^t\rmd\kappa)_x^{-1}(\xi) \big); (x;\xi)\in \Gamma\Big\},
$$
we have the norm
\begin{eqnarray*}
\Vert \Lambda\Vert_{N,V,{\varphi},\kappa} = \sup_{\xi\in V} \big(1+\vert \xi\vert\big)^N \big\vert\widehat{ (\kappa_*\Lambda){\varphi}}(\xi)\big\vert;
\end{eqnarray*}
\item the seminorms of the strong topology of distributions
\begin{eqnarray*}
\sup_{{\varphi}\in \mcB}\vert\langle \Lambda,{\varphi}\rangle\vert,
\end{eqnarray*}
where $\mcB$ is a bounded set of $C^\infty_c(\mcX)$ which means that there is some compact $K$ such that $\text{supp}(\varphi)\subset K$ for all $\varphi\in\mcB$, and for any differential operator $Q$ one has $\sup_{\varphi\in B}\Vert Q{\varphi}\Vert_{L^\infty(K)}<\infty$.
\end{itemize}
To be bounded in $\mathcal{D}^\prime_\Gamma(U)$ will always mean 
that all the above seminorms are bounded.  
\end{defn}

We refer to \cite[p.~823]{DangAHP} and \cite{BDH16} for results about why this topology is well-behaved with respect to some natural operations on distributions.

Many wave front sets of interest for us are contained in {some} conormal bundles that we need to recall.

\ssk

\begin{defn}[Conormal bundles]
Let $Y\subset X$ be a closed submanifold in some smooth manifold $X$. We will denote by $N^*\left(Y\subset X \right)$ the \textbf{conormal bundle of $Y$} which is defined
as 
$$
\big\{ (y;\eta) \in T^*X ; y\in Y, \eta(v)=0, \forall v\in T_yY \big\}.
$$
\end{defn}

\ssk

Let us give several examples of distributions and the corresponding wave front sets. In all these examples, the wave front sets will be {some} conormal bundles. We start by looking at indicator functions.

\ssk

\begin{Example}--
The indicator function $\mathbf 1_D\in L^2(\mathbb{R}^2)$ of some smooth planar domain $D\subset \mathbb{R}^2$ has wave front set contained in the conormal bundle $N^*(\partial D)$ of the boundary of the domain $D$ (in fact the conormal bundle minus the zero section $\mathbf{0}$).    
\end{Example}

\ssk

The next example of distributions together with their wave front sets is given by $\delta$--like distributions: 

\ssk

\begin{Example}--
The delta distribution $\delta_{\{0\}}^{\mathbb{R}^n} \in \mathcal{D}^\prime(\mathbb{R}^n)$ has wave front set equal to the {punctured} conormal bundle $N^*(\{0\}\subset \mathbb{R}^n) \setminus \mathbf{0} $ which happens to coincide with the cotangent fibre $T_0^*\mathbb{R}^n$ over the origin $0\in \mathbb{R}^n$.  
Now if we consider the distribution $\delta_{\{0\}}^{\mathbb{R}^n}(x_1-x_2) \in \mathcal{D}^\prime(\mathbb{R}^n\times \mathbb{R}^n)$ which is nothing but the Schwartz kernel of the identity map, then either by some direct calculation relying on translation invariance or by some application of the pull--back Theorem, its wave front set will be contained in
the conormal $N^*\left(\{x_1=x_2\}\subset (\mathbb{R}^n)^2 \right)$ which is the conormal bundle of the diagonal.
\end{Example}

\ssk

The last example is also fundamental for both quantum field theory and stochastic PDEs. It describes the wave front set of {the Green function} for the Laplacian and also wave fronts of heat propagators.

\ssk

\begin{Example}--
Let {$(\mathcal{X},g)$} be a closed compact Riemannian manifold {with metric tensor $g$, and let $\Delta_g$ stand for} the Laplace Beltrami operator on $\mathcal{X}$. Then the Schwartz kernel $\mathbf{G}\in \mathcal{D}^\prime(\mathcal{X}\times \mathcal{X})$ of $(1-\Delta_g)^{-1}$, sometimes called Green function, is a distribution on $\mathcal{X}\times \mathcal{X}$ whose wave front set lies in the conormal bundle $N^*(d_2\subset \mathcal{X}\times \mathcal{X})$ where $d_2=\{(x,x)\in \mathcal{X}\times \mathcal{X} \}$ is the diagonal of $\mathcal{X}\times \mathcal{X}$. For the heat kernel $K_{\vert t-s\vert}(x,y)1_{\mathbb{R}_{\geqslant 0}}(t-s)$ viewed as a distribution in $\mathbb{R}\times \mathbb{R}\times \mathcal{X}\times \mathcal{X}$, the wave front set lies in $N^*\left( d_2\subset (\mathbb{R}\times \mathcal{X})^2 \right)$ where $d_2$ is now the space--time diagonal.
\end{Example}

\ssk

\begin{lem} \label{LemLinearFlow}
The family of distribution
$$
\delta\big(z'-e^{-\ell\rho}z\big) \qquad (1\leqslant \ell\leqslant +\infty)
$$
on $\bbR^k\times\bbR^k$ is bounded in $\mcD'_{\Gamma_\rho}(\bbR^k\times\bbR^k)$, where
\begin{equation*} 
\Gamma_\rho = \bigcup_{1\leqslant \ell\leqslant +\infty} \Big\{ \big((z,e^{-\ell\rho}z),(\lambda,e^{\ell\rho}\lambda)\big) \,;\, (z,\lambda)\in T^*\bbR^k \Big\} \subset T^*(\bbR^k\times\bbR^k).
\end{equation*}
\end{lem}

\medskip

This estimate can also be used to give an upper bound on the wave front set of the resolvent $\left(\rho+z\right)^{-1}$ which implies the radial type estimates for $\rho$. This is very similar in spirit to the radial estimates from the works of Melrose \cite{Mel94}, Vasy \cite{Vasy} or Dyatlov \& Zworski \cite{DZ16}.

\medskip

\begin{Dem}
Note that the distributions $\delta(z'-e^{-\ell\rho}z)\in \mathcal{D}^\prime(\mathbb{R}^k\times \mathbb{R}^k)$ are nothing but the Schwartz kernels of the transfer operator $\varphi\in C^\infty(\mathbb{R}^k)\mapsto e^{-\ell\rho*}\varphi\in C^\infty(\mathbb{R}^k) $, so we will use the identification $[e^{-\ell\rho*}]=\delta(z'-e^{-\ell\rho}z)$. The fact that this family of distributions is bounded (weak boundedness implies strong boundedness by uniform boundedness) automatically follows from the continuity of the pull-back of a distribution by a smooth family of diffeomorphisms and the strong convergence of $\delta(z'-e^{-\ell\rho}z)$ to $\delta(z' - (x,0,0))$ when $\ell$ goes to infinity, for $z=(x,y,t)$. 
 
Fix an arbitrary compact subset $K\subset \mathbb{R}^{d_1+d_2+d_3}$ that is stable by the scaling maps
$$
(x,y,{\sf t})\mapsto \big(x,e^{-\ell}y,e^{-2\ell}{\sf t}\big) \quad (\ell\geqslant 0)\,.
$$ 
Then we shall restrict the Schwartz kernel $[e^{-\ell\rho}]$ to $K\times K$. It means we estimate this wave front set near the diagonal but for arbitrary large $\ell$. Choose some test functions ${g}_1,{g}_2$ in $C^\infty_K(\mathbb{R}^{d_1+d_2+d_3})$, supported in $K$. In local coordinates we have

\makebox[\textwidth][c]{
\begin{minipage}{\dimexpr\textwidth+10cm}
\begin{align*}
{}&\int_{\mathbb{R}^{d_1+d_2+d_3}} e^{i\xi_2\cdot x+i\eta_2\cdot y+i\tau_2\cdot t} {g}_2(x,y,t) e^{-t\rho*}({g}_1 e^{i\xi_1\cdot x+i\eta_1\cdot y+i\tau_1\cdot {\sf t}}) \, \rmd x\rmd yd{\sf t}  \\
&= e^{-\ell(d_2+2)}\int_{\mathbb{R}^{d_1+d_2+d_3}} e^{i\xi_2\cdot x+i\eta_2\cdot y+i\tau_2 t} {g}_2(x,y,t) \big({g}_1 (x,e^{-\ell}y,e^{-2\ell}t)e^{i\xi_1\cdot x+ie^{-\ell}\eta_1\cdot y+e^{-2\ell}\tau_1\cdot t}\big) \, \rmd x\rmd y\rmd{\sf t}   \\
&= \widehat{{g}_\ell}\left(\xi_1+\xi_2,\eta_2+e^{-\ell}\eta_1,\tau_2+e^{-2\ell}\tau_1 \right)
\end{align*}
\end{minipage}
}
\esp
\noindent
where 
$$
{g}_\ell(x,y,t) \defeq {g}_2(x,y,t) {g}_1\big(x, e^{-\ell}y, e^{-2\ell} t\big)
$$ 
is a bounded family of smooth compactly supported functions (this is crucial) when $\ell\in [0,+\infty)$. We then have for any $N\geqslant 1$ the upper bound
\begin{eqnarray}
\big\vert \widehat{{g}_\ell}(\xi,\eta,\tau) \big\vert\leqslant C_N \big(1+\vert\xi\vert+\vert\eta\vert+\vert\tau\big\vert\big)^{-N}
\end{eqnarray}
where the constant $C_N$ does not depend on $\ell\in [0,+\infty)$. Hence in any closed conic set $V$ which does not meet the subset
$$
\Lambda = \Big\{ \Big(\xi,-\xi, \eta, -e^{-\ell}\eta, \tau, -e^{-2\ell}\tau\Big)\in (\bbR^k\times\bbR^k)^*, \ell\geqslant 1 \Big\}, 
$$
there exists some $\varepsilon>0$ such that for all $\big(\xi_1,\xi_2,\eta_1,\eta_2,\tau_1,\tau_2\big)\in V\subset (\bbR^k\times\bbR^k)^*$ we have for all $\ell\geqslant 1$ the inequality 
$$
\Big\vert \big(\xi_1+\xi_2,e^{-s}\eta_1+\eta_2,e^{-2s}\tau_1+\tau_2\big) \Big\vert \geqslant \varepsilon \Big( \vert\xi_1\vert+\vert\xi_2\vert+\vert\eta_1\vert+\vert\eta_2\vert+\vert\tau_1\vert+\vert\tau_2\vert \Big).
$$ 
This implies the following Fourier bound
\begin{equation*} \begin{split}
\Big\vert \int_{\mathbb{R}^k} e^{i\xi_2\cdot x+i\eta_2\cdot y+i\tau_2 \cdot t} {g}_2(x,y,t) & e^{-\ell\rho*}({g}_1 e^{i\xi_1\cdot x+i\eta_1\cdot y+i\tau_1\cdot t}) \, \rmd x\rmd y \rmd t \Big\vert   \\
&\leqslant C_N \Big(1+\vert\xi_1+\xi_2\vert+\vert\eta_2+e^{-s}\eta_1\vert+\vert\tau_2+e^{-2s}\tau_1\vert \Big)^{-N}   \\
&\leqslant C_N \varepsilon^{-N}\Big(1+\vert\xi_1\vert+\vert\xi_2\vert+\vert\eta_1\vert+\vert\eta_2\vert+\vert\tau_1\vert+\vert\tau_2\vert \Big)^{-N}
\end{split} \end{equation*}
for all $\ell\geqslant 1$ and $(\xi_1,\xi_2,\eta_1,\eta_2,\tau_1,\tau_2)\in V\subset (\bbR^k\times\bbR^k)^*$. The previous bound analyses the wave front set of the family 
$\delta(z'-  e^{-\ell\rho}(z))$ near $T^*\left(K\times K\right)\subset T^*(\bbR^k\times\bbR^k)$. Since $K$ is arbitrary the family of distributions $\delta(z'-  e^{-\ell\rho}(z))$ is bounded in $\mathcal{D}^\prime_{\Gamma_\rho}(\bbR^k\times\bbR^k)$, with $\Gamma_\rho \subset T^*(\bbR^k\times\bbR^k)$ given by   
\begin{equation*} \begin{split}
\Gamma_\rho = \; &\underset{\textrm{the radial set which is the conormal of the singular set of }\rho}{ \underbrace{ \Big\{ \big(x,x, 0,0,0,0; \xi,-\xi,0,\eta_2,0,\tau_2 \big) ;  (\xi,\eta_2,\tau_2)\neq (0,0,0) \Big\} }}   \\
&\cup \Big\{ \Big(x,y,t, x,e^{-\ell}y,e^{-2\ell} t;\xi,\eta,\tau,-\xi,-e^\ell\eta,-e^{2\ell}\tau \Big); \ell\geqslant 1 , (\xi,\eta,\tau)\neq (0,0,0) \Big\}.
\end{split} \end{equation*}
This concludes the proof.
\end{Dem}


Set 
\begin{equation} \label{EqSmallPi}
\pi(z)\defeq (x,0,0).
\end{equation}
Lemma \ref{LemLinearFlow} is useful to give a description of the Taylor subtraction operation. The Taylor subtractors of order $0$ and $1$ read
\begin{equation*} \begin{split}
&R_0: \varphi \mapsto \varphi - \varphi\circ\pi,   \\
&R_1: \varphi\mapsto \varphi-\varphi(x,0,0) - y\cdot\partial_y\varphi(x,0,0)-t \partial_t\varphi(x,0,0),
\end{split} \end{equation*} 
We call these operators $R_0$ and $R_1$, with the letter $R$ chosen for `{\sl remainder}'. Denote generically by $[\Lambda]$ the Schwartz kernel of an operator $\Lambda$.

\ssk

\begin{prop} \label{PropTaylorRemainder}
The operators $R_0,R_1$ have Schwartz kernel
\begin{equation*} \begin{split}
[R_0] &= \int_0^\infty \big[\rho e^{-a\rho*}\big] \rmd a,   \\
[R_1] &= \int_0^\infty \Big[\big(1-e^{a}+e^{a}\rho\big)\rho e^{-a\rho*}\Big] \rmd a - [\rho]
\end{split} \end{equation*} 
and 
\begin{equation*} \begin{split}
[e^{-\ell\rho*}R_0] &= \int_0^\infty \big[\rho e^{-(\ell+a)\rho*}\big]\,\rmd a,   \\ 
[e^{-\ell\rho*}R_1] &= \int_0^\infty \big[(1-e^{a}+e^{a}\rho)\rho e^{-(\ell+a)\rho*}\big]\rmd a - [e^{-\ell\rho*}\rho].
\end{split} \end{equation*} 
and the families of distributions $\big(\big[(e^{-\ell\rho})^*R_0\big]\big)_{0\leqslant \ell\leqslant +\infty}$ and  $\big(\big[(e^{-\ell\rho})^*R_1\big]\big)_{0\leqslant \ell\leqslant +\infty}$ are bounded in $\mcD'_{\Gamma_\rho}(\bbR^k\times\bbR^k)$.
\end{prop}

\ssk

\begin{Dem}
We write a detailed proof for $R_0$; the proof for $R_1$ is very similar and left to the reader. For a test function ${o}$ with compact support on $\bbR^k\times\bbR^k$ write $\rho_1{o}$ for the action of the vector field on the first component of ${o}$. We have
$$
\big\langle \big[\rho(e^{-\ell\rho})^*\big] \,,\, {o} \big\rangle = -\int_{\bbR^k} (\rho_1{o})\big(z,e^{-\ell\rho}(z)\big)\,\rmd z
$$
and since $(\rho_1{o})(z,e^{-\ell\rho}z)$ vanishes along the singular set $\{y=0, t=0\}$ the integrand is of order $e^{-\ell}$, so the integral is converging. The wavefront bound follows from the wave front bound on the propagator $[(e^{-\ell\rho})^*]$ and the fact that the wave front of a distribution is stable under the action the vector field $\rho$.
\end{Dem}

\ssk

\subsubsection{Distributions with some regularity and scaling properties.}

We assume from now on that $\mcX$ has a Riemannian structure and we denote by $K_r(x,y)$ its heat kernel ($r>0$). We come back to the general setting of an open subset $U\subset \mcX$ and assume we are given a closed conic set $\Gamma$ in $T^*U\backslash\{0\}$. It is a classical fact that for $\alpha<0$ the Besov space $C^\alpha(\mcX) = B^\alpha_{\infty,\infty}(\mcX)$ can be characterized as the set of distributions $\Lambda$ on $\mcX$ such that 
$$
\underset{x\in\mcX}{\sup}\,\underset{0<r\leqslant 1}{\sup}\, r^{-\frac{\alpha}{2}} \big\vert \langle \Lambda , K_r(x,\cdot)\rangle \big\vert < \infty.
$$
{(See for instance Theorem 2.34 of \cite{BCD}.)} A distribution $\Lambda\in\mcD'(U)$ is an element of $\mcD'_\Gamma(U)$ iff for all pseudodifferential operators $Q$
with 
Schwartz kernel compactly supported in $U\times U$ and which is microlocally smoothing near $\Gamma$ (this means that for any $(x_0;\xi_0)\in \Gamma$, there exists a conic neighborhood $U\times V$ of $(x_0;\xi_0)$ s.t. the symbol $a\in S(T^*\mcX)$ of the operator $Q$ satisfies the estimate $ \vert a(x;\xi)\vert \leqslant C_N \left(1+\vert \xi\vert\right)^{-N}, \forall N$ for all $(x;\xi)\in U\times V$), one has for every compact subset $C$ of $\mcX$, 
\begin{equation} \label{EqGammaWaveFrontSet}
\underset{x\in C}{\sup}\,\underset{0<r\leqslant 1}{\sup}\, \big\vert \langle  \Lambda , (QK_r)(x,\cdot)\rangle \big\vert < \infty.
\end{equation}
One can describe an element of $C^\alpha_{\textrm{loc}}(U)$, with $\alpha<0$, with wave front set in $\Gamma$ in terms similar to \eqref{EqGammaWaveFrontSet} as the set of distributions $\Lambda\in \mcC^\alpha_{\textrm{loc}}(U)$ {such that} for all pseudodifferential operators $Q$ with Schwartz kernel compactly supported in $U\times U$ and whose symbol is microlocally smoothing near $\Gamma$, one has for every compact subset $C$ of $\mcX$
$$
\underset{x\in C}{\sup}\,\underset{0<r\leqslant 1}{\sup}\Big( r^{-\frac{\alpha}{2}} \big\vert \big\langle \Lambda , K_r(x,\cdot)\big\rangle \big\vert + \big\vert \big\langle \Lambda , (QK_r)(x,\cdot)\big\rangle \big\vert \Big) < \infty.
$$ 
The term $r^{-\frac{\alpha}{2}} \big\vert \big\langle \Lambda , K_r(x,\cdot)\big\rangle \big\vert$ ensures that $\Lambda$ is H\"older whereas the pseudodifferential operator $Q$ probes the smoothness of $\Lambda$ outside of the wave front set.

We now come back to the parabolic setting $\mcY\subset\mcX$ of \eqref{EqModelParabolicSituation} in Section \ref{SubsectionScalingFields} and denote by $\rho$ a parabolic scaling field for this embedding. Let $U$ stand for an open set of $\mcX$ which is stable by the backward semiflow of $\rho$, in the sense that $e^{-\ell\rho}\left(U \right)\subset U$ for all $\ell\geqslant 0$.

\ssk

\begin{defn}
Given some closed conic set $\Gamma\subset T^*U\setminus \{{\bf 0}\}$ we denote by $(e^{-\ell\rho})^*\Gamma$ the set defined as
\begin{align*}
(e^{-\ell\rho})^*\Gamma \defeq \Big\{ \left( e^{\ell\rho}x; (^t\rmd e^{\ell\rho})_x^{-1}(\xi)  \right) ; (x;\xi)\in \Gamma \Big\}\,.
\end{align*}
Then we assume that the lifted flow of $e^{-\ell\rho}$ leaves the conic set $\Gamma$ stable i.e.
$$
(e^{-\ell\rho})^*\Gamma\subset\Gamma
$$
for all $\ell\geqslant 0$.
\end{defn}

\ssk

 Last recall that for a distribution $\Lambda$ on $U$ and a smooth diffeomorphism $\varphi$ of $U$, the distribution $\varphi^*\Lambda$ is defined for any test function $f$ by $(\varphi^*\Lambda,f)=(\Lambda, \vert \det \rmd\varphi \vert^{-1}f\circ\varphi^{-1})$. Having described the necessary geometric framework, we are ready to state the definition of the scaling spaces that will be used extensively in the present work.

\ssk

\begin{defn*}
For $\alpha<0$ and $a\in\bbR$ we define the \textbf{\textsf{scaling space}} $\mcS^{a,\rho;\alpha}_\Gamma(U)$ as the space of distributions $\Lambda\in\mcD'(U)$ with the following property. For every pseudodifferential operators $Q$ with  
Schwartz kernel compactly supported in $U\times U$ and whose symbol is microlocally smoothing\footnote{That is denoting by $a$ the symbol of $Q$, for all $(x_0,\xi_0)\in \Gamma$, for all $N>0$, we have $\vert a(x,\xi)\vert \lesssim \left(1+\vert \xi\vert \right)^{-N}$ for every $(x,\xi)$ in some conical neighbourhood of $(x_0,\xi_0)$} near $\Gamma$ for each compact set $C\subset U$, we have
$$
\underset{\ell\geqslant 1}{\sup}\, \underset{x\in C}{\sup}\,\underset{0<r\leqslant 1}{\sup} \Big( r^{-\frac{\alpha}{2}} \big\vert \big\langle \Lambda , K_r(x,\cdot) \big\rangle\big\vert + e^{a\ell}  \big\vert \big\langle (e^{-\ell\rho})^*\Lambda , QK_r(x,\cdot) \big\rangle\big\vert \Big) < \infty\,.
$$
\end{defn*}

\ssk

 For $a\in\bbR$ fixed and $\rho$ a parabolic scaling field for the inclusion $\mcY\subset\mcX$ whose lifted backward flow on the cotangent stabilizes $\Gamma$: $e^{-\ell\rho*}\Gamma\subset \Gamma, \forall \ell\geqslant 0$ (which implies in particular that $N^*(\mcY\subset\mcX)\subset\Gamma$), we define 
$$
\mcS^a_\Gamma(U) = \bigcup_{\alpha\in \bbR} \mcS^{a,\rho;\alpha}_\Gamma(U).
$$ 
The letter `$\mcS$' is chosen for {\sl scaling}. The exponent $a$ retains the scaling property and $\Gamma$ retains some information on the wavefront set.  We do not retain the vector field $\rho$ in the notation $\mcS^a_\Gamma(U)$. Most of the time it comes implicitely together with the data of the closed conic set $\Gamma$. Moreover, in the particular case where $\Gamma=T^\bullet U$ or $\Gamma=N^*\left(\mcY\subset \mcX\right)\setminus \{{\bf 0}\}$, it can indeed be proved that the spaces $\mcS^{a,\rho;\alpha}_\Gamma(U)$ do not depend on which particular scaling field $\rho$ for the inclusion $\mcY\subset\mcX$ is used. A similar statement will be given and proved below in Proposition \ref{Prop:intrinsicscalingspaces}. Note that the space $\mcS^a_\Gamma(U)$ is a priori larger than the space of conormal distributions with wavefront set in $N^*\left(\mcY\subset U \right)$ since {some} elements in $\mcS^a_\Gamma(U)$ might have some wavefront set contained in the cone $\Gamma$ which is not necessarily included in $N^*\left(\mcY\subset U \right)$. {\it The notation $\mcS^a_\Gamma(U)$ does not emphasize the dependence of this space on the inclusion $\mcY\subset\mcX$; this will always be clear for us from the context. Note the fact that for all element $\Lambda\in \mcS^a_\Gamma(U)$ the family of scaled distributions $\left( e^{a\ell} e^{-\ell\rho*}\Lambda \right)_{\ell\geqslant 0}$ is bounded in $\mcD^\prime_\Gamma$.} 

\ssk

\begin{Example}--
An elementary example is given by the principal value of $1/\vert x\vert$ in $\bbR$, where $\mcY=\{0\}\subset\bbR$ and it has scaling exponent $a=-1$ and wavefront set $\Gamma=T_0^*\bbR$.
\end{Example}

\ssk

The next proposition gives an example of an element of some space $\mcS^{a,\rho ; 0}(U)$ for some scaling exponent $a$ and some scaling field $\rho$. For $n\geqslant 2$, denote by ${\bf d}_n$ the diagonal of $M^n$. Denote by $\rho_n$ a scaling field on $M^n$ for the inclusion ${\bf d}_n\subset M^n$ and define on $M^2\times \bbR^2$ the parabolic scaling field
$$
\rho = 2(t-s)\partial_s + \rho_2.
$$
Denote also by 
$$
\pi_{\leqslant2} : (x_1,\dots,x_n)\in M^n\rightarrow (x_1,x_2)\in M^2
$$ 
the canonical projection on the first two components. 

\ssk

\begin{prop} \label{l:topologytwopointfunctiong}
Let $M$ be a closed manifold and $A_r(x,y)$ be a smooth kernel on $M^2\backslash{\bf d}_2$ such that one can associate to any small enough open set $U$ a coordinate system in which one has for all multiindices $\alpha,\beta$
\begin{equation} \label{EqHomogeneityBoundsKernels}
\big\vert\partial^\alpha_{s,t}\partial^\beta_{x,y} A_{\vert t-s\vert}(x,y)\big\vert \lesssim \big(\sqrt{t-s}+\vert y-x\vert\big)^{a-2\vert\alpha\vert - \vert\beta\vert}.
\end{equation}
Then the family 
$$
\Big(e^{\ell a}(e^{-\ell\rho})^* \pi_{\leqslant2}^* A_{\vert t-\cdot\vert}(\cdot,\cdot)\Big)_{\ell\geqslant 0}
$$
is bounded in 
$$
\mcD'_{N^*(\{s=t\})}\Big((M^n\times\bbR) \backslash \big(\pi_{\leqslant2}^*{\bf d_2}\cap\{s=t\}\big)\Big),
$$
that is 
$$
\pi_{\leqslant2}^* A_{\vert t-\cdot\vert}(\cdot,\cdot)\in\mcS^{a,\rho ; 0}_{N^*(\{s=t\})}\Big((M^n\times\bbR)\backslash\big( \pi_{\leqslant2}^*{\bf d_2}\cap\{s=t\}\big)\Big).
$$
\end{prop}

\ssk

In the sequel we denote by $\mathcal{K}^a$ the $C^\infty$--module of kernels $A_r(x,y)$ as above depending on two variables endowed with the weakest topology containing the $C^\infty\big({ [0,+\infty)_{1/2}}\times \big( M^2\backslash{\bf d}_2\big)\big)$ topology and which makes all the seminorms defined by the estimates (\ref{EqHomogeneityBoundsKernels}) continuous.

\ssk

\begin{Dem}
We first localize in a neighbourhood $U\times U$ of the diagonal since $K$ is smooth off--diagonal. It is enough to prove the claim for $A_{\vert t-s\vert}(x,y){\varphi}_1(y){\varphi}_2(x)$ where ${\varphi}_i\in C_c^\infty(U)$ and use a partition of unity to get the global result. In $U\times U$ we pull-back everything to the configuration space, which we write with a slight abuse of notations
$$ 
\pi_{\leqslant2}^*(A{\varphi}_1{\varphi}_2)(t,s,x_1,\dots,x_n) = A_{\vert t-s\vert}(x_1,x_2){\varphi}_1(x_1){\varphi}_2(x_2).
$$
We already know that this kernel satisfies some bound of the form
\begin{eqnarray*}
\big\vert A(t,s,x_1,x_2){\varphi}_1(x_1){\varphi}_2(x_2) \big\vert \lesssim \left(\sqrt{\vert t-s\vert}+\vert x_1-x_2\vert \right)^{-a}.
\end{eqnarray*}
Somehow we would like to flow both sides of the inequality by the parabolic dynamics $(e^{-\ell\rho})^*$ and bound the term $e^{-\ell\rho*}\left(\sqrt{\vert t-s\vert}+\vert x_1-x_2\vert \right)^{-a}$ asymptotically when $\ell$ goes to $+\infty$. We use for that purpose the Normal Form Theorem for the space part of the isotropic scaling fields
$$
\rho_{[n]} \overset{\textrm{def}}{=} \sum_{k=2}^n h_k\cdot \partial_{h_k},
$$
for some new coordinates $(h_k)_{k=2}^n$ that vanish at order $1$ along the deep space diagonal ${\bf d}_n$. The fact that $x_1-x_2$ vanishes at first order along ${\bf d}_n$ implies by Taylor expansion at first order that
\begin{eqnarray}
x_1-x_2 = L(h)+\mathcal{O}(\vert h\vert^2)
\end{eqnarray}
where $L(h)$ is a linear function of $(h_k)_{k=2}^n$. One then has
$$
(e^{-\ell\rho_{[n]}})^*\left(x_1-x_2\right) = (e^{-\ell\rho_{[n]}})^* L(h) + \mathcal{O}(e^{-2\ell}\vert h\vert^2) = L(e^{-\ell}h) + \mathcal{O}(e^{-2\ell}\vert h\vert^2), 
$$
and an exponential lower bound of the form
$$
e^{-\ell}\vert x_1-x_2\vert\lesssim \big\vert (e^{-\ell\rho_{[n]}})^*(x_1-x_2) \big\vert
$$
which yields the desired bound 
$$
\Big\vert  \partial_{t}^\alpha \partial_x^\beta e^{-\ell\rho*}  \pi^*(A{g}_1{g}_2)\big(t,s,x_1,\dots,x_n\big)\Big\vert \lesssim e^{\ell a}\left(\sqrt{\vert t-s\vert}+\vert x_1-x_2\vert \right)^{-a-2\vert \alpha\vert-\vert \beta\vert}
$$
and proves the claim. The above bound allows for instance to justify that the singularities when $x_1\neq x_2$ are conormal along the equal time region $t=s$ since we are smooth on each half region $t\geqslant s$ and $s>t$.
\end{Dem}

{
\subsubsection{Invariance properties of scaling spaces.}

We end this section by stating some invariance results which will be useful later, in the proof of Theorem \ref{ThmBlackBoxFeynmanGraphs}, when we need to make a particular choice of scaling field in the proof and then use the fact that the blow-up does not depend on this choice. In the next proposition all the scaling fields are relative to the closed embedding $\mcY\subset \mcX$. }

\ssk

\begin{prop} \label{Prop:intrinsicscalingspaces}
Assume $a\in \bbR$, $\mcY\subset \mcX $ and the distribution $\Lambda\in \mcD^\prime(\mcX)$, respectively $\mcD^\prime(\mcX\setminus \mcY)$, is \textbf{weakly homogeneous of degree $a$} for some scaling field $\rho$, in the sense that the family 
\begin{align*}
\left(e^{\ell a}e^{-\ell\rho*}(\varphi\Lambda)\right)_{\ell\geqslant 0}
\end{align*}
is bounded in $\mcD^\prime(\mcX)$, respectively  $\mcD^\prime(\mcX\setminus \mcY)$, for some scaling field $\rho$ whose critical manifold is $\mcY$ and some {function} ${\varphi}$ such that ${\varphi}=1$ near $\mcY$ and $\rho$ is well-defined near $\text{supp}({\varphi})$. Then $\Lambda$ is weakly homogeneous of degree $a$ for every scaling fields. More precisely, for any scaling field $\widetilde{\rho}$ with respect to  $\mcY $, every function $\widetilde{{\varphi}}$ supported in the domain of $\widetilde{\rho}$ such that $\widetilde{{\varphi}}=1 $ near $\mcY$, 
\begin{align}\label{eq:eqdefS^a}
\text{the family of distributions}\;
\left(e^{ta}e^{-t\widetilde{\rho}*}(\widetilde{{\varphi}}\Lambda)\right)_{t\geqslant 0}\;\text{is bounded in}\;\mcD^\prime(\mcX),\,\text{respectively  $\mcD^\prime(\mcX\setminus \mcY)$}.
\end{align}
\end{prop}

In the sequel, we shall denote by $\mcS^a_{\mcY}(\mcX)$ (respectively  $\mcS^a_{\mcY}(\mcX\setminus \mcY)$) the space of distributions such that \eqref{eq:eqdefS^a} holds.  Consequently, assume we are given an open subset $\mcU\subset \mcX$ such that $\mcU$ is stable by $\rho_1$ and $\rho_2$. Then every $\Lambda\in \mcD^\prime(\mcU)$ which is weakly homogeneous of degree $a$ for $\rho_1$ is also weakly homogeneous of the same degree $a$ for $\rho_2$. In the sequel, we shall denote by $\mcS^a_{\mcY}(\mcU)$ the space of such distributions.

We note that the space $\mcS^a_{\mcY}(\mcX)$, respectively  $\mcS^a_{\mcY}(\mcX\setminus \mcY)$, is intrinsic, defined independently from the choice of scaling fields. These spaces do not contain microlocal informations hence the topologies of the spaces $\mcS^a$ are weaker than the topologies of the spaces $\mcS^a_\Gamma$ except when $\Gamma$ is the whole cotangent space $T^*\mcX\setminus {\bf 0}$. 

\ssk

\begin{Dem}
The proof can be found in~\cite[Thm 3.3 p.~828]{DangAHP} and relies on the linearization Theorem for scaling fields proved in ~\cite{DangAHP} and also in the beautiful exposition of Meinrenken~\cite[Lemma 2.1 p.~226]{Meinrenkensurvey}.
We sketch a proof for completeness. 
We localize near $x\in \mcY$, we choose some chart $\kappa:U\ni x \mapsto \bbR^{d_1+d_2+d_3}$ in which 
$\kappa( \mcY \cap U)= (\bbR^{d_1}\times \{0\}) \cap \kappa(U) $.
Then once we push $\rho$ by the linear chart $\kappa$ we still get a scaling field that we abusively denote by $\rho$. In these coordinates the vector field $\rho$ has the form $\rho=2t\cdot\partial_t+y\cdot\partial_y + R(t,y,\partial_t)+H(t,y,\partial_y)$ where $R$ and $H$ vanish at order $2$ when $(t,y)$ go to $(0,0)\in \bbR^{d_2+d_3}$. Choose ${o}$ supported in the open chart $U$, $\kappa_*(\Lambda{\varphi})$ is a distribution in $\bbR^{d_1+d_2+d_3}$ weakly homogeneous of degree $a$ under scaling by $\rho$. With no loss of generality it suffices to prove the weak homogeneity of $\kappa_*(\Lambda{\varphi}) $ when we scale with a different scaling field $\rho_2$, the general claim can be deduced by localising plus gluing with a partition of unity without problem. Then the linearization proof \cite[Prop 2.3 p.~826]{DangAHP} tells us that  we have an equation of the form
$$ 
e^{-\ell\rho}\circ U(\ell) = e^{-\ell\rho_2}   
$$
where $U(\ell)$ is a family of diffeomorphism germs near $(x,0,0)$ which has a limit when $\ell\uparrow\infty$, the limit is still a diffeomorphism germ. So for every test function $\varphi\in C^\infty_c(\kappa(U))$ one has 
$$ 
e^{\ell a} \left\langle e^{-\ell\rho_2*}\kappa_*(\Lambda{\varphi}), \varphi  \right\rangle = e^{\ell a} \left\langle U(\ell)^* e^{-\ell\rho*}\kappa_*(\Lambda{\varphi}), \varphi  \right\rangle = e^{\ell a} \left\langle  e^{-\ell\rho*}\kappa_*(\Lambda{\varphi}), U(\ell)^{-1*}\varphi  \right\rangle,
$$
where the last quantity on the right hand side is bounded when $\ell\geqslant 0$ since $(U(\ell)^{-1*}\varphi)_{\ell\geqslant 0}$ is a bounded family of test functions and $\big( e^{\ell a} e^{-\ell\rho*}\kappa_*(\Lambda{\varphi})\big)_{\ell\geqslant 0} $ is a bounded family of distributions by assumption. Then we just proved that $\big( e^{\ell a} e^{-\ell\rho_2*}\kappa_*(\Lambda{\varphi})\big)_{\ell\geqslant 0}$ is a weakly, hence strongly, bounded  family of distributions which concludes the proof.
\end{Dem}

\medskip

\subsection{The canonical extension}
\label{SubsectionCanonicalExtension}

\subsubsection{The basic extension result$\mathbf{.}$}
Let us use a unique notation ${\bf 0}$ for the zero section of any vector bundle.

\ssk

\begin{defn*}
Let $\mcX$ be a smooth manifold and $\mcY\subset\mcX$. A closed conic set $\Gamma\subset T^*(\mcX\backslash\mcY)\backslash{\bf 0}$ is said to satisfy the \textbf{\textsf{conormal landing condition}} if its closure $\widetilde\Gamma$ in $T^*(\mcX)\backslash{\bf 0}$ satisfies $\widetilde\Gamma\subset\big(\Gamma\cup N^*(\mcY)\big)$.
\end{defn*}

\ssk
In what follows, we will use the terminology \emph{smooth weighted manifolds} 
for smooth weighted manifolds of product type as introduced in Section~\ref{SubsectionScalingFields}.

\ssk

\begin{thm} \label{ThmCanonicalExtension}
Let $\mcX$ be a smooth weighted manifold and $\mcY\subset\mcX$ and $\Gamma\subset T^*(\mcX\backslash\mcY)\backslash{\bf 0}$ be a closed conic set that satisfy the conormal landing condition. Assume we are given a family $(\Lambda_\epsilon)_{0<\epsilon\leqslant 1}$ of distributions on $\mcX$ that converge in $\mcD'(\mcX\backslash\mcY)$ as $\epsilon$ goes to $0$ to an element $\Lambda\in\mcS^a_\Gamma(\mcX\backslash\mcY)$.
\begin{enumerate}
	\item[\textsf{\textbf{(a)}}] If 
	$$
	-\textrm{\emph{codim}}_w(\mcY\subset\mcX) < a
	$$ 
	then $\Lambda$ has a unique extension into a distribution over $\mcX$ such that the convergence of $\Lambda_\epsilon$ to $\Lambda$ occurs in $\mcS^a_{\Gamma\cup N^*(\mcY)}(\mcX)$.   \vspace{0.15cm}
	
	\item[\textsf{\textbf{(b)}}] If
	$$
	-\textrm{\emph{codim}}_w(\mcY\subset\mcX) - 1 \leqslant a < -\textrm{\emph{codim}}_w(\mcY\subset\mcX)
	$$ 
	there exists a family $\Lambda_{\mcY,\epsilon}$ of distributions supported on $\mcY$, with wavefront set in $N^*(\mcY)$ such that $\Lambda_\epsilon - \Lambda_{\mcY,\epsilon}$ has a limit in $\mcD'(\mcX)$ and the convergence occurs in $\mcS^{a'}_{\Gamma\cup N^*(\mcY)}(\mcX)$ for all
	$$
	a' < a.
	$$
\end{enumerate}
\end{thm}

\ssk

 The particular case where $\mcY=\{p\}$ is a point will be used in Theorem \ref{Thm:extensiontwosteps} below. We loose any information on the wavefront set of the extension at $p$ in that case, so the convergence of $\Lambda_\epsilon$ to $\Lambda$ happens in $\mcS^b_{\Gamma\cup T_p^*M}(\mcX)$, for $b\in\{a,a'\}$, depending on the situation.

\ssk

\begin{Dem}
We follow the proof of similar results proved in an elliptic setting in \cite{DangAHP} -- see Theorem 1.10, Theorem 4.4 and Section 6 therein. We give here the main arguments to emphasize the differences with \cite{DangAHP}  that come from our parabolic setting. The main idea of the proof is to start from a continuous partition of unity which approximates the constant function $1$ but vanishes near $\mcY$. Then we multiply the distribution $\Theta$ by this partition of unity to product an approximation of $\Theta$ outside $\mcY$ and use the assumption on the weak homogeneity of $\Theta$ near $\mcY $ to conclude that the approximation genuinely converges to some given distribution $ \Theta^+ $ which is the desired extension.

\ssk

Let $\rho$ be a scaling field for the inclusion $\mcY\subset\mcX$ such that $\Lambda\in\mcS^{a,\rho;\alpha}_{\Gamma\cup N^*(\mcY)}(\mcX)$ for some $\alpha\in\bbR$, and let ${\varphi_\mcY}$ be a smooth function equal to $1$ in a neighborhood of $\mcY$ stable by the backward semiflow of $\rho$ and such that ${\varphi_\mcY}$ vanishes outside some larger neighborhood.   \vspace{0.1cm}

{\it (1)} We first use the normal form theorem to reduce our problem to the model case of a distribution on $\bbR^k$, with $k=d_1+d_2+d_3$ with coordinates $(x,y,t)$, the scaling field $\rho = y\partial_y + 2t\partial_t$ is linear and globally defined, and the extension is done with respect to the linear subspace $\bbR^{d_1}\subset\bbR^k$. We work in that setting in the remainder of the proof. Let then $(\Theta_\epsilon)_{0<\epsilon\leqslant 1}$ be a family of distributions on $\bbR^k$. We assume that the $\Theta_\epsilon$ converge as $\epsilon$ goes to $0$ to an element $\Theta\in\mcS^a_\Gamma(\bbR^k\backslash\bbR^{d_1})$ where $a > -\textrm{codim}_w(\mcY\subset\mcX)$.   \vspace{0.1cm}

{\it (2)} Pick $0<\ell_0$ and think of it as being large. We use the continuous partition of unity
\begin{eqnarray*}
\textrm{Id} - (e^{\ell_0\rho})^*{\varphi_\mcY} = \textrm{Id} - {\varphi_\mcY} + \int_0^{\ell_0} (e^{\ell\rho})^*\left(-\rho {\varphi_\mcY}\right)\rmd \ell
\end{eqnarray*}
to define an extension of our distribution $\Theta$. We set for convenience $\overline{\varphi_\mcY} \defeq -\rho{\varphi_\mcY}$; its support does not meet $\mcY$. We have for any test function $f\in\mcD(\bbR^k)$
\begin{eqnarray*}
\big\langle \Theta (1 - (e^{\ell_0\rho})^*{\varphi_\mcY}),f \big\rangle = \big\langle \Theta(1-{\varphi_\mcY}),f \big\rangle + \int_0^{\ell_0} e^{-\ell(d_2+2d_3)} \left\langle \overline{\varphi_\mcY} (e^{-\ell\rho})^*\Theta , (e^{-\ell\rho})^* f  \right\rangle\,\rmd \ell.
\end{eqnarray*}
The exponential factor $e^{-\ell(d_2+2d_3)}$ comes from the Jacobian of the flow of $e^{-\ell\rho}$. Note that $d_2+2d_3 = \textrm{codim}_w(\mcY\subset\mcX)$. If $\Gamma\subset T^*\bbR^k\backslash{\bf 0}$ stands for a closed conic set invariant by the lifted dynamics of $(e^{-\ell\rho})^*$ such that $\Gamma\cap T^*\bbR^{d_1} \subset N^*(\bbR^{d_1})$, our choice of scaling exponent $a$ ensures that the family 
$$
(\Theta^{(\ell)})_{\ell\geqslant 0} \defeq (e^{a\ell}(e^{-\ell\rho})^*\Theta)_{\ell\geqslant 0}
$$
is bounded in $\mcD'_\Gamma(\bbR^k)$. One then has for the Schwartz kernels
$$
\big[\Theta\left(1-(e^{\ell_0\rho})^*{\varphi_\mcY}\right)\big](z,z') = \big[\Theta (1-{\varphi_\mcY})\big](z,z') + \int_0^{\ell_0} e^{-\ell(a+d_2+2d_3)} \big(\overline{\varphi_\mcY} \, \Theta^{(\ell)}\big)(z)  \delta\big(z'- e^{-\ell\rho}(z)\big) \,\rmd \ell.
$$
\noindent We know from the hypocontinuity theorem on the H\"ormander product of distributions \cite[Thm 6.1 p.~219]{BDH16} that the family
$$
(\overline{\varphi_\mcY} \, \Theta^{(\ell)})(z) \delta\big(z'- e^{-\ell\rho}(z)\big)
$$
with $\ell\geqslant \ell_1$ large enough, is bounded in $\mcD'_{\overline\Gamma}(\bbR^k)$, where
$$
\overline\Gamma \defeq (\Gamma\times{\bf 0}) \cup \Gamma_\rho \cup \big((\Gamma\times{\bf 0}) + \Gamma_\rho\big).
$$
For the moment this means that the $\ell_0$-dependent family of distributions associated with the kernels
\begin{eqnarray*}
(\star) \defeq \int_0^{\ell_0} e^{-\ell(a+d_2+2d_3)} \big(\overline{\varphi_\mcY} \, \Theta^{(\ell)}\big)(z) \delta\big(z' - e^{-\ell\rho}(z)\big) \, \rmd \ell
\end{eqnarray*}
is bounded in $\mathcal{D}'_{\overline\Gamma}(\bbR^k\times\bbR^k)$ uniformly in $\ell_0\geqslant \ell_1$. In particular the integral converges in $\mathcal{D}'_{\overline\Gamma}(\bbR^k\times\bbR^k)$ when $\ell_0$ goes to $+\infty$. Now we interpret the integral over the variable $z$ as a push-forward along the fibers of the linear projection 
$$
p_2 : (z,z')\mapsto z'.
$$
The pushforward Theorem yields that $p_{2*}(\cdot)$ is bounded in $\mathcal{D}'_{p_{2*}\overline\Gamma}(\bbR^k)$ where
\begin{eqnarray*}
p_{2*}\overline\Gamma =  (p_{2*}\Gamma_\rho) \cup p_{2*} \big(\Gamma\times{\bf 0} + \Gamma_\rho \big) 
\end{eqnarray*}
and
\begin{equation*} \begin{split}
p_{2*}\Gamma_\rho &= \big\{\big((x,0,0),(0,\eta,\tau)\big)\big\}   \\
p_{2*}\left(\Gamma\times{\bf 0} + \Gamma_\rho \right) &= \Big\{ \Big(e^{-\ell\rho}(z) , (e^{-\ell\rho})^*(\lambda) \Big) ; (z,\lambda) \in \Gamma, 0\leqslant \ell\leqslant +\infty\Big\} \subset \Gamma \cup N^*(\mcY)
\end{split} \end{equation*}
since the cone $\Gamma$ is invariant by the lifted flow of $e^{-\ell\rho}$ provided $\ell<+\infty$ and the limit points of the form $\lim_{\ell\uparrow \infty}\big(e^{-\ell\rho}(z) , (e^{-\ell\rho})^*(\lambda) \big) $ for $ (z,\lambda) \in \Gamma$ must belong to the conormal $N^*(\mcY)$ by the conormal landing condition on $\Gamma$. It is at this precise place that we are using the conormal landing condition assumption on $\Gamma$. The distributions $\Theta(1-(e^{\ell\rho})^*{\varphi_\mcY})$ are thus converging in $\mcD'_{\Gamma\cup N^*\bbR^{d_1}}(\bbR^k)$ to
$$
\langle \Theta^+ , f\rangle = \langle \Theta(1-{\varphi_\mcY}) , f\rangle + \int_0^\infty e^{-\ell(d_2+2d_3)} \langle \overline{\varphi_\mcY} \, (e^{-\ell\rho})^*\Theta , f\circ e^{-\ell\rho}\rangle\,\rmd \ell.
$$ 
The uniqueness of the extension $\Theta^+$ follows from the continuity of all the operations involved in above. To see the scaling property of the extension we note that the family 
$$
\big(\Theta^{(\ell)} \big)_{\ell\geqslant 0}=\big( e^{a\ell}(e^{-\ell \rho})^*\Theta\big)_{\ell\geqslant 0}
$$ 
is bounded in $\mcD'_\Gamma(\bbR^k)$, since $\Theta\in \mcS_\Gamma^a(\bbR^k)$ means that the family  $\left( e^{a\ell}(e^{-\ell \rho})^*\Theta\right)_{\ell\geqslant 0}$ is bounded in $\mcD^\prime_\Gamma(\bbR^k)$ by definition of $\mcD_\Gamma^\prime(\bbR^k)$ and $\mcS_\Gamma^\prime(\bbR^k)$, and observe that
$$ 
e^{\ell a}(e^{-\ell\rho})^*\Theta^+ = p_{2*}\left(\int_0^\infty e^{-\ell(a+d_2+2d_3)} \left((\Theta^{\ell+a})' \overline{\varphi_\mcY} \otimes 1\right)[e^{-\ell\rho}] \, \rmd \ell\right).
$$

{\it (3)} In the borderline case our proof follows closely~\cite[Prop 4.9 p.~841]{DangAHP} except we work in a parabolic setting. We proceed as above with $(e^{-\ell\rho})^*$ replaced by $(e^{-\ell\rho})^*R_0$ if $-\textrm{codim}_w(\mcY\subset\mcX) -1 <a\leqslant -\textrm{codim}_w(\mcY\subset\mcX)  $ and $(e^{-s\rho})^*R_1$ if $a=-\textrm{codim}_w(\mcY\subset\mcX) -1 $. So our extension reads
$$
\langle \Theta^+ , f\rangle = \langle \Theta(1-{\varphi_\mcY}) , f\rangle + \int_0^\infty e^{-\ell(d_2+2d_3)} \big\langle \overline{\varphi_\mcY} (e^{-\ell\rho})^*\Theta , \big(R_if \big) \circ e^{-\ell\rho}\big\rangle\,\rmd \ell,
$$
where $R_i f, i=0,1$ is obtained from $f$ by Taylor subtraction. The integral converges absolutely since $[e^{-\ell\rho}R_i]=\mathcal{O}_{\mathcal{D}'_\Gamma}(e^{-\ell(1+i)})$ and the map
$$
\Theta\mapsto \Theta^+
$$
is continuous from $ \mcS^a_\Gamma(\mcX\setminus \mcY) $ to $\mcS^{a'}_{\Gamma\cup N^*(\mcY)}(\mcX)$ for all $a'<a$ as we will see below when we check the weak homogeneity of the extension $\Theta^+$. This shows that when 
$$
-\textrm{codim}_w(\mcY\subset\mcX) -1 <a\leqslant -\textrm{codim}_w(\mcY\subset\mcX)
$$
one can take
$$
\Lambda_{\bbR^{d_1},\epsilon}(f) = \int_0^\infty e^{-\ell(d_2+2d_3)} \big\langle \Theta_\epsilon , \overline{\varphi_\mcY}\circ e^{-\ell\rho} \, \Pi(f) \big\rangle \, \rmd \ell
$$
and when 
$$
a=-\textrm{codim}_w(\mcY\subset\mcX) -1
$$
one can take
$$
\Lambda_{\bbR^{d_1},\epsilon}(f) = \int_0^\infty e^{-\ell(d_2+2d_3)} \big\langle \Theta_\epsilon \,,\, \overline{\varphi_\mcY}\circ e^{-\ell\rho} \, \Pi(f)+t(\partial_tf)(\cdot,0,0)+y\cdot(\partial_yf)(\cdot,0,0) \big\rangle \, \rmd \ell.
$$
For simplicity, in the remainder of the proof we shall specialize to the case $-\textrm{codim}_w(\mcY\subset\mcX) -1 <a\leqslant -\textrm{codim}_w(\mcY\subset\mcX) $. By the wave front set condition on $\Theta_\epsilon$ one can always decompose $\Lambda_{\bbR^{d_1},\epsilon}$ under the product form
$$
\Lambda_{\bbR^{d_1},\epsilon} = c_\epsilon \Lambda_{\bbR^{d_1}}
$$
where
$$
\Lambda_{\bbR^{d_1}} = \Pi
$$
is a distribution independent of $\epsilon$, supported on $\bbR^{d_1}$, with wavefront set contained in $N^*(\bbR^{d_1})$, and the function $c_\epsilon$ is given by
$$
c_\epsilon(x) = \int_0^\infty e^{-\ell(d_2+2d_3)} \Theta_\epsilon (x,y,t) \overline{\varphi_\mcY} \, \big(e^{-\ell\rho}(x,y,t)\big) \, \rmd \ell \rmd y\rmd t.
$$
To check the weak homogeneity bound write
\begin{eqnarray*}
e^{\ell'a}\left\langle e^{-\ell'\rho}\Theta^+ , f\right\rangle = \int_0^\infty  \left\langle \Theta^{(\ell)} \overline{\varphi_\mcY} , e^{-(\ell-\ell')\rho}f - \Pi(f)\right\rangle \rmd \ell
\end{eqnarray*}
and observe that the support of $e^{-(\ell-\ell')\rho}f$ meets the support of $\overline{\varphi_\mcY}$ only if $\ell\geqslant C+\ell'$ for a constant $C$ that depends only on the support of $\overline{\varphi_\mcY}$. So the integral can be split in
\begin{eqnarray*}
- \int_0^{C+\ell'} \left\langle \Theta^{(\ell)} \overline{\varphi_\mcY} , \Pi(f)\right\rangle \rmd \ell + \int_{C+\ell'}^{\infty} \left\langle \Theta^{(\ell)} \overline{\varphi_\mcY} , e^{-(\ell-\ell')\rho}f - \Pi(f)\right\rangle \rmd \ell
\end{eqnarray*}
A change of variable shows that the second term  is uniformly bounded in $\ell'$ whereas the first term is bounded above by $(C+\ell')\Vert f\Vert_{C^0}$. This concludes the proof that $e^{\ell a}e^{-t\rho}\overline{U} = \mathcal{O}_{\mathcal{D}^\prime}(a)$.
\end{Dem}

\medskip

In the case (b), note that the extension is no longer unique. Any two extensions differ by some conormal distribution supported on $\mathcal{Y}$ whose order is $0$ if $-\textrm{codim}_w(\mcY\subset\mcX) -1 <a\leqslant -\textrm{codim}_w(\mcY\subset\mcX) $ and of order $1$ if $a=-\textrm{codim}_w(\mcY\subset\mcX) -1$.

\medskip

\subsubsection{Two step canonical extensions.}
We need the following refinement of the canonical extension. It involves a two step extension procedure.

\ssk

\begin{thm} \label{Thm:extensiontwosteps}
Let $\mcY\subset \mcX$ be a closed embedded submanifold in the ambient smooth manifold $\mcX$, $p\in \mcY$ a point in $\mcY$, $a_1,a_2$ some real numbers. Assume we are given a smooth function ${\varphi_\mcY}$ such that ${\varphi_\mcY}=1$ near $\mcY$ and ${\varphi_\mcY}=0$ outside some larger neighbourhood. Set 
$$
\Omega\defeq \Big\{ m\in \mcX: \mathrm{dist}(m,p) > 2\,\mathrm{dist}(m,\mcY)\Big\}.
$$
{Assume} there exists a scaling field $\rho_1$ with respect to $\{p\}$ and a scaling field $\rho_2$ with respect to $\mcY$, such that both flows of $\rho_1$ and $\rho_2$ preserve the inclusion $p\in \mcY$. {Let $\Lambda$ be a distribution in $\mcD^\prime(\mcX\setminus \mcY )$ such that}
\begin{itemize}
    \item[--] for every test function $\varphi\in C^\infty_c(\mcX\setminus \mcY)$, 
\begin{align} \label{EqOnePullBacks}
e^{\ell_1 a_1} \left\langle  (e^{-\ell_1\rho_1})^* \left( {\varphi_\mcY}\Lambda\right) \,,\, \varphi \right\rangle \qquad (\ell_1\geqslant 0);
\end{align}
is bounded.

    \item[--] for every test function $\varphi\in C^\infty_c(\Omega\setminus \mcY)$, 
\begin{align} \label{EqTwoPullBacks}
e^{\ell_2a_2 + \ell_1a_1} \left\langle (e^{-\ell_2\rho_2})^* (e^{-\ell_1\rho_1})^* \left( {\varphi_\mcY}\Lambda\right) \,,\, \varphi \right\rangle \qquad (\ell_1, \ell_2\geqslant 0)   
\end{align}
is bounded.
\end{itemize} 
If
\begin{equation*} \begin{split}
a_1 &> -\mathrm{dim}_w(\mcX),   \\
a_2 &> -\mathrm{codim}_w(\mcY\subset \mcX),
\end{split} \end{equation*}
then there exists a canonical extension of $\Lambda$ in $\mcD^\prime(\mcX)$.
\end{thm}

\ssk

 Note that in \eqref{EqTwoPullBacks} we first pullback $\varphi_\mcY\Lambda$ by the map $e^{-\ell_1\rho_1}$ and then pull back the resulting distribution by $e^{-\ell_2\rho_2}$. Note also that we no longer need any microlocal control in the above statement {that only deals with the} weak topology. Let us emphasize that {we ask \eqref{EqTwoPullBacks} to hold for some} test functions supported in the domain $\Omega\setminus \mcY$ whereas the {first} scaling assumption applies when we scale toward the point $p$. In the Feynman amplitude context of Section \ref{SubsectionConfigurationSpace} both scaling fields will be admissible in the sense of Definition~\ref{def:explicitEuler} hence the fact that the inclusion $p\in \mcY$ is preserved by both dynamics will be immediate. 

\ssk

 To lighten the notations we will write below $e^{-\rho*}\Lambda$ for $(e^{-\rho})^*\Lambda$. The structure of the proof is simple and is sketched as follows in the model situation of $\bbR^d$ and some linear subspace for $\mcY$. We first extend $\Lambda$ to $\Omega$ using \eqref{EqOnePullBacks} and the canonical extension theorem, Theorem \ref{ThmCanonicalExtension}, and the assumption on $a_2$. It can be seen from \eqref{EqTwoPullBacks} that this extension $\mcR\Lambda$ is weakly $a_1$-homogeneous with respect to $\{p\}$. Then choose a $\rho_1$-radial function $0\leqslant\varphi_\Omega\leqslant1$ that is smooth outside the point $p$, equal to $1$ in a neighbourhood of $\mcY$ in $\Omega$ and is $0$ outside $\Omega$. The distribution $\Lambda = \varphi_\Omega\Lambda + (1-\varphi_\Omega)\Lambda$ has $\varphi_\Omega\mcR\Lambda + (1-\varphi_\Omega)\Lambda$ as an extension to $\mcX\backslash\{p\}$. We can then use \eqref{EqTwoPullBacks} and the canonical extension theorem to extend the distribution to $\mcX$.   

\ssk

\begin{Dem}
 Let us say that a function $f$ is $\rho_1$-radial if $f(e^{c\rho_1}(x))=f(x)$ for all $c\in\bbR$ and $x\in \mcX$. 

\ssk

We reduce the proof to some normal form. We cover the whole of $\mcX$ by some locally finite open cover, use charts and a subordinated partition of unity
$\sum {\varphi}_i=1$. It suffices to prove the same claim for $\Lambda{\varphi} $ where ${\varphi}$ supported in some open chart $\kappa: \mcX\supset U \rightarrow \bbR^k $ containing $p$, where $ \kappa(\mcY\cap U)\subset \bbR^{d_1+d_2}\times \{0,0\}$ and $\kappa(p)=0\in \bbR^{k}$. So we are reduced to study the distribution $\kappa_*\left(\Lambda{\varphi} \right)$ which is compactly supported and satisfies the assumption of our Theorem on
$\mcX=\bbR^k$ with $k=d_1+d_2+d_3+d_4$ with coordinates $(x,t,y,s)$ such that
\begin{align*}
&\{p\} = \big\{x=0,t=0,y=0,s=0 \big\},\quad \text{and} \quad \mcY=\{y=0,s=0 \}=\bbR^{d_1+d_2}\times \{0\}, \\ 
&\Omega = \Big\{ \sqrt{\vert x\vert^2+\vert t\vert^2+\vert y\vert^2+\vert s\vert^2} > 2 \sqrt{\vert y\vert^2+\vert s\vert^2}  \Big\}.
\end{align*}
Near $p$ we have from the Linearization Theorem proved in~\cite[Prop 2.3]{DangAHP} or \cite{Meinrenkensurvey} the identities
$$ 
e^{-\ell E_1} = e^{-\ell\rho_1}\circ U_1(\ell)\,,\;\text{and} \;e^{-\ell E_2} =  e^{-\ell\rho_2}\circ U_2(\ell)
$$
where $E_1,E_2 $ are the linear scaling fields reading
\begin{align*}
E_1 &= 2t\cdot\partial_{t}+x\cdot\partial_{x} + 2s\cdot\partial_{s}+y\cdot\partial_{y},   \\
E_2 &= 2s\cdot\partial_{s}+y\cdot\partial_{y},
\end{align*}
and $U_1(\ell)$ and $U_2(\ell)$ are two smooth germs of diffeomorphisms near $p$ depending smoothly on $e^{-\ell}$ when $\ell\uparrow \infty$, and both $U_1,U_2$ have limits when $\ell\uparrow \infty$ that also are smooth germs of diffeomorphisms near $p\in \kappa(U)$. Then, for every test function $\varphi\in C^\infty_c( \Omega\setminus \mcY )$, we have
\begin{equation*} \begin{split}
\left\langle e^{-\ell E_1*}e^{-rE_2*} (\Lambda{\varphi}), \varphi  \right\rangle &= \big\langle U_1(\ell)e^{-\ell\rho_1*}  e^{-rE_2*} (\Lambda{\varphi}), \varphi  \big\rangle   \\
&= \big\langle e^{-\ell\rho_1*}  e^{-rE_2*} (\Lambda{\varphi}) \,,\, U_1(\ell)^{-1*}\varphi \big\rangle   \\
&= \big\langle e^{-\ell\rho_1*} U_2(r)^{*}  e^{-r\rho_2*} (\Lambda{\varphi}) \,,\, U_1(\ell)^{-1*}\varphi \big\rangle \\
&= \big\langle \left( e^{-\ell\rho_1*} U_2(r)^{*} e^{u\rho_1*}\right) e^{-\ell\rho_1*}  e^{-r\rho_2*}(\Lambda{\varphi}) \,,\, U_1(\ell)^{-1*}\varphi \big\rangle   \\
&= \big\langle  e^{-\ell\rho_1*}  e^{-r\rho_2*}(\Lambda{\varphi}) \,,\, \Psi_{\ell,r}^{-1*} U_1(\ell)^{-1*}\varphi \big\rangle,\Lambda
\end{split} \end{equation*}
where 
$$
\Psi_{\ell,r} \defeq e^{-\ell\rho_1*} U_2(r)^{*} e^{\ell\rho_1*}.
$$

At this stage we make the observation that the families $\left(\Psi_{u,r}^{-1*} U_1(u)^{-1*}\varphi\right)_{ u\geqslant 0 ,r\geqslant 0 } $
are bounded families of test functions since $(\Psi_{u,r})_{u\geqslant 0,r\geqslant 0}$ is a bounded family of smooth germs of diffeomorphisms.
By the Banach-Steinhaus Theorem the family of distributions 
$$
\big(e^{\ell a_1+ra_2}   e^{-\ell\rho_1*}  e^{-r\rho_2*} \Lambda{\varphi}\big)_{\ell,r\geqslant 0}
$$ 
is weakly, hence strongly, bounded. From this we deduce that
$$
e^{\ell a_1+ra_2} \Big\langle e^{-\ell\rho_1*}  e^{-r\rho_2*} \Lambda{\varphi} \,,\, \Psi_{\ell,r}^{-1*} U_1(\ell)^{-1*}\varphi \Big\rangle
$$
is bounded for all test functions $\varphi \in C^\infty_c(\Omega\setminus \mcY)$. Therefore the family $e^{\ell a_1+ra_2} e^{-\ell E_1*} e^{-rE_2*}(\Lambda{\varphi})$ is also bounded in $\mcD^\prime(\Omega\setminus \mcY)$. From the Linearization Theorem, without loss of generality, we may thus choose some linear scaling fields $\rho_1$ and $\rho_2$ with respect to $\{p\}$ and $\mcY$, respectively, that read
\begin{align*}
\rho_1 &= 2t\cdot\partial_{t}+x\cdot\partial_{x} + 2s\cdot\partial_{s}+y\cdot\partial_{y},   \\
\rho_2 &= 2s\cdot\partial_{s}+y\cdot\partial_{y}\,.
\end{align*}
Since $\Lambda$ is a distribution on $\mcD'(\Omega\setminus \mcY)$ with scaling degree $a_2$ with respect to $\mcY$ and since $a_2>-\mathrm{codim}_w(\mcY\subset \mcX)$, the extension theorem \ref{ThmCanonicalExtension} implies that we can extend $\Lambda$ to a distribution $\mcR\Lambda$ on $\Omega$, recall that $p\notin \Omega$. 
The first extension $\mcR\Lambda$ for $\Lambda$ extending from $\Omega\setminus \mcY$ to the larger space $\Omega$ satisfies 
\begin{align*}
\mcR\Lambda=\Lambda(1-{\varphi_\mcY})+\underbrace{\int_0^\infty  \left( e^{r\rho_2*}\psi\right)\Lambda\rmd r}
\end{align*}
for every smooth function ${\varphi_\mcY}$ such that ${\varphi_\mcY}=1$ near $\mcY$, ${\varphi_\mcY}=0$ outside some larger neighbourhood (recall that here $\psi=-\rho_2{\varphi_\mcY}$) where the integral underbraced converges in $\mcD^\prime( \Omega)$ by the assumption on the scaling degree of $\Lambda$ with respect to  $\mcY$. Indeed for any test function $\varphi\in C^\infty_c(\Omega)$ one has
$$
\left\langle\int_0^\infty  \left( e^{r\rho_2*}\psi\right)\Lambda\rmd r ,\varphi\right\rangle= \int_0^\infty e^{-r(d_3+2d_4+a_2)}\left\langle \left( e^{ra_2} e^{-r\rho_2*} \Lambda \right) ,\psi e^{-r\rho_2*}\varphi  \right\rangle \rmd r 
$$
where it is immediate that the integrand converges since $d_3+2d_4+a_2>0 $, the family $\left(\psi e^{-r\rho_2*}\varphi\right)_{r\geqslant 0}$ forms a bounded family of test functions in $C^\infty_c(\Omega\setminus \mcY)$ and $\big(e^{ra_2} e^{-r\rho_2*}\Lambda \big)_{r\geqslant 0}$ is a bounded family of distributions in $\mcD^\prime(\Omega\setminus \mcY)$. We also have from the identity for $T>0$
$$ 
1 - {\varphi_\mcY}+\int_0^T  e^{r\rho_2*}\psi\rmd r=1- e^{T\rho_2*}{\varphi_\mcY}, 
$$
and the absolute convergence of the integral $\int_0^\infty  \left( e^{r\rho_2*}\psi\right)\Lambda\,\rmd r$, that $\mcR\Lambda$ can be defined as the limit
$$
\mcR\Lambda=\lim_{T\uparrow \infty} \Lambda\left( 1- e^{T\rho_2*}{\varphi_\mcY}\right). 
$$
Here we make the observation that for every isomorphism $f:\bbR^{k}\mapsto \bbR^k$ that stabilizes $\mcY$ and commutes with $\rho_2$ we have the identity
\begin{align*}
\mcR\Lambda=\Lambda(1-f^*{\varphi_\mcY})+{\int_0^\infty  \left( e^{r\rho_2*}f^*\psi\right)\Lambda \, \rmd r}.
\end{align*}
Since $f^*{\varphi_\mcY}$ still satisfies the same technical assumptions as ${\varphi_\mcY}$ and $-\rho_2 f^*{\varphi_\mcY}=-f^*\rho_2{\varphi_\mcY}=f^*\psi$, the integral underbraced also converges in $\mcD^\prime(\Omega)$. The key observation is the continuous partition of unity identity for $T>0$ 
$$ 
1-{\varphi_\mcY}+\int_0^T  e^{r\rho_2*}\psi \, \rmd r = 1- e^{T\rho_2*}{\varphi_\mcY}, 
$$
which implies
\begin{equation*} \begin{split}
1-{\varphi_\mcY}+\int_0^T  &e^{r\rho_2*}\psi\rmd r - \bigg(1-f^*{\varphi_\mcY} + \int_0^T  e^{r\rho_2*}f^*\psi \, \rmd r \bigg)   \\
&= 1- e^{T\rho_2*}{\varphi_\mcY} -\left(1- e^{T\rho_2*}f^*{\varphi_\mcY}  \right)  =e^{T\rho_2*}f^*{\varphi_\mcY}-e^{T\rho_2*}{\varphi_\mcY}   = \int_T^\infty e^{r\rho_2*}(\psi-f^*\psi) \, \rmd r,
\end{split} \end{equation*}
where the last equality holds true in $C^\infty(\Omega\setminus\mcY)$. Then, we can control the difference
$$ 
\Lambda (1- e^{T\rho_2*}{\varphi_\mcY}  )- \Lambda (1- e^{T\rho_2*}f^*{\varphi_\mcY}  )=\Lambda( e^{T\rho_2*}(f^*{\varphi_\mcY}-{\varphi_\mcY}) ) =  \int_T^\infty e^{r\rho_2*}(\psi-f^*\psi) \Lambda \, \rmd r.
$$
Repeating the above estimate, for every test function $\varphi\in C^\infty_c(\Omega\setminus \mcY)$
$$
\bigg\vert \left\langle \int_T^\infty e^{r\rho_2*}(\psi-f^*\psi) \Lambda \rmd r \,,\, \varphi\right\rangle \bigg\vert \lesssim \int_T^\infty e^{-r(d_2+2d_3+a_2)} \, \rmd r\longrightarrow 0 
$$
as $T\uparrow \infty$.

\ssk

The whole point is that we start from the information that $\Lambda$ defined on $\Omega\setminus \mcY$ is weakly homogeneous with respect to $\{p\}$ of degree $a_1$ and we need to make sure that the same property still holds true for the first extension $\mcR\Lambda$. Therefore we need to check that the partial extension $\mcR\Lambda\in \mcD^\prime(\Omega)$ (recall $p\notin \Omega$) is still weakly homogeneous of degree $a_1$ when we scale with respect to $\{p\}$.

We scale the renormalized $\mcR\Lambda$ using $e^{-t\rho_1}$ which gives
\begin{align*}
e^{\ell a_1} e^{-\ell\rho_1*} \mcR\Lambda &= e^{\ell a_1} e^{-\ell\rho_1*} \left(\Lambda(1-f^*{\varphi})+{\int_0^\infty  \left( e^{r\rho_2*}f^*\psi\right)\Lambda \, \rmd r}\right)   \\
&= e^{\ell a_1}\left( e^{-\ell\rho_1*}\Lambda \right) (1-{\varphi})+\underbrace{\int_0^\infty  \left( e^{r\rho_2*}\psi\right)\left( e^{\ell a_1} e^{-\ell\rho_1*}\Lambda \right)\rmd r},
\end{align*}
choosing $f=e^{\ell\rho_1*}$, using the fact the flows $e^{-r\rho_2}$ and $e^{-\ell\rho_1}$ commute, and the absolute convergence of the underbraced term. Then we make the second observation that the family $\left(e^{ta_1} e^{-\ell\rho_1*} \Lambda\right)_{\ell\geqslant 0}$ satisfies the assumptions of the extension Theorem~\ref{ThmCanonicalExtension} applied to the scaling with respect to  $\{p\}$ \textbf{uniformly} in the parameter $\ell\geqslant 0$.
Using the boundedness of the extension map $\mcR$, this concludes that $\mcR \Lambda\in \mcD^\prime(\Omega)$ is weakly homogeneous 
of degree $a_1$ with respect to  $\{p\}$.

\ssk

We can now conclude, choose a partition of unity $1={\varphi}_\Omega+(1-{\varphi}_\Omega)$ where ${\varphi}_\Omega\in C^\infty(\bbR^k\setminus \{0\})$ is scale invariant with respect to $\rho_1$, ${\varphi}_\Omega|_\Omega=1$ near $\mcY$ and ${\varphi}_\Omega=0$ outside $\Omega$, then note that the two families 
$$
\big( e^{\ell a_1} e^{-\ell\rho_1*} {\varphi}_\Omega\mcR\Lambda \big)_{\ell\geqslant 0}
$$
and 
$$
\big(e^{\ell a_1} e^{-\ell\rho_1*} \left(1-{\varphi}_\Omega\right) \Lambda \big)_{\ell\geqslant 0} 
$$ 
are bounded in $\mcD^\prime(\bbR^k\setminus \{0\})$. Hence applying the extension Theorem~\ref{ThmCanonicalExtension} to the sum ${\varphi}_\Omega\mcR\Lambda+ (1-{\varphi}_\Omega)\Lambda $ yields that $\Theta={\varphi}_\Omega\mcR\Lambda + (1-{\varphi}_\Omega)\Lambda\in \mcD^\prime(\bbR^k\setminus \{0\})$ has a canonical extension in $\mcD^\prime(\bbR^k)$ which is the result we wanted.
\end{Dem}

\ssk

\begin{Examples}
On {the Euclidean plane} $\mathbb{R}^2$, fix $p=(p_1,p_2)\in \mathbb{R}^2$ and let 
$$
\varphi(x_1, x_2) \defeq |x_1-x_2|^{a_2} \big( |x_1-p_1|+|x_2-p_2|\big)^{a_1-a_2}.
$$
It is weakly homogeneous of degree $a_2$ with respect to the diagonal $\{x_1=x_2\}$ and of degree $a_1$ with respect to the point $p$. Theorem \ref{Thm:extensiontwosteps} therefore applies to $\varphi$ provided $a_2>-1$ and $a_1>-2$.
\end{Examples}

\section{Kolmogorov-Chentsov and Feynman graphs}
\label{SectionKolmogorovFeynman}

We use a variant of the classical Kolmogorov-Chentsov regularity and convergence theorems to prove the convergence of the regularized and renormalized enhanced noise $\widehat{\xi}_r$. In its abstract form, given some Banach space $(E,\vert\cdot\vert)$, this statement ensures the convergence as $r>0$ goes to $0$ in $L^p(\Omega,\mathcal{F},\bbP)$ of a $C([0,T],E)$-valued family $(\Lambda_r)_{0<r\leqslant1}$ of random processes under the assumption that 
\begin{equation} \label{EqKolmogorovChentsov}
\big\Vert (\Lambda_{r_1}-\Lambda_{r_2})(t) - (\Lambda_{r_1}-\Lambda_{r_2})(s)\big\Vert_{L^p} \lesssim c(r_1,r_2)\,\vert t-s\vert^a
\end{equation}
for some positive finite exponents $p,a$ with $a>1/p$, and some positive constants $c(r_1,r_2)$ that tend to $0$ as $r_1+r_2$ goes to $0$. We will benefit in our setting from the fact that the different components of $\widehat{\xi}_r$ live in the Wiener chaoses generated by $\xi$ of degree at most $5$, so hypercontractivity can be used to trade some high moments of some real-valued random variables in these chaoses for some power of some second moment of these random variables. These second moments take the form of some finite sums of Feynman graphs, that is iterated integrals $\int_{D(s,t)} \prod_e K^e_{(r_1,r_2)}$ with regularization parameters $(r_1,r_2)$, where $ K^e_{(r_1,r_2)}$ are some distributional kernels indexed by the edges $e$, and where the product runs over the edges $e$ of some given graph and a domain of integration $D(s,t)$ that depends on $(s,t)$. An elementary telescopic sum argument shows that it suffices to prove some continuity bounds of the form
\begin{equation} \label{EqBoundsKernels}
\Big\vert \int_{D(s,t)} \prod_e K^e_{(r_1,r_2)}\Big\vert \lesssim \vert t-s\vert^b \prod_e \Vert K^e_{(r_1,r_2)}\Vert_e 
\end{equation}
for some positive exponent $b$, for some $e$-dependent natural notion of size $\Vert\cdot\Vert_e$ of each kernel, to obtain \eqref{EqKolmogorovChentsov} as a consequence of \eqref{EqBoundsKernels}. The constant $c(r_1,r_2)$ in \eqref{EqKolmogorovChentsov} is essentially the size in $\Psi^{\kappa}(M)$ for some small $\kappa>0$ of the difference of operators $e^{r_1\Delta}\circ e^{r_2\Delta}-\text{Id}$ viewed as pseudodifferential operators of order $\kappa>0$ acting on $M$, quantifying the convergence of the covariance of the space regularized spacetime white noise $\xi_r$ to its limit $\xi$. We will thus concentrate in the sequel on proving some bounds of the form \eqref{EqBoundsKernels}.

Similarly, a telescopic sum decomposition allows to bring back the estimate of some quantities of the form $\Vert\Lambda(t)-\Lambda(s)\Vert_{L^2}$ to some continuity estimate $\Big\vert \int \prod_e K^e \Big\vert \lesssim \prod_e \Vert K^e\Vert_e $ on some Feynman integral as one of the kernels will be of the form $K^e(t,\cdot)-K^e(s,\cdot)$, for which we will have an estimate $\Vert K^e(t,\cdot)-K^e(s,\cdot)\Vert_e \lesssim \vert t-s\vert^b$, for some $b$. 

This type of manipulations is classical, and almost all works only give the full details of the $(r,t)$-uniform estimates on $\Lambda_r(t)$ in $L^p$, as in Mourrat, Weber \& Xu's reference work \cite{MWX}. We will proceed similarly here. We illustrate this mechanics in this section on the example of the Wick monomials $\X, \begin{tikzpicture}[scale=0.3,baseline=0cm] \node at (0,0) [dot] (0) {}; \node at (0.3,0.6)  [noise] (noise1) {}; \node at (-0.3,0.6)  [noise] (noise2) {}; \draw[K] (0) to (noise1); \draw[K] (0) to (noise2); \end{tikzpicture}$ and $\IXthree$.

\medskip

\subsection{A Kolmogorov-Chentsov type argument}
\label{subsectionKolmo}

We start by giving the definitions of the Besov spaces  that we need for our analysis.

\ssk

\begin{defn} \label{def:besov}
    In the sequel, we fix $I$ a finite set and $(U_i,\kappa_i)_{i\in I}$ a cover of $M$ with $\kappa_i:U_i\rightarrow \bbR^3$, along with a quadratic partition of unity $({\varphi}_i)_{i\in I}$ on the closed manifold $M$ subordinated to $(U_i)_{i\in I}$.
    For any $\alpha\in\bbR$ and $p,q\in[1,\infty]$ we define the Besov space $B^\alpha_{pq}(M)$ on the closed manifold $M$ as the completion of $C^\infty(M)$ with the norm
\begin{align} \label{eq:def1Besovnorm}
    \|f\|_{B^\alpha_{pq}(M)}\defeq \max_{i\in I}\|\kappa_{i*}({\varphi_i} f)\|_{B^\alpha_{pq}(\bbR^3)}\defeq
    \max_{i\in I}\Big\|2^{k\alpha } \|\Delta_k\big(\kappa_{i*}({\varphi_i} f) \big)\|_{L^p(\bbR^3)}\Big\|_{\ell^q_k},
\end{align}
where the operators $\Delta_k$ for $k\geqslant-1$ are Littlewood-Paley blocks on the flat space $\bbR^3$.
\end{defn}

\ssk

We can rewrite this norm as follows. We fix a collection $(\psi_i)_{i\in I}$ of functions on $\bbR^3$ such that $\mathrm{supp}(\psi_i)\subset \kappa_i(U_i)$ and $\psi_i|_{\kappa_i(\mathrm{supp}({\varphi_i}))}=1$; there is enough room, since the support of ${\varphi_i}$ does not fill $U_i$. We will see below that we can deduce from results in the companion paper \cite{BDFTCompanion}
 that the norm defined in \eqref{eq:def1Besovnorm} is equivalent to 
\begin{align} \label{eq:def2Besovnorm}
     \max_{i\in I}\Big\|2^{k\alpha } \|\psi_i\Delta_k\big(\kappa_{i*}({\varphi_i} f) \big)\|_{L^p(\bbR^3)}\Big\|_{\ell^q_k}.
\end{align}
We therefore define for $i\in I$ and $k\geqslant-1$ some Littlewood-Paley blocks on the manifold $M$ by
\begin{align}  \label{eq:defLPblock}  
P^i_k(f)\defeq\kappa_i^*\big[\psi_i\Delta_k\big(\kappa_{i*}({\varphi_i} f) \big)\big].
\end{align}
With this notation in hand $\|f\|_{B^\alpha_{pq}(M)}$ is therefore equivalent to
\begin{align} \label{eq:def3Besovnorm}
     \max_{i\in I}\Big\|2^{k\alpha } \|P^i_k(\bigcdot)\|_{L^p(M)}\Big\|_{\ell^q_k}.
\end{align}
The equivalence of \eqref{eq:def1Besovnorm} with \eqref{eq:def2Besovnorm} is deduced from some argument in the proof of  \cite[Lemma 2.10]{BDFTCompanion}, where we control on $\bbR^3$ the commutator of a Littlewood-Paley block with multiplication by a smooth compactly supported function. Let us explain in more detail. For all smooth function $a$, $M_a$ denotes the multiplication operator by $a$. On $\bbR^3$, we would like to compare $M_\psi \Delta_k M_{\kappa_{i*}{\varphi_i}} $ with $\Delta_k M_{\kappa_{i*}{\varphi_i}}  $. Note that $\Delta_k M_{\kappa_{i*}{\varphi_i}}=\Delta_k M_\psi M_{\kappa_{i*}{\varphi_i}}  $ since $\psi=1$ on the support of $\kappa_{i*}{\varphi_i}$ hence the difference $M_\psi \Delta_k M_{\kappa_{i*}{\varphi_i}} -\Delta_k M_{\kappa_{i*}{\varphi_i}}  $ can be written as a commutator 
$  
[M_{\psi},\Delta_k ] M_{\kappa_{i*}{\varphi_i}} $.
By the argument in the proof of \cite[Lemma 2.10]{BDFTCompanion}, this composition is a smoothing operator in the semiclassical sense where $\hbar=2^{-k}$.
It follows that
$$ 
\Vert \Delta_k \kappa_{i*}({\varphi_i} f)- \psi_i \Delta_k \kappa_{i*}({\varphi_i} f)\Vert_{L^p} \lesssim 2^{-kN} \Vert f\Vert_{C^{-N}} 
$$
for all $N$. So we can absorb this error term in the semi norms which shows the equivalence of the two norms. In the sequel, to lighten the notation, we often let $B^\alpha_{p,q}$ stand for $B^\alpha_{pq}(M)$ and set 
$$
C^\alpha \defeq B^\alpha_{\infty \infty}(M).
$$
Propositions \ref{PropContinuityParaproductResonance}, \ref{PropLeibniz} and \ref{PropInterpolation} give some elementary properties of these spaces that we use.

\ssk

Similarly, for any vector field $X\in \Gamma(TM)$,  we can define
\begin{align} \label{eq:def1Besovnorm_vector}
    \|X\|_{B^\alpha_{pq}(M)}\defeq \max_{i\in I}\|\kappa_{i*}({\varphi_i} X)\|_{B^\alpha_{pq}(\bbR^3)}\defeq
    \max_{i\in I} \max_{\ell=1,2,3}\Big\|2^{k\alpha } \|\Delta_k\big(\kappa_{i*}({\varphi_i} X)\big)_\ell \|_{L^p(\bbR^3)}\Big\|_{\ell^q_k},
\end{align}
then define $\mcB_{p,q}^s(TM)$ and $C^s(TM)$. By definition, we have $\|\nabla f\| _{C^\beta(TM)}\lesssim  \| f\|_{C^{\beta+1}(M)} $ for all $\beta\in \mathbb{R}$.

\ssk

In the sequel, we denote by $\psi(|\bigcdot|)\in\mcS(\bbR)$ the Schwartz function defining the Schwartz kernel of $\Delta_0$ in $\bbR^3$: $\Delta_0(x-y):=\psi(|x-y|) $; the Fourier transform $\widehat{\Delta_0}$ is supported in an annulus. Note that for any $k\geqslant0$, the Schwartz kernel of $\Delta_k$ is given by
\begin{align*}
    \Delta_k(x-y)=2^{3k}\psi\big(2^k|x-y|\big)\,.
\end{align*}

\ssk

We are now ready to state our Kolmogorov-Chentsov type lemma. It turns out that due to the lack of stationarity of the stochastic objects on manifolds (as opposed to the flat case, or the sphere which are Riemannian symmetric spaces), the inequalities stated in Lemma~\ref{lem:Kolmo} below involve a point-wise bound {that is required} to be uniform in some base-point $x\in M$. The mechanics involved here is classical; we give the details for the reader's convenience.

\ssk

\begin{lem} \label{lem:Kolmo}
Fix $T>0$ and let $\Lambda$ be a random distribution in $\mathcal{D}^\prime([0,T]\times M)$ which belongs to the direct sum of finitely many Wiener chaos. 
\begin{itemize}
    \item[--] Assume one has 
\begin{eqnarray}\label{eq:kolmo1}
 \mathbb{E}\big[ (P^i_k\Lambda)(t,x)^2 \big] \lesssim2^{-2k \gamma_0}  
\end{eqnarray}
uniformly in $x\in M$, $i\in I$ and $k\geqslant0$ for some $\gamma_0\in\bbR$, for some time $0\leqslant t\leqslant T$. Then for every $\gamma<\gamma_0$ and $p>1$ we have
\begin{eqnarray*}
 \mathbb{E}[ \Vert \Lambda(t)\Vert_{C^\gamma}^p]\lesssim_p 1.   
\end{eqnarray*}

    \item[--] If, in addition to \eqref{eq:kolmo1}, the distribution $\Lambda$ further satisfies 
\begin{eqnarray}\label{eq:kolmo2}
\mathbb{E}\Big[ \big\vert (P^i_k\Lambda)(t_1,x) - (P^i_k\Lambda)(t_2,x)\big\vert^2 \Big] \lesssim 2^{-2k \gamma_0} \vert t_1-t_2\vert^{2\alpha_0}   
\end{eqnarray}
uniformly in $(t_1,t_2,x)\in[0,T]^2\times M$ and $(i,k)\in I\times\bbN$ for some $\gamma_0 \in \mathbb{R}, \alpha_0>0$ then for every $\gamma<\gamma_0$ and $\alpha<\alpha_0$ we have
\begin{eqnarray*}
\Lambda\in C^{\alpha}([0,T],C^\gamma(M)). 
\end{eqnarray*}    
\end{itemize}
\end{lem}

\ssk

\begin{Dem}    
We concentrate on the second item as the first item is proved by a similar reasoning. By Sobolev embedding by have
\begin{equation*} \begin{split}
\bbE\Big[ \big\| \Lambda(t_1)-\Lambda(t_2)\big\|^p_{C^\gamma} \Big] &\lesssim \bbE\Big[\big\|\Lambda(t_1)-\Lambda(t_2)\big\|^p_{B_{p,p}^{\gamma+3/p}}\Big]   \\
&\lesssim \max_{i\in I} \sum_{k\geqslant0}2^{pk(\gamma+3/p)}\int_M \bbE\Big[\big\vert(P^i_k \Lambda)(t_1,x) - (P^i_k \Lambda)(t_2,x)\big\vert^p \Big] \rmd x.
\end{split} \end{equation*}
Then the Gaussian hypercontractivity estimate implies that
\begin{align*}
      \bbE\Big[\big\|\Lambda(t_1)-\Lambda(t_2)\big\|^p_{C^\gamma}\Big] &\lesssim \max_{i\in I}\sum_{k\geqslant0}2^{pk(\gamma+3/p)}\int_M \bbE\Big[ \big\vert(P^i_k \Lambda)(t_1,x) - (P^i_k \Lambda)(t_2,x)\big\vert^2 \Big]^{\frac{p}{2}} \rmd x
        \\
    &\lesssim
    \max_{i\in I}\sum_{k\geqslant0}\sup_{x\in M}\Big( 2^{2k(\gamma+3/p)}\bbE\Big[\big\vert (P^i_k \Lambda)(t_1,x) - (P^i_k \Lambda)(t_2,x)\big\vert^2\Big] \Big)^{\frac{p}{2}}.
\end{align*}  
Therefore using the hypothesis \eqref{eq:kolmo2} we end up with
\begin{align*}
\bbE\Big[\big\|\Lambda(t_1)-\Lambda(t_2)\big\|^p_{C^\gamma}\Big]
\lesssim
\sum_{k\geqslant0} 2^{pk(\gamma+3/p-\gamma_0)}|t_1-t_2|^{p\beta}\lesssim |t_1-t_2|^{p \beta },
\end{align*}
which holds provided we choose $p$ large enough to have $\gamma<\gamma_0-3/p$. The proof now follows using a classical Kolmogorov argument in time.

\end{Dem}

\ssk

As an echo to the introduction of Section \ref{SectionKolmogorovFeynman}, we add an important remark. Assume we have a family $(\Lambda_r)_{r>0} $ of random distributions that belongs to a fixed finite sum of Wiener chaoses, for which there exists some positive constants $c(r_1,r_2)$ that go to $0$ as $r_1+r_2$ goes to $0$, and such that 
\begin{itemize}
	\item[--] one has $\mathbb{E}\big[ (P^i_k(\Lambda_{r_1}-\Lambda_{r_2}))(t,x)^2 \big] \lesssim c(r_1,r_2) \, 2^{-2k \gamma_0}$, uniformly in $x\in M$, $i\in I$ and $k\geqslant0$ for some $\gamma_0\in\bbR$, for some time $0\leqslant t\leqslant T$,   \vspace{0.1cm} 
	\item[--] one has $\mathbb{E}\big[\big( P^i_k(\Lambda_{r_1}-\Lambda_{r_2})(t_1,x) - P^i_k(\Lambda_{r_1}-\Lambda_{r_2})(t_2,x)\big)^2 \big] \lesssim c(r_1,r_2) \,  2^{-2k \gamma_0} \vert t_1-t_2\vert^{2\alpha_0}$, uniformly in $(t_1,t_2,x)\in[0,T]\times[0,T]\times M$, $(i,k)\in I\times\bbN$ for some $\alpha_0>0$.
\end{itemize}  
Then for every $\gamma<\gamma_0$ and $\alpha<\alpha_0$, we have that the family $(u_r)_{r>0} $ is Cauchy and converges in $L^p(\Omega,\mathcal{F},\bbP)$ to some limit random variable $u \in C^\alpha([0,T],C^\gamma(M))$; the proof is almost verbatim what we wrote above.

\ssk

We denote by $P^i_k(x,y)$ the smooth Schwartz kernel of the operator $P^i_k$.

\ssk

\begin{defn} \label{def:Qgamma}
 We introduce a bilinear kernel depending on a {base} point $x\in M$ and $\gamma\in\bbR$ allowing to probe the regularities of some objects {defined on $M$} as follows
\begin{eqnarray} \label{EqDefnQGamma}
\mcQ^\gamma_x (y_1,y_2) \defeq \sum_{i\in I} \sum_{k\geqslant-1} 2^{2k\gamma} P_k^i (x,y_1)     P_k^i (x,y_2)  \,.
\end{eqnarray}
\end{defn}

\ssk

Note that if there exists $\gamma\in \mathbb{R}$ such that one has
$$ 
\mathbb{E}\Big[\mcQ_x^\gamma\big(\Lambda(t,\cdot) , \Lambda(t,\cdot)\big)\big] = \mathbb{E}\bigg[ \sum_{k\geqslant-1} 2^{2k\gamma} (P^i_k\Lambda)(t,x)^2\bigg] = \sum_{k\geqslant-1} 2^{2k\gamma} \mathbb{E}\big[(P^i_k\Lambda)(t,x)^2\big] < \infty
$$
uniformly in $x\in M$, then we have {the estimate \ref{eq:kolmo1}:}
$$ 
\mathbb{E}\big[ (P^i_k \Lambda) (t,x)^2 \big] \lesssim 2^{-2k\gamma}
$$
uniformly in $k\geqslant-1$. The microlocal properties of the kernel $\mcQ^\gamma_x$ are stated in the following lemma.

\ssk

\begin{lem}
Fix $x\in M$ and $\gamma\in\bbR$. The series defining $\mcQ^\gamma_x$ converges in $C^\infty(M^2\setminus \{(x,x)\})$ and in the space $\mathcal{D}^\prime_{\Gamma_x}(M\times M)$ of distributions with wave front set 
$$
\Gamma_x = T^*_{(x,x)} M^2.
$$
Moreover it is weakly homogeneous of degree $-6-2\gamma$ with respect to the scaling towards $\mcY = (x,x)$, that is to say we have $\mcQ^\gamma_x\in\mcS_{\Gamma_x}^{-6-2\gamma}$ with a control uniform in $x$.
\end{lem}

\ssk

\begin{Dem}
For the convergence in  $C^\infty\big(M^2\setminus \{(x,x)\}\big)$, in a local chart indexed by $i\in I$, the kernel of $\mcQ^\gamma_x$ is given by
\begin{eqnarray*}
\sum_{k\geqslant0}2^{k(6+2\gamma)}  \psi_i^2(\widetilde x) (\kappa_{i*}{\varphi_i})^{\otimes2}(y_1,y_2) \, \psi\big(2^k| y_1-\widetilde x |\big) \, \psi\big(2^k| y_2-\widetilde x |\big), 
\end{eqnarray*}
where $\psi\circ|\bigcdot|$ is a Schwartz function on $\bbR$ with Fourier transform $\eta$ supported on an annulus, and $\widetilde x\defeq \kappa_i(x)$. For $y_i\neq \widetilde x$  we have for any $N\geqslant0$ and any $\alpha\in\bbN^3$
$$ 
\big\vert \partial_{y_i}^\alpha \psi\big(2^k| y_i-\widetilde x |\big) \big\vert \lesssim 2^{k\vert\alpha\vert} \big(1+2^k\vert y_i-\widetilde x \vert \big)^{-N} \lesssim 2^{k\vert\alpha\vert-Nk}.
$$
Therefore choosing $N$ large enough ensures the convergence of the series in $C^\infty$.

We now turn to the proof of the convergence of the series in the sense of distributions in  $\mathcal{D}_{\Gamma_x}^\prime(M\times M)$. We use different arguments depending on the sign of $\gamma$. 

$\bullet$ Suppose that $\gamma<0$. Then we have $\big\Vert  2^{3k}\psi\big(2^k| \bigcdot-\widetilde x |\big) \big\Vert_{L^1}=\Vert \psi \Vert_{L^1} $ and the series \eqref{EqDefnQGamma} converges absolutely in $L^1$ since
\begin{align*}
\bigg\Vert \sum_{k\geqslant0} 2^{k(6+2\gamma)} \psi_i^2(\widetilde x) (\kappa_{i*}{\varphi_i})^{\otimes2}(y_1,y_2) \, \psi\big(2^k| y_1-\widetilde x |\big) \, &\psi\big(2^k|y_2-\widetilde x|\big) \bigg\Vert_{L^1_{y_1,y_2}}   \\
&\lesssim \sum_{k\geqslant0} 2^{k\gamma} \Vert \psi\Vert_{L^1}^2  \Vert \psi_i\Vert^2_{L^\infty}\Vert {\varphi_i}\Vert_{L^\infty}^2\lesssim1\,.
\end{align*}

$\bullet$ Suppose now that $\gamma\geqslant 0$. We must prove that the series converges in the distributional sense. For every test function $\varphi\in C^\infty_c(\mathbb{R}^3\times \mathbb{R}^{3})$ we must prove the convergence of the series
\begin{equation*} \begin{split}
&\sum_{k\geqslant0} 2^{k(6+2\gamma)} \int_{(\mathbb{R}^3)^2} \varphi(y_1,y_2) \psi_i^2(\widetilde x) (\kappa_{i*}{\varphi_i})^{\otimes2}(y_1,y_2) \psi\big(2^k| y_1-\widetilde x |\big) \psi\big(2^k| y_2-\widetilde x|\big) \rmd y_1\rmd y_2   \\
&= \sum_{k\geqslant0} 2^{2k\gamma} \int_{(\mathbb{R}^3)^2} \varphi\big( 2^{-k} (y_1-\widetilde x)+\widetilde x, 2^{-k}(y_2-\widetilde x)+\widetilde x\big)   \\
&\hspace{2cm}\times \psi_i^2(\widetilde x) (\kappa_{i*}{\varphi_i})^{\otimes2}\big( 2^{-k} (y_1-\widetilde x)+\widetilde x,2^{-k} (y_2-\widetilde x)+\widetilde x\big) \, \psi\big(| y_1-\widetilde x| \big) \, \psi\big(|y_2-\widetilde x|\big) \rmd y_1\rmd y_2.
\end{split} \end{equation*}
We rename by $\widetilde\varphi$ the test function 
$$
\widetilde\varphi(y_1,y_2)\defeq\varphi(y_1,y_2) (\kappa_{i*}{\varphi_i})^{\otimes2}(y_1,y_2).
$$ 
The series therefore reads
\begin{align*}
\sum_{k\geqslant0} 2^{2k\gamma} \int_{(\mathbb{R}^3)^2} \widetilde{\varphi} \big( 2^{-k} (y_1-\widetilde x)+\widetilde x, 2^{-k}(y_2-\widetilde x)+\widetilde x\big) \psi\big(| y_1-\widetilde x| \big) \, \psi\big(|y_2-\widetilde x|\big) \, \psi_i^2(\widetilde x) \rmd y_1\rmd y_2\,.
\end{align*}
Note that the product $\psi\big( |y_1-\widetilde x| \big) \, \psi\big(|y_2-\widetilde x|\big) \, \psi_i^2(\widetilde x)$ is a Schwartz function in $y_1,y_2$ uniformly in $\widetilde x$. The idea is to Taylor expand 
$\widetilde\varphi$ about $(\widetilde x,\widetilde x)$ at order $N$ with integral remainder. This yields
$$ \widetilde\varphi=P_N+ R_N\,,\; R_N(y_1,y_2)=\mathcal{O}\big( (|y_1-\widetilde x|\vee|y_2-\widetilde x|)^{N+1}\big)\,, $$
where $P_N$ is a polynomial function of $(y_1,y_2)$ centered at $(\widetilde x,\widetilde x)$. The rescaled $\widetilde\varphi$ reads
\begin{align*}
\widetilde\varphi&\Big( 2^{-k} (y_1-\widetilde x)+\widetilde x, 2^{-k}(y_2-\widetilde x)+\widetilde x\Big)   \\
&=P_N\Big( 2^{-k} \big(y_1-\widetilde x)+\widetilde x, 2^{-k}(y_2-\widetilde x)+\widetilde x\Big) + R_N\Big( 2^{-k} (y_1-\widetilde x)+\widetilde x, 2^{-k}(y_2-\widetilde x)+\widetilde x\Big),
\end{align*}  
where $R_N=\mathcal{O}\big(2^{-k(N+1)}\big)$ depends on the jets of order $N+1$ of $\widetilde\varphi$. Injecting the above decomposition in our series yields
$$
\sum_{k\geqslant0} 2^{2k\gamma} \int_{(\mathbb{R}^3)^2} (P_N+R_N) \Big( 2^{-k} (y_1-\widetilde x)+\widetilde x, 2^{-k}(y_2-\widetilde x)+\widetilde x\Big) \, \psi\big(|y_1-\widetilde x|\big) \, \psi\big(|y_2-\widetilde x|\big) \, \psi_i^2(\widetilde x) \rmd y_1\rmd y_2.
$$
Now we make use of the fact that we have
$$
\int_{(\mathbb{R}^3)^2} P_N\Big( 2^{-k} (y_1-\widetilde x)+\widetilde x, 2^{-k}(y_2-\widetilde x)+\widetilde x\Big) \, \psi\big( |y_1-\widetilde x| \big) \, \psi\big(|y_2-\widetilde x|\big) \, \psi_i^2(\widetilde x) \rmd y_1\rmd y_2 = 0, 
$$
since the support of $\eta$ is included in an annulus, which implies that its integral against all polynomials vanish. The series therefore simplifies as
\begin{equation*} \begin{split}
\sum_{k\geqslant0} 2^{2k\gamma} \int_{(\mathbb{R}^3)^2} &\underset{\mathcal{O}(2^{-k(N+1)})} {\underbrace{ R_N  \big( 2^{-k} (y_1-\widetilde x)+\widetilde x, 2^{-k}(y_2-\widetilde x)+\widetilde x\big)} } \, \psi\big( |y_1-\widetilde x|\big) \, \psi\big(|y_2-\widetilde x|\big) \, \psi_i^2(x)\, \rmd y_1\rmd y_2   \\ 
&\lesssim \sum_{k=0}^\infty 2^{k(2\gamma-N-1)} \sum_{\vert\alpha\vert =N+1} \Vert \partial^\alpha \widetilde\varphi \Vert_{L^1},
\end{split} \end{equation*}
where the series on the right hand side converges as soon as $N+1>2\gamma$ and we control the convergence  from the $N+1$ derivatives of $\widetilde\varphi$ hence $\eta$. We just proved that $\mcQ^\gamma_x$ converges as distribution of order $N+1$ for all $N+1>2\gamma$.  

Finally we need to control the weak homogeneity when we scale towards $(x,x)$. Working in a chart $i\in I$ this reduces to 
estimate the weak homogeneity near $(x,x)$ of the series
$$ 
U \defeq  \sum_{k\geqslant 0} 2^{k(6+2\gamma)} \psi\big(2^k|y_1-\widetilde x|\big)\psi\big(2^k|y_2-\widetilde x|\big) \in \mathcal{S}^\prime(\mathbb{R}^3\times \bbR^3).
$$
By duality it suffices to estimate the weak homogeneity at $\infty$ of its Fourier transform
\begin{equation*} \begin{split}
\widehat U(\xi_1,\xi_2) &= \sum_{k\geqslant 0} 2^{k(6+2\gamma)} \mcF\big[\psi\big(2^k|\bigcdot-\widetilde x|\big)\big](\xi_1) \; \mcF\big[ \psi\big(2^k|\bigcdot-\widetilde x|\big)\big](\xi_2)   \\
&= \sum_{k\geqslant 0} 2^{2k\gamma} e^{\imath \widetilde x\cdot(\xi_1+\xi_2)} \, \eta (2^{-k}\xi_1) \, \eta(2^{-k} \xi_2).    
\end{split} \end{equation*}
Here we can use the fact that $\eta (2^{-k}\xi)$ is zero unless $2^{k-1}\leqslant \vert \xi\vert \leqslant 2^{k+1}$, so that both $\xi_1$ and $\xi_2$ are localized at scale $2^j$ for some $j\geqslant0$, which yields the bound
$$ 
\bigg\vert \sum_{k\geqslant 0} 2^{2k\gamma}  e^{\imath \widetilde{x}\cdot(\xi_1+\xi_2)} \eta (2^{-k}\xi_1) \eta (2^{-k} \xi_2) \bigg\vert = \mathcal{O}\big(\max(|\xi_1|^{2\gamma} , |\xi_2|^{2\gamma})\big)
$$
uniform in $\widetilde{x}$, hence the Fourier transform $\widehat U$ is such that the family of tempered distributions $\big(\lambda^{-2\gamma} \widehat U(\lambda \,\bigcdot)\big)_{\lambda\geqslant 1}$ is bounded. By Plancherel this entails that the family of distributions 
$$
\big(\lambda^{-2\gamma-6}U(\lambda\bigcdot + (\widetilde{x},\widetilde{x}))\big)_{\lambda \in (0,1]}
$$ 
is bounded too, which concludes the proof.
\end{Dem}

\subsection{The Wick monomials}
\label{SubsectionWick}

\subsubsection{The elementary term $\X$.}

As an example of direct application of the Kolmogorov-Chentsov lemma \ref{lem:Kolmo}, we discuss below the regularity of the dynamical free field.

\ssk

\begin{lem}\label{lem_X}
For any $T>0$ and $\epsilon>0$ we have $\X\in C_T\mcC^{-1/2-\epsilon}$.
\end{lem}

\ssk

\begin{Dem}
We start by performing the estimate at some fixed time $t>0$. By stationary in time of $\X$ one has for any $x\in M$ the $t$-independent bound
$$
\bbE\big[\mcQ_x^\gamma\big(\X(t,\bigcdot),\X(t,\bigcdot)\big)\big] \lesssim \int_{M^2} \mcQ^\gamma_{x} (y_1,y_2)P^{-1}(y_1,y_2) \rmd v(y_1)\rmd v(y_2)\,.
$$
Next we use the representation \eqref{eq:defLPblock} of the Schwartz kernel of $P_k^i$, which implies that pushing the estimate in the chart $\kappa_i(U_i)$ we have

$$
\bbE\big[ (P_k^i \X)(t,x)^2 \big] \lesssim 2^{6k}\int_{(\bbR^3)^2} \psi_i(\widetilde x)^2 \psi\big(2^k|y_1-\widetilde x|\big) \psi\big(2^k|y_2-\widetilde x|\big)(\kappa_{i*}{\varphi_i})^{\otimes2}(y_1,y_2) \kappa^{\otimes2}_{i*}P^{-1} (y_1,y_2) \rmd y_1\rmd y_2
$$
where we denote $\widetilde x\defeq \kappa_i(x)$. Also, denoting by $\kappa^{\otimes2}_{i*}P^{-1}=:P_{i,i}^{-1}$ the pulled--back Green function, we obtain
\begin{equation*} \begin{split}
\bbE\big[ (P_k^i \X) (t,x)^2 \big]  \lesssim \int_{(\mathbb{R}^3)^2} \psi_i(\widetilde x)^2 \, \psi\big( |y_1-\widetilde x|\big) \, &\psi\big( |y_2-\widetilde x|\big) \, (\kappa_{i*}{\varphi_i})^{\otimes2}\Big(2^{-k}(y_1-\widetilde x)+\widetilde x,2^{-k}(y_2-\widetilde x)+\widetilde x\Big)   \\
&\qquad\times  P_{i,i}^{-1}\big(2^{-k}(y_1-\widetilde x)+\widetilde x,2^{-k}(y_2-\widetilde x)+\widetilde x\big)  \rmd y_1\rmd y_2\,.
\end{split} \end{equation*}
Since $\widetilde x$ is chosen uniformly in a compact set, $\psi\big( |y_1-\widetilde x|\big),\psi\big( |y_2-\widetilde x|\big)$ are Schwartz in $y_1,y_2$ and $P_{i,i}^{-1}\big(2^{-k}(y_1-\widetilde x)+\widetilde x,2^{-k}(y_2-\widetilde x)+\widetilde x\big)$ is integrable in $y_1,y_2$ near $y_1=y_2$, so the second integral is well-defined. Using the bound on the rescaled Green function
$$
P_{i,i}^{-1}\Big(2^{-k}(y_1-\widetilde x)+\widetilde x,2^{-k}(y_2-\widetilde x)+\widetilde x\Big) \lesssim 2^{k} \vert y_1-y_2 \vert^{-1}, 
$$
we can now conclude that we have 
\begin{eqnarray*}
\bbE\big[ (P_k^i \X)(t,x)^2 \big]\lesssim 2^k 
\int_{(\mathbb{R}^3)^2} \big\vert \psi\big( |y_1-\widetilde x|\big)\psi\big( |y_2-\widetilde x|\big)\big\vert \, \vert y_1-y_2\vert^{-1} \, \rmd y_1\rmd y_2  \lesssim 2^k.
\end{eqnarray*}
-- We now show that $\X$ is indeed continuous in time. To do so, let us first introduce the notation
$$
G^{(1)}_{i,i}(t_1-t_2,x,y)\defeq \kappa_{i*}^{\otimes2} \bigg( \frac{ e^{-\vert t_1-t_2 \vert P} }{P} \bigg) (x,y).  
$$
The quantity $\mathbb{E}\big[\big(P^i_k \X(t_1,x)-P^i_k \X(t_2,x)\big)^2\big]$ is bounded above by
\begin{equation} \label{EqCalculIntermediaire} \begin{split}
\bigg\vert   
&\int_{(\mathbb{R}^3)^2} \psi_i(\widetilde x)^2 \,\psi\big( |y_1-\widetilde x|\big) \, \psi\big( |y_2-\widetilde x|\big)  (\kappa_{i*}{\varphi_i})^{\otimes2}\Big(2^{-k}(y_1-\widetilde x)+\widetilde x,2^{-k}(y_2-\widetilde x)+\widetilde x\Big)   \\
&\times
\Big(P_{i,i}^{-1}\big(2^{-k}(y_1-\widetilde x)+\widetilde x,2^{-k}(y_2-\widetilde x)+\widetilde x\big)-G^{(1)}_{i,i}\big(t_1-t_2,2^{-k}(y_1-\widetilde x)+\widetilde x,2^{-k}(y_2-\widetilde x)+\widetilde x\big)\Big) \bigg\vert
\end{split} \end{equation}
for an integral with respect to $\rmd y_1\rmd y_2$. Our goal is to bound the difference of kernels $P^{-1}_{i,i}-G^{(1)}_{i,i}(r,\bigcdot)$ for $r$ small. 
To do so, we use the identity
$$ 
\frac{1-e^{-rP}}{P}  =  \int_0^r e^{-sP}  \rmd s    
$$
which expresses the difference as an integral of the heat operator. Then we can inject in the integral expression the bound on the heat kernel on the product chart 
$$ 
\big| e^{-sP} (y_1,y_2) \big| \lesssim s^{-3/2}e^{-s} e^{ -C\vert y_1-y_2\vert^2/s }, 
$$  
where $C>0$. We obtain the estimate for any $\beta\in(0,1)$ {and any $0<r\leqslant1$}
\begin{align}\label{eq:estimate_time_continuity}\nonumber
\Big\vert P^{-1}_{i,i}(y_1,y_2)-G^{(1)}_{i,i}(r,y_1,y_2)  \Big\vert &\lesssim    
\int_0^r s^{-3/2} e^{ -C\vert y_1-y_2\vert^2/s -s}  \rmd s\lesssim\int_{r^{-1}}^\infty e^{ -Cs\vert y_1-y_2\vert^2 }e^{-\frac{1}{s}} s^{-\frac{1}{2}}\rmd s \\ \nonumber
&\lesssim \vert y_1-y_2\vert^{-1} \int_{r^{-1}\vert y_1-y_2\vert^2}^\infty e^{ -Cs-\frac{1}{s} } s^{-\frac{1}{2}}\rmd s
\lesssim \vert y_1-y_2\vert^{-1}  (e^{-\frac{\vert y_1-y_2\vert^2}{r}}) \\ &\lesssim  \vert y_1-y_2\vert^{-1} (1+\frac{\vert y_1-y_2\vert^2}{r})^{-N}\lesssim  \vert y_1-y_2\vert^{-1-\beta} r^{\frac{\beta}{2}}.
\end{align}
Replacing in the difference estimate \eqref{EqCalculIntermediaire} finally yields
$$
\mathbb{E}\Big[\big((P^i_k \X)(t_1,x) - (P^i_k \X)(t_2,x)\big)^2 \Big] \lesssim 2^{k(1+\beta)} \vert t_1-t_2\vert^{\frac{\beta}{2}},
$$
{so we can complete the proof with Lemma \ref{lem:Kolmo}.}
\end{Dem}

\ssk

Together with Kolmogorov's classical regularity theorem what we said above justifies that $\begin{tikzpicture}[scale=0.3,baseline=0cm] \node at (0,0)  [dot] (1) {}; \node at (0,0.8)  [noise] (2) {}; \draw[K] (1) to (2); \end{tikzpicture}$ takes almost surely its values in $C^\alpha([0,T],C^{-1/2-\epsilon}(M))$, for all small enough $\alpha>0$ and all $\epsilon>0$, and its norm in the corresponding space has moments of any finite order.

\ssk

\begin{Rem}
In the sequel we will not dwell on the proof of the estimate \eqref{eq:kolmo2} for larger trees. However it is by now classical that when computing in charts the covariance of $(P^i_ku)(t_1,x) - (P^i_ku)(t_2,x)$ the amplitudes appearing can be reorganised by means of some telescopic sums in order to introduce difference terms of the form $P^{-1}_{i,i}-G^{(1)}_{i,i}(r,\bigcdot)$. In view of \eqref{eq:estimate_time_continuity} this term always yields a good factor in time modulo a small loss in spatial regularity.
\end{Rem}

\subsubsection{Locally covariant Wick renormalization.}
\label{ssslocallycovariant}

In the sequel we will systematically first estimate regularities using Wick renormalization. Then we shall compare the usual Wick renormalization which is not {\em locally covariant} with a  {\em locally covariant} renormalization in which we only subtract {\sl universal quantities} at the cost of producing objects which do not belong to homogeneous Wiener chaoses.  Let us recall  from Definition \ref{def:localcounterterms} the notion of {\em local covariance} with a simple example which also explains why the usual Wick renormalization fails to be locally covariant.
In the usual Wick renormalization for some massive Gaussian free field $\phi$ on some Riemannian $3$-manifold $(M,g)$ with covariance $P^{-1}$, one first mollifies $\phi$ via heat regularization. This yields a random smooth function $\phi_r\defeq e^{-rP}\phi$. To renormalize the square $\phi_r^2$, one subtracts to the square $(\phi_r)^2(x)$ of the mollified field at $x$, the counterterm $a_r(x)=e^{-2rP}P^{-1}(x,x)$ and it is well-known that the difference $(\phi_r)^2(\cdot)- a_r(\cdot)$ will converge as random distribution when $r\downarrow 0$. However, the counterterm $a_r(x)=e^{-2rP}P^{-1}(x,x)$ that we subtracted is \textbf{nonlocal} in the metric $g$ at $x$, it depends on the global Riemannian geometry of $(M,g)$ and not on finite jets of the metric $g$ at $x$. Hence such $a_r$ is not {\em locally covariant} in the above sense.
Now we observe that the diagonal value $e^{-2rP}P^{-1}(x,x)$ has an asymptotic expansion of the form:
$$ e^{-2rP}P^{-1}(x,x)\sim \frac{1}{4\pi^{\frac{3}{2}}r^{\frac{1}{2}}}+\mathcal{O}(1).$$
If instead of subtracting the diagonal value of $e^{-2rP}P^{-1}$ one subtracted its \textbf{singular part}: $(\phi_r)^2(\bigcdot)- \frac{1}{4\pi^{\frac{3}{2}}r^{\frac{1}{2}}}$, then one would still get a random distribution at the limit when $r\downarrow 0$ but the covariantly renormalized Wick square $:\phi^2:$ would no longer have zero expectation.
So one may wonder why is it so important to subtract only locally covariant quantities?
The answer lies in the deep notion of locality in quantum field theory.
It is a folklore result in quantum fields on curved backgrounds that subtracting non locally covariant counterterms is incompatible with locality in the sense of Atiyah-Segal. Let us quote the beautiful discussion on the regularization of tadpoles and the relation with Atiyah-Segal gluing in~\cite[1.2 p.~1852]{KandelMnevWernli}:
"{\em In various treatments of scalar theory, tadpole diagrams were set to zero
(this corresponds to a particular renormalization scheme – in flat space,
this is tantamount to normal ordering, \dots ). However, in our
framework this prescription contradicts locality in Atiyah-Segal sense,\dots One good solution is to prescribe to the tadpole diagrams the
zeta-regularized diagonal value of the Green’s function. We prove that assigning to a surface its zeta-regularized tadpole is compatible with locality,\dots
However there are other consistent prescriptions (for
instance, the tadpole regularized via point-splitting and subtracting the singular term, \dots). This turns out to be related to Wilson’s idea
of RG flow in the space of interaction potentials,\dots
}"
We refer the interested reader to~\cite[Section 5 p.~1885]{KandelMnevWernli} for further details on this central topic of quantum fields on curved backgrounds.
So in the present paper, we follow a similar strategy as in the previous example and try to subtract only locally covariant quantities, in fact we shall see that we subtract universal quantities that do not even depend on the metric $g$.

\subsubsection{The quadratic term.}
\label{subsect_quadratic}

The argument used for the study of $\begin{tikzpicture}[scale=0.3,baseline=0cm] \node at (0,0)  [dot] (1) {}; \node at (0,0.8)  [noise] (2) {}; \draw[K] (1) to (2); \end{tikzpicture}$, without regularization, works for the study of $\begin{tikzpicture}[scale=0.3,baseline=0cm] \node at (0,0) [dot] (0) {}; \node at (0.3,0.6)  [noise] (noise1) {}; \node at (-0.3,0.6)  [noise] (noise2) {}; \draw[K] (0) to (noise1); \draw[K] (0) to (noise2); \end{tikzpicture}_r$, with regularization. Since $\xi_r$ is regularized in space, $\begin{tikzpicture}[scale=0.3,baseline=0cm] \node at (0,0)  [dot] (1) {}; \node at (0,0.8)  [noise] (2) {}; \draw[K] (1) to (2); \end{tikzpicture}_r$ is here a function on $[0,T]\times M$. Set
$$
a_r(z) \defeq \bbE\big[\begin{tikzpicture}[scale=0.3,baseline=0cm] \node at (0,0)  [dot] (1) {}; \node at (0,0.8)  [noise] (2) {}; \draw[K] (1) to (2); \end{tikzpicture}_r^2(z)\big], \qquad
\overline{\begin{tikzpicture}[scale=0.3,baseline=0cm] \node at (0,0) [dot] (0) {}; \node at (0.3,0.6)  [noise] (noise1) {}; \node at (-0.3,0.6)  [noise] (noise2) {}; \draw[K] (0) to (noise1); \draw[K] (0) to (noise2); \end{tikzpicture}}_r(z) \defeq \begin{tikzpicture}[scale=0.3,baseline=0cm] \node at (0,0)  [dot] (1) {}; \node at (0,0.8)  [noise] (2) {}; \draw[K] (1) to (2); \end{tikzpicture}_r^2(z) - a_r(z).
$$
To treat the regularity of {$\overline\Xtwo_r$}, we will follow a similar strategy as for $\X$ and try to control
$$ 
\mathbb{E}\big[ (P_k^i {\overline\Xtwo_r})(t,x)^2\big]
$$
when $i$ becomes large and uniformly in $x\in M$.
 Denote by $G_r^{(2)}(t-s)$ the operator with kernel 
$$
\bigg(\frac{e^{-(2r+\vert t-s\vert)P}}{P}(x,y)\bigg)^2.
$$ 
The operators $G_r^{(2)}(0)$ take values in $\Psi^{-1}(M)$ and $G_r^{(2)}(t)$ is of order $\vert t\vert^{\gamma/2}$ in $\Psi^{-1 +\gamma}(M)$ since $e^{-\vert t\vert P}P^{-1}$ is of order $\vert t\vert^\gamma$ in $\Psi^{-2 + 2\gamma}(M)$. This holds uniformly in $r\in [0,1]$. By definition, and using the Wick formula, we can compute the quantity $\mathbb{E}[ P_k^i {\overline\Xtwo_r}(t,x)^2  ]$, which is equal to

\makebox[\textwidth][c]{
\begin{minipage}{\dimexpr\textwidth+10cm}
\begin{align*} 
\mathbb{E}\big[(P_k^i{\overline\Xtwo_r})(t,x)^2 \big] &= \int_{M^2{ \times [0,t]^2}} { G_r^{(2)}}(t_1-t_2,y_1,y_2)P_k^i(x,y_1)P_k^i(x,y_2) \, \rmd v(y_1)\rmd v(y_2) \rmd t_1\rmd t_2   \\
&\lesssim \int_{(\bbR^3)^2{ \times [0,t]^2}} { G^{(2)}_{r,ii}}(t_1-t_2,y_1,y_2)\psi_i(\widetilde x)^2 \psi\big( |y_1-\widetilde x|\big)\psi\big( |y_2-\widetilde x|\big)   \\
&\hspace{2.2cm}\times (\kappa_{i*}{\varphi_i})^{\otimes2}\big(2^{-k}(y_1-\widetilde x)+\widetilde x,2^{-k}(y_2-\widetilde x)+\widetilde x\big)   \\
&\lesssim 2^{2k} \int_{(\mathbb{R}^3)^2}  \vert y_1-y_2 \vert^{-2}\psi_i(\widetilde x)^2 \psi\big( |y_1-\widetilde x|\big)\psi\big( |y_2-\widetilde x|\big)   \\
&\hspace{2.2cm}\times   (\kappa_{i*}{\varphi_i})^{\otimes2}\big(2^{-k}(y_1-\widetilde x)+\widetilde x,2^{-k}(y_2-\widetilde x)+\widetilde x\big),
\end{align*}
\end{minipage}
}
\esp

\noindent for some integrals over $(\bbR^3)^2$ with respect to $\rmd y_1\rmd y_2$, and where again $\widetilde x=\kappa_i(x)$ and ${ G^{(2)}_{r,ii}} = \kappa_{i*}^{\otimes2}{G^{(2)}_r}$. We used the {$r$ and $t_1,t_2$-uniform} estimate 
$$ 
\big\vert G^{(2)}_{r,ii}(t_1-t_2,x_1,x_2)\big\vert \lesssim \vert x_1-x_2 \vert^{-2}
$$
and the changes of variables 
$$
y_1\mapsto 2^{-i}(y_1-\widetilde x)+\widetilde x\,,\; y_2\mapsto 2^{-i}(y_2-\widetilde x)+\widetilde x.
$$
We can now conclude since the integral 
\begin{align*}
\int_{(\mathbb{R}^3)^2}  \vert y_1-y_2 \vert^{-2} \, \psi_i(\widetilde x)^2 \, \psi\big( |y_1-\widetilde x|\big) \, \psi\big( |y_2-\widetilde x|\big) \, (\kappa_{i*}{\varphi_i})^{\otimes2}\big(2^{-k}(y_1-\widetilde x)+\widetilde x,2^{-k}(y_2-\widetilde x)+\widetilde x\big)
\end{align*}
with respect to $\rmd y_1\rmd y_2$ is bounded uniformly in $x\in \text{supp}(\psi_i)$ because $\psi\big(|y-\widetilde x|\big)$ is a Schwartz function of $y$.

Some similar estimate are proved to conclude from Kolmogorov-Chentsov regularity/convergence theorem that $\overline{\begin{tikzpicture}[scale=0.3,baseline=0cm] \node at (0,0) [dot] (0) {}; \node at (0.3,0.6)  [noise] (noise1) {}; \node at (-0.3,0.6)  [noise] (noise2) {}; \draw[K] (0) to (noise1); \draw[K] (0) to (noise2); \end{tikzpicture}}_r \in C_T^\alpha C^{-1-\epsilon}(M)$ for some $0<\alpha\leqslant1$ and all $\epsilon>0$, uniformly in $r\in(0,1]$, and we further get from the $r$-uniform and continuity of the different quantities as functions of $r$ the convergence of $\overline{\begin{tikzpicture}[scale=0.3,baseline=0cm] \node at (0,0) [dot] (0) {}; \node at (0.3,0.6)  [noise] (noise1) {}; \node at (-0.3,0.6)  [noise] (noise2) {}; \draw[K] (0) to (noise1); \draw[K] (0) to (noise2); \end{tikzpicture}}_r $ in $C_T^\alpha C^{-1-\epsilon}(M)$ to a limit which we denote by $\overline{\begin{tikzpicture}[scale=0.3,baseline=0cm] \node at (0,0) [dot] (0) {}; \node at (0.3,0.6)  [noise] (noise1) {}; \node at (-0.3,0.6)  [noise] (noise2) {}; \draw[K] (0) to (noise1); \draw[K] (0) to (noise2); \end{tikzpicture}}$.

\ssk

We emphasized in Section \ref{ssslocallycovariant} that we define in this work the \textbf{renormalization in a locally covariant way} with respect to the Riemannian metric $g$. Therefore, we shall subtract from $\X^2_r$ only the \textbf{singular part} of the {\sl function} $a_r(z)=\mathbb{E}[\X^2_r (z)]$. We actually prove that this singular part $a_r$ is a universal constant. The {\sl function} $a_r(z)$ differs from the {\sl constant} $a_r$, but for $z=(t,x)$ we have
$$
a_r(z) = \frac{e^{-2rP}}{P}(x,x) = \int_{2r}^1 e^{-sP}(x,x)\,\rmd s + b(x) \,,
$$
where $b$ is a smooth function. The small time asymptotics of the heat kernel then tells us that $e^{-sP}(x,x)=\frac{1}{(4\pi s)^{\frac{3}{2}}}+\mathcal{O}(s^{-\frac{1}{2}}) $, so that the function $a_r(\cdot)-a_r$ is indeed bounded, uniformly in $r\in(0,1]$.
To prove that the difference $a_r(\cdot)-a_r$ is smooth, we rely on the description of the heat kernel of~\cite[def 2.1 p.~6]{Grieser}. In local coordinates in some open subset $U$, the heat kernel can be represented as $s^{-\frac{3}{2}}A(a,\frac{x-y}{\sqrt{s}},y)$ where $A\in C^\infty([0,+\infty)_{\frac{1}{2}}\times \mathbb{R}^3\times U)$. Hence 
$a_r(x)-a_r=\int_{2r}^1 s^{-\frac{3}{2}}\left( A(\sqrt{s},0,x)-A(0,0,x)\right) \rmd s $ converges together with all its derivatives in $x$ since 
$ \partial_x^\beta s^{-\frac{3}{2}}\left( A(\sqrt{s},0,x)-A(0,0,x)\right) =\mathcal{O}(s^{-\frac{1}{2}})$ for all multi-indices $\beta$. Hence the convergent integral defining $a_r(x)-a_r$ depends smoothly on $x\in U$.

The convergence of $\Xtwo_r\defeq \X_r^2-a_r $ in $C_T^\eta C^{-1-\epsilon}(M)$ to a limit which we denote by $\begin{tikzpicture}[scale=0.3,baseline=0cm] \node at (0,0) [dot] (0) {}; \node at (0.3,0.6)  [noise] (noise1) {}; \node at (-0.3,0.6)  [noise] (noise2) {}; \draw[K] (0) to (noise1); \draw[K] (0) to (noise2); \end{tikzpicture}$ follows as a consequence. We note that as a random variable, $\begin{tikzpicture}[scale=0.3,baseline=0cm] \node at (0,0) [dot] (0) {}; \node at (0.3,0.6)  [noise] (noise1) {}; \node at (-0.3,0.6)  [noise] (noise2) {}; \draw[K] (0) to (noise1); \draw[K] (0) to (noise2); \end{tikzpicture}$ is not a homogeneous element in the chaos of order $2$ of the Gaussian noise since we did not subtract the full expectation. However it differs from a homogeneous element by a deterministic smooth function hence $\begin{tikzpicture}[scale=0.3,baseline=0cm] \node at (0,0) [dot] (0) {}; \node at (0.3,0.6)  [noise] (noise1) {}; \node at (-0.3,0.6)  [noise] (noise2) {}; \draw[K] (0) to (noise1); \draw[K] (0) to (noise2); \end{tikzpicture} $ has moments of any order $1\leqslant r<\infty$ that are equivalent to its second moment. In the sequel, we will always prove stochastic estimates for homogeneous elements in Wiener chaoses and then justify why the locally covariant renormalization, subtracting only universal local quantities, still yields a stochastic object with the same analytic properties.

\subsubsection{The cubic term.}\label{subsect_cubic}

For the stochastic term $\IXthree_r$, we do not bound the regularity by hands any more and use the microlocal machinery.
Recall that our aim is to find the range of $\gamma\in \mathbb{R}$ so that
\begin{align*}
\bbE\big[\mcQ_x^\gamma\big({\IXthree_r}(t,\bigcdot),{\IXthree_r}(t,\bigcdot)\big)\big] < \infty     
\end{align*}
uniformly in $x\in M$. By application of Wick's Theorem, we have
\begin{align*} 
\bbE\big[\mcQ_x^\gamma\big({\IXthree_r}(t,\bigcdot) \,&,\, {\IXthree_r}(t,\bigcdot)\big)\big]   \\
&= \int_{(-\infty,t]^2\times M^4} {G_r^{(3)}}(s_1-s_2,z_1,z_2 ) e^{-\vert t-s_1\vert P}(y_1,z_1)  e^{-\vert t-s_2\vert P}(y_2,z_2) \mcQ_x^\gamma (y_1,y_2 ),
\end{align*}
for an integral with respect to $\rmd v(y_1, y_2, z_1, z_2) \rmd s_1\rmd s_2$, where we denote by $G_r^{(3)}(t_1-t_2)$ the operator with kernel 
$$
\bigg(\frac{e^{-(2r+\vert t_1-t_2\vert)P}}{P}(x,y)\bigg)^3.
$$ 
This operator belongs to $\mcS^{-3}_\Gamma(\bbR^2\times M^2)$ with
$$
\Gamma = N^*\Big(\{t=s\}\times{\bf d}_2\subset \bbR^2\times M^2\Big) \cup N^*\Big(\{t=s\}\subset \bbR^2\times M^2\Big).
$$
By the general theory described above, the only irreducible subgraph of the Feynman graph is the graph itself, therefore the Feynman amplitude is well-defined on 
$(-\infty,t]^2\times M^4\setminus \Sigma $ where the singular locus $\Sigma$ corresponds to the generalized diagonal
\begin{align*}
\underset{ \text{codim}_w = 12  }{ \underbrace{ \Big\{ y_1=y_2=z_1=z_2=x \Big\} } } \cap
\underset{ \text{codim}_w = 4 }{ \underbrace{  \{s_1=s_2=t\}  } }
 \subset (-\infty,t]^2\times M^4
\end{align*}
which has weighted codimension $16$. The sum of the weak homogeneities of the kernels in the Feynman amplitude is given by
\begin{align*}
    -3+2(-3)-6-2\gamma\,,
\end{align*}
so that is finite for $\gamma<\frac12$ by Theorem \ref{ThmBlackBoxFeynmanGraphs}.

To estimate time regularity of the random object ${\IXthree_r}$, we need to control time increments $\bbE\big[\mcQ_x^\gamma\big({ \IXthree_r}(t_1,\bigcdot)-{\IXthree_r}(t_2,\bigcdot),{ \IXthree_r}(t_1,\bigcdot)-{\IXthree_r}(t_2,\bigcdot)\big)\big]$ for given $\gamma<\frac{1}{2}$. By application of Wick's Theorem and symmetry arguments, this equals
\begin{align*}
\bbE\Big[\mcQ_x^\gamma&\Big({ \IXthree_r}(t_1,\bigcdot)-{\IXthree_r}(t_2,\bigcdot),{ \IXthree_r}(t_1,\bigcdot)-{ \IXthree_r}(t_2,\bigcdot)\Big)\Big]   \\
&=2\int_{\mathbb{R}^2\times M^4} G_r^{(3)}(s_1-s_2,z_1,z_2 ) e^{-\vert t_1-s_1\vert P}(y_1,z_1)H(t_1-s_1)   \\
&\quad\times\underbrace{\left( e^{-\vert t_1-s_2\vert P}(y_2,z_2)H(t_1-s_2) -e^{-\vert t_2-s_2\vert P}(y_2,z_2)H(t_2-s_2)\right)} \mcQ_x^\gamma (y_1,y_2 )
\end{align*}
where $H$ is the Heaviside function and the term underbraced is a difference of heat kernels taken at closed times $(t_1,t_2)$. Now observe that we need to treat $t_2$ as an extra parameter and enlarge the previous scaling to the generalized diagonal $\{y_1=y_2=z_1=z_2=x\}\cap \{s_1=s_2=t_1\}\subset \mathbb{R}^2\times M^4$ with an extra scaling which contracts the extra time variable $t_2$ on $t_1$. In practice, the enlarged scaling flow maps $\big(s_1,s_2,t_2,z_1,z_2,y_1,y_2\big)\in \mathbb{R}^3\times M^4$ to

$$ 
\Big(e^{-2t}(s_1-t_1)+t_1,e^{-2t}(s_2-t_1)+t_1,e^{-2t}(t_2-t_1)+t_1,e^{-t\rho_{M^4}}(z_1,z_2,y_1,y_2  ) \Big) \in \mathbb{R}^3\times M^4
$$
where $\rho_{M^4}\in C^\infty(TM^4)$ is any scaling field associated to the deepest diagonal $\{ z_1=z_2=y_1=y_2 \}\subset M^4 $. Then for $\Gamma$ some suitable conical neighborhood of the conormal bundle
$$
N^*\Big( \big\{y_1=y_2=z_1=z_2=x\big\}\cap \{s_1=s_2=t_1\}\subset \mathbb{R}^2\times M^4 \Big),
$$
we have an estimate of the form
$$
\left( e^{-\vert t_1-s_2\vert P}(y_2,z_2)H(t_1-s_2) -e^{-\vert t_2-s_2\vert P}(y_2,z_2)H(t_2-s_2)\right)=\mathcal{O}_{\mathcal{S}_\Gamma^{-3-\varepsilon}(\mathbb{R}\times M^2)}(\vert t_1-t_2\vert^{\frac{\varepsilon}{2}} )
$$ 
for all $\varepsilon\in (0,1)$ where we trade some space--time regularity to get a small power $\vert t_1-t_2\vert^{\frac{\varepsilon}{2}}$ in factor. Therefore, for every $\gamma<\frac{1}{2}$, if we choose $\varepsilon\in (0,1)$ small enough so that the weak homogeneity $-3+2(-3-\varepsilon)-6-2\gamma >-16$ which yields the $r$-uniform bound
$$
 \bbE\Big[\mcQ_x^\gamma\Big({ \IXthree_r}(t_1,\bigcdot)-{ \IXthree_r}(t_2,\bigcdot),{\IXthree_r}(t_1,\bigcdot)-{ \IXthree_r}(t_2,\bigcdot)\Big)\Big]=\mathcal{O}\big(\vert t_1-t_2\vert^{\frac{\varepsilon}{2}} \big)
 $$ 
 by the fact that Theorem \ref{ThmBlackBoxFeynmanGraphs} holds in a parameter version ($t_2$ is the new parameter) and since our extension procedures are linear continuous in the relevant topologies.
 
 We obtain the $L^p(\Omega,\mathcal{F},\bbP)$ convergence of $\IXthree_r$ to some limit $\IXthree$ in $C_T\mcC^{1/2-\epsilon}(M)$ using the telescopic sum argument described in the introduction of Section \ref{SectionKolmogorovFeynman}. The computations are similar to the preceding computations, except that they involve difference kernels $\bigg(\frac{e^{-(r_1+r_2+\vert t_1-t_2\vert)P}}{P}(x,y)\bigg)^3-\bigg(\frac{e^{-(2r_2+\vert t_1-t_2\vert)P}}{P}(x,y)\bigg)^3$, $\bigg(\frac{e^{-(r_1+r_2+\vert t_1-t_2\vert)P}}{P}(x,y)\bigg)^3-\bigg(\frac{e^{-(2r_1+\vert t_1-t_2\vert)P}}{P}(x,y)\bigg)^3 $ with norm $\mathcal{O}((r_1+r_2)^{\frac{\varepsilon}{2}}) $ in the space $\mathcal{S}_\Gamma^{-3-\varepsilon}(\mathbb{R}^2\times M^2 )$ for $\Gamma = N^*\Big(\{t=s\}\times{\bf d}_2\subset \bbR^2\times M^2\Big) \cup N^*\Big(\{t=s\}\subset \bbR^2\times M^2\Big)$.

\medskip

Rather than keep proceeding in an ad hoc way, we develop in the next section a systematic machinery to deal with the convergence of the other components of the enhanced noise
$$
\widehat{\xi}_r \defeq \Big(\xi_r, \begin{tikzpicture}[scale=0.3,baseline=0cm]
\node at (0,0) [dot] (0) {};
\node at (0.3,0.6)  [noise] (noise1) {};
\node at (-0.3,0.6)  [noise] (noise2) {};
\draw[K] (0) to (noise1);
\draw[K] (0) to (noise2);
\end{tikzpicture}_r, \; 
\IXthree_r\,,\;
\begin{tikzpicture}[scale=0.3,baseline=0cm]
\node at (0,0) [dot] (0) {};
\node at (0,0.5) [dot] (1) {};
\node at (-0.4,1)  [noise] (noise1) {};
\node at (0,1.2)  [noise] (noise2) {};
\node at (0.4,1)  [noise] (noise3) {};
\draw[K] (0) to (1);
\draw[K] (1) to (noise1);
\draw[K] (1) to (noise2);
\draw[K] (1) to (noise3);
\end{tikzpicture}_r \odot \begin{tikzpicture}[scale=0.3,baseline=0cm]
\node at (0,0)  [dot] (1) {};
\node at (0,0.8)  [noise] (2) {};
\draw[K] (1) to (2);
\end{tikzpicture}_r, \; 
\begin{tikzpicture}[scale=0.3,baseline=0cm]
\node at (0,0) [dot] (0) {};
\node at (0,0.5) [dot] (1) {};
\node at (-0.4,1)  [noise] (noise1) {};
\node at (0.4,1)  [noise] (noise2) {};
\draw[K] (0) to (1);
\draw[K] (1) to (noise1);
\draw[K] (1) to (noise2);
\end{tikzpicture}_r \odot \begin{tikzpicture}[scale=0.3,baseline=0cm]
\node at (0,0) [dot] (0) {};
\node at (0.3,0.6)  [noise] (noise1) {};
\node at (-0.3,0.6)  [noise] (noise2) {};
\draw[K] (0) to (noise1);
\draw[K] (0) to (noise2);
\end{tikzpicture}_r - \frac{b_r}{3}, \; 
\big\vert\nabla \begin{tikzpicture}[scale=0.3,baseline=0cm]
\node at (0,0) [dot] (0) {};
\node at (0,0.5) [dot] (1) {};
\node at (-0.4,1)  [noise] (noise1) {};
\node at (0.4,1)  [noise] (noise2) {};
\draw[K] (0) to (1);
\draw[K] (1) to (noise1);
\draw[K] (1) to (noise2);
\end{tikzpicture}_r\big\vert^2 - \frac{b_r}{3}, \;  
\begin{tikzpicture}[scale=0.3,baseline=0cm]
\node at (0,0) [dot] (0) {};
\node at (0,0.5) [dot] (1) {};
\node at (-0.4,1)  [noise] (noise1) {};
\node at (0,1.2)  [noise] (noise2) {};
\node at (0.4,1)  [noise] (noise3) {};
\draw[K] (0) to (1);
\draw[K] (1) to (noise1);
\draw[K] (1) to (noise2);
\draw[K] (1) to (noise3);
\end{tikzpicture}_r \odot \begin{tikzpicture}[scale=0.3,baseline=0cm]
\node at (0,0) [dot] (0) {};
\node at (0.3,0.6)  [noise] (noise1) {};
\node at (-0.3,0.6)  [noise] (noise2) {};
\draw[K] (0) to (noise1);
\draw[K] (0) to (noise2);
\end{tikzpicture}_r - b_r \begin{tikzpicture}[scale=0.3,baseline=0cm]
\node at (0,0)  [dot] (1) {};
\node at (0,0.8)  [noise] (2) {};
\draw[K] (1) to (2);
\end{tikzpicture}_r \,\Big).
$$ 
It provides a convergence statement for Feynman graphs similar to Weinberg's power counting theorem. A reader who is primarily interested in its application to the study of the $\Phi^4_3$ equation can safely skip this part and read only Theorem \ref{ThmBlackBoxFeynmanGraphs} after they get acquainted with the notations in Section \ref{SubsectionConfigurationSpace}. Then they can go directly to Section \ref{SectionStochasticBounds}.

\section{Induction on Feynman amplitudes}
\label{SectionInductionGraphs}

In the present section we describe a general induction to control analytically the Feynman amplitudes which appear when we study the regularities of the stochastic trees appearing in our equation. This is the content of Theorem \ref{ThmBlackBoxFeynmanGraphs}. We need to introduce several layers of formalism before giving this statement; they are interesting on their own. We first describe correctly in Section \ref{SubsectionConfigurationSpace} the configuration spaces on which we work, and give the definition of the Feynman graphs in Section \ref{subsectionfeynmangraphsdefi}. {Working in a non-homogeneous manifold, it turns out to be useful to introduce some} pointed scaling space in Section \ref{subsubsectionpointedscaling}. {The inductive proof of Theorem \ref{ThmBlackBoxFeynmanGraphs} is given in Section \ref{SectionProofGeneralWeinberg}. It} involves a double scaling, where we need to control the growth of the amplitude when all points collide and also when all points collide on a given $x\in M$, uniformly in $x\in M$. This makes the proof much more involved than the usual Weinberg convergence like theorems in usual Quantum Field Theories. It is also because of these double scalings that we need the double extension theorem from Theorem \ref{Thm:extensiontwosteps}.

\subsection{Configuration space}
\label{SubsectionConfigurationSpace}

The Feynman amplitudes which arise from the stochastic estimates involve the analysis of products of distributional kernels on space-time configuration spaces. In the present section we introduce the necessary general formalism for analysing these amplitudes.

\subsubsection{Diagonals and scaling fields.}
In the sequel we write 
$$
\mcM\defeq \bbR\times M
$$ 
and work on 
$$
\bbR^p\times M^{p+q}\simeq\mcM^p \times M^{q}
$$
for $p, q\geqslant0$. Given a fixed $t\in\bbR$ we always view some elements of $\bbR^p\times M^{p+q}$ as some elements of $\mcM^{p+q}$ using the mapping
\begin{align}\label{eq:sametimes}
 \mcM^p \times M^{q}\simeq  \mcM^p \times \big(\{t\}\times M\big)^{q}\subset\mcM^{p+q},
\end{align}
that is to say attributing the time $t$ to the purely spatial points. The configuration space $\mcM^n$ of $n=p+q$ points in $\mcM$ will play a particular role in the sequel. Given a distinguished point $(t,x)\in\mcM$, writing $m_i=(t_i,x_i)\in\mcM$, for $I\subset\{1,\dots,n\}$ we denote by 
\begin{equation} \label{DefnDiagonals} \begin{split}
{\bf d}_I &\defeq \Big\{m=(m_1,\dots,m_n)\,;\,x_i=x_j\;\text{and}\;t_i=t_j=t\; \textrm{ for }i\neq j \textrm{ if }(i,j)\in I^2\Big\},   \\
&\hspace{1cm}\text{when space-time points labelled by $I$ collide;}  \\
{\bf d}_{I,(t,x)} &\defeq \Big\{m=(m_1,\dots,m_n)\,,\,m_i=(t,x)\;  \textrm{ if }i\in I\Big\}, \quad \text{the marked diagonal}  \\
&\hspace{1cm} \text{where all space-time points labelled by $I$ collide to $(t,x)$,   }   \\
\mcT_I &\defeq \Big\{m=(m_1,\dots,m_n)\,,\,t_i=t_j=t\; \textrm{ for } i\neq j \textrm{ if }(i,j)\in I^2\Big\},   \\
& \hspace{1cm}\text{when only time components labelled by $I$ coincide with $t$  } \\ 
\end{split} \end{equation}
the corresponding diagonals in the product spaces. Let us make several important observations. First the marked diagonal ${\bf d}_{[n],(t,x)}$ is in reality just \textbf{one point} $\big((t,x),\dots,(t,x)\big)$ in the configuration space $\mcM^n$. Second one has the natural inclusion relations 
$$
{\bf d}_I\subset\mcT_I \textrm{ and } {\bf d}_I\subset {\bf d}_J \textrm{ and } \mcT_I\subset \mcT_J \textrm{ if } J\subset I.
$$
We denote by
$$
{\bf d}_n \defeq \Big\{ \big((t,x_1),\dots,(t,x_1)\big)\in\mcM^n\,:\,x_1\in M \Big\},
$$
and
$$
{\bf d}_{n,(t,x)} \defeq \Big\{\big((t,x),\dots,(t,x)\big)\Big\}\subset\mcM^n.
$$
In the sequel, to take \eqref{eq:sametimes} into account, we will work on some submanifold $ \mcM^p \times (\{t\}\times M)^{q}=\mcT_J\subset \mcM^n$ for some fixed $J\subset \{1,\dots,n\}$,  $|J|=q$, where all time variables indexed by $J$ coincide and equal $t$. 

\ssk

\begin{Examples}
We give two examples of stochastic estimates we will meet in the sequel so that the reader can see in what type of space-time domains we need to integrate our Feynman amplitudes.

In the sequel we will calculate $\mathbb{E}\big[ \mcQ_x^\gamma( \IXthree(t),\IXthree(t) ) \big]$ which can be represented as a Feynman amplitude (underbraced below) tested against the constant function $1$ that reads
\begin{equation*} \begin{split}
&\mathbb{E}\big[ \mcQ^\gamma_{x}( \IXthree (t) ,\IXthree (t) )\big]   \\
&= \int_{(-\infty,t]^2\times M^4 } \underbrace{ G^{(3)}\big(s_a-s_b,x_a,x_b\big) \, 
                                   \underline{\mcL}^{-1}\big(t-s_a,y_*,x_a\big) \,
                                   \underline{\mcL}^{-1}\big(t-s_b,z_*,x_b\big) \,
                                   \mcQ^\gamma_{x}(y_*,z_*)}
\end{split} \end{equation*}
for some function $G^{(3)}$. The integration is with respect to $\rmd v(x_a,x_b,y_*,z_*)\rmd s_a \rmd s_b$. In this case we integrate a certain Feynman amplitude underbraced on some domain $ \mcT_{\{c,d \}}\subset \mcM^4 $  with two space-time variables $ (s_a,x_a)$, $(s_b,x_b)$ and two space variable $(s_c,y_*)$, $(s_d,z_*)$ promoted to space-time variables by writing $s_c=s_d=t$. On the other hand, when we want to prove the continuity in time using a Kolmogorov type argument we will also consider 
\begin{equation*} \begin{split}
&\mathbb{E}\big[ \mcQ^\gamma_{x}( \IXthree (t_1) ,\IXthree (t_2) ) \big]   \\
&= \int_{(-\infty,t]^2\times M^4 } \underbrace{ G^{(3)}\big(s_a-s_b,x_a,x_b\big) \,
    \underline{\mcL}^{-1}\big(t_1-s_a,y_*,x_a\big) \,
    \underline{\mcL}^{-1}\big(t_2-s_a,z_*,x_b\big) \,
    \mcQ^\gamma_{x}(y_*,z_*)}.
\end{split} \end{equation*}
In this case we integrate over $\mcT_{\{c,d\}} \subset\mcM^4 $ where the last two time variables are taken equal
to some fixed times $t_1$, $t_2$. With this example we see that we also need to fix the time variables.
\end{Examples}

\ssk

We will also use a \textbf{particular class of scaling fields that will leave all $\mcT_J$ and ${\bf d}_I $ stable}, such class of scaling fields will be called \textbf{admissible}.

\ssk

\begin{defn} \label{def:explicitEuler}
\textsf{\textbf{(Admissible scaling fields)}} Pick some open chart $U\subset M,\kappa:U\mapsto \mathbb{R}^d$ such that $\kappa(U)\subset \mathbb{R}^d$ is an open convex ball -- this is always possible up to making things smaller. 

We define a scaling field $\rho_{[p]}$ in $U$ from its flow given by for any $(x_1,\dots,x_p)\in \kappa(U)^p$ by
\begin{eqnarray*}
(x_1,\dots,x_p)\in U^p\mapsto \Big(x_1,e^{-t}(x_2-x_1)+x_1,\dots, e^{-t}(x_p-x_1)+x_1\Big)\in \kappa(U)^p;
\end{eqnarray*}
it means $\rho_{[p]}$ reads $\sum_{j=2}^p (x_j-x_1)\partial_{x_j} $ in the above coordinate chart. We define the local scaling field on $(\bbR\times U)^p$, with local coordinates $(s_i,x_i)_{1\leqslant i\leqslant p}$, setting
\begin{eqnarray*}
\rho = 2 \bigg(\sum_{j=2}^p(s_j-s_1)\partial_{s_j} \bigg) + \rho_{[p]} \defeq 2\rho_{times} + \rho_{[p]}\,.
\end{eqnarray*}
We obtain global scaling fields by gluing together the above local objects. Consider a cover $\cup_i  U_i^p$ of some neighborhood of the space diagonal and choose ${o}\in C^\infty_c(  \cup_i U_i^p)$ such that ${o}=1$ near the space diagonal. For a subordinated partition of unity $\sum{\varphi_i}=1$ of $\text{supp}\left({o}\right)$, we set $\rho=2\rho_{times}+{o}\sum_i {\varphi_i}\rho_i$ where each local scaling field $\rho_i\in C^\infty(T\left(U_i^p\right))$ is constructed in some charts as above.

\ssk

The same construction also works when we scale towards some marked points, in all charts $\kappa:U\mapsto M$ containing $x$, we decide that we scale with the flow
\begin{eqnarray*}
(x_1,\dots,x_p)\in U^p\mapsto \Big(e^{-t}(x_1-x)+x,e^{-t}(x_2-x)+x,\dots, e^{-t}(x_p-x)+x\Big)\in \kappa(U)^p,
\end{eqnarray*}
it means $\rho_{[p]}$ reads $\sum_{j=1}^p (x_j-x)\partial_{x_j} $ in the above coordinate chart.
If the chart does not contain $x$, we decide our flow is trivial (the generator of the flow is the zero vector field) and $\rho_{[p]}=0$ in such a chart. We define the local scaling field on $(\bbR\times U)^p$, with local coordinates $(s_i,x_i)_{1\leqslant i\leqslant p}$, setting
\begin{eqnarray*}
\rho_{(t,x)} = 2 \bigg(\sum_{j=1}^p(s_j-t)\partial_{s_j} \bigg) + \rho_{[p]} \defeq 2\rho_{times} + \rho_{[p]}.
\end{eqnarray*}
We obtain global scaling fields by gluing together the above local objects. Consider a cover $\cup_i  U_i^p$ of some neighborhood of the space diagonal and choose ${\varphi}\in C^\infty_c(  \cup_i U_i^p)$ such that ${\varphi}=1$ near the space diagonal. For a subordinated partition of unity $\sum{\varphi_i}=1$ of $\text{supp}(\varphi)$, we set $\rho=2\rho_{times}+{\varphi}\sum_i {\varphi_i}\rho_i$ where each local scaling field $\rho_i\in C^\infty(T\left(U_i^p\right))$ is constructed in charts as above.
\end{defn}

\ssk

We make the observation that in all the above situations the constructed vector fields generate some dynamics which preserve all the diagonals $\textbf{d}_I,\mathcal{T}_I$ or all the marked diagonals $\textbf{d}_{I,(t,x)}$ in the case of scaling fields $\rho_{(t,x)}$ on marked points \textbf{by construction} (they preserve these diagonals in the charts locally, hence globally by gluing). The above definition shows that \emph{admissible} scaling fields are abundant. Admissible scaling fields enjoy another remarkable property. The cotangent lift of $e^{-u\rho}$ stabilizes all the conormal bundles of all the partial diagonals.

\ssk

\begin{lem}\label{lem:conormstability}
For all $I\subset \{1,\dots,p\}$ and $\ell\geqslant 0$, for all admissible scaling fields $\rho$, one has
$$
e^{-\ell\rho*}\left( N^*(\textbf{d}_I\subset \mcM^p) \right)\subset N^*(\textbf{d}_I\subset \mcM^p).
$$
\end{lem}

\ssk

\begin{Dem}
If $v\in T_x\mathbf{d}_I$, $\rmd e^{\ell\rho}_x(v)\in T_{e^{\ell\rho}(x)}\mathbf{d}_I$ (since the flow stabilizes the diagonals) implies that if $\xi\in T_{e^{\ell\rho}(x)}\mathbf{d}_I^\perp$ then $\left\langle\xi, \rmd e^{\ell\rho}_x(v) \right\rangle=0$ for all $v\in T_x\mathbf{d}_I$ hence
$^t\rmd e^{\ell\rho}_x(v)(\xi) \in T_x\mathbf{d}_I^\perp=N_x^*(\mathbf{d}_I) $, which concludes our proof.
\end{Dem}

\ssk

We will typically be given a family $(\Lambda_\epsilon)_{0<\epsilon\leqslant 1}$ of distributions on $\mcT_J\setminus \left(\cup_{I\subset \{\{1,\dots,p\}\}} {\bf d}_I \right)$ that converge to a limit as a distribution outside all the diagonals of $\mcT_J$. We will use Theorem \ref{ThmCanonicalExtension} to extend it to the whole of $\mcT_J$ by an inductive procedure under some scaling-type  assumptions. The inductive structure of the extension procedure will come from the geometric form of {\sl Popineau \& Stora's lemma}, which we recall here. We associate to $I\subset\{1,\dots, p\}$, 
the open set:
$$
\mcO_I \defeq \Big\{m=(m_1,\dots,m_p)\in\mcM^p\,;\,m_i\neq m_j\ \forall\,(i,j)\in I\times I^c\Big\} \subset \mcM^p.
$$

\ssk
In the next lemma and throughout the paper, whenever one considers a Cartesian product of Manifolds $\mcM^p$, we will always call \textit{deepest diagonal} the diagonal $\mathbf{d}_p\defeq\{x_1=\dots=x_p  \}\subset\mcM^p$. Moreover, we will always call \textit{larger diagonal} any diagonal of the form $\{x_{i_1}=\dots=x_{i_{|I|}}: I=\{i_1,\dots ,i_{|I|}\}\subsetneq [p] \}\subset\mcM^p$.

\begin{lem}\label{lem:PopineauStora}
One has 
$$
\mcM^p\backslash {\bf d}_p = \bigcup_{I\subset\{1,\dots,p\}} \mcO_I
$$
and there is an associated smooth partition of the unity, $1=\sum_{I\subset\{1,\dots,p\}} \eta_I\in C^\infty(\mcM^p\setminus {\bf d}_p)$, with the family $(\eta_I\circ e^{-\ell\rho})_{\ell \geqslant 0}$ bounded in $C^\infty\left(\mcM^p\backslash {\bf d}_p\right) $ for every admissible scaling field $\rho$  with respect to the deepest diagonal ${\bf d}_p\subset\mcM^p$.
\end{lem}

\ssk 

The proof is simple and can be found in \cite[Lemma 6.3]{DangHers}. The proof of the claim on the family $(\eta_I\circ e^{-\ell\rho})_{\ell\geqslant 0}$ can be found in \cite[Lemma 6.3.1 p.~131]{Dangthese}. Since $\mcT_J\subset \mcM^p$ for $J\subset \{1,\dots,p\}$, the above partition of unity induces naturally a partition of unity on $\mcT_J\setminus {\bf d}_p$ with the same properties.


\subsubsection{H\"ormander product of distributions.}
\label{SubsectionHormander}
In the simplest cases the distributions $\Lambda_\epsilon$ will be given as some products of distributions, with each factor depending possibly only on a subset of the variables $\mcM^p$. The easiest case in which to make sense of such products relies on H\"ormander's product theorem \cite[Thm 6.1 p.~219]{BDH16} and gives the following statement.

\ssk

\begin{lem} \label{LemSimpleProduct}
If $\Lambda_1\in\mcD'(\mcM^p)$ depends only on the first $1\leqslant k<p$ components of $\mcM^p$ and $\Lambda_2\in\mcD'(\mcM^p)$ depends only on the last $p-(k-1)$ components, so they have only one component in common, and
\begin{equation*} \begin{split}
WF(\Lambda_1) \subset \bigcup_{I\subset\{1,\dots, k\}}N^*({\bf d}_I)\cup N^* \left(\mcT_I\right),   \\
WF(\Lambda_2) \subset \bigcup_{J\subset\{k,\dots, p\}}N^*({\bf d}_J)\cup N^* \left(\mcT_J\right),
\end{split} \end{equation*}
then the product $\Lambda_1\Lambda_2$ is well-defined in $\mcD'(\mcM^p)$ and
$$
WF(\Lambda_1\Lambda_2) \subset \big(WF(\Lambda_1)+WF(\Lambda_2)\big)\cup WF(\Lambda_1)\cup WF(\Lambda_2)\,.
$$
\end{lem}

\ssk

\begin{Dem}
Denote by $\lambda$ a generic element of $T^*\mcM$. If $(\lambda_1,\dots,\lambda_k,0,\dots,0)$ and \\$(0,\dots,0,\mu_k,\mu_{k+1},\dots,\mu_p)$ stand for some non-null elements of $T^*(\mcM^p)$ such that
$$
\sum \lambda_i = 0\,, \quad \sum \mu_j =0\,,
$$
then the convex sum
$$
\big(\lambda_1,\dots,\lambda_k,0,\dots,0\big) + \big(0,\dots,,0,\mu_k,\mu_{k+1},\dots,\mu_p\big) = \big(\lambda_1,\dots,\lambda_k+\mu_k,\mu_{k+1},\dots,\mu_p\big)
$$
cannot vanish. This implies that $WF(\Lambda_1)+WF(\Lambda_2)$ does not meet the zero section $\{0\}$ and one can apply H\"ormander's Theorem \cite[Thm 6.1 p.~219]{BDH16} which yields the existence of the distributional product $\Lambda_1\Lambda_2$ together with a bound on the wave front set $WF\left(\Lambda_1\Lambda_2 \right)$ of the product.
\end{Dem}

\ssk

We give another important consequence of Theorem \ref{ThmCanonicalExtension} before talking about Feynman amplitudes. 

\ssk

\begin{prop} \label{PropProductThm}
Let $\mcY\subset \mcX$ be a closed embedding and let $\rho$ stand for a parabolic scaling field for the inclusion $\mcY\subset \mcX$. Assume we are given some closed conic sets $\Gamma_1, \Gamma_2$ in $T^*\mcX\setminus \mcY$ such that
$$
{\bf 0} \notin (\Gamma_1+\Gamma_2).
$$ 
and such that $e^{-\ell\rho*}(\Gamma_i)\subset \Gamma_i, \, \forall \ell\geqslant 0$ and $i\in \{1,2\}$. Assume also that we are given two distributions 
$$
\Lambda_1 \in \mcS^{s_1,\rho}_{\Gamma_1}(\mcX\setminus \mcY) \cap \mcD^\prime(\mcX), \qquad \Lambda_2 \in \mcS^{s_2,\rho}_{\Gamma_2}(\mcX)\cap \mcD^\prime(\mcX)
$$
so the product $\Lambda_1\Lambda_2$ is well-defined on $ \mcU\subset \mcX\backslash \mcY$. If 
$$
s_1+s_2>-\textrm{\emph{codim}}_w(\mcY)
$$ 
then this product has a unique extension as an element of $\mcS^{s_1+s_2,\rho}_\Gamma(\mcX)$ with
$$
\Gamma = \Gamma_1 \cup \Gamma_2\cup \big(\Gamma_1+\Gamma_2\big) \cup N^*( \mcY\subset \mcX).
$$
\end{prop}

\ssk

The condition ${\bf 0} \notin (\Gamma_1+\Gamma_2)$ ensures that the distributional product $\Lambda_1\Lambda_2$ is well-defined at least on $\mcX\setminus \mcY$ and the stability of $\Gamma_1,\Gamma_2$ under the lifted cotangent flow ensures that the convex sum $\big(\Gamma_1+\Gamma_2\big)\subset T^*\mcX\setminus \mcY$ satisfies the conormal landing condition for the inclusion $\mcY \subset \mcX$. The statement of the above proposition means that for any mollification $\Lambda_1^\epsilon, \Lambda_2^\epsilon$ of these distributions which converge in the respective functional spaces the product $\Lambda_1^\epsilon\Lambda_2^\epsilon$ is converging in $\mcS^{s_1+s_2,\rho}_\Gamma(\mcM^p)$ to a limit independent of the mollification.

\bigskip

\subsection{Feynman graphs and Feynman amplitudes}
\label{subsectionfeynmangraphsdefi}
We are now ready to define the Feynman graphs we are using in order to control the H\"older-Besov norms of the tree appearing in the construction of the $\Phi^4_3 $ measure. For an oriented finite graph $(V,E)$ with vertex set $V$ and edge set $E$ we denote by  $v(e)_-, v(e)_+$ its two vertices, according to its orientation.

\ssk

\begin{defn*}
    A \textsf{\textbf{Feynman graph for}} $\Phi^4_3$ is an oriented finite graph $\mcG_{(t,x)}=(V,E)$ with $p$ vertices in $V$ and an edge set $E$ with no two edges joining a given pair of vertices, along with
   \begin{itemize}
       \item[--]  a distinguished edge $e_{\textrm{ref}}\in E$,
       \item[--]  a subset $J$ of $ V$ which indicates which times are set equal to the fixed $t\in\bbR$,
       \item[--] for each vertex $v\in V$, a variable $z_v=(t_v,x_v)\in \mathcal M$, with the restriction that for the two vertices $v(e_{\textrm{ref}})_-,v(e_{\textrm{ref}})_+$ attached to $e_{\textrm{ref}}$, one has $z_{v(e_{\textrm{ref}})_\pm}=(t,x_{v(e_{\textrm{ref}})_\pm})$ for some $x_{v(e_{\textrm{ref}})_\pm}\in M$.
   \end{itemize} 
We set 
$$
\widetilde E\defeq E\setminus \{e_{\textrm{ref}}\}.
$$
We furthermore assume that the following facts hold.
\begin{itemize}
    \item[--] The set $V$ of vertices can be partitioned as
$$
V = V'\sqcup V_A
$$
where $V'$ is a disjoint union of singletons and $V_A$ is a disjoint union of 
$$
n_\mcG\defeq |A|
$$ 
triples of vertices indexed by a finite set $A$, with each triple made up of a distinguished pair of vertices and another vertex. For $j\in[n_\mcG]$ such a triple in $V_A$ reads $\big(v^j_*,(v^j_1,v^j_2)\big)$ where $(v^j_1,v^j_2)$ is the distinguished pair and $v^j_*$ the remaining vertex.

    \item[--] For every $j\in[n_\mcG]$ there is no edge in the graph relating $v_1^j$ to $v_2^j$ or one of these points to $v^j_*$.

    \item[--] We are given for each edge $e\in E$ a kernel $K_e\in \mathcal{K}^{a_e}$ for some scaling exponent $a_e\in \mathbb{R}$. (The space $\mathcal{K}^{a_e}$ was defined in Lemma \ref{l:topologytwopointfunctiong}). Moreover for the distinguished edge $e_{\textrm{ref}}\in E$ the corresponding kernel is given by $K_{e_{\textrm{ref}}} = \mcQ_x^\gamma$ and $e_{\textrm{ref}}$ is the only edge in $E$ whose kernel $K_e$ is of the form $\mcQ^\gamma_y$ for some $y\in M$. 
\end{itemize}
\end{defn*}

\ssk

\begin{figure}[h!] \label{Fig1}
\begin{center}
\includegraphics[scale=1.9]{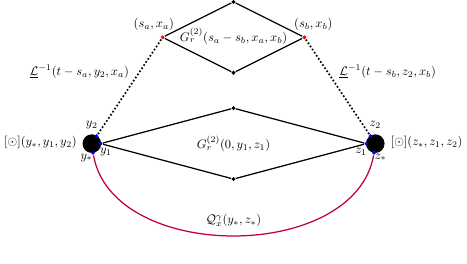}
\caption{An example of Feynman graph $\mcG(=\mcG_{24})$ with $n_\mcG=2$ along with the contributions to $\mcA_\mcG$ of its edges and vertices. Space-time vertices are pictured by \textcolor{red}{red} nodes, purely space vertices (with time set to the value $t$) are pictured by \textcolor{blue}{blue} nodes, and black nodes represent the noises.}
\end{center}
\end{figure}

See figure \ref{Fig1} below for an illustration. In the sequel we often omit the base point $(t,x)$, writing $\mcG$ instead of $\mcG_{(t,x)}$. Since renormalization also involves the analysis of singularities of Feynman subgraphs we also need a notion of Feynman subgraphs adapted to our specific setting.

\ssk

\begin{defn*}
A \textsf{\textbf{Feynman subgraph}} $\mathcal{G}_1=(V_1,E_1)\subset \mathcal{G}_{(t,x)}=(V,E)$ is the data of
\begin{itemize}
    \item[--] some subset $V_1$ of the vertices $V$ of $\mathcal{G}$, 
    \item[--] some subset $E_1$ of the edges $E$ of $\mathcal{G}$ such that
$V_1$ can be partitioned as $V_1=V^\prime_1\sqcup V^\prime_A$ and $V^\prime_A\subset V_A$ respects the partitioning of $V_A$, any triple $\big(v^j_*,(v^j_1,v^j_2)\big),\, j\in n_{\mathcal{G}_1}$ in $V^\prime_A$ corresponds to a triple in $V_A$;
\end{itemize} 
along with the conditions that every edge $e\in E_1$ has its bounding vertices $v(e)_-,v(e)_+ $ in $V_1$, and that the subgraph $\mathcal{G}_1$ does not necessarily contain the distinguished edge $e_{\textrm{ref}}$ of $\mathcal{G}$. A (sub)graph $\mathcal{G}$ is said to be $\mathrm{irreducible}$ if it cannot be disconnected by removing exactly one edge $e\in E$. 
\end{defn*}

\ssk

We also need to recall the notion of loops for the Feynman graphs we consider, this is given by the usual Euler formula.

\ssk

\begin{defn*}
Given a Feynman graph $\mcG$ as above, we define the \textsf{\textbf{number of loops of the Feynman graph}} $\mcG$ as 
$$
b_1(\mcG)\defeq \vert E(\mcG)\vert - \vert V^\prime(\mcG)\vert - n_{\mcG} + 1.
$$
\end{defn*}

\ssk

We are given some distributions 
\begin{align*}
    [\odot]\big(\cdot,\cdot,\cdot\big) &\in \mcS^{-6}_{N^*({\bf d}_3) }(M^3)\,,\\
    \mcQ^\gamma_x &\in \mcS^{-6-2\gamma}_{ T^*_{(x,x)}M^2  }(M^2)\,,
\end{align*}
where the scaling of $[\odot]$ is with respect to the deepest diagonal ${\bf d}_3 =\{(y,y,y): y\in M\}\subset M^3$ and the scaling of $\mcQ^\gamma_x$ is with respect to the point $(x,x)$. Here $[\odot]$ is a general notation for the kernel $[{\odot}_{i}]$ of the resonant product in some chart $i\in I$ which is defined by
\begin{align*}
    [\odot_i]\big(x,y,z\big) \defeq \sum_{\vert k-\ell\vert\leqslant 1} P^i_k(x,y)  \widetilde{P}^i_\ell(x,z)\,,
\end{align*}
where $P^i_k,\widetilde{P}^i_k$ are the Littlewood-Paley blocks introduced in in Appendix~\ref{SectionLPProjectors}, and $\mcQ^\gamma_x$ is the kernel probing the regularity of the trees, introduced below in Definition~\ref{def:Qgamma}. The weak homogeneity exponent $-6$ for $ [\odot_i]$ comes from the fact that $[\odot_i]\in \mathcal{D}^\prime(M^3)$ is the Schwartz kernel of the resonant product and that our manifold $M$ has dimension $3$. 

\ssk

{ As a last piece of notation let us introduce a function space $\mcR^\gamma$ adapted to the kernels varying with a parameter $x\in M$, such as the kernel $\mcQ^\gamma_x$. Note that we already know that given $x\in M$ the kernel $\mcQ^\gamma_x\in \mcD^\prime(M\times M) $ is singular at $(x,x)$ and smooth everywhere else. Also, the singular locus is moving with $x\in M$. Our goal is to define a correct functional space $\mcR^\gamma$ that measures the singularities of $\mcQ^\gamma_x$ uniformly in $x\in M$. This is done in the following definition.

\ssk

\begin{defn}
\label{def:functionspaceQ}
Fix $x_0\in M$, along with a pair of closed neighbourhoods $U_{x_0}\Subset \Omega_{x_0} \subset M$ of $x_0$, and define a
conical neighbourhood 
$$
\mathcal{C}_{x_0}\defeq\bigcup_{x\in \Omega_{x_0}} T^*_{(x,x)}M^2 
$$ 
of $T^*_{(x_0,x_0)}M^2$. Then we define $\mcR^\gamma$ as the space of all families $(\mcQ_x)_{x\in U_{x_0}}$ of distributions 
$$
\mcQ_x\in\mcD'(M^2)\cap C^\infty\big(M^2\setminus\{(x,x)\}\big)
$$
such that
\begin{align*}
\left(e^{-(6+2\gamma)\ell}e^{-\ell\rho_{x}*}\mcQ_x \right)_{\ell\geqslant 0, x\in U_{x_0}}    
\end{align*}
is bounded in $\mcD^\prime_{\mathcal{C}_{x_0}}(M^2)$ for every family of scaling fields $\rho_x$ scaling with respect to $(x,x)$.
\end{defn}
}

\ssk

We can now define the Feynman amplitudes. 

\ssk

\begin{defn}
We view $[\odot]$ and $\mcQ^\gamma_x$ as some distributions on $\mcM^3$ and $\mcM^2$ by pull-back by the canonical projection from $\mcM^p$ to $M^p$. Denote by ${\bf d}_{V'}$ the diagonals of $\mathcal M^p$, for $V'\subset V$. The amplitude $\mcA^{\mcG}$ associated with the graph $\mcG_{(t,x)}$ is the distribution on $\mcT_J\backslash \bigcup_{V'\subset V}{\bf d}_{V'}$ defined by the product
$$
\mcA_{\mcG}(z_1,\dots,z_p) \defeq \mcQ^\gamma_x\big(x_{v(e_{\textrm{ref}})_-},x_{v(e_{\textrm{ref}})_+}\big) \prod_{e\in \widetilde E} K_e\big(z_{v(e)_-}, z_{v(e)_+}\big) \prod_{1\leqslant j\leqslant n_\mcG} [\odot]\big(
x_{v^j_*},
x_{v^j_1},x_{v^j_2} \big)
$$
with the second product corresponding to all of $V_A$, see Figure \ref{Fig1} for a fully detailed example. We talk of $\mcA_\mcG$ as the \textbf{\textsf{Feynman amplitude associated with $\mcG$}}.
\end{defn}

\ssk

The following fact is a direct consequence of Lemma \ref{LemSimpleProduct}. It allows two things in the analysis of Feynman amplitudes
\begin{enumerate}
    \item If a Feynman graph is a tree, i.e. it contains no loops, then the corresponding Feynman amplitude $\mathcal{A}_\mcG$ is always well-defined as distribution on the corresponding configuration space.
    \item We can reduce the analysis to \textbf{irreducible graphs} which contain at least one loop since joining two subgraphs by a bridge is always well-defined microlocally.
\end{enumerate}

These two facts are detailed in the next two statements. The following Lemma states that Feynman trees are always well defined.

\ssk

\begin{lem}\label{LemWellPosedTrees}
If $\mcG=(V,E)$ is a tree, for every $e\in \widetilde{E}$, each two point kernel $K_e$ belongs to the module $\mathcal{K}^{a_e}, a_e\in \mathbb{R}$ endowed with the topology of Lemma \ref{l:topologytwopointfunctiong}, each three point kernel 
$$
[\odot]\big(\cdot,\cdot,\cdot\big) \in \mcS^{-6}_{N^*({\bf d}_3\subset M^3) }(M^3)
$$ 
where the scaling is with respect to  ${\bf d}_3\subset M^3$ and the eventual marked edge $e_{\textrm{ref}}$ is associated with the propagator $\mcQ^\gamma_x$ which belongs to the topological space $\mcR^\gamma$ from Definition \ref{def:functionspaceQ}. Then the multilinear map 
$$
\Big([\odot] , (K_e)_{e\in E}, \mcQ^\gamma_x\Big) \in   \mcS^{-6}_{N^*\left({\bf d}_3\subset M^3\right) }(M^3) \times \prod_{e\in E} \mathcal{K}^{a_e}\times \mathcal{R}^\gamma  \longmapsto\mcA_\mcG \in \mcD'_{\Gamma}\big(\mcT_J \big)
$$ 
where 
$$
\Gamma = \bigcup_{V'\subset V} N^*({\bf d}_{V'})\cup N^*\left(\mcT_{V'}\right)
$$
is continuous.
\end{lem}

\ssk

The simplest example which illustrates the above claim is the composition of  pseudo-differential kernels. If we represent each kernel by an edge then composition can be interpreted in terms of Feynman rules as gluing the edges at one common vertex and this is always perfectly well defined, the diagonal singularities of the kernels do not matter. We now state a useful corollary of Lemma~\ref{LemWellPosedTrees} which allows to restrict the analysis of Feynman amplitudes to \textit{connected irreducible subgraphs}. 

\ssk

\begin{cor}
Let $\mathcal{A}_\mcG$ be a Feynman amplitude which is obtained by joining two irreducible amplitudes $\mcA_{\mcG_1},\mcA_{\mcG_2}(z_j, {\bf z}_2)$ by a propagator $K\in \mathcal{D}^\prime(M^2)$ whose wave front set is in the conormal $N^*\left({\bf d}_2  \subset M^2\right)$ to the diagonal
\begin{align*}
\mcA_{\mcG_1}({\bf z}_1,z_i)K(z_i,z_j) \mcA_{\mcG_2}(z_j, {\bf z}_2)    
\end{align*}
then if $\mcA_{\mcG_1}$ and $\mcA_{\mcG_2}(z_j, {\bf z}_2)$ are some well defined distributions with wave front set in 
\begin{equation} \label{EqConditionWaveFront} \begin{split}
WF(\mcA_{\mcG_1}) &\subset \bigcup_{V_1'\subset V_1} N^*({\bf d}_{V_1'})\cup N^*\left(\mcT_{V_1'}\right),   \\
WF(\mcA_{\mcG_2}) &\subset \bigcup_{V_2'\subset V_2} N^*({\bf d}_{V_2'})\cup N^*\left(\mcT_{V_2'}\right),
\end{split} \end{equation}
then the global amplitude $\mcA_{\mcG}$ is a well-defined distribution with wave front set included in 
\begin{align*}
\bigcup_{V'\subset V} N^*({\bf d}_{V'})\cup N^*\left( \mcT_{V} \right).    
\end{align*}
The weak homogeneity of $\mcA_\mcG$ is then the sum of the weak homogeneities of the subamplitudes and of the kernel $K$.
\end{cor}

\ssk

\begin{Dem}
Since the amplitude of a reducible graph reads
\begin{eqnarray*}
\mcA_{\mcG_1}({\bf z}_1,z_i)K(z_i,z_j) \mcA_{\mcG_2}(z_j, {\bf z}_2)
\end{eqnarray*}
for some collective variables $({\bf z}_1, z_i), (z_j, {\bf z}_2)$ partitioning $\{z_v\}_{v\in V}$ and corresponding to a partition $V=V_1\cup V_2$ of $V$, where $K(z_i,z_j)$ has wave front contained in $N^*\big({\bf d}_2\subset \mathcal M^2\big)\cup N^*\big( \{t_i=t_j\} \big) $ and \eqref{EqConditionWaveFront} holds, one can apply Lemma \ref{LemSimpleProduct} twice to 
$$
\mcA_{\mcG_1}({\bf z}_1,z_i)K(z_i,z_j)  
$$
then to
$$
\big(\mcA_{\mcG_1}({\bf z}_1,z_i)K(z_i,z_j) \big) \mcA_{\mcG_2}(z_j, {\bf z}_2).
$$
It shows that the product is well defined so the only difficulty is to treat the amplitudes $\mcA_{\mcG_1}$ and $\mcA_{\mcG_2}$.
\end{Dem}

\ssk

\subsubsection{Pointed scaling spaces.}
\label{subsubsectionpointedscaling}
We next define a parametrized version of the functional scaling spaces $\mathcal{S}_\Gamma^a$ in configuration space which generalizes Definition \ref{def:functionspaceQ}. The elements in our new functional space depend on some space time point $z=(t,x)\in \mcM$. We will use these functional spaces when we will scale the whole Feynman amplitude $\mcA_\mcG$ with respect to the marked diagonal $(z,\dots,z)\subset \mcM^p$ and test that everything is uniform in $z\in \mcM$. We start by describing the the geometrical setting. We are given:

\begin{itemize}
    \item[--] For every $z_0=(t_0,x_0)$ an open neighborhood $U_{z_0}$ of $z_0$ in $\mcM$ so that $U_{z_0}^p $
     is a neighborhood of $(z_0,\dots,z_0)\in \mcM^p$;      
       
    \item[--] a continuous family of scaling fields $\rho_{z}\in C^\infty(T(\mcM^p)), z\in U_{z_0}$ on $\mcM^p$ such that $\rho_{z}$ scales with respect to $\{(z,\dots,z)\}\subset U_{z_0}^p$ and $e^{-\ell\rho_{z}}U^p_{z_0}\subset U^p_{z_0}$ for every $z_0\in U_{z_0}$, $\ell\geqslant 0$ and the flow of $\rho_{z}$ preserves all the diagonals $\textbf{d}_I$ and marked diagonals $\textbf{d}_{I,z}$.
\end{itemize}

We need to give an important example which shows that such geometric setting is non--empty and that one can always produce such setting.

\ssk

\begin{Example}
In a local product chart $(a,b)\times U$, $U\subset M$, for every $z=(t,x)\in (a,b)\times U$ in the chart, the typical example of such a vector field reads
$$ 
\rho_z = 2  (t_1-t)\partial_{t_1}+\dots + 2(t_p-t)\partial_{t_p} +   (x_1-x)\cdot\partial_{x_1}+\dots+(x_p-x)\cdot\partial_{x_p}.
$$    
\end{Example}

\ssk

Then we define the functional data. As above we define the functional spaces associated to distributions which are weakly homogeneous at a space-time diagonal $(z,\dots,z) \in \mcM^I $, in a way which is uniform in $z\in \mcM$. Here is an example that may help fixing the setting.

\ssk

\begin{Example}
The function
$$ 
(\bbR^3)^p\ni (x_1,\dots,x_p)\mapsto \Big(\vert x_1-x \vert \cdots \vert x_p-x\vert\Big)^{-\frac{1}{2}} 
$$
is a function on the configuration space $(\bbR^3)^p$ with a singular locus $\big\{x_1=\dots=x_p=x\big\}$ which is moving with $x$.
\end{Example}

\ssk

Recall from \eqref{DefnDiagonals} the definitions of the sets ${\bf d}_I$ and $\mcT_I$.  

\ssk

\begin{defn} \label{def:functionspaceSuni}
Fix $a\in \mathbb{R}$, $J\subset \{1,\dots,p\}$ and $z_0=(t_0,x_0)\in\mcM$. Choose some neighbourhood $U_{z_0}$ of $z_0$. Then define $\mathcal{S}^a(\mcT_J ; U_{z_0})$ as the set of families $(T_{z})_{z\in U_{z_0}}$ of distributions $T_{z}\in\mcD'(\mcT_J\setminus \mathbf{d}_{p,z})$ such that 
\begin{align*}
\Big(e^{\ell a}e^{-\ell\rho_{z}*}T_z\Big)_{\ell\geqslant 0, z\in U_{z_0}}    
\end{align*}
is bounded in $\mcD^\prime(\mcT_J)$ for every $\rho_z$ as above.
\end{defn}

 Implicitly, in the definition of $\mcT_J$ there is a time variable $t$ that we view as parameter where some time variables are taken to be equal to this time parameter $t$, $t$ is not equal to $t_0$ but $(t,x)\in U_{t_0,x_0}$.

\ssk

The next statement makes explicit a construction of such families; it is concerned with recentering.

\ssk

\begin{prop}\label{Prop:recentering}
Assume we are given a distribution $T\in \mcS^a_{\Gamma}( \mcT_J)$ with $\Gamma=N^*\left(\textbf{d}_I \subset \mcT_J\right)$ for some $I\subset \{1,\dots,p\}$. Then for any admissible $\rho_z$ scaling with respect to $\textbf{d}_{V(\mcG),z}$ the family 
$$ 
\left( e^{\ell a} e^{-\ell\rho_z*}T\right)_{\ell\geqslant 0}
$$
is bounded in $\mcD^\prime(\mcT_J)$ uniformly in $z$.

 Let $I\subset \{1,\dots, p\}$ and denote by $\pi:\mcM^p\mapsto \mcM^I$ the canonical projection map. Given  $T\in \mcS^a_{\Gamma}(\mcM^I)$ where $\Gamma=\cup_{I'\subset I} N^*\textbf{d}_{I'}$ and we scale with respect to  $\textbf{d}_I$. Then 
$\pi^*T\in \mcS^a_{\pi^*\Gamma}(\mcM^p)$ where we scale with respect to  the deepest diagonal $\textbf{d}_p$ using admissible scaling fields.
\end{prop}

\ssk

Let us comment on the above statement. We are given a subamplitude $\mcA_{\mcG^\prime}$ coming with a certain scaling degree with respect to its 
deepest diagonal $\textbf{d}_{V(\mcG^\prime)}$. 
The question is how this subamplitude scales inside a bigger graph. To answer this question, we need to lift 
the subamplitude to a bigger configuration space $\mcM^{V(\mcG)}$, and we would like to examine the scaling degree 
with respect to the deepest diagonal $\textbf{d}_{V(\mcG)}$ of the larger graph or the marked diagonal 
$\textbf{d}_{V(\mcG),z}$ of the larger graph. We need to ensure the scaling degree is unaffected under changes of the scaling dynamics.

\ssk

\begin{Dem}
The proof is very similar to the proof of~\cite[Lemma 6.4.5 p.~144]{Dangthese}.
By the invariance results under scalings from Proposition~\ref{Prop:intrinsicscalingspaces}, 
the proof reduces to the comparison of linear scalings with respect to  different points.  
So we are reduced to the following generic situation,
we work on $(\bbR^{1+d})$ and we would like to compare
the two linear scaling flows~:
$$ e^{-s\rho_{z_0}}: ( t,x )\in \bbR^{1+d} \mapsto ( e^{-2s} (t-t_0)+t_0, e^{-s}(x-x_0)+x_0)\in \bbR^{1+d}  $$
which scales with respect to  $z_0=(t_0,x_0)$
and the second scaling
$$ e^{-s\rho_{z_1}} ( t,x )\in \bbR^{1+d} \mapsto ( e^{-2s} (t-t_1)+t_1, e^{-s}(x-x_1)+x_1)\in \bbR^{1+d} $$
which scales with respect to  a different point $z_1=(t_1,x_1)$.

Then just observe the identity
$ e^{-s\rho_{z_1}}=(e^{-s\rho_{z_1}} e^{s\rho_{z_0}}) e^{ -s\rho_{z_0}} $ where the 
composition $ \Phi(s)\defeq e^{-s\rho_{z_1}} e^{s\rho_{z_0}}$ has a smooth linear limit when $s\rightarrow +\infty$ as can be easily inspected by a direct calculation: $ \Phi(s):(t,x)\mapsto ( t + (e^{-2s}-1)(t_1-t_0) , x+ (e^{-s}-1)(x_1-x_0) ) $.
\end{Dem}

In spite of the fact that these spaces are rather ad hoc and not really intrinsic, they are sufficient to capture uniformity in $x\in M$ and therefore to control the size of the Feynman amplitudes uniformly in $x$.

\ssk

\subsubsection{The inductive Theorem for convergent amplitudes.}
\label{SectionProofGeneralWeinberg}

Our next goal is to describe a recursive algorithm that controls the convergence as a distribution {\it over the space} $\mcT_J$. For a given graph $\mcG$ with marked edges, given a fixed $J\subset V(\mcG)$, we prove in the next statement that for every $x_0\in M$ and every compact neighborhood  $U_{x_0}$ of $x_0$, the following multilinear Feynman map is continuous under suitable conditions on the weak homogeneity $a_e$ of the two point kernels $(K_e)_{e\in E}$ 
\begin{equation}\label{eq:Feynmanmap}
\Big([\odot] , (K_e)_{e\in E}, \mcQ_x^\gamma\Big) \in \mcS^{-6}_{N^*\left(\mathbf{d}_3 \subset M^3 \right) }(M^3) \times \prod_{e\in E} \mathcal{K}^{a_e} \times \mcR^\gamma \longmapsto\mcA_{\mcG,x} \in \mcS^{a_1}_{\Gamma}\big(\mcT_J \big)\cap  \mcS^{a_2}(\mcT_J ; U_{x_0})
\end{equation} 
where 
\begin{equation*} \begin{split}
\Gamma &= \bigcup_{V'\subset V(\mcG)} \Big(N^*({\bf d}_{V'}\subset \mcT_J)\cup N^*\left(\mcT_{V'}\subset \mcT_J\right)\Big)   \\
a_1 &= - 6 n_\mcG - \sum_{e\in E\setminus \{e_{\textrm{\it ref}}\}}a_e   \\
a_2 &= a_1 - 6 - 2\gamma.
\end{split} \end{equation*}
The space $\mcS^{a_2}(\mcT_J ; U_{x_0})$ which appears in Definition \ref{def:functionspaceSuni} accounts for the fact that our estimates should be uniform in $x\in U_{x_0}$ when we scale with respect to the marked diagonal $\mathbf{d}_{V(\mcG),x}$. Recall also that we work on $\mcT_J\subset \mcM^{V(\mcG)}$ for some $J\subset V(\mcG)$ because we take into account that our amplitudes are integrated on regions where certain time variables coincide. If the graph $\mcG$ has no distinguished edge $e_{\textrm{\it ref}}$, then we do not need to test the regularity in the space 
$ \mcS^{a_2}(\mcT_J ; U_{x_0})$ and the target functional space is just $ \mcS^{a_1}_{\Gamma}\big(\mcT_J \big)$ for $a_1=-6n_\mcG-\sum_{e\in E}a_e$.

\ssk

We only consider below some subgraphs $\mcG'=(V',E')$ of $\mcG$ which contain all the points of a given triple if ever they contain one of them. Recall all our analysis takes place in the submanifold $\mcT_J$ of $\mcM^p$. With a slight abuse of notation we will also denote by ${\bf d}_{V'}$ the diagonal ${\bf d}_{V'}\cap \mcT_J$. 

\ssk

For any marked subgraph $\mcG'$ we set 
\begin{equation*} \begin{split}
a_{\mcG',1} &\defeq -6n_{\mcG'}-\sum_{e\in E'\setminus \{e_{\textrm{ref}}\}}a_e   \\
a_{\mcG',2} &\defeq a_{\mcG',1}-6-2\gamma,
\end{split} \end{equation*}
while for any subgraph $\mcG'$ with no marked/reference edge we set $a_{\mcG'} \defeq -6n_{\mcG'}-\sum_{e\in E'}a_e $ and 
$$
\Gamma_{\mcG'} \defeq \bigcup_{V''\subset V'} \Big\{ N^*\big({\bf d}_{V''}\subset \mcT_J\big)\cup N^*\big(\mcT_{V''} \subset \mcT_J \big) \Big\}.
$$

\ssk

A given Feynman graph $\mcG$ will be said to \textit{have a loop} if $b_1(\mcG)>0$.

\ssk

\begin{thm} \label{ThmBlackBoxFeynmanGraphs}
The following holds.
\begin{enumerate}
	\item[\textsf{\textbf{(a)}}] If every connected irreducible subgraph $\mcG'=(V',E')$ of $\mcG$ with a loop satisfies
	\begin{equation} \label{EqWeinbergCondition} \begin{split}
	&a_{\mcG^\prime, 1} + \textrm{\emph{codim}}_w({\bf d}_{V'}) > 0,   \\
    &a_{\mcG^\prime, 2} + \textrm{\emph{codim}}_w({\bf d}_{V',(t,x)}) > 0,
	\end{split} \end{equation}
	then the Feynman map defined by Equation \ref{eq:Feynmanmap} is continuous.   \vspace{0.1cm}
	
	\item[\textsf{\textbf{(b)}}] If $\mcG=(V,E)$ is a graph that contains no distinguished edge $q$, every connected irreducible strict subgraph of $\mcG$ with a loop satisfies Condition \eqref{EqWeinbergCondition} and $\mcG$ satisfies 
	$$
	a_{\mcG} + \textrm{\emph{codim}}_w({\bf d}_V { \subset \mcT_J}) > -1, 
 	$$
	then there exists a family $\Lambda_{\mcY,\epsilon}$ of distributions supported on ${\bf d}_V$, with wavefront set in $N^*({\bf d}_V\subset \mcT_J)$ such that $\Lambda_\epsilon - \Lambda_{\mcY,\epsilon}$ has a limit in $\mcD'\big(\mcT_J\big)$ and the convergence occurs in $\mcS^{a'}_{\Gamma_\mcG\cup N^*({\bf d}_V)}\big(\mcT_J\big)$ for all
	$$
	a' < -\textrm{\emph{codim}}_w\big({\bf d}_V\subset \mcT_J\big)\,.
	$$
\end{enumerate}	
\end{thm}

\ssk

\noindent Before we give the proof of Theorem \ref{ThmBlackBoxFeynmanGraphs} an important Remark is in order.

\ssk

\begin{Rem}
    We aim to point an important fact. The above result is general, but relies on the fact that the irreducible subgraphs are convergent. It turns out that we took care to implement the fact that for any of the Feynman graphs that we need to study, in particular those indexing the covariances of $\Xtwo\odot\IXtwo$ and $\Xtwo\odot\IXthree$, the most divergent subgraph is the graph itself. Note that this would not be the case if we were to replace all the cubic kernels $[\odot](x,y,z)$ by $\delta_x(y)\delta_x(z)$. Indeed  while $[\odot]$ and the product of two $\delta$s have the same weak homogeneity ($-6$) with respect to the deep diagonal $\{x=y=z\}$ they do not share the same microlocal properties. We will further elaborate on this point later in Remark \ref{rem:resonant}. In particular when we deal with the resonant products $\Xtwo\odot\IXtwo$ and $\Xtwo\odot\IXthree$ in Section~\ref{SectionStochasticBounds}, this will imply that the subgraph containing the covariance $G^{(2)}(z_1,z_2)$ of $\Xtwo$ and the probe operator $\mcQ_x^\gamma(y_1,y_2)$ is very convergent, since these two operators are linked by $[\odot](y_1,z_1,w_1)$ and { $[\odot](y_2,z_2,w_2)$} for $w_i$ far from $y_i, z_i$. Conversely, replacing $[\odot](y_i,z_i,w_i)$ by $\delta_{y_i}(z_i)\delta_{y_i}(w_i)$, then this subgraph would be more divergent that the graph itself, which corresponds to the fact that while $\Xtwo\odot\IXtwo$ and $\Xtwo\odot\IXthree$ are of regularities $0^-$ and $(-1/2)^-$, the products $\Xtwo\times\IXtwo$ and $\Xtwo\times\IXthree$ are both of regularity $(-1)^-$.
\end{Rem}

\begin{Dem}
Our proof proceeds by induction on subgraphs for the inclusion relation. For a subset $V_1=V_1^\prime\cup V_A^\prime  \subset V(\mcG) $ of vertices and $E^\prime\subset E(\mcG)$ of edges, we denote by $\mcG_{V_1,E^\prime}$ the corresponding subgraph. 

The initialization of the induction is immediate, the simplest subgraphs are just the propagators or the kernel of the resonant product which have their respective wave front sets on the conormals of the diagonals.

Given a graph $\mcG$ constructed according to our rules, assume that all subgraphs $\mcG^\prime$ of $\mcG$ have well-defined Feynman amplitudes $\mcA_{\mcG^\prime}$ with wave front set in $\Gamma_{\mcG^\prime}$, without loss of generality, we may assume that $\mcG$ has $p$ vertices which means that $V(\mcG)$ is in bijection with $\{1,\dots,p\}$ hence $\mcM^{V(\mcG)}\simeq\mcM^p$. For every $I\subsetneq V(\mcG) \simeq  \{1,\dots,p\} $ we consider the open subset $\mcU_I$ defined as follows
\begin{align}
\mcU_I \defeq \Big\{ (m_1,\dots,m_p)\in \mcM^{V(\mcG)}; i\in I, j\in V(\mcG)\setminus I, m_i\neq m_j \Big\}.
\end{align}
By the Popineau-Stora covering Lemma \ref{lem:PopineauStora} we have the covering $\mcT_J\setminus \textbf{d}_{V(\mcG)}=\cup_{I\subsetneq V(\mcG)} \mcU_I$. The idea is to restrict $\mcA_\mcG|_{\mcU_I}$, then to factor the restricted amplitude as a product of subamplitudes which are well-defined by the inductive assumption and a product of smooth kernels.

For given a subset $I\subsetneq V(\mcG)$ of vertices, we set 
$$
I'\defeq V(\mcG)\setminus I, \qquad (I,I')\defeq (I\times I')\cup (I'\times I).
$$
If $K$ is another subset of $V(\mcG)$, we write $K\not\subset I^{(')}$ to say that $K$ is neither a subset of $I$ nor of $I'$. With this notation the amplitude $\mcA_\mcG$ factors as
\begin{align*}
\mcA_\mcG=\mcA_{\mcG_I} \mcA_{\mcG_{I^\prime}} \left( \prod_{e : (v(e)_-, v(e)_+)\in (I,I^\prime) } K_e\big(z_{v(e)_-},z_{v(e)_+}\big) \prod_{j:\{v^j_1,v^j_2,v^j_*\}\not\subset I^{(')} } [\odot](x_{v^j_*},x_{v^j_1},x_{v^j_2})  \right)     \,,
\end{align*}
where the subgraph $\mcG_I$, respectively  $\mcG_{I^\prime}$, is some subgraph of $\mcG$ with only vertices in $I$, respectively  $I^\prime$, and edges $e\in E$ such that both vertices $v(e)_-,v(e)_+ \in I$, respectively  $v(e)_-,v(e)_+ \in I^{\prime}$, bounding $e$ belong to $I$, respectively  $I'$. More importantly, the product on the resonant kernels indexed by integers $j$ runs over the kernels $[\odot]$ whose vertices $\big(v^j_*,(v^j_1,v^j_2) \big)$ are not entirely contained neither in $I$ nor $I^\prime$: some vertices amongst them are in $I$ and some others are in $I^\prime$. Note the fact that the product on resonant kernels might well be empty, for instance if all distinguished triples are contained either in $I$ or $I^\prime$.

\ssk

Now observe that when we restrict the amplitude $\mcA_\mcG$ on the open subset $\mcU_I$, this yields
\begin{align*}
\mcA_\mcG|_{\mcU_I}= \underset{ \text{exterior product} }{ \underbrace{  \mcA_{\mcG_I} \mcA_{\mcG_{I^\prime}} }} \; 
\underset{\in C^\infty}{\underbrace{\left( \prod_{e : (v(e)_-, v(e)_+)\in(I, I^\prime) } K_e\big(z_{v(e)_-} , z_{v(e)_+}\big) \prod_{j:\{v^j_1,v^j_2,v^j_*\}\not\subset I^{(')} } [\odot]\big(x_{v^j_*} , x_{v^j_1} , x_{v^j_2}\big) \right)}}, 
\end{align*}
where the restriction to open subset $\mcU_I$ of the product inside the parenthesis is smooth, and the product $\mcA_{\mcG_I} \mcA_{\mcG_{I^\prime}}$ denotes an exterior tensor product of distributions which is always well-defined. We know by induction that the two amplitudes $\mcA_{\mcG_I}\big((z_j)_{j\in I}\big)$ and $\mcA_{\mcG_{I'}}\big((z_j)_{j\in I'}\big)$ are well-defined distributions with wave front sets
$$
WF(\mcA_{\mcG_{I}}) \subset \Gamma_I\defeq \bigcup_{K\subset I} N^*({\bf d}_{K}\subset \mcT_J)\cup N^*\left(\mcT_{K}\subset \mcT_J  \right)\,,
$$
and 
$$
WF(\mcA_{\mcG_{ I'}}) \subset \Gamma_{I'}\defeq \bigcup_{K'\subset I'} N^*({\bf d}_{K'} \subset \mcT_J )\cup N^*\left(\mcT_{K'}\subset \mcT_J  \right).
$$

Analysing the graph $\mcG$, there are two situations, either $\mcG$ contains some
distinguished edge $\mcQ_x^\gamma$ in which case we should study the scaling behaviour in two steps, first scale on the partial diagonals $\mathbf{d}_{ V(\mcG)}$ where all points labelled by $ V(\mcG)$ collide, then study the scaling on the marked diagonals $\mathbf{d}_{ V(\mcG),(t,x)}$ (recall the marked diagonal is just one point in $\mcM^{\vert V(\mcG)\vert}$) where all points labelled by $ V(\mcG)$ collide uniformly on $(t,x)$, we use the spaces of Definition~\ref{def:functionspaceSuni} to control the uniformity in the parameter $(t,x)$. If the graph contains no distinguished edge, then we just scale in one step with respect to  $\mathbf{d}_{ V(\mcG)}$. 

\ssk

Assume we are in the first situation where $\mcG$ contains a distinguished edge, the second case is simpler to handle.
Choosing any admissible scaling field $\rho$ scaling with respect to the deep diagonal $\textbf{d}_{V(\mcG)}$  and using the stability of $\mcU_I$ by such scaling field (a consequence of admissibility of $\rho$, we made an essential use of the fact that the flow by scaling fields $\rho$ from Definition \ref{def:explicitEuler} preserves both time and space-time diagonals $\mathcal{T}_J,{\bf d}_I$ hence every $\mcU_I, I\subsetneq V(\mcG)$). The weak homogeneity of each term in factor does not depend on the choice of scaling field with respect to  $\textbf{d}_{V(\mcG)}$ by Proposition \ref{Prop:intrinsicscalingspaces}. Moreover, we also use the property that the symplectic lifts of these scaling fields preserve the conormals of all time and space-time diagonals $N^*\left(\mathcal{T}_I\right),N^*\left({\bf d}_I\right)$ so that the cones $\Gamma_I,\Gamma_{I'}$ containing microsingularities of $\mcA_{\mcG_I},\mcA_{\mcG_{I'}}$ are stable by the lifted flow: $e^{-\ell\rho*}\Gamma_I\subset \Gamma_I$, $e^{-\ell\rho*}\Gamma_{I'}\subset \Gamma_{I'}$ by Lemma \ref{lem:conormstability}. Using the fact that each element $\eta_I\in C^\infty(M^{d_{V(\mcG)}}\setminus \textbf{d}_{V_\mcG})$ of the partition of unity belongs to $\mcS_{\emptyset}^0( M^{V(\mcG)}\setminus \textbf{d}_{V(\mcG)} )$, it has scaling degree $0$ in $ C^\infty(M^{d_{V(\mcG)}})\setminus \textbf{d}_{V_\mcG}$ in the sense that 
$$ 
\left( e^{-\ell\rho*}\eta_I\right)_{\ell\geqslant 0} 
$$
is a bounded family in $C^\infty(M^{V(\mcG)}\setminus \textbf{d}_{V(\mcG)})$. We can thus define the renormalized product
$$ 
\eta_I \mcA_{\mcG_I} \mcA_{\mcG_{I^\prime}} \left( \prod_{e : (v(e)_-, v(e)_+)\in (I,I^\prime) } K_e(z_{v(e)_-},z_{v(e)_+}) \prod_{j:\{v^j_1,v^j_2,v^j_*\}\not\subset I^{(')} } [\odot](x_{v^j_*},x_{v^j_1},x_{v^j_2})  \right)     
$$
on $\mcT_J\setminus \textbf{d}_{V(\mcG),(t,x)}$ by applying Proposition \ref{PropProductThm} twice and we find that the product is well-defined since the weak homogeneity of the above product is $>-\text{codim}_w(\textbf{d}_{V(\mcG)})$ by assumption. We insist that when we scale with respect to  $\textbf{d}_{V(\mcG)}$ the weak homogeneity of $\mcQ^\gamma_x$ is $0$ by Lemma \ref{Lem:Qgammasmooth}. By summation over $I$ we conclude that $\mcA_{\mcG}$ is well-defined on $\mcT_J\setminus \textbf{d}_{V(\mcG),(t,x)}$.

To define an extension on the whole configuration space $\mcT_J$ we need the two step extension from Theorem \ref{Thm:extensiontwosteps}.
We now need to apply Proposition~\ref{PropProductThm} and now Theorem \ref{Thm:extensiontwosteps} a second time
to the expression
$$ 
\eta_I \mcA_{\mcG_I} \mcA_{\mcG_{I^\prime}} \left( \prod_{e : (v(e)_-, v(e)_+)\in (I,I^\prime) } K_e\big(z_{v(e)_-} , z_{v(e)_+}\big) \prod_{j:\{v^j_1,v^j_2,v^j_*\}\not\subset I^{(')} } [\odot]\big(x_{v^j_*} , x_{v^j_1} , x_{v^j_2}\big) \right)
$$
but for the scaling fields with respect to  the distinguished diagonal $\textbf{d}_{V(\mcG),(t,x)}$. {\it It is at this point that we are using our second inductive assumption on the weak homogeneity of all subgraphs containing the distinguished edge}, equivalently all subamplitudes containing the propagator $\mcQ_x^\gamma$. We make an induction on all marked subgraphs whose amplitude contains $\mcQ^\gamma_x$, the induction starts with $\mcQ^\gamma_x$ itself. We use the fact that the scaling degree of $\mcQ^\gamma_x$ with respect to $\textbf{d}_{V(\mcG)}\setminus \textbf{d}_{V(\mcG),(t,x)}$ equals $0$, while when scaling with respect to $\textbf{d}_{V(\mcG),(t,x)}$, the scaling degree of $\mcQ^\gamma_x$ is now given by $-6-2\gamma$. This last claim is proven in Lemma~\ref{Lem:Qgammasmooth} below.

\ssk

At this stage given a graph $\mcG$, assume by induction that all subgraphs $\mcG^\prime$ containing the distinguished edge have scaling degree with respect to  the marked diagonal $> -\text{codim}_w(\textbf{d}_{V(\mcG^\prime),(t,x)})$. Then every product defined on $\mcT_J\setminus \textbf{d}_{V(\mcG)}$
$$ 
\eta_I \mcA_{\mcG_I} \mcA_{\mcG_{I^\prime}} \left( \prod_{e : (v(e)_-, v(e)_+)\in (I,I^\prime) } K_e\big(z_{v(e)_-} , z_{v(e)_+}\big) \prod_{j:\{v^j_1,v^j_2,v^j_*\}\not\subset I^{(')} } [\odot]\big(x_{v^j_*} , x_{v^j_1} , x_{v^j_2}\big) \right)
$$
will be weakly homogeneous of degree $a_{\mcG,2}=\sum_{e} a_e- 6n_\mcG  -6-2\gamma $ when we scale with respect to  the marked diagonal $\textbf{d}_{V(\mcG),(t,x)}$ where we need to include the scaling degree of $\mcQ_x^\gamma$ which equals $-6-2\gamma$.
So each piece above satisfies the assumptions of Theorem~\ref{Thm:extensiontwosteps} hence extends to $\mcT_J$.

The uniform estimates in $x$ essentially follow from the stability assumptions of the cones by change of scaling in Proposition~\ref{Prop:recentering}.
\end{Dem}

\ssk

\begin{lem}\label{Lem:Qgammasmooth}
Fix $d\geqslant2$ and denote by $\pi_{\leqslant 2} : \mcM^d\rightarrow\mcM^2$ the canonical projection. For all values of $\gamma\in \mathbb{R}$ the kernel $\pi_{\leqslant 2}^*\mcQ^\gamma_x$ is a smooth germ near $\mathbf{d}_d\setminus \mathbf{d}_{d,x}$ in $\bbR^2\times M^d\setminus \mathbf{d}_{d,x}$. We define
\[
\mcO \defeq \Big\{m\in\bbR^2\times M^d:\mathrm{dist}(m, {\bf d}_{d,x})> 2\mathrm{dist}(m,{\bf d}_{d}) \Big\}.
\]
Then $\pi_{\leqslant 2}^*\mcQ^\gamma_x$ is weakly homogeneous of degree $0$ when scaling with respect to $\mathbf{d}_d\setminus \mathbf{d}_{d,x}$ in
$\Omega \setminus {\bf d}_d$ in the following sense: For every compact $K \Subset \Omega$, for all scaling field $\rho$ scaling with respect to  $\textbf{d}_d$, there exists $\ell_K>0$ such that the family
$$ 
\Big(\big\{e^{-\ell\rho*}\pi_{\leqslant 2}^*\mcQ^\gamma_x\big\}|_K\Big)_{\ell\geqslant \ell_K}
$$
is bounded in $C^\infty(K)$.
\end{lem}

\ssk

\begin{Dem}
For any element $m\in\mcO$ we know that the limit point $m_\infty = \lim_{t\uparrow \infty} e^{-\ell\rho}(m) \in \textbf{d}_d\setminus \textbf{d}_{d,x}$ has the form $m_\infty=(y,\dots,y,t,t)$ for $y\neq x$ and therefore $\pi_{\leqslant 2}^*\mcQ^\gamma_x(\cdot)$ is smooth near the limit point $m_\infty=(y,\dots,y,t,t)$ since $\pi_{\leqslant 2}^*\mcQ^\gamma_x(m_\infty)=\mcQ^\gamma_x(y,y)$ for $y\neq x$. Now by a continuity argument, we know that for any element $m\in\mcO$ there exists a neighbourhood $V_m\subset \Omega$ and $\ell_m>0$, such that for all $\ell\geqslant \ell_m$ one has $e^{-\ell\rho}(V_m)\subset \Omega$ and $\textrm{dist}\left( \pi_{\leqslant 2}(e^{-\ell\rho}(V_m)), (x,x) \right)\geqslant \delta>0 $ hence $\pi_{\leqslant 2}^*\mcQ^\gamma_x|_{e^{-\ell\rho}(V_m)} $ is smooth uniformly in $\ell\geqslant\ell_m$ which yields the claim by compactness.
\end{Dem}

\section{Random fields from renormalization}
\label{SectionStochasticBounds}

{We are finally in a position to prove the convergence} of the enhancement $\widehat{\xi}_r$ of the regularized noise $\xi_r$
$$
\widehat{\xi}_r \defeq \Big(\xi_r, \begin{tikzpicture}[scale=0.3,baseline=0cm]
\node at (0,0) [dot] (0) {};
\node at (0.3,0.6)  [noise] (noise1) {};
\node at (-0.3,0.6)  [noise] (noise2) {};
\draw[K] (0) to (noise1);
\draw[K] (0) to (noise2);
\end{tikzpicture}_r, \; 
\IXthree_r\,,\;
\begin{tikzpicture}[scale=0.3,baseline=0cm]
\node at (0,0) [dot] (0) {};
\node at (0,0.5) [dot] (1) {};
\node at (-0.4,1)  [noise] (noise1) {};
\node at (0,1.2)  [noise] (noise2) {};
\node at (0.4,1)  [noise] (noise3) {};
\draw[K] (0) to (1);
\draw[K] (1) to (noise1);
\draw[K] (1) to (noise2);
\draw[K] (1) to (noise3);
\end{tikzpicture}_r \odot \begin{tikzpicture}[scale=0.3,baseline=0cm]
\node at (0,0)  [dot] (1) {};
\node at (0,0.8)  [noise] (2) {};
\draw[K] (1) to (2);
\end{tikzpicture}_r, \; 
\begin{tikzpicture}[scale=0.3,baseline=0cm]
\node at (0,0) [dot] (0) {};
\node at (0,0.5) [dot] (1) {};
\node at (-0.4,1)  [noise] (noise1) {};
\node at (0.4,1)  [noise] (noise2) {};
\draw[K] (0) to (1);
\draw[K] (1) to (noise1);
\draw[K] (1) to (noise2);
\end{tikzpicture}_r \odot \begin{tikzpicture}[scale=0.3,baseline=0cm]
\node at (0,0) [dot] (0) {};
\node at (0.3,0.6)  [noise] (noise1) {};
\node at (-0.3,0.6)  [noise] (noise2) {};
\draw[K] (0) to (noise1);
\draw[K] (0) to (noise2);
\end{tikzpicture}_r - \frac{b_r}{3}, \; 
\big\vert\nabla \begin{tikzpicture}[scale=0.3,baseline=0cm]
\node at (0,0) [dot] (0) {};
\node at (0,0.5) [dot] (1) {};
\node at (-0.4,1)  [noise] (noise1) {};
\node at (0.4,1)  [noise] (noise2) {};
\draw[K] (0) to (1);
\draw[K] (1) to (noise1);
\draw[K] (1) to (noise2);
\end{tikzpicture}_r\big\vert^2 - \frac{b_r}{3}, \;  
\begin{tikzpicture}[scale=0.3,baseline=0cm]
\node at (0,0) [dot] (0) {};
\node at (0,0.5) [dot] (1) {};
\node at (-0.4,1)  [noise] (noise1) {};
\node at (0,1.2)  [noise] (noise2) {};
\node at (0.4,1)  [noise] (noise3) {};
\draw[K] (0) to (1);
\draw[K] (1) to (noise1);
\draw[K] (1) to (noise2);
\draw[K] (1) to (noise3);
\end{tikzpicture}_r \odot \begin{tikzpicture}[scale=0.3,baseline=0cm]
\node at (0,0) [dot] (0) {};
\node at (0.3,0.6)  [noise] (noise1) {};
\node at (-0.3,0.6)  [noise] (noise2) {};
\draw[K] (0) to (noise1);
\draw[K] (0) to (noise2);
\end{tikzpicture}_r - b_r \begin{tikzpicture}[scale=0.3,baseline=0cm]
\node at (0,0)  [dot] (1) {};
\node at (0,0.8)  [noise] (2) {};
\draw[K] (1) to (2);
\end{tikzpicture}_r \,\Big)
$$ 
{in the space
$$
\mcC^{-5/2-\eps}([0,T]\times M)\times C_T\mcC^{-1-2\epsilon}(M) \times C_T\mcC^{1/2-3\epsilon}(M) \times C_T\mcC^{-4\epsilon}(M)^3 \times C_T\mcC^{-1/2-5\epsilon}(M)\,.
$$ 
Here is the general strategy that we use to deal with the convergence of the non-Wick product random fields.} First, a general object $\tau_m$ lying in the $p$-th inhomogeneous Gaussian chaos can be expended onto the $p$ first homogeneous chaoses, so that its covariance can be controlled by the covariances of the projections and subsequently by some amplitudes indexed by Feynman graphs, denoted $\mcA_{mn}$ for $n=p,p-2,$ etc. The next step is to localize the amplitudes in local charts using the partition of unity $1=\sum_{i\in I}{\varphi_i}$ subordinated to the cover $M=\cup_{i\in I} U_i$, and to control for every $i\in I$ the localized version $\mcA^i_{mn}$. The amplitudes $\mcA_{mn}$ correspond to products of kernels, and the aim is to identify the range of $\gamma$ for which they are well defined using item \textbf{\textsf{(a)}} of Theorem~\ref{ThmBlackBoxFeynmanGraphs}. Two difficulties may occur:

\begin{enumerate}
    \item Because they contain a divergent subgraph (the term coming from the lowest chaos), the amplitudes $\mcA_{m0}$ and $\mcA_{m1}$ cannot be directly handled using item \textbf{\textsf{(a)}} of Theorem \ref{ThmBlackBoxFeynmanGraphs}, and rather requires the use of item \textbf{\textsf{(b)}}: this amounts to subtracting a local counterterm.   \vspace{0.15cm}
    
    \item The amplitudes $\mcA_{mn}$ coming from higher chaoses do not contain any divergent subgraph and could therefore be defined using item \textbf{\textsf{(a)}} of Theorem~\ref{ThmBlackBoxFeynmanGraphs}. However, due to the fact that $\mcA_{mn}$ contains a subamplitude including $\mcQ_x^\gamma$ whose weak homogeneity is worse than that of the whole amplitude, it may happen that the value of $\gamma$ could a priori be deteriorated. This last case requires a special treatment leveraging the precise microlocal properties of $[\odot]$ -- this more careful analysis is performed in Section~\ref{subsubsec:shielding}.
\end{enumerate}
{The analysis is completed in Section \ref{subsection_completing}. Note here that we devote Section \ref{SectionDivergentPartTau20} to the exact calculation of the counterterms from the divergent subamplitude; this uses some subtle arguments involving heat asymptotics, stationary phase and the borderline case of our extension Theorem \ref{ThmCanonicalExtension} where we need a renormalization.}

\medskip

\subsection{Diagrammatic notation for higher chaoses} 
\label{subsubsec:diagramnotation}

To handle the quartic and quintic terms, we first need to introduce some notations. 

\ssk

From Appendix \ref{SectionLPProjectors}, we define a family $(P^i_k,\widetilde{P}^i_k)_{k\in \mathbb{N},i\in I}$ of generalized Littlewood-Paley-Stein projectors indexed by the frequency $2^k$ and the chart index $i\in I$.
The localized resonant product $\odot_i$, where $i$ is the chart index, is defined 
in the appendix as $u\odot_iv=\sum_{\vert k-\ell\vert\leqslant 1} P^i_k(u)\widetilde{P}^i_\ell(v) $.
The goal of the present subsection is to deal with the localized objects 
$$
\tau_1=
\begin{tikzpicture}[scale=0.3,baseline=0cm]
\node at (0,0) [dot] (0) {};
\node at (0,0.5) [dot] (1) {};
\node at (-0.4,1)  [noise] (noise1) {};
\node at (0,1.2)  [noise] (noise2) {};
\node at (0.4,1)  [noise] (noise3) {};
\draw[K] (0) to (1);
\draw[K] (1) to (noise1);
\draw[K] (1) to (noise2);
\draw[K] (1) to (noise3);
\end{tikzpicture}_r \odot_i \begin{tikzpicture}[scale=0.3,baseline=0cm]
\node at (0,0)  [dot] (1) {};
\node at (0,0.8)  [noise] (2) {};
\draw[K] (1) to (2);
\end{tikzpicture}_r\,,  \quad 
\overline\tau_2 = \begin{tikzpicture}[scale=0.3,baseline=0cm]
\node at (0,0) [dot] (0) {};
\node at (0,0.5) [dot] (1) {};
\node at (-0.4,1)  [noise] (noise1) {};
\node at (0.4,1)  [noise] (noise2) {};
\draw[K] (0) to (1);
\draw[K] (1) to (noise1);
\draw[K] (1) to (noise2);
\end{tikzpicture}_r \odot_i \begin{tikzpicture}[scale=0.3,baseline=0cm]
\node at (0,0) [dot] (0) {};
\node at (0.3,0.6)  [noise] (noise1) {};
\node at (-0.3,0.6)  [noise] (noise2) {};
\draw[K] (0) to (noise1);
\draw[K] (0) to (noise2);
\end{tikzpicture}_r - {\varphi_i}\frac{b_r}{3}\,,  \quad
\overline\tau_3 = {\varphi_i}\big\vert\nabla \begin{tikzpicture}[scale=0.3,baseline=0cm]
\node at (0,0) [dot] (0) {};
\node at (0,0.5) [dot] (1) {};
\node at (-0.4,1)  [noise] (noise1) {};
\node at (0.4,1)  [noise] (noise2) {};
\draw[K] (0) to (1);
\draw[K] (1) to (noise1);
\draw[K] (1) to (noise2);
\end{tikzpicture}_r\big\vert^2 - {\varphi_i}\frac{b_r}{3}\,,
\quad
\overline{\tau}_4 = \begin{tikzpicture}[scale=0.3,baseline=0cm]
\node at (0,0) [dot] (0) {};
\node at (0,0.5) [dot] (1) {};
\node at (-0.4,1)  [noise] (noise1) {};
\node at (0,1.2)  [noise] (noise2) {};
\node at (0.4,1)  [noise] (noise3) {};
\draw[K] (0) to (1);
\draw[K] (1) to (noise1);
\draw[K] (1) to (noise2);
\draw[K] (1) to (noise3);
\end{tikzpicture}_r \odot_i \begin{tikzpicture}[scale=0.3,baseline=0cm]
\node at (0,0) [dot] (0) {};
\node at (0.3,0.6)  [noise] (noise1) {};
\node at (-0.3,0.6)  [noise] (noise2) {};
\draw[K] (0) to (noise1);
\draw[K] (0) to (noise2);
\end{tikzpicture}_r -{\varphi_i} b_r \begin{tikzpicture}[scale=0.3,baseline=0cm]
\node at (0,0)  [dot] (1) {};
\node at (0,0.8)  [noise] (2) {};
\draw[K] (1) to (2);
\end{tikzpicture}_r\,,
$$
where ${\varphi_i}\in C^\infty_c(U_i)$ is the cut--off function used to define $\odot_i$, and find for each of them the range of exponents $\gamma$ for which our test criterion is verified.

Note that the trees are defined with the localized resonant term $\odot_i$ made with these projectors and recall from Appendix~\ref{SectionLPProjectors} that these resonant terms are not commutative, in the sense that $A\odot_i B\neq B\odot_i A$. However, we do not have to worry about the definitions of $\tau_1$ and $\overline{\tau}_2$, since the analytic properties of $P^i_k$ and $\widetilde{P}^i_k$ are similar, which entails that for every $A$, $B$, the construction of the renormalized part of $B\odot_i A$ is totally equivalent to that of $A\odot_i B$ which we provide here.   

\ssk

Denote by $[\odot_i](x,y,z)$ the kernel of the localized resonant operator $\odot_i$ on the chart $U_i\subset M$. Recall that we write

$$
\underline{\mathcal{L}}^{-1}(t-s,x,y) = {\bf 1}_{(-\infty,t]}(s) \, e^{-(t-s)P}(x,y)
$$
and

\begin{align*} 
G_r^{(p)}(t-s,x,y) &=  \left(\Big\{e^{-(\vert t-s\vert+2r) P}P^{-1}\Big\}(x,y)\right)^p,\hspace{0.31cm} (1\leqslant p\leqslant 3)   \\
[\odot_i](x,y,z) &=  \sum_{\vert k-\ell\vert\leqslant 1} P^i_k(x,y) \, \widetilde{P}^i_\ell(x,z),\hspace{1.3cm} (i\in I)   \\
\mcQ^\gamma_x(y,z) &= \sum_{i\in I} \sum_{k\geqslant-1} 2^{2k\gamma} P_k^i (x,y) P_k^i (x,z),\hspace{0.42cm} (\gamma\in\bbR, x\in M).
\end{align*}
These kernels are in some spaces of the form $\mcS^a_\Gamma$ for different ambiant spaces, scaling exponents $a$ and wavefront sets as follows. 
\begin{itemize}
	\item[--] The kernel $\underline{\mathcal{L}}^{-1}(t-s,x,y)$ has scaling exponent $-3$ and wavefront set 
	$$
	N^*\left(\{t=s\}\times {\bf d}_2\subset \mathbb{R}^2\times M^2 \right).
	$$
	\item[--] The kernel $G_r^{(p)}(t-s,x,y)$ have scaling exponent $-p$ and wavefront set 
	$$
	N^*\left(\{t=s\}\subset \mathbb{R}^2\times M^2 \right)\cup N^*\left(\{t=s\}\times {\bf d}_2 \subset \mathbb{R}^2\times M^2 \right).
	$$
	\item[--] The kernel $[\odot_i](x,y,z)$ has scaling exponent $-6$ and wavefront set
	$$
	 N^*\left(\{x=y=z\}\subset M^3 \right)\,.
	$$
	\item[--] The kernel $\mcQ_x^\gamma(y_1,y_2)$ has scaling exponent $-6-2\gamma$ and wavefront set
	$$
	  T^*_{(x,x)}  (M^2) \,.
	$$
\end{itemize}

Beware that in the case of $\mcQ^\gamma_x$, the {wavefront set} depends on the marked point $x\in M$. In this list, the kernels on $\mathcal M^2$ satisfy a local diagonal bound of the form
$$
\big\vert\partial_{\sqrt{t},\sqrt{s},y_1,y_2}^\alpha K\big\vert \lesssim \left(\sqrt{\vert t-s \vert}+\vert y_1-y_2\vert \right)^{-a-\vert\alpha\vert}
$$
for the corresponding scaling exponent $a$. We also often see the kernels $\underline{\mcL}^{-1}$, $\mcQ_x^\gamma$ and $G_r^{(1)}$ as some time-dependent space operators $X(t_1-t_2)$ whose kernels are then given by $X(t_1-t_2,y_1,y_2)$. The proof of the above microlocal bounds is done in detail in our companion work~\cite[Theorem 1.1]{BDFTCompanion}, in Section 4 thereof.

\ssk

We use a pictorial representation of the $\tau_i$ in which the black dot $\bullet$ represents a resonant operator and the noises are coloured circles. In a given graph, noises of the same colour are integrated outside all diagonals of the corresponding set of variables. In the present subsection, it will be convenient to first discuss stochastic estimates for Wick ordered elements which live in homogeneous Wiener chaoses, then explain why our locally covariant renormalization yields stochastic elements that differ from the {corresponding} Wick renormalized elements only up to {some} higher regularity elements. The Wiener chaos decomposition of the $\tau_i$ is

\makebox[\textwidth][c]{
\begin{minipage}{\dimexpr\textwidth+10cm}
\begin{align*}
\tau_1 &= \begin{tikzpicture}[scale=0.6,baseline=0cm]
\node at (0,0) [blackdot] (0) {};
\node at (-0.4,0.7) [dot] (1) {};
\node at (0.4,0.7)  [noiseblue] (noise1) {};
\node at (0,1.2)  [noisegray] (noise2) {};
\node at (-0.4,1.4)  [noisegray] (noise3) {};
\node at (-0.8,1.2)  [noisegray] (noise4) {};
\draw[DK] (0) to (1);
\draw[K] (0) to (noise1);
\draw[K] (1) to (noise2);
\draw[K] (1) to (noise3);
\draw[K] (1) to (noise4);
\end{tikzpicture} 
=
\begin{tikzpicture}[scale=0.6,baseline=0cm]
\node at (0,0) [blackdot] (0) {};
\node at (-0.4,0.7) [dot] (1) {};
\node at (0.4,0.7)  [noisegray] (noise1) {};
\node at (0,1.2)  [noisegray] (noise2) {};
\node at (-0.4,1.4)  [noisegray] (noise3) {};
\node at (-0.8,1.2)  [noisegray] (noise4) {};
\draw[DK] (0) to (1);
\draw[K] (0) to (noise1);
\draw[K] (1) to (noise2);
\draw[K] (1) to (noise3);
\draw[K] (1) to (noise4);
\end{tikzpicture} 
+
3\begin{tikzpicture}[scale=0.6,baseline=0cm]
\node at (0,0) [blackdot] (0) {};
\node at (-0.4,0.7) [dot] (1) {};
\node at (0.4,0.7)  [dot] (noise2) {};
\node at (-0.4,1.4)  [noisegray] (noise3) {};
\node at (-0.8,1.2)  [noisegray] (noise4) {};
\draw[DK] (0) to (1);
\draw[K] (0) to (noise2);
\draw[K] (1) to (noise2);
\draw[K] (1) to (noise3);
\draw[K] (1) to (noise4);
\end{tikzpicture} \,,\qquad\qquad\quad\;
\tau_2= \begin{tikzpicture}[scale=0.6,baseline=0cm]
\node at (0,0) [blackdot] (0) {};
\node at (0.1,0.1) [dot] (0prime) {};
\node at (-0.4,0.7) [dot] (1) {};
\node at (-0.7,1.2)  [noisegray] (noise1) {};
\node at (-0.1,1.2)  [noisegray] (noise2) {};
\node at (0.4,0.7)  [noiseblue] (noise3) {};
\node at (0.75,0.2)  [noiseblue] (noise4) {};
\draw[DK] (0) to (1);
\draw[K] (1) to (noise1);
\draw[K] (1) to (noise2);
\draw[K] (0prime) to (noise3);
\draw[K] (0prime) to (noise4);
\end{tikzpicture}  
=
\begin{tikzpicture}[scale=0.6,baseline=0cm]
\node at (0,0) [blackdot] (0) {};
\node at (0.1,0.1) [dot] (0prime) {};
\node at (-0.4,0.7) [dot] (1) {};
\node at (-0.7,1.2)  [noisegray] (noise1) {};
\node at (-0.1,1.2)  [noisegray] (noise2) {};
\node at (0.4,0.7)  [noisegray] (noise3) {};
\node at (0.75,0.2)  [noisegray] (noise4) {};
\draw[DK] (0) to (1);
\draw[K] (1) to (noise1);
\draw[K] (1) to (noise2);
\draw[K] (0prime) to (noise3);
\draw[K] (0prime) to (noise4);
\end{tikzpicture}  
+ 2
\begin{tikzpicture}[scale=0.6,baseline=0cm]
\node at (0,0) [blackdot] (0) {};
\node at (0.1,0.1) [dot] (0prime) {};
\node at (-0.4,0.7) [dot] (1) {};
\node at (-0.7,1.2)  [noisegray] (noise1) {};
\node at (0.4,0.7)  [dot] (noise3) {};
\node at (0.75,0.2)  [noisegray] (noise4) {};
\draw[DK] (0) to (1);
\draw[K] (1) to (noise1);
\draw[K] (1) to (noise3);
\draw[K] (0prime) to (noise3);
\draw[K] (0prime) to (noise4);
\end{tikzpicture} 
+ 2
 \begin{tikzpicture}[scale=0.6,baseline=0cm]
\node at (0,0) [blackdot] (0) {};
\node at (0.1,0.1) [dot] (0prime) {};
\node at (-0.4,0.7) [dot] (1) {};
\node at (0.4,0.7)  [dot] (noise3) {};
\node at (1,0.7)  [dot] (noise4) {};
\draw[DK] (0) to (1);
\draw[K] (1) to (noise3);
\draw[K] (1) to[bend left=+50] (noise4);
\draw[K] (0prime) to (noise3);
\draw[K] (0prime) to (noise4);
\end{tikzpicture}\,,\\
\tau_3 &= \begin{tikzpicture}[scale=0.6,baseline=0cm]
\node at (0,0) [blackdot] (0) {};
\node at (-0.4,0.7) [dot] (1) {};
\node at (0.4,0.7) [dot] (1prime) {};
\node at (-0.6,1.2)  [noisegray] (noise1) {};
\node at (-0.2,1.2)  [noisegray] (noise2) {};
\node at (0.2,1.2)  [noiseblue] (noise3) {};
\node at (0.6,1.2)  [noiseblue] (noise4) {};
\draw[DK] (0) to (1);
\draw[DK,green] (0) to (1prime);
\draw[K] (1) to (noise1);
\draw[K] (1) to (noise2);
\draw[K] (1prime) to (noise3);
\draw[K] (1prime) to (noise4);
\end{tikzpicture}
=
\begin{tikzpicture}[scale=0.6,baseline=0cm]
\node at (0,0) [blackdot] (0) {};
\node at (-0.4,0.7) [dot] (1) {};
\node at (0.4,0.7) [dot] (1prime) {};
\node at (-0.6,1.2)  [noisegray] (noise1) {};
\node at (-0.2,1.2)  [noisegray] (noise2) {};
\node at (0.2,1.2)  [noisegray] (noise3) {};
\node at (0.6,1.2)  [noisegray] (noise4) {};
\draw[DK] (0) to (1);
\draw[DK,green] (0) to (1prime);
\draw[K] (1) to (noise1);
\draw[K] (1) to (noise2);
\draw[K] (1prime) to (noise3);
\draw[K] (1prime) to (noise4);
\end{tikzpicture}
+ 2 
\begin{tikzpicture}[scale=0.6,baseline=0cm]
\node at (0,0) [blackdot] (0) {};
\node at (-0.4,0.7) [dot] (1) {};
\node at (0.4,0.7) [dot] (1prime) {};
\node at (-0.6,1.2)  [noisegray] (noise1) {};
\node at (0,1.2)  [dot] (noise2) {};
\node at (0,1.2)  [dot] (noise3) {};
\node at (0.6,1.2)  [noisegray] (noise4) {};
\draw[DK] (0) to (1);
\draw[DK,green] (0) to (1prime);
\draw[K] (1) to (noise1);
\draw[K] (1) to (noise2);
\draw[K] (1prime) to (noise3);
\draw[K] (1prime) to (noise4);
\end{tikzpicture}
+2
\begin{tikzpicture}[scale=0.6,baseline=0cm]
\node at (0,0) [blackdot] (0) {};
\node at (-0.4,0.7) [dot] (1) {};
\node at (0.4,0.7) [dot] (1prime) {};
\node at (0,1.7)  [dot] (noise1) {};
\node at (0,1.1)  [dot] (noise2) {};
\node at (0,1.1)  [dot] (noise3) {};
\node at (0,1.7)  [dot] (noise4) {};
\draw[DK] (0) to (1);
\draw[DK,green] (0) to (1prime);
\draw[K] (1) to (noise1);
\draw[K] (1) to (noise2);
\draw[K] (1prime) to (noise3);
\draw[K] (1prime) to (noise4);
\end{tikzpicture}  \,,\qquad
\tau_4 = \begin{tikzpicture}[scale=0.6,baseline=0cm]
\node at (0,0) [blackdot] (0) {};
\node at (0.1,0.1) [dot] (0prime) {};
\node at (-0.4,0.7) [dot] (1) {};
\node at (-0.8,1.2)  [noisegray] (noise1) {};
\node at (-0.4,1.4)  [noisegray] (noise2) {};
\node at (0,1.2)  [noisegray] (noise3) {};
\node at (0.4,0.7)  [noiseblue] (noise4) {};
\node at (0.75,0.2)  [noiseblue] (noise5) {};
\draw[DK] (0) to (1);
\draw[K] (1) to (noise1);
\draw[K] (1) to (noise2);
\draw[K] (1) to (noise3);
\draw[K] (0prime) to (noise4);
\draw[K] (0prime) to (noise5);
\end{tikzpicture} = 
\begin{tikzpicture}[scale=0.6,baseline=0cm]
\node at (0,0) [blackdot] (0) {};
\node at (0.1,0.1) [dot] (0prime) {};
\node at (-0.4,0.7) [dot] (1) {};
\node at (-0.8,1.2)  [noisegray] (noise1) {};
\node at (-0.4,1.4)  [noisegray] (noise2) {};
\node at (0,1.2)  [noisegray] (noise3) {};
\node at (0.4,0.7)  [noisegray] (noise4) {};
\node at (0.75,0.2)  [noisegray] (noise5) {};
\draw[DK] (0) to (1);
\draw[K] (1) to (noise1);
\draw[K] (1) to (noise2);
\draw[K] (1) to (noise3);
\draw[K] (0prime) to (noise4);
\draw[K] (0prime) to (noise5);
\end{tikzpicture}  
+ 6
\begin{tikzpicture}[scale=0.6,baseline=0cm]
\node at (0,0) [blackdot] (0) {};
\node at (0.1,0.1) [dot] (0prime) {};
\node at (-0.4,0.7) [dot] (1) {};
\node at (-0.8,1.2)  [noisegray] (noise1) {};
\node at (-0.4,1.4)  [noisegray] (noise2) {};
\node at (0.4,0.7)  [dot] (noise4) {};
\node at (0.75,0.2)  [noisegray] (noise5) {};
\draw[DK] (0) to (1);
\draw[K] (1) to (noise1);
\draw[K] (1) to (noise2);
\draw[K] (1) to (noise4);
\draw[K] (0prime) to (noise4);
\draw[K] (0prime) to (noise5);
\end{tikzpicture}
+ 6
\begin{tikzpicture}[scale=0.6,baseline=0cm]
\node at (0,0) [blackdot] (0) {};
\node at (0.1,0.1) [dot] (0prime) {};
\node at (-0.4,0.7) [dot] (1) {};
\node at (-0.8,1.2)  [noisegray] (noise1) {};
\node at (0.4,0.7)  [dot] (noise4) {};
\node at (0.8,1.4)  [dot] (noise5) {};
\draw[DK] (0) to (1);
\draw[K] (1) to (noise1);
\draw[K] (1) to (noise4);
\draw[K] (0prime) to (noise4);
\draw[K] (0prime) to[bend right=+50] (noise5);
\draw[K] (1) to (noise5);
\end{tikzpicture}\,.
\end{align*}
\end{minipage}
}
\esp

\noindent where black edges stand for the kernel of $\underline\mcL^{-1}$ and \textcolor{green}{green edges} stand for the kernel of $\Delta\underline\mcL^{-1}$. The reason for the presence of the Laplacian operators in $\tau_3$ will be made clear in Section~\ref{subsec:tau3}.

For $m\in\{1,2,3\}$ {we write} $\tau_m = \tau_{m4} + \tau_{m2} + \tau_{m0}$ with $\tau_{mn}$ in the homogeneous chaos of degree $n\in \{0,2,4\}$ and accordingly $\tau_4= \tau_{45} + \tau_{43} + \tau_{41}$. By orthogonality 
$$
\bbE\Big[\mcQ_x^\gamma\big(\tau_m(t,\bigcdot),\tau_m(t,\bigcdot)\big)\Big] =\sum_{n\in\bbN}\bbE\Big[\mcQ_x^\gamma\big(\tau_{mn}(t,\bigcdot),\tau_{mn}(t,\bigcdot)\big)\Big].
$$
Remember that we are not interested in the $\tau_m$ themselves but in their renormalized versions generically written $\overline\tau_m = \tau_m - c_{m,r}$, with $c_{m,r}$ for `counterterm', for which we still have the orthogonality relation
\begin{equation} \label{EqChaosDecomposition} \begin{split}
\bbE\Big[\mcQ_x^\gamma\big(\overline\tau_m(t,\bigcdot) , &\overline\tau_m(t,\bigcdot)\big)\Big]   \\
&= \sum_{n\geqslant2} \bbE\Big[\mcQ_x^\gamma\big(\tau_{mn}(t,\bigcdot),\tau_{mn}(t,\bigcdot)\big)\Big] + \bbE\Big[\mcQ_x^\gamma\big((\tau_{mp}(t,\bigcdot)-c_{{ m},r}),(\tau_{mp}(t,\bigcdot)-c_{m,r})\big)\Big],
\end{split} \end{equation}
where $p=0$ for $m\in\{1,2,3\}$ and $p=1$ for $m=4$. Also, note that each element $\tau_{mn}$ in the Wiener chaos can be written as $F_n(:\xi^{\otimes n}:) $ where each $F_n:C^\infty(\mathcal M)^{\otimes n}\rightarrow \mathcal{D}^\prime(\mathcal M)$ is a multilinear functional valued in distributions involving iterated integrals of the heat operator, products represented by the trees. The following elementary well-known result tells us that we only need to bound some very symmetric Feynman diagrams to bound the preceding expectations.

\begin{lem} \label{LemHairerTrick}
Let $F\in L^2(M^n)$ be a function of $n$ variables on some compact Riemannian manifold $(M,g)$, $\xi$ the white noise on $(M,g)$. Then:
\begin{eqnarray*}
\mathbb{E}\Big[ \left\langle F,:\xi^{\otimes n}:\right\rangle^2 \Big] \leqslant \Vert F\Vert^2_{L^2(M^n)}.
\end{eqnarray*} 
\end{lem}

\begin{Dem}
We define the symmetrization operator $S_n$ as 
$$
S_n(\varphi)(x_1,\dots,x_n)=\frac{1}{n!}\sum_{\sigma\in S_n}\varphi(x_{\sigma(1)},\dots,x_{\sigma(n)}).
$$ 
Note the important fact that $S_n$ is self--adjoint on $L^2(M^n)$, so $S_n$ is the orthogonal projector on the closed subspace of symmetric $L^2$ functions. By the It\^o isometry property we have
$$
\mathbb{E}\left( \left\langle F,:\xi^{\otimes n}:\right\rangle^2 \right)=\mathbb{E}\left( \left\langle S_nF,:\xi^{\otimes n}:\right\rangle^2 \right)=\Vert S_nF\Vert^2_{L^2(M^n)};
$$
hence to prove the {statement} it suffices to note that $\Vert S_nF \Vert_{L^2(M^n)}\leqslant \Vert F\Vert_{L^2(M^n)}$, which is obvious since $S_n$ is an orthogonal projector.
\end{Dem}

Each expectation {in \eqref{EqChaosDecomposition}} can therefore be bounded by a quantity of the form
$$
\int_{\mathcal M^{n+2}} F_n(x_1;y_1,\dots,y_n)F_n(x_2;y_1,\dots,y_n) \mcQ_x^\gamma(x_1,x_2) \, \rmd y\rmd x \,,
$$
which can be represented by some mirror symmetric Feynman diagram. Using a {\color{purple} purple edge} for the kernel of the operator $\mcQ_x^\gamma$, a dotted edge for $\underline\mcL^{-1}$, a black edge for $G_r^{(1)}$, and a \textcolor{green}{green edge} for $\Delta\underline\mcL^{-1}$, we have

\begin{align*}
\bbE\Big[\mcQ_x^\gamma\big(\overline\tau_1(t,\bigcdot) \,,\, \overline\tau_1(t,\bigcdot)\big)\Big] &\lesssim 
\begin{tikzpicture}[scale=0.6,baseline=-0.1cm]
\node at (0,-0.3) [dot] (1) {};
\node at (0,0.3) [dot] (2) {};
\node at (0,0.6) [dot] (3) {};
\node at (0,0.9) [dot] (4) {};
\node at (-0.6,0.6) [dot] (5) {};
\node at (0.6,0.6) [dot] (6) {};
\node at (-0.8,0) [blackdot] (7) {};
\node at (0.8,0) [blackdot] (8) {};
\draw[K] (1) to (7);
\draw[K] (1) to (8);
\draw[K] (2) to (5);
\draw[K] (2) to (6);
\draw[K] (3) to (5);
\draw[K] (3) to (6);
\draw[K] (4) to (5);
\draw[K] (4) to (6);
\draw[DK] (7) to (5);
\draw[DK] (8) to (6);
\draw[K,purple] (7) to[bend right=+80] (8);
\end{tikzpicture}
+ 
\begin{tikzpicture}[scale=0.6,baseline=-0.1cm]
\node at (0,0.4) [dot] (1) {};
\node at (0,1) [dot] (2) {};
\node at (-0.4,0.7) [dot] (3) {};
\node at (-0.4,0.2) [dot] (4) {};
\node at (0.4,0.7) [dot] (5) {};
\node at (0.4,0.2) [dot] (6) {};
\node at (-0.8,0) [blackdot] (7) {};
\node at (0.8,0) [blackdot] (8) {};
\draw[K] (1) to (3);
\draw[K] (1) to (5);
\draw[K] (2) to (3);
\draw[K] (2) to (5);
\draw[DK] (3) to (7);
\draw[K] (3) to (4);
\draw[K] (4) to (7);
\draw[K] (5) to (6);
\draw[DK] (5) to (8);
\draw[K] (6) to (8);
\draw[K,purple] (7) to[bend right=+80] (8);
\end{tikzpicture}=:\mcG_{14}+\mcG_{12}\,,  
\end{align*}
\begin{align*}
\bbE\Big[\mcQ_x^\gamma\big(\overline\tau_2(t,\bigcdot) \,,\, \overline\tau_2(t,\bigcdot)\big)\Big] &\lesssim 
\begin{tikzpicture}[scale=0.6,baseline=-0.1cm]
\node at (0,-0.2) [dot] (1) {};
\node at (0,0.2) [dot] (2) {};
\node at (0,0.4) [dot] (3) {};
\node at (0,0.8) [dot] (4) {};
\node at (-0.4,0.6) [dot] (5) {};
\node at (0.4,0.6) [dot] (6) {};
\node at (-0.8,0) [blackdot] (7) {};
\node at (0.8,0) [blackdot] (8) {};
\draw[K] (1) to (7);
\draw[K] (1) to (8);
\draw[K] (2) to (7);
\draw[K] (2) to (8);
\draw[K] (3) to (5);
\draw[K] (3) to (6);
\draw[K] (4) to (5);
\draw[K] (4) to (6);
\draw[DK] (7) to (5);
\draw[DK] (8) to (6);
\draw[K,purple] (7) to[bend right=+80] (8);
\end{tikzpicture}
+ 
\begin{tikzpicture}[scale=0.6,baseline=-0.1cm]
\node at (0,-0.4) [dot] (1) {};
\node at (0,0.8) [dot] (2) {};
\node at (-0.6,0.4) [dot] (3) {};
\node at (-0.3,0) [dot] (4) {};
\node at (0.6,0.4) [dot] (5) {};
\node at (0.3,0) [dot] (6) {};
\node at (-0.9,0) [blackdot] (7) {};
\node at (0.9,0) [blackdot] (8) {};
\draw[K] (1) to (7);
\draw[K] (1) to (8);
\draw[K] (2) to (3);
\draw[K] (2) to (5);
\draw[DK] (3) to (7);
\draw[K] (3) to (4);
\draw[K] (4) to (3);
\draw[K] (4) to (7);
\draw[K] (5) to (6);
\draw[DK] (5) to (8);
\draw[K] (6) to (8);
\draw[K,purple] (7) to[bend right=+80] (8);
\end{tikzpicture}
+
\bbE\Big[\mcQ_x^\gamma\big((\tau_{20}(t,\bigcdot)-{\varphi_i}\frac{b_r}{3}) \,,\, (\tau_{20}(t,\bigcdot)-{\varphi_i}\frac{b_r}{3})\big)\Big]   \\
&\eqdef \mcG_{24} + \mcG_{22} + \mcG_{20}\,,  
\end{align*}

\begin{align*}
\bbE\Big[\mcQ_x^\gamma\big(\overline\tau_3(t,\bigcdot) \,,\, \overline\tau_3(t,\bigcdot)\big)\Big] &\lesssim 
\begin{tikzpicture}[scale=0.6,baseline=-0.1cm]
\node at (0,-0.4) [dot] (1) {};
\node at (0,0) [dot] (2) {};
\node at (0,0.2) [dot] (3) {};
\node at (0,0.6) [dot] (4) {};
\node at (-0.3,0.4) [dot] (5) {};
\node at (0.3,0.4) [dot] (6) {};
\node at (-0.3,-0.2) [dot] (7) {};
\node at (0.3,-0.2) [dot] (8) {};
\node at (-0.9,0) [blackdot] (9) {};
\node at (0.9,0) [blackdot] (10) {};
\draw[K] (1) to (7);
\draw[K] (1) to (8);
\draw[K] (2) to (7);
\draw[K] (2) to (8);
\draw[K] (3) to (5);
\draw[K] (3) to (6);
\draw[K] (4) to (5);
\draw[K] (4) to (6);
\draw[DK] (9) to (5);
\draw[DK, green] (9) to (7);
\draw[DK] (10) to (6);
\draw[DK, green] (10) to (8);
\draw[K,purple] (9) to[bend right=+80] (10);
\end{tikzpicture}
+ 
\begin{tikzpicture}[scale=0.6,baseline=-0.1cm]
\node at (0,-0.55) [dot] (1) {};
\node at (0,0.55) [dot] (2) {};
\node at (-0.5,0.3) [dot] (3) {};
\node at (-0.5,-0.3) [dot] (4) {};
\node at (-0.2,0) [dot] (5) {};
\node at (0.5,0.3) [dot] (6) {};
\node at (0.5,-0.3) [dot] (7) {};
\node at (0.2,0) [dot] (8) {};
\node at (-1,0) [blackdot] (9) {};
\node at (1,0) [blackdot] (10) {};
\draw[K] (1) to (4);
\draw[K] (1) to (7);
\draw[K] (2) to (3);
\draw[K] (2) to (6);
\draw[K] (3) to (5);
\draw[K] (4) to (5);
\draw[K] (6) to (2);
\draw[K] (6) to (8);
\draw[K] (7) to (8);
\draw[DK] (9) to (3);
\draw[DK,green] (9) to (4);
\draw[DK] (10) to (6);
\draw[DK, green] (10) to (7);
\draw[K,purple] (9) to[bend right=+80] (10);
\end{tikzpicture}
+
\bbE\Big[\mcQ_x^\gamma\big((\tau_{30}(t,\bigcdot)-{\varphi_i}\frac{b_r}{3}) \,,\, (\tau_{30}(t,\bigcdot)-{\varphi_i}\frac{b_r}{3})\big)\Big]   \\
&\eqdef \mcG_{34}+\mcG_{32} +\mcG_{30}\,,
\end{align*}

\begin{align*}   
 \bbE[\mcQ_x^\gamma\big(\overline\tau_4(t,\bigcdot) \,,\, \overline\tau_4(t,\bigcdot)\big)] &\lesssim 
\begin{tikzpicture}[scale=0.6,baseline=-0.1cm]
\node at (0,-0.3) [dot] (1) {};
\node at (0,-0.1) [dot] (2) {};
\node at (0,0.3) [dot] (3) {};
\node at (0,0.5) [dot] (4) {};
\node at (0,0.8) [dot] (5) {};
\node at (-0.4,0.6) [dot] (6) {};
\node at (0.4,0.6) [dot] (7) {};
\node at (-0.8,0) [blackdot] (8) {};
\node at (0.8,0) [blackdot] (9) {};
\draw[K] (1) to (8);
\draw[K] (1) to (9);
\draw[K] (2) to (8);
\draw[K] (2) to (9);
\draw[K] (3) to (6);
\draw[K] (3) to (7);
\draw[K] (4) to (6);
\draw[K] (4) to (7);
\draw[K] (5) to (6);
\draw[K] (5) to (7);
\draw[DK] (8) to (6);
\draw[DK] (9) to (7);
\draw[K,purple] (8) to[bend right=+80] (9);
\end{tikzpicture}
+ 
\begin{tikzpicture}[scale=0.6,baseline=-0.1cm]
\node at (0,-0.4) [dot] (1) {};
\node at (0,0.35) [dot] (2) {};
\node at (0,0.85) [dot] (3) {};
\node at (-0.4,0.6) [dot] (4) {};
\node at (-0.4,0) [dot] (5) {};
\node at (0.4,0.6) [dot] (6) {};
\node at (0.4,0) [dot] (7) {};
\node at (-0.9,0) [blackdot] (8) {};
\node at (0.9,0) [blackdot] (9) {};
\draw[K] (1) to (8);
\draw[K] (1) to (9);
\draw[K] (2) to (4);
\draw[K] (2) to (6);
\draw[K] (3) to (4);
\draw[K] (3) to (6);
\draw[DK] (4) to (8);
\draw[K] (4) to (5);
\draw[K] (5) to (8);
\draw[K] (6) to (7);
\draw[DK] (6) to (9);
\draw[K] (7) to (9);
\draw[K,purple] (8) to[bend right=+80] (9);
\end{tikzpicture}
+ 
\bbE\Big[\mcQ_x^\gamma\big((\tau_{41}(t,\bigcdot)-{\varphi_i}b_r\X_r) \,,\, (\tau_{41}(t,\bigcdot)-{\varphi_i}b_r\X_r)\big)\Big]   \\
&\eqdef \mcG_{45}+\mcG_{43} +\mcG_{41}.
\end{align*}

\esp

\noindent We will use the notation $\mcG_{mn}$ for {the} positive quantity represented by the mirror graph associated with $\tau_{mn}$ and the notation $\mcA_{mn}$ for the associated distribution on the corresponding configuration space ; we use Theorem \ref{ThmBlackBoxFeynmanGraphs} to prove their convergence. We use the word `{\sl amplitude}' to talk about any of the $\mcA_{mn}$. The terms involving $\tau_{20}$ and $\tau_{30}$ are treated differently from the mirror graphs. Recall that the vertex set is partitioned as $V=V'\sqcup V_A$. We always denote by $x_a,x_b,$ etc. the space points associated with the vertices in $V'$, while in our setting $V_A$ will always contain two triplets $((v^j_1,v^j_2),v^j_*)$ whose associated space points are denoted $((y_1,y_2),y_*)$ and $((z_1,z_2),z_*)$ for $n_\mcG=2$.

Our aim is now to control all of the eleven amplitudes $\mcA_{mn}$ using Theorem~\ref{ThmBlackBoxFeynmanGraphs}. It turns out that this naive approach will only work for $\mcA_{12}$ and, after renormalization, $\mcA_{20}$ and $\mcA_{30}$. All the remaining eight amplitudes need a special care for one of their subamplitudes. This analysis is performed in Section \ref{subsubsec:shielding}. Once the problematic subamplitudes are constructed, it will then suffice to plug this knowledge inside the recursion of the proof of Theorem~\ref{ThmBlackBoxFeynmanGraphs}, and this will yields the existence of $\mcA_{mn}$.

\medskip

\subsection{Modification of the inductive proof}
\label{subsubsec:shielding}

As anticipated, while Theorem~\ref{ThmBlackBoxFeynmanGraphs} turns out to be sufficient to handle some graphs and subgraphs of the graphs introduced in Section \ref{subsubsec:diagramnotation}, yet most of the total amplitudes need a particular care in order to be controlled.

\begin{enumerate}
	\item The amplitudes of the graphs $\mcG_{14}$, $\mcG_{24}$, and $\mcG_{45}$ are given for $(m,n)\in\{(1,4 ),(2,4),(4,5)\}$ by
\begin{align*}
    \mathcal A_{mn}= &[\odot](y_*,y_1,y_2) \,  [\odot](z_*,z_1,z_2) \,  \mcQ_x^\gamma(y_*,z_*) \,  G^{(k)}_r(0,y_1,z_1)  \, \underline{\mcL}^{-1}(t-s_a,y_2,x_a)  \, \underline{\mcL}^{-1}(t-s_b,z_2,x_b)   \\
    &G^{(\ell)}_r(s_a-s_b,x_a,x_b),
\end{align*}
where $(k,\ell)=(1,3)$ for $\mcG_{14}$, $(k,\ell)=(2,2)$ for $\mcG_{24}$, and $(k,\ell)=(2,3)$ for $\mcG_{45}$, and have a subgraph with amplitude reading
\begin{align*}
 \mcC^{(k)}(y_*,y_1,y_2,z_*,z_1,z_2)\defeq [\odot](y_*,y_1,y_2)  \,  [\odot](z_*,z_1,z_2) \,  \mcQ_x^\gamma(y_*,z_*) \, G^{(k)}_r(0,y_1,z_1) 
\end{align*}
which is \textit{less convergent}, in the sense of the power counting from Theorem \ref{ThmBlackBoxFeynmanGraphs}, than the whole graph: since this is a one loop subgraph, naively applying the power counting criterion from Theorem~\ref{ThmBlackBoxFeynmanGraphs}) on the deepest diagonal $\{ y_*=y_1=y_2=z_*=z_1=z_2=x \}$ would result in a potential loss in the regularity $\gamma$. The exact same phenomenon also occurs for the amplitude \eqref{eq:amplitude34} of the graph $\mcG_{34}$ whose subamplitude
\begin{align*}
  \widetilde\mcC^{(2)}\big(y_*,y_1,y_2,&z_*,z_1,z_2,x_c,x_d,s_c,s_d\big) \defeq [\odot](y_*,y_1,y_2) \, [\odot](z_*,z_1,z_2)   \\
  &\times \mcQ_x^\gamma(y_*,z_*) \, \Delta\underline\mcL^{-1}(t-s_c,y_1,x_c) \, \Delta\underline\mcL^{-1}(t-s_d,z_1,x_d) \, G^{(2)}_r(s_c-s_d,x_c,x_d)
\end{align*}
is also \textit{less convergent} than the whole amplitude, which equally forbids the use of a naive power counting argument.   \vspace{0.15cm}

	\item The amplitudes of the graphs $\mcG_{22}$ and $\mcG_{43}$ are given for $(m,n)\in\{(2,2),(4,3)\}$ by
\begin{align*}
    \mathcal A_{mn}= &[\odot](y_*,y_1,y_2)  \,  [\odot](z_*,z_1,z_2) \, \mcQ_x^\gamma(y_*,z_*) \, G^{(1)}_r(0,y_1,z_1)    \underline{\mcL}^{-1}(t-s_a,y_2,x_a) \, \underline{\mcL}^{-1}(t-s_b,z_2,x_b)   \\
    & G^{(1)}_r(t-s_a,y_1,x_a) \,  G^{(1)}_r(t-s_b,y_2,x_b) \, G^{(\ell)}_r(s_a-s_b,x_a,x_b),
\end{align*}
where $\ell=1$ for $\mcG_{22}$ and $\ell=2$ for $\mcG_{43}$, and have a subgraph with amplitude $\mcC^{(1)}$ which is \textit{less convergent} than the whole graph (marginally for $\mcG_{43}$). Again the same phenomenon occurs for the amplitude \eqref{eq:amplitude32} of the graph $\mcG_{32}$ that has a subamplitude $\widetilde \mcC^{(1)}$ defined as $\widetilde \mcC^{(2)}$ above with $G^{(2)}_r$ replaced with $G^{(1)}_r$ which is \textit{less convergent} than the whole amplitude.   \vspace{0.15cm}

	\item The amplitude of the graph $\mcG_{41}$, which is given by
\begin{align}\label{eq:defA41}
    \mathcal A_{41}= &   \mcQ_x^\gamma(y_*,z_*)  \, \mcB(y_*,y_1,y_2,x_a,s_a) \, \mcB(z_*,z_1,z_2,x_b,s_b)   G^{(1)}_r(s_a-s_b,x_a,x_b),
\end{align}
where
\begin{align} \label{eq:defB}
\mcB(y_*,y_1,y_2,x_a,s_a) \defeq [\odot](y_*,y_1,y_2)G^{(2)}_r( t-s_a,y_1,x_a) \, \underline{\mcL}^{-1}(t-s_a,y_1,x_a)-b_r\delta_{\mathbf{d}_4}\,,
\end{align}
requires a renormalization to define the subamplitudes 
$$
\mcB(y_*,y_1,y_2,x_a,s_a) \textrm{ and } \mcB(z_*,z_1,z_2,x_b,s_b).
$$
\end{enumerate}

Fortunately, for the items $\textsf{(a)}$ and $\textsf{(b)}$, the presence of the kernel $[\odot]$ does prevent this loss in the value of $\gamma$ from occurring, due to a subtle mechanism which is not taken into account by the recursion of Theorem~\ref{ThmBlackBoxFeynmanGraphs}. Indeed, while the kernel $[\odot](y_*,y_1,y_2)$ is divergent on the deep diagonal $\{y_*=y_1=y_2\}$, it is \textit{smooth} as long as one of the three points is distinct from the two others and applying any smoothing operator at one of its input variables, indexed by $(y_1,y_2)$, makes the \textit{whole kernel smooth}. In particular, provided the kernel $[\odot]$ is tested against a smooth kernel in one of its two input variables $(y_1,y_2)$, then it is smooth in the two remaining variables -- see Lemma~\ref{Lemshield}.
 In terms of multilinear operators, this translates the important fact that 
\begin{center}
    for every $f\in\mcD'(M)$ and $g\in C^\infty(M)$, $f\odot g\in C^\infty(M)$.
\end{center}
To take both this effect and the renormalization of the divergent subamplitudes of $\mcA_{41}$ into account, we introduce in this section a minor modification of the induction in the proof of Theorem~\ref{ThmBlackBoxFeynmanGraphs}. More precisely, the strategy is to first construct the subamplitudes by hand, and then to inject our knowledge of the subamplitudes into the induction yielding Theorem~\ref{ThmBlackBoxFeynmanGraphs}. Indeed, if an amplitude $\mcA$ contains the subamplitude $\mcB$, $\mcC^{(k)}$ or $\widetilde\mcC^{(k)}$, then we will add the estimates on the subamplitude to the hypothesis of Theorem~\ref{ThmBlackBoxFeynmanGraphs} when applying it to $\mcA$, which will yield control over $\mcA$ avoiding the issue which would naively be caused by the subamplitude.

\subsubsection{The smoothing effect from the kernel of the resonant operator.} 
\label{subsect_smoothing_effect}

This section is devoted to items $\textsf{(a)}$ and $\textsf{(b)}$ above, that is to say with the construction of the subamplitudes $\mcC^{(k)}$ and $\widetilde\mcC^{(k)}$ for $k\in\{1,2\}$. Note that by a naive application of Theorem~\ref{ThmCanonicalExtension}, it would be possible to construct these subamplitudes for every $\gamma<-k/2$. However, to avoid loosing some regularity, the aim is to actually construct them for any $\gamma\in\bbR$. This is possible thanks to the lemma below, which expresses the \textit{smoothing property of the kernel $[\odot](x,y,w)$}: outside the diagonal $\{x=y=z\}$, $[\odot](x,y,w)$ is smooth. Therefore, in sequel, the presence of this kernel will guarantee that several subamplitudes containing it actually have an arbitrary positive power counting.

\ssk

\begin{lem}\label{Lemshield}
Fix a chart label $i\in I$, and define a collection of scaling fields $(\rho_{i,x})_{x\in M}$ as follows: for $x\in U_i$, the scaling field $\rho_{i,x}$ is such that $\kappa_i(e^{-t\rho_{i,x}}z )= e^{-t}(\kappa_i(z)-\kappa_i(x))+\kappa_i(x)$, while if $x\notin U_i$ we choose for $\rho_{i,x}$ any scaling field with respect to $x\in M\setminus U_i$. Also fix $\Xi_i\in C^\infty_c(M^3)$ such that $\Xi_i(y_*,y_1,y_2)$ has support in $y_2$ contained in the open chart $U_i$. Then, the following holds: for any $x\in M$, $\varphi \in \mcD (M)$, and any scaling vector field $\rho_{i,x}$ scaling with respect to $x$ chosen as above, the partial integration
\begin{align*}
I^i_{u}(y_*,y_1)\defeq  e^{-6u}\int_{M}  \left(e^{-u\rho_{i,x}*}  [\odot] \right) (y_*,y_1,y_2) \varphi(y_2) \Xi_i(y_*,y_1,y_2)  \rmd v(y_2) 
\end{align*}
is an element of $ C^\infty (M^2)$ bounded \textbf{uniformly} in $u\in [0,+\infty)$ and $x\in M$.
\end{lem}

\ssk

\begin{Dem}
Recall that in the chart $i\in I$, we have
\begin{align*}
[\odot_i ](y_*,y_1,y_2)&=\sum_{|k-l|\leqslant 1} 2^{3k+3l}\frac{\psi_i(y_*)\widetilde\psi_i(y_*)}{ \vert g_i\vert(y_1) \vert g_i\vert(y_2) } \psi\big(2^k|y_1-y_*|\big)   \psi\big(2^l|y_2-y_*| \big) \kappa_{i*}^{\otimes2}({\varphi_i}\otimes\widetilde{\varphi_i})(y_1,y_2) \,,
\end{align*} 
where $\vert g_i\vert\defeq \sqrt{\det (\kappa_{i*}g) } $, and that the family
$$ e^{-6u}  \left(e^{-u\rho_{i,x}*}  [\odot_i] \right)\,,\; u\in [0,+\infty)\,,\; x\in M\,, $$
is bounded in  $  \mcD^\prime_{N^*(\mathbf{d}_3)} (M^3)$. 
We first assume without loss of generality that $x\in U_i$ then we choose a \textbf{very specific scaling vector field}
adapted to our chart structure as
$$ e^{-u\rho_{i,x}}(y_*,y_1,y_2  )\defeq (e^{-u}(y_*-x)+x,e^{-u}(y_1-x)+x,e^{-u}(y_2-x)+x)\,,$$
so the scaling is linear in the chart $\kappa_i:U_i\rightarrow \bbR^3$. Here and throughout the proof, with a slight abuse of notation, for any $a\in M$ we also use the notation $a$ to denote $\kappa_i(a)$. With the same above of notation, we identify $I^i_u$ with its localization on $U_i$, which thus reads

\makebox[\textwidth][c]{
\begin{minipage}{\dimexpr\textwidth+10cm}
\begin{align*}
I_{u}^i(y_*,y_1)=&\sum_{|k-l|\leqslant 1} 2^{3k+3l}e^{-6u} \int_{\bbR^3} \psi_i(e^{-u}( y_*-x)+x)\widetilde{\psi}_i(e^{-u}(y_*-x)+x) \\
&\times \frac{\psi\big(2^ke^{-u}\vert y_1-y_* \vert\big) \psi\big(2^le^{-u}\vert y_2-y_* \vert\big)}{\vert g_i\vert ( e^{-u}(y_1-y_*)+y_*)  \vert g_i\vert  ( e^{-u}(y_2-y_*)+y_*)  }    \\
& \times \kappa^{\otimes 2}_{i*}({\varphi_i}\otimes\widetilde{\varphi_i})(e^{-u}(y_1-x)+x,e^{-u}(y_2-x)+x)
 \varphi (y_2)\Xi_i(y_*,y_1,y_2)  \rmd y_2\, ,
\end{align*}
\end{minipage}
}

\noindent 

Our goal is to prove that the above function $I_{u}^i$ is smooth in the variables $(y_*,y_1)$ 
uniformly in the scale parameter $u\geqslant0$ and in $x\in\bbR^3$. At this point in our proof, it is crucial that the scaling by $\rho_{i,x}$ is the same scaling as the one used to define the Littlewood-Paley blocks. We consider  the series  with $k=l$ in $I_{u}^i$, other series are treated similarly, which reads
\begin{align*}
\sum_{k=0}^\infty 2^{6k}e^{-6u}\psi\big(2^ke^{-u}\vert y_1-y_* \vert\big) \psi\big(2^ke^{-u}\vert y_2-y_* \vert\big)\in \mcS^\prime(\bbR^3\times \bbR^3 \times \bbR^3)  \,, 
\end{align*}
and write $e^{-u}=2^{-\ell}$. The above family can be re-expressed as
\begin{align*}
&\sum_{k=0}^\infty 2^{6(k-\ell)}\psi\big(2^{k-\ell}\vert y_1-y_* \vert\big) \psi\big(2^{k-\ell}\vert y_2-y_* \vert\big)  
=\sum_{k=-\ell}^\infty 2^{6k}\psi\big(2^{k}\vert y_1-y_* \vert\big) \psi\big(2^{k}\vert y_2-y_* \vert\big)\,,
\end{align*}
the summation getting shifted. Moreover, introduce the family of smooth functions 
$$
F_{\ell,x}(y_*,y_1,y_2) 
$$
defined by the formula
\begin{equation*} \begin{split}
&\frac{\psi_i(2^{-\ell}( y_*-x)+x)\widetilde{\psi}_i(2^{-\ell}(y_*-x)+x) \kappa^{\otimes 2}_{i*}({\varphi_i}\otimes\widetilde{\varphi_i})\big(2^{-\ell}(y_1-x)+x,2^{-\ell}(y_2-x)+x\big)}{\vert g_i\vert(2^{-\ell}(y_1-y_* ) +y_* ) \vert g_i\vert(2^{-\ell}(y_2-y_* ) +y_*)}   \\
&\times \varphi(y_2)\Xi_i(y_*,y_1,y_2)
\end{split} \end{equation*}
They are bounded in $C^\infty$ as functions of the variables $y_*,y_1,y_2 $ uniformly in $\ell,x$, and have compact support in the variable $y_2$ in some compact $K\Subset \bbR^3$. By Plancherel's theorem applied to the variable $y_2$, we have
\begin{equation*} \begin{split} 
\sum_{k=-\ell}^\infty 2^{6k} \int_{\bbR^3} &\psi\big(2^{k}\vert y_1-y_* \vert\big) \psi\big(2^{k}\vert y_2-y_* \vert\big)   F_{\ell,x}(y_*,y_1,y_2)   \rmd y_2   \\
&=\sum_{k=-\ell}^\infty 2^{3k} \int_{\bbR^3} e^{\imath\xi_2\cdot y_*} \psi\big(2^{k}\vert y_1-y_* \vert\big) \eta\big(2^{-k}\xi_2\big)   \mathcal{F}_{y_2} F_{\ell,x}(y_*,y_1,\xi_2) \rmd\xi_2,
\end{split} \end{equation*}
where we recall that $\eta$ denotes the Fourier transform of $\psi\circ|\bigcdot|$. Using the fact that $F_{\ell,x}$ is smooth, which entails the bound $ \partial_{y_*,y_1}^\alpha \mathcal{F}_{y_2} F_{\ell,x}(y_*,y_1,\xi_2) = \mathcal{O}\big(\langle\xi_2\rangle^{-2N}\big) $, we obtain
$$\sum_{k=1}^\infty 2^{3k}  \int_{\bbR^3} e^{\imath\xi_2\cdot y_*} \psi\big(2^{k}\vert y_1-y_* \vert\big)\eta\big(2^{-k}\xi_2\big)   \mathcal{F}_{y_2} F_{\ell,x}(y_*,y_1,\xi_2)   \rmd\xi_2 \lesssim \sum_{k=1}^\infty 2^{(3-N)k}  \int_{\bbR^3}  \langle\xi_2\rangle^{-N}  \rmd\xi_2\,,$$
which choosing $N$ large enough is absolutely convergent. Uniform in $x\in\bbR^3$ bounds on the derivatives can be obtained in a similar fashion. Regarding the sum over $k$ negative we have
\begin{equation*} \begin{split}
   \bigg|\sum_{k=-\ell}^0  2^{3k}  \int_{\bbR^3} e^{\imath\xi_2\cdot y_*} \psi\big(2^{k}\vert y_1-y_* \vert\big) \eta\big(2^{-k}\xi_2\big)   \mathcal{F}_{y_2} F_{\ell,x}(y_*,y_1,\xi_2)\rmd\xi_2\bigg|
   \lesssim   \sum_{k=0}^\ell 2^{-3k} \int_{\bbR^3}\eta\big(2^{k}\xi_2\big)  \langle\xi_2\rangle^{-N} \rmd\xi_2 \lesssim 1,
\end{split} \end{equation*}
uniformly in $\ell\geqslant 0$. Indeed $\sum_{k=0}^\ell 2^{-3k}\eta(2^{k}\xi_2) \leqslant \sum_{k=0}^\ell \eta(2^{k}\xi_2)\lesssim 1 $ by definition the Littlewood-Paley blocks. This concludes the proof of the claim.
\end{Dem}

\ssk

\begin{rem} \label{rem:resonant}
We note that this smoothing effect is second microlocal in nature and cannot be captured by classical (``conical'')
wave front set analysis. In fact, it can be only captured semiclassically. Let us just explain things in the flat case.
Indeed, the wave front set of both $[\odot](y_*,y_1,y_2)$ and $\delta_{y_*}(y_1)\delta_{y_*}(y_2)$ are 
contained in the conormal of the deepest diagonal
$$ 
\Big\{ y_1=y_2=y_*; \eta_1+\eta_2+\eta_*=0 \Big\} 
$$
but we see that the Fourier support of $[\odot]$ is smaller
than the hyperplane $\{ \eta_1+\eta_2+\eta_*=0 \}$: indeed,
it is contained in the subset $\{ \eta_1+\eta_2+\eta_*=0, \vert \eta_1\vert\simeq \vert \eta_2\vert\}$ which is not a closed conical subset, since $\{ \eta_1+\eta_2+\eta_*=0,  \eta_1\neq 0, \eta_2\neq 0\}$ is not a closed conical subset. This is this precise Fourier support of the kernel $[\odot]$ which is responsible for the smoothing effect. The purpose of the above lemma is therefore to capture such phenomenon on manifolds using local charts. 
\end{rem}

\ssk

With the above lemma expressing the smoothing property of the kernel $[\odot]$ at hand, we are ready to construct the subamplitudes $\mcC^{(k)}$ and $\widetilde\mcC^{(k)}$. Note that since these amplitudes contain $\mcQ^\gamma_x$, we aim to construct them both on $\mathbf{d}_6\setminus\mathbf{d}_{6,x}$ and $\mathbf{d}_{6,x}$, where it has a different weak homogeneity.

\ssk

\begin{lem}\label{LemC^k}
Fix $k\in\{1,2\}$. For any $\gamma\in \bbR$, the distribution $\mcC^{(k)}$ is well-defined on $M^6$. More precisely, we have
\begin{align*}
    \mcC^{(k)}\in\mcS^{-18-k-2\gamma} (M^6;U_{x_0})\cap \mcS_{\Gamma}^{-12-k}
\end{align*}
uniformly in $x\in U_{x_0}$, where

\makebox[\textwidth][c]{
\begin{minipage}{\dimexpr\textwidth+10cm}
\begin{align*}
\Gamma =&N^*(\mathbf{d}_{6})\cup N^*\{ y_*=y_1=y_2 \} \cup N^*\{ z_*=z_1=z_2 \} 
\cup N^*\{ y_1=z_1 \}
\\
& \cup N^*\{ y_*=y_1=y_2=z_1  \}\cup  N^*\{ z_*=z_1=z_2=y_1  \} \,.
\end{align*}
\end{minipage}
}

\noindent Moreover, a similar statement holds for $\widetilde\mcC^{(k)}$, which is well-defined as a distribution on $\bbR^2\times M^8$ and belongs to $\mcS^{-18-k-2\gamma} (\bbR^2\times M^8;U_{x_0})\cap \mcS_{\widetilde\Gamma}^{-12-k}$ where $\widetilde\Gamma$ is a union of conormals defined accordingly.

\end{lem}

\ssk

\begin{Rem}
    With this lemma in hand, we will be able to inject our knowledge on $\mcC^{(k)}$ and $\widetilde\mcC^{(k)}$ inside the recurrence of the proof of Theorem~\ref{ThmBlackBoxFeynmanGraphs}, when applied to an amplitude $\mcA$ containing these subamplitudes. Since they exist for every $\gamma\in\bbR$, they will therefore not contributes to the upper bound on $\gamma$, which will actually come from the scaling of the whole amplitude $\mcA$ itself as desired. 
\end{Rem}

\ssk

\begin{Dem}
We prove the statement for $\mcC^{(k)}$, the proof for $\widetilde\mcC^{(k)}$ being a straightforward modification (\eqref{eq:DeltaL-1} shows that $\Delta\underline\mcL^{-1}$ actually scales like a Dirac, so that its presence does not affect the argument). Recall that this subamplitude reads
\begin{align*}
\mcC^{(k)}(y_*,y_1,y_2,z_*,z_1,z_2)= [\odot](y_*,y_1,y_2)   [\odot](z_*,z_1,z_2)      \mcQ_x^\gamma(y_*,z_*)   G^{(1)}_r(0,y_1,z_1)\,. 
\end{align*}
We first claim that $\mcC^{(k)}$ is well-defined as a distribution on $M^6$ for all $\gamma\in\bbR$ (and not just $\gamma<-1/2$). Indeed, when testing it against some $\varphi\in\mcD(M^6)$, one can first perform the partial integration 
\begin{align*}
  \int_M  [\odot](y_*,y_1,y_2)\varphi(y_*,y_1,y_2,z_*,z_1,z_2)\rmd v(y_2)\,.
\end{align*}
An application of Lemma~\ref{Lemshield} shows that this expression is a smooth function on $M^6$, and testing it against $ [\odot](z_*,z_1,z_2)      \mcQ_x^\gamma(y_*,z_*)   G^{(1)}_r(0,y_1,z_1)$ proves the claim. 

Regarding the scaling on $\Gamma$, observe that 

\makebox[\textwidth][c]{
\begin{minipage}{\dimexpr\textwidth+10cm}
\begin{align*}
&\underbrace{\int_M e^{-6u} \left(e^{-u\rho_{i}*} [\odot_i]  \right)(y_*,y_1,y_2)\varphi(y_*,y_1,y_2,z_*,z_1,z_2)\rmd v(y_2)}   \\
&\times e^{-(6+k)u} e^{-u\rho_i*}\left( [\odot_i](z_*,z_1,z_2) \, \mcQ_x^\gamma(y_*,z_*) \, G^{(k)}_r(0,y_1,z_1)\right)
\end{align*}
\end{minipage}
}
\noindent is bounded since $ [\odot](z_*,z_1,z_2)      \mcQ_x^\gamma(y_*,z_*)   G^{(k)}_r(0,y_1,z_1)$ is a tree and the underbraced term is bounded by Lemma~\ref{Lemshield}. Note that we used the fact that $\mcQ^\gamma_x$ is weakly homogeneous of degree 0, which is proven in Lemma~\ref{Lem:Qgammasmooth}. It follows that
$$ e^{-(12+k)u}e^{-u\rho_i*} \left( [\odot_i](y_*,y_1,y_2)   [\odot_i](z_*,z_1,z_2)      \mcQ_x^\gamma(y_*,z_*)   G^{(k)}_r(0,y_1,z_1)\right) $$ 
is bounded in $\mcD^{\prime}_{\Gamma}(M^6)$. Then, by Proposition~\ref{Prop:intrinsicscalingspaces}, the same holds with any scaling field  
$\rho$ with respect to  $\textbf{d}_6$, which concludes after summation over $i\in I$
that $\mcC^{(k)}\in  \mcS_{\Gamma}^{-12-k}$.

It remains to scale on the deep marked diagonal $\mathbf{d}_{6,x}$. This directly stems from Theorem~\ref{Thm:extensiontwosteps}, and from an easy modification of the above discussion, since the only propagator whose scaling degree changes when we scale with respect to  $\textbf{d}_6$ or $\textbf{d}_{6,x}$
is $\mcQ^\gamma_x$, as discussed in Lemma~\ref{Lem:Qgammasmooth}. Here we need to rely on Proposition~\ref{Prop:recentering} which allows to keep the same scaling degrees for the new scaling field $\rho_x$.
\end{Dem}

\subsubsection{Renormalising the divergent subamplitude.}\label{subsect_rernorm_subamplitude}

This section is devoted to the construction of the subamplitude $\mcB$, which is performed in the following Lemma.

\ssk

\begin{lem}\label{LemB}
 For any $\gamma<0$, the distribution $\mcB$ is well-defined on $M^4\times\bbR$. More precisely, we have
\begin{align*}
    \mcB\in \mcS_{\Gamma}^{-11}\,,
\end{align*}
where $\Gamma$ is equal to
\begin{equation*} \begin{split} 
&N^*\big(\{s_a=t\}\big) \cup N^*\big(\{s_a=t,y_1=x_a\}\big) \cup N^*\big(\big\{s_a=t,y_2=x_a\big\}\big) \cup N^*\big(\{y_*=y_1=y_2\}\big)   \\
&\cup N^*\Big(\big\{s_a=t,y_1=y_2=x_a\big\}\Big) \cup N^*\Big(\big\{s_a=t,y^*=y_1=y_2\big\}\Big) \cup N^*\Big(\big\{s_a=t,y^*=y_1=y_2=x_a\big\}\Big).
\end{split} \end{equation*}
\end{lem}

\ssk

\begin{Rem}
    Again, with this lemma in hand, we will be able to inject our knowledge on $\mcB$ inside the recurrence of the proof of Theorem~\ref{ThmBlackBoxFeynmanGraphs}, when applied to an amplitude $\mcA_{41}$ which contains two copies of this subamplitude.  
\end{Rem}

\ssk

\begin{Dem}
Recall the definition \eqref{eq:defB} of $\mcB$ and note that in \eqref{eq:defB}, $\delta_{\mathbf{d}_4}$ is the unique distribution on $M^4\times\bbR$ supported on $\mathbf{d}_4=\{s_a=t, y_*=y_1=y_2=x_a\} $ and such that for all $\varphi\in C^\infty_c( M^4\times\bbR)$, one has
\begin{equation*} \begin{split} 
\langle \delta_{{\bf d}_4},\varphi\rangle_{\mathcal M^4} &\defeq \int_{M^4\times \bbR} \delta_{{\bf d}_4} (y_*,y_1,y_2,x_a,t,s_a)\varphi(y_*,y_1,y_2,x_a,t,s_a)  \rmd v_{M^4\times \bbR}   \\
&= \int_{y_*\in M} \varphi(y_*,y_*,y_*,y_*,t,t) \rmd v_M(y),
\end{split} \end{equation*}
where both $\rmd v_{M^4\times \bbR}$ and $\rmd v_{M} $ are the natural volume forms on the respective Riemannian manifolds, so that $\delta_{{\bf d}_4}$ only depends on the choice of Riemannian volume form on $M$. We denote by $\widetilde\mcB$ the unrenormalized amplitude of $\mcB$ which reads
\begin{align*}
\widetilde\mcB(y_*,y_1,y_2,x_a,s_a)\defeq    [\odot](y_*,y_1,y_2)G^{(2)}_r( t-s_a,y_1,x_a)\underline{\mcL}^{-1}(t-s_a,y_1,x_a)\,,
\end{align*}
and verifies $\mcB=\widetilde\mcB-b_r\delta_{\mathbf{d}_4}$. Moreover, to simplify the notation, we write $y_a$ rather than $x_a$, and therefore focus on $\widetilde\mcB(y_*,y_1,y_2,y_a,s_a)$. Finally, throughout the proof, given $y_*\in M$ we use the short-hand notation $z=(y_*,y_*,y_*,y_*,t)\in M^4\times\bbR$.

\vspace{6pt}

The definition of $\mcB$ is a consequence of item \textbf{\textsf{(b)}} in Theorem \ref{ThmBlackBoxFeynmanGraphs}, and of the discussion performed in Section~\ref{SectionDivergentPartTau20}. Indeed, naively, we know that $\widetilde\mcB$ belongs to $\mcS_\Gamma^{-11}(M^4\times \bbR^2\setminus {\bf d}_4)$. But the weighted codimension of ${\bf d}_4$ also equals $11$, so we are in case \textbf{\textsf{(b)}} of Theorem \ref{ThmCanonicalExtension} and our extension requires a renormalization as in Hadamard's finite parts. We thus aim to show that one can renormalize $\widetilde\mcB$ by subtracting some explicit distributional counterterm ${\varphi_i} c_r$ proportional to $\delta_{{\bf d}_4}$. A priori $c_r=c_r(y_*,t)$ is a function on $M\times\bbR$, but we also want to show that one can actually take $c_r$ constant equal to $b_r$.

\vspace{6pt}

The first step in our proof is to implement in real conditions the abstract extension Theorem~\ref{ThmCanonicalExtension} with counterterms, also controlling the wave front of the extension. The second step is to explicitly compute the abstract counterterm $c_r$ whose existence is given by Theorem~\ref{ThmCanonicalExtension} in terms of trace densities of some operators, this computation is similar to the one which is done for the quartic term $\overline\tau_2$ in Section~\ref{SectionDivergentPartTau20}. 

\vspace{6pt}

Let us now study the renormalization problem for $\widetilde\mcB(y_*,y_1,y_2,y_a,s_a)$ locally in $U_i^4\times \bbR$ for $i\in I$, since the diagonal ${\bf d}_4=\{y_*=\cdots=y_a,s_a=t\}$ can be covered by such sets, so that we can recover the global extension 
just from working with the local extensions. Fix ${\varphi_i} : M^4\rightarrow\bbR_{\geqslant0}$ with support in $U_i^4$, identically equal to $1$ in a neighbourhood of $\{y_*=\cdots=x_a\}\cap U_i^4$. Here and from now on, abusing notation, use the same notation to denote both elements of $U_i$ and their coordinates. As in the extension Theorem~\ref{ThmCanonicalExtension}, we use the parabolic scaling defined by the scaling field $\rho=2(t-s)\partial_{s}+\sum_{j\in\{1,2,a\}}(y_j-y_*)\cdot\partial_{y_j} $ whose semiflow $(e^{-t\rho})_{t\geqslant 0}$ leaves $U_i^4$ stable. Then we have the continuous partition of unity formula ${\varphi_i}=\int_0^\infty e^{u\rho*}w_i \rmd u$ where we set $w_i\defeq-\rho{\varphi_i}\in C^\infty(U_i^4)$ which vanishes near $\{y_*=\dots=x_a\}\cap U^4_i$. Moreover, we denote for $k\in\bbN_{\geqslant1}$ by $$\dens_k(y_1,\dots,y_k)\defeq\prod_{j=1}^k\sqrt{\det(g)_{y_j}}$$ the density of the Riemannian volume on $U_i^k$ endowed with the product metric with respect to the Lebesgue measure $\prod_{j=1}^k\rmd y_j$. 
 
Fix a test function $\varphi\in C^\infty_c(U_i^4\times \bbR)$. We decompose the pairing $\left\langle \widetilde{\mcB},\varphi\right\rangle$ of $\widetilde{\mcB}$ with $\varphi$ as
\begin{align*}
&\left\langle \widetilde{\mcB}(1-{\varphi_i}),\varphi\right\rangle + \int_0^\infty\int^t_{-\infty} \int_{U_i^4}  \Big(w_i e^{-u\rho*}\widetilde\mcB e^{-u\rho*}\big(\Pi_z(\varphi \dens_4)\big)\Big)(y_*,y_1,y_2,y_a,s_a) e^{-11u}\prod_{j\in\{*,1,2,a\}}\rmd y_{j}\,\rmd s_a \rmd u \, +
\end{align*}
\begin{align*}
\int_0^\infty \int_{-\infty}^t\int_{U_i^4}  (\widetilde{\mcB} e^{u\rho*}w_i)(y_*,y_1,y_2,y_a,s_a) (\varphi\dens_4)(z) \prod_{j\in\{*,1,2,a\}}\rmd y_j\, \rmd s_a \rmd u\,,
\end{align*}
where we used the notation $\Pi_z(\varphi \dens_4)\defeq\varphi \dens_4-(\varphi \dens_4)(z)$ to denote the recentering at the point $z=(y_*,y_*,y_*,y_*,t)$, and the factor $e^{-11u}$ comes from the Jacobian of the semi-flow generated by $\rho$. Moreover, observe that $e^{-u\rho*}\widetilde\mcB=\mathcal{O}_{\mathcal{D}^\prime}(e^{11u}) $ and that $\Pi_z(\varphi\dens_4)$ vanishes on $\mathbf{d}_4\cap (U_i^4\times\bbR)$, which implies that
$$
e^{-u\rho*}\big(\Pi_z(\varphi\dens_4)\big)=\mathcal{O}(e^{-u})\,.
$$
The integral over $u$ in the second term of the right hand side  is therefore absolutely convergent (in fact there is a subtlety related to compactness, but it is immediate to check that the product $\psi_i e^{-u\rho*}\big(\Pi_z(\varphi\dens_4) \big)$ forms a bounded family of test functions). We have thus proved that the first two terms of the right hand side  have well-defined limits as $r\downarrow 0$. It remains to identify the counterterm as the divergent part of the third term, which we denote by $\mathfrak D(\bigcdot)$. Note that all the quantities $f=f(r)$ that we compute and which depend on the cut-off $r$ are contained in some algebra $\mathcal{O}_{>0}$ of functions of the cut-off $r$ which are smooth functions of $r\in (0,1]$ and have log-polyhomogeneous expansions as $r\downarrow0$, $f(r)=a{r}^{-1/2}+b\log r^{-1}+c+o(1)$ where $(a,b,c)\in \bbR^3$. We define the divergent part of any element $f \in \mathcal{O}_{>0}$ as $\mathfrak{D}(f)\defeq  a{r}^{-1/2}+b\log r^{-1}$.

With this definition in hand, the divergent part of the third term verifies

\begin{align*}
\mathfrak{D}\Big(\int_0^\infty &\int_{-\infty}^t\int_{U_i^4}  (\widetilde\mcB e^{u\rho*}w_i)(y_*,y_1,y_2,y_a,s_a) (\varphi\dens_4)(z) \prod_{j\in\{*,1,2,a\}}\rmd y_j\, \rmd s_a \rmd u\Big)\\ 
&= \int_{U_i} \mathfrak{D}\Big( \int_{-\infty}^t\int_{U_i^3}  (\widetilde\mcB{\varphi_i})(y_*,y_1,y_2,y_a,s_a) \dens_3(y_*,y_*,y_*) \prod_{j\in\{1,2,a\}}\rmd y_j\,\rmd s_a\Big) \varphi(z)  \rmd v( y_*)   \\
&= \left\langle c_r \delta_{{\bf d}_4},\varphi\right\rangle\,,
 \end{align*}
 which implies by definition of $\delta_{\textbf{d}_4}$ that the counterterm reads
\begin{align}\label{eq:countertermusefull}
c_r(y_*,t)&=\mathfrak{D}\Big(  \int_{-\infty}^t\int_{U_i^3}  (\widetilde\mcB{\varphi_i})(y_*,y_1,y_2,y_a,s_a) \dens_3(y_*,y_*,y_*) \prod_{j\in\{1,2,a\}} \rmd y_j\,\rmd s_a \Big)\nonumber\\
&=\mathfrak{D}\Big(  \int_{-\infty}^t\int_{U_i^3}  (\widetilde\mcB{\varphi_i})(y_*,y_1,y_2,y_a,s_a) \dens_3(y_1,y_2,y_a) \prod_{j\in\{1,2,a\}} \rmd y_j\,\rmd s_a \Big)\,.
\end{align} 
To go from the first to the second line of this last equality, again we used the fact that $$\dens_3(y_1,y_2,y_a)-\dens_3(y_*,y_*,y_*) $$ vanishes on $\{y_*=\dots=y_a\}$, so that it does not contribute to the singular part of the above integral, which gives us the freedom to change the evaluation point of the volume elements. The reader might also wonder why we could interchange the extraction of singular parts with the integration: this is due to the fact that the integrand admits an asymptotic expansion. To conclude about the expression of $c_r(y_*,t)$, since in \eqref{eq:countertermusefull} we reconstructed the volume form $ \prod_{j\in\{1,2,a\}} \rmd v(y_j)=\dens_3(y_1,y_2,y_a) \prod_{j\in\{1,2,a\}} \rmd y_j$, it remains to plug the expression of $\widetilde\mcB$, and we obtain

\makebox[\textwidth][c]{
\begin{minipage}{\dimexpr\textwidth+10cm}
\begin{align*}
c_r(y_*,t)&=\mathfrak{D}\Big(\int_{-\infty}^{t} 
\sum_{\vert k-\ell\vert\leqslant 1}  P_k^i(y_*,\cdot) \circ \widetilde{{\varphi}}_i\circ\left( G_r^{(2)}(t-s)\circ e^{-(t-s)P}\right)\circ\widetilde{{\varphi}}_i \circ \widetilde{P}_\ell^i(\cdot,y_*)\, \rmd s\Big)\,,
\end{align*}
\end{minipage}
}
\esp

where the $P_k^i,\widetilde{P}_\ell^i$ are the Littlewood-Paley-Stein projectors, the functions $\widetilde{{\varphi}}_i\in C^\infty_c(U_i)$ are arbitrary test functions such that $\widetilde{{\varphi}}_i=1$ near $y_*$ and where we used the explicit definition of $[ \odot_i ]$ in terms of the Littlewood-Paley-Stein projectors. This is indeed the desired expression \eqref{EqTau20}, since this last term is  precisely a trace density whose singular part matches the divergent part of $\tau_{20}$ up to a combinatorial factor (see Section~\ref{SectionDivergentPartTau20}). In particular, $c_r$ can ultimately be chosen independent of $y_*,t$, and factors out of the pairing.

It remains to prove why the renormalized
amplitude 
\begin{eqnarray*}
\mcB_0\defeq
\lim_{r\downarrow 0} \big(\widetilde\mcB_r-c_r(y_*,t)\delta_{{\bf d}_4}\big)
\end{eqnarray*}
has the correct wave front set in $\Gamma$. To do so, it suffices to prove the property for $\mcB{\varphi_i} R$ where $R$ is the Taylor subtraction operator from Proposition \ref{PropTaylorRemainder}. This follows from the proof of Theorem \ref{ThmCanonicalExtension} where one replaces the wave front bound on $[e^{-u\rho}]$ by the identical wave front bound on $[e^{-u\rho}R]$ which coincide by Proposition~\ref{PropTaylorRemainder}. 
\end{Dem}

\subsection{Completing the diagrammatic estimates}
\label{subsection_completing}

\subsubsection{Bounds for $\tau_1$.}

The amplitude $\mathcal{G}_{14}=\begin{tikzpicture}[scale=0.6,baseline=-0.1cm]
\node at (0,-0.3) [dot] (1) {};
\node at (0,0.3) [dot] (2) {};
\node at (0,0.6) [dot] (3) {};
\node at (0,0.9) [dot] (4) {};
\node at (-0.6,0.6) [dot] (5) {};
\node at (0.6,0.6) [dot] (6) {};
\node at (-0.8,0) [blackdot] (7) {};
\node at (0.8,0) [blackdot] (8) {};
\draw[K] (1) to (7);
\draw[K] (1) to (8);
\draw[K] (2) to (5);
\draw[K] (2) to (6);
\draw[K] (3) to (5);
\draw[K] (3) to (6);
\draw[K] (4) to (5);
\draw[K] (4) to (6);
\draw[DK] (7) to (5);
\draw[DK] (8) to (6);
\draw[K,purple] (7) to[bend right=+80] (8);
\end{tikzpicture}$ contains the subgraph with subamplitude $\mcC^{(1)}$, and thus needs the special treatment performed in Section~\ref{subsubsec:shielding} (see Lemma~\ref{LemC^k}), using the smoothing effect of the kernel of $\odot$. To complete the argument, we verify the scalings of all the other subgraphs. $\mcA_{14}$ has four closed, irreducible, connected subgraphs, for each of which we need to compute the weak homogeneity. We do the computation for  $\mcG_{14}$ itself and let the reader deal with its three subgraphs. The amplitude $\mathcal A_{14}$ of $\mathcal G_{14}$ is

\makebox[\textwidth][c]{
\begin{minipage}{\dimexpr\textwidth+10cm}
\begin{align*}
[\odot_i](y_*,y_1,y_2) \, [\odot_i](z_*,z_1,z_2) \, &\mcQ_x^\gamma(y_*,z_*) \, G^{(1)}_r( 0,y_1,z_1)   \\
&\times \underline{\mcL}^{-1}(t-s_a,y_2,x_a) \, \underline{\mcL}^{-1}(t-s_b,z_2,x_b) \, G^{(3)}_r(s_a-s_b,x_a,x_b).
\end{align*}
\end{minipage}
}
\esp

\noindent By summing the weak homogeneity of all analytical objects  appearing in the amplitude $\mcA_{14}$, the kernels $[\odot_i],G_r^{(1)},\underline{\mcL}^{-1},  G_r^{(3)},\mcQ_x^\gamma$, we get for any $\gamma<0$
\begin{align*}
\sum \text{weak homogeneities}&=2(-6)+(-1)+2(-3) -3-6-2\gamma =-28-2\gamma   \\&>-\text{codim}_w\big(\{s_a=s_b=t, y_*=\dots=x_b=x\}\big) = -4 -24=-28\,,
\end{align*} 
which is sharp. One get the same condition on $\gamma$ when checking the condition on the subgraphs of $\mcG_{14}$ so Theorem \ref{ThmBlackBoxFeynmanGraphs} entails that the contribution of $\mcG_{14} $ is finite for all $\gamma<0$. 

The verification for the graph $\mcG_{12}=\begin{tikzpicture}[scale=0.6,baseline=-0.1cm]
\node at (0,0.4) [dot] (1) {};
\node at (0,1) [dot] (2) {};
\node at (-0.4,0.7) [dot] (3) {};
\node at (-0.4,0.2) [dot] (4) {};
\node at (0.4,0.7) [dot] (5) {};
\node at (0.4,0.2) [dot] (6) {};
\node at (-0.8,0) [blackdot] (7) {};
\node at (0.8,0) [blackdot] (8) {};
\draw[K] (1) to (3);
\draw[K] (1) to (5);
\draw[K] (2) to (3);
\draw[K] (2) to (5);
\draw[DK] (3) to (7);
\draw[K] (3) to (4);
\draw[K] (4) to (7);
\draw[K] (5) to (6);
\draw[DK] (5) to (8);
\draw[K] (6) to (8);
\draw[K,purple] (7) to[bend right=+80] (8);
\end{tikzpicture}  $ is similar. We have
\begin{align*}
    \mathcal A_{12}= &[\odot_i](y_*,y_1,y_2)   [\odot_i](z_*,z_1,z_2)      \mcQ_x^\gamma(y_*,z_*)   G^{(1)}_r( t-s_a,y_1,x_a)  G^{(1)}_r( t-s_b,z_1,x_b)  \underline{\mcL}^{-1}(t-s_a,y_2,x_a)    \\&\times\underline{\mcL}^{-1}(t-s_b,z_2,x_b) G^{(2)}_r(s_a-s_b,x_a,x_b)  \,.
\end{align*}
For instance the subamplitude $[\odot_i](y_*,y_1,y_2)G_r^{(1)}(t-s_a,y_1,x_a)\underline{\mcL}^{-1}(t-s_a,z_1,x_a)$ associated with a triangle attached to a $\bullet$ is weakly homogeneous of degree 
$$
-6-3-1=-10>-\text{codim}_w\big(\{s_a=t, y_*=y_1=z_1=x_a\}\big) = -2+3(-3) = -11.
$$ 
It yields the same range of Sobolev regularity $\gamma<0$. We invite the reader to check all the subgraphs by themselves. In the locally covariant renormalization picture, the random distribution $\tau_1$ will differ from the Wick renormalized one by a quantity of the form $$f\left(\left( \underline{\mcL}^{-1}\begin{tikzpicture}[scale=0.3,baseline=0cm]
\node at (0,0)  [dot] (1) {};
\node at (0,0.8)  [noise] (2) {};
\draw[K] (1) to (2);
\end{tikzpicture}\right) \odot_i \begin{tikzpicture}[scale=0.3,baseline=0cm]
\node at (0,0)  [dot] (1) {};
\node at (0,0.8)  [noise] (2) {};
\draw[K] (1) to (2);
\end{tikzpicture}\right)+
f\bigg(\int[\odot_i]\big(y_*,y_1,y_2\big)\underline{\mcL}^{-1}(t-s_a,y_2,x_a)G_r^{(1)} (t-s_a,y_1,x_a)\rmd s_a\rmd v( y_1,\rmd y_2,\rmd x_a)\bigg)
 $$ where $f$ is some deterministic smooth function. The resonant product between $\left(\underline{\mcL}^{-1} \begin{tikzpicture}[scale=0.3,baseline=0cm]
\node at (0,0)  [dot] (1) {};
\node at (0,0.8)  [noise] (2) {};
\draw[K] (1) to (2);
\end{tikzpicture}\right)\in \mcC^{\frac{3}{2}-\varepsilon}, \forall \varepsilon>0$ and $\begin{tikzpicture}[scale=0.3,baseline=0cm]
\node at (0,0)  [dot] (1) {};
\node at (0,0.8)  [noise] (2) {};
\draw[K] (1) to (2);
\end{tikzpicture}\in \mcC^{-\frac{1}{2}-\varepsilon}$, $\forall \varepsilon>0$ is well-defined in $\mcC^{1-\varepsilon}$, $\forall \varepsilon >0$. The second term, which can be rewritten as $f\bigg(\sum_{\vert k-\ell\vert\leqslant 1}\int_{-\infty}^t \Big(P^i_k\,\circ\, e^{-(t-s)P} \circ G^{(1)}(t-s)\circ \widetilde{P}^i_\ell \Big)(y_*,y_*) \,\rmd s\bigg)$ is a well-defined smooth function that we can control by the following argument. The operators verify $\underline\mcL^{-1}\in \Psi_H^{-1}$ and $ G^{(1)}\in \Psi_P^{-2}$, where the heat calculus $\Psi_H$ is defined in Grieser's note \cite{Grieser} -- or see Definition \ref{def:heatcalculus}) and the parabolic calculus $\Psi_P$ in Definition \ref{defi:flatparabolic}. Then the composition $\left(\int_{-\infty}^t \underline\mcL^{-1}(t-s) \circ G^{(1)}(t-s)\rmd s \right)$ belongs to $\Psi_P^{-3}$, hence it belongs to $\Psi^{-4}(M)$ uniformly in $t$ and is trace class on $M$. The series of pseudodifferential operators
\begin{equation*} \begin{split}
&\sum_{\vert k-\ell\vert\leqslant 1}\int_{-\infty}^t \Big(P^i_k\,\circ\, e^{-(t-s)P} \circ G^{(1)}(t-s)\circ \widetilde{P}^i_\ell \Big)(y_*,y_*)\,\rmd s =  \\
&\sum_{\vert k-\ell\vert\leqslant 1} \int_{-\infty}^t \Big\{\Big(P^i_k\circ \widetilde{P}^i_\ell \circ e^{-(t-s)P} \circ G^{(1)}(t-s) \Big)  + \Big(P^i_k\circ  [e^{-(t-s)P} \circ G^{(1)}(t-s),\widetilde{P}^i_\ell] \Big)\Big\}(y_*,y_*)  
\end{split} \end{equation*}
will also converge in $\Psi^{-4}(M)$ since the series  $\sum_{\vert k-\ell\vert\leqslant 1}P^i_k\circ \widetilde{P}^i_\ell  $ converges in $\Psi^0_{0,1}(M)$ and the commutator $ [e^{-(t-s)P} \circ G^{(1)}(t-s),\widetilde{P}^i_\ell]$ is bounded in $\Psi^{-5}(M)$ uniformly in $(\ell,i)$ since the sequence $(\widetilde{P}_\ell^i)_{\ell,i} $ is bounded in $\Psi^0(M)$. Therefore the second series involving the commutator term converges in $\Psi^{-5}(M)$. We use for that purpose a commutator identity that says that for every pseudodifferential operator $A\in \Psi^m(M)$, the series
\begin{eqnarray*}
\sum_{\vert k-\ell\vert\leqslant 1}\left( P^i_k A \widetilde{P}^i_\ell - A P^i_k \widetilde{P}^i_\ell \right)
\end{eqnarray*}
converges as a pseudodifferential operator of order $m-1$ that we prove in our companion work~\cite[
Proposition 2.11]{BDFTCompanion}.   Finally this implies that the term 
$$
\int[\odot_i]\big(x_*,x_1,x_2\big)\underline{\mcL}^{-1}\big((t,x_2),(s,x_a)\big)G_r^{(1)} \big((t,x_1),(s,x_a)\big)\rmd s\rmd x_1\rmd x_2\rmd x_a
$$ 
is a well-defined smooth function.

\medskip

\subsubsection{Elementary bounds for $\tau_2$.}\label{sss:elementary_est_tau_2}
Again, $\mcA_{24}$ and $\mcA_{22}$ need the special treatment from Section~\ref{subsubsec:shielding} since they respectively have the subamplitudes $\mcC^{(2)}$ and $\mcC^{(1)}$: Lemma~\ref{LemC^k} states that these subamplitudes are indeed well-defined for every $\gamma\in\bbR$.

\ssk

To complete the argument, we check for the reader's convenience the weak homogeneities of the amplitudes of all the other subgraphs. The amplitude $\mathcal A_{24}$ of $\mcG_{24}=\begin{tikzpicture}[scale=0.6,baseline=-0.1cm]
\node at (0,-0.2) [dot] (1) {};
\node at (0,0.2) [dot] (2) {};
\node at (0,0.4) [dot] (3) {};
\node at (0,0.8) [dot] (4) {};
\node at (-0.4,0.6) [dot] (5) {};
\node at (0.4,0.6) [dot] (6) {};
\node at (-0.8,0) [blackdot] (7) {};
\node at (0.8,0) [blackdot] (8) {};
\draw[K] (1) to (7);
\draw[K] (1) to (8);
\draw[K] (2) to (7);
\draw[K] (2) to (8);
\draw[K] (3) to (5);
\draw[K] (3) to (6);
\draw[K] (4) to (5);
\draw[K] (4) to (6);
\draw[DK] (7) to (5);
\draw[DK] (8) to (6);
\draw[K,purple] (7) to[bend right=+80] (8);
\end{tikzpicture}$
is 
\begin{equation*} \begin{split}
[\odot_i](y_*,y_1,y_2) \, [\odot_i](z_*,z_1,z_2) \, &\mcQ_x^\gamma(y_*,z_*) \, G^{(2)}_r( 0 ,y_1,z_1)   \\
&\times \underline{\mcL}^{-1}(t-s_a,y_2,x_a) \, \underline{\mcL}^{-1}(t-s_b,z_2,x_b) \, G^{(2)}_r(s_a-s_b,x_a,x_b).
\end{split} \end{equation*}
Here again this graph has four closed, irreducible, connected subgraphs and we calculate the weak homogeneity of $\mcA_{24}$ itself by summing the weak homogeneity of all analytical objects, the kernels $G_r^{(2)}, [\odot_i],\underline{\mcL}^{-1},\mcQ^\gamma$, appearing in the amplitude
\begin{align*}
\sum \text{weak homogeneities} &= 2(-2)+2(-6)+2(-3)-6-2\gamma =-28-2\gamma \\ 
&> - \text{codim}_w\big(\{s_a=s_b=t, y_*=\dots=x_b\}\big) = -4 -24 = -28\,,
\end{align*} 
hence $\gamma<0$. Repeating this verification for all subgraphs yields the result that $\mcG_{24}<\infty$ for all $\gamma<0$, by using again Theorem \ref{ThmBlackBoxFeynmanGraphs}.

The verification for the graph $\mcG_{22}=\begin{tikzpicture}[scale=0.6,baseline=-0.1cm]
\node at (0,-0.4) [dot] (1) {};
\node at (0,0.8) [dot] (2) {};
\node at (-0.6,0.4) [dot] (3) {};
\node at (-0.3,0) [dot] (4) {};
\node at (0.6,0.4) [dot] (5) {};
\node at (0.3,0) [dot] (6) {};
\node at (-0.9,0) [blackdot] (7) {};
\node at (0.9,0) [blackdot] (8) {};
\draw[K] (1) to (7);
\draw[K] (1) to (8);
\draw[K] (2) to (3);
\draw[K] (2) to (5);
\draw[DK] (3) to (7);
\draw[K] (3) to (4);
\draw[K] (4) to (3);
\draw[K] (4) to (7);
\draw[K] (5) to (6);
\draw[DK] (5) to (8);
\draw[K] (6) to (8);
\draw[K,purple] (7) to[bend right=+80] (8);
\end{tikzpicture}$ of amplitude

\makebox[\textwidth][c]{
\begin{minipage}{\dimexpr\textwidth+10cm}
\begin{align*}
  \mathcal A_{22}= &[\odot_i](y_*,y_1,y_2)   [\odot_i](z_*,z_1,z_2)      \mcQ_x^\gamma(y_*,z_*)   G^{(1)}_r( t-s_a ,y_1,x_a)  G^{(1)}_r(t-s_b  ,z_1,x_b)\\
    &\times   \underline{\mcL}^{-1}(t-s_a,y_2,x_a)    \underline{\mcL}^{-1}(t-s_b,z_2,x_b) G^{(1)}_r(s_a-s_b,x_a,x_b)  
\end{align*}
\end{minipage}
}
\esp

\noindent is similar, for instance the subamplitude 
$$
[\odot_i](y_*,y_1,y_2) \, G_r^{(1)}(t-s_a,y_1,x_a)
$$ 
is weakly homogeneous of degree $-6-1=-7>\text{codim}_w\big(\{s_a=t, y_*=y_1=y_2=x_a\}\big) = -2 -9 = -11$ and yields the same range of regularity exponent. In a similar way as what we did for $\tau_1$, the
locally covariant renormalized $\tau_2$ will differ from the Wick renormalized $\tau_2$ by homogeneous terms in Wiener chaoses of order $2$ and $0$ which have the form $P_1\begin{tikzpicture}[scale=0.3,baseline=0cm]
\node at (0,0) [dot] (0) {};
\node at (0,0.5) [dot] (1) {};
\node at (-0.4,1)  [noise] (noise1) {};
\node at (0.4,1)  [noise] (noise2) {};
\draw[K] (0) to (1);
\draw[K] (1) to (noise1);
\draw[K] (1) to (noise2);
\end{tikzpicture}+P_2\begin{tikzpicture}[scale=0.3,baseline=0cm]
\node at (0,0) [dot] (0) {};
\node at (0.3,0.6)  [noise] (noise1) {};
\node at (-0.3,0.6)  [noise] (noise2) {};
\draw[K] (0) to (noise1);
\draw[K] (0) to (noise2);
\end{tikzpicture}+f_3$ where $P_1,P_2$ are smoothing operators and $f_3$ is a smooth function. We used the fact that both $\begin{tikzpicture}[scale=0.3,baseline=0cm]
\node at (0,0) [dot] (0) {};
\node at (0,0.5) [dot] (1) {};
\node at (-0.4,1)  [noise] (noise1) {};
\node at (0.4,1)  [noise] (noise2) {};
\draw[K] (0) to (1);
\draw[K] (1) to (noise1);
\draw[K] (1) to (noise2);
\end{tikzpicture}$ and $\begin{tikzpicture}[scale=0.3,baseline=0cm]
\node at (0,0) [dot] (0) {};
\node at (0.3,0.6)  [noise] (noise1) {};
\node at (-0.3,0.6)  [noise] (noise2) {};
\draw[K] (0) to (noise1);
\draw[K] (0) to (noise2);
\end{tikzpicture}$ differ from their Wick renormalized version by a smooth function and also that for any smooth function $f\in C^\infty(M)$, the multiplication operator by the localized resonant product $u\in \mathcal{D}^\prime(M)\mapsto f\odot_i u\in C^\infty(M)$ is smoothing.

\subsubsection{The divergent part of $\tau_{20}$.}
\label{SectionDivergentPartTau20}

Recall that we denote by $[\odot_i]$ the kernel of $\odot_i$. 
An immediate calculation yields for $z=(t,y_*)$
\begin{align} \nonumber
\tau_{20}(z) = \mathbb{E}[\tau_2(z)] &= 2\int_{-\infty}^t \int_{M^3} [\odot_i](y_*,y_1,y_2) \, {\mcL}^{-1}(t-s_a,y_2,x_a) \, G_r^{(2)}(t-s_a,y_1,x_a) \rmd v(y_1,y_2,x_a)\rmd s_a   \\
&= 2\sum_{\vert k-\ell\vert\leqslant 1} \int_{-\infty}^t P^i_k(y_*,\bigcdot)\circ G_r^{(2)}(s-t)\circ e^{-(t-s)P} \circ \widetilde{P}^i_\ell(\bigcdot,y_*)\rmd s\,, \label{EqTau20}
 \end{align}
with the shorthand notation $\rmd v(y_1,y_2,x_a)\defeq \rmd v(y_1)\rmd v(y_2)\rmd v(x_a)$ for the Riemannian volume form on $M^3$, with $\rmd v$ the corresponding volume form on $M$, the factor $2$ in front comes from Wick's Theorem.

\ssk

\noindent {\it \S1. Preliminary analysis.} 
Recall that 
$$
G_r^{(2)}(t-s,x,y) = \left(\Big\{e^{-(\vert t-s\vert+2r) P} P^{-1}\Big\}(x,y)\right)^2\,.
$$
We reformulate the integral 
\begin{equation} \label{EqTIndexedComposition}
\int_{-\infty}^t  G_r^{(2)}(t-s) \circ e^{-(t-s)P}\,\rmd s 
\end{equation}
as the composition of two operators in the parabolic calculus $(\Psi_P^\alpha)_{\alpha\in\bbR}$ defined in  Definition \ref{defi:flatparabolic}. This calculus extends the heat calculus $(\Psi_H^\alpha)_{\alpha\in\bbR}$ as defined in Grieser's note \cite{Grieser} in order to include the kernels $G_r^{(p)}, p\in \{1,2,3\}$ needed to define $\varphi^4_3$ amplitudes. Since $G_r^{(2)}\in \Psi^{-\frac{3}{2}}_P $ uniformly in $r>0$ and $e^{-tP} \in \Psi^{-1}_H$, we prove in \cite[Theorem 4.17]{BDFTCompanion} that the composition $\int_{-\infty}^t  G_r^{(2)}(s,t) \circ e^{-(t-s)P} \, \rmd s$ defines an element in $\Psi_P^{-\frac{5}{2}+\gamma}$ for any $\gamma>0$, uniformly in $r>0$. Hence it is a pseudodifferential operator depending continuously on $t$ of order $-3+2\gamma$, by the comparison Theorem in our companion work~\cite[Proposition 4.14]{BDFTCompanion} viewing parabolic operators as parameter dependent pseudodifferential operators. It follows that the $t$-indexed family of operators \eqref{EqTIndexedComposition} is bounded in $\Psi^{-3+2\gamma}(M)$ and fails to be trace class when the regularization parameter $r$ tends to $0$. Our goal in the sequel of this section is to extract the singular part of this operator.

First we need to disentangle this operator in \eqref{EqTau20} from the Littlewood-Paley-Stein projectors $P^i_k, \widetilde{P}^i_\ell$. We use for that purpose a commutator identity proved in \cite[Proposition 2.11]{BDFTCompanion} states that for every pseudodifferential operator $A\in \Psi^m(M)$, the series
\begin{eqnarray*}
\sum_{\vert k-\ell\vert\leqslant 1}\left( P^i_k A \widetilde{P}^i_\ell - A P^i_k \widetilde{P}^i_\ell \right)
\end{eqnarray*}
converges as a pseudodifferential operator of order $m-1$. The operator
\begin{equation} \label{EqCommutator}
\sum_{\vert k-\ell \vert\leqslant 1} \left( P^i_k G_r^{(2)}(t-s)e^{-(t-s)P} \widetilde{P}^i_\ell - G_r^{(2)}(t-s)e^{-(t-s)P} P^i_k \widetilde{P}^i_\ell\right)
\end{equation}
is in particular in $\Psi^{-4+2\gamma}(M)$ uniformly in $r\in [0,1]$, for any $0<\gamma<1/2$, so it is trace class.

Denote by $\delta_{y_*}$ is the unique distribution depending on the Riemannian volume form $\rmd v(y)$ such that 
$$
\int_M \delta_{y_*}(z)f(z)\,\rmd v(z) = f(y_*)\,,
$$ 
for all bounded measurable functions $f$. With this notation one has
$$
\tau_{20}(z) = 2\int_{-\infty}^t \sum_{\vert k-\ell\vert\leqslant 1} \left\langle P^i_k\left(\delta_{y_*}\right), G_r^{(2)}(t-s)e^{-(t-s)P} \widetilde{P}^i_\ell(\delta_{y_*})\right\rangle  \rmd s
$$
and we see from the preceding regularity result for the commutator \eqref{EqCommutator} that the singular part $\mathfrak D \tau_{20}(z)$ of $\tau_{20}(z)$ coincides with the singular part of 
\begin{align}\label{eq:equdontjaibesoin}
    2\left\langle \Big( \Big\{\int_{-\infty}^t G_r^{(2)}(t-s)e^{-(t-s)P}\rmd s\Big\}\delta_{y_*}\Big)\odot_i \delta_{y_*},1 \right\rangle \,.
\end{align}
Given that the localized paraproduct of any two distributions is always well-defined, and that (cf Appendix~\ref{SectionLPProjectors})

$$
{\varphi_i}(uv)=u\odot_i v+ u\prec_iv+u\succ_iv,
$$
this localized singular part coincides with the singular part of 
$$
2{\varphi_i}({y_*})\Big\{\int_{-\infty}^t G_r^{(2)}(t-s)e^{-(t-s)P}\rmd s\Big\}({y_*},{y_*})\,.
$$

\ssk

\noindent {\it \S2. Explicit computation of the divergent part.} We proceed in two steps by localising first in space and time and then by using the heat kernel asymptotics.   \vspace{0.1cm}

\ssk

{\it \S2.1. Space and time localisation.} Fix $z=(t,y_*)$. We start by localising in space near $y_*$. Let ${o}$ stand for a smooth indicator function of a neighbourhood $U_{y_*}$ of $y_*$ in $M$ -- without loss of generality the domain of a chart. Since the quantity
$$
2\int_{-\infty}^t \int_M G_r^{(2)}(t-s)(y_*,y) \big(1-{o}(y) \big) e^{-(t-s)P}(y,y_*)\rmd v(y)\rmd s
$$
has a well-defined limit when $r\downarrow0$, we concentrate on 
\begin{equation*} \begin{split}
(\star) &\defeq 2\int_{-\infty}^t \int_{U_{y_*}} G_r^{(2)}(t-s)(y_*,y) e^{-(t-s)P}(y,y_*)  {o}(y)\rmd v(y)\rmd s   \\
&= 2\int_0^\infty \left(\int_{[a+2r,\infty)^2} \int_{U_{y_*}} e^{-s_1P}(y_*,y)e^{-s_2P}(y_*,y)   e^{-aP}(y_*,y){o}(y)\rmd v(y)\rmd s_1\rmd s_2 \right) \rmd a \,.
\end{split} \end{equation*}

{\it \S2.2. Use the heat kernel asymptotic.} We have near $y_*$
$$
e^{-sP}(y_*,y) = \frac{1}{(4\pi s)^{\frac{3}{2}}} e^{-\frac{\vert y_*-y\vert_{g(y_*)}^2}{4s}} e^{-s} + R\big(s,y_*, \frac{y_*-y}{\sqrt{s}}\big) =: K^0(s,y_*,y) + R\Big(s,y_*, \frac{y_*-y}{\sqrt{s}}\Big)
$$
for a remainder term $R\in \Psi^{-2}_H(M)$ in the heat calculus as defined in~\cite{Grieser}, that is $R\big(s,y_*, \frac{y_*-y}{\sqrt{s}}\big)$ has the same estimates as $s$ times the heat kernel itself. It follows from that fact that replacing any of the heat kernels $e^{-s_1P}, e^{-s_2P}, e^{-aP}$ by a remainder term $R$ gives a contribution to $(\star)$ that remains uniformly bounded for $r\in [0,1]$. We can therefore keep in our computations only the leading term of the heat expansion. Integrating first with respect to $y$ the stationary phase in $U_{y_*}$ gives the asymptotics
\begin{equation*} \begin{split}
2\int_{U_{y_*}} K^0(s_1,y_*,y) &K^0(s_2,y_*,y) {o}(y) K^0(a,y_*,y)\,\rmd v(y)    \\
&=\frac{2}{(4\pi)^{9/2}}\times\frac{e^{-(s_1+s_2+a)}}{(s_1s_2a)^{{3}/{2}}}\int_{U_{y_*}} e^{-\frac{\vert y_*-y\vert^2_{g(y_*)}}{ 4}(s_1^{-1}+s_2^{-1}+a^{-1})  }{o}(y) \sqrt{\det(g)_y} \, \rmd y   \\
&=\frac{2}{(4\pi)^3} \times\frac{e^{-(s_1+s_2+a)}{o}(y_*) \sqrt{\det(g)_{y_*}} }{ (s_1^{-1}+s_2^{-1}+a^{-1})^{{3}/{2}} (s_1s_2a)^{{3}/{2}} \sqrt{\det\mathrm{Hess} (|\bigcdot-y_*|^2_{g(y_*)} )(y_*) }  }    \\
&\qquad+ \mathcal{O}\Big((s_1s_2a)^{-{3}/{2}}(s_1^{-1}+s_2^{-1}+a^{-1})^{-{5}/{2}}\Big)\\
\end{split} \end{equation*}
\begin{equation*} \begin{split}
&= \frac{2}{(4\pi)^3}\times\frac{e^{-(s_1+s_2+a)}}{ (s_1^{-1}+s_2^{-1}+a^{-1})^{{3}/{2}} (s_1s_2a)^{{3}/{2}}}   \\
&\qquad+ \mathcal{O}\Big((s_1s_2a)^{-{3}/{2}}(s_1^{-1}+s_2^{-1}+a^{-1})^{-{5}/{2}}\Big)\,,
\end{split} \end{equation*}
with an error term $\mathcal{O}\big((s_1s_2a)^{-{3}/{2}}(s_1^{-1}+s_2^{-1}+a^{-1})^{-{5}/{2}}\big) $ bounded uniformly in the $y_*$ variable. Note that to go from the second to the last line we used ${o}(y_*)=1$, along with the fact that the determinant of the Hessian equals the determinant of $g$ evaluated at $y_*$.\\
Here, note that the divergence of the integral in $a$ only takes place when $a\downarrow0$ so that, in order to seek the divergent part, one can restrict the integration over $a$ to $[0,1]$. Moreover, the exponential decay $e^{-(s_1+s_2+a)}$ will not help, so that we can discard it when computing the divergent part. In the end, we are therefore left with computing the singular part of
\begin{align*}
\frac{2}{(4\pi)^3}&\int_0^1 \int_{[a+2r,\infty)^2}\frac{ \rmd s_1\rmd s_2\rmd a  }{(s_1a+s_2a+s_1s_2)^{3/2}}   \\
&= \frac{2}{(4\pi)^3} \int_0^1\frac{\rmd a}{a+2r}\int_1^\infty\int_1^\infty\frac{\rmd\alpha\rmd\beta}{(\alpha+\beta+\alpha\beta-\frac{2r}{a+2r}(\alpha+\beta))^{3/2}}   \\
&=  \frac{2}{(4\pi)^3}
\int_0^{1/2r}\frac{\rmd b}{1+b}\int_1^\infty\int_1^\infty\frac{\rmd\alpha\rmd\beta}{(\alpha+\beta+\alpha\beta-\frac{1}{1+b}(\alpha+\beta))^{3/2}}
\\&=
 \frac{2}{(4\pi)^3}
\int_0^{1/2r}\frac{\rmd b}{1+b}\int_1^\infty\int_1^\infty\Big(1-\frac{1}{1+b}\Big(\frac{\alpha+\beta}{\alpha+\beta+\alpha\beta}\Big)\Big)^{-3/2}\frac{\rmd\alpha\rmd\beta}{(\alpha+\beta+\alpha\beta)^{3/2}}
 \,,
\end{align*}
Here, we first did the changes of variable $s_1=(a+2r)\alpha$, $s_2=(a+2r)\beta$, and then $b=a/2r$. Using the fact that $\frac{1}{1+b}\big(\frac{\alpha+\beta}{\alpha+\beta+\alpha\beta}\big)$ is smaller than one, we can expand 
\begin{align*}
    \Big(1-\frac{1}{1+b}\Big(\frac{\alpha+\beta}{\alpha+\beta+\alpha\beta}\Big)\Big)^{-3/2}=1-\frac{3(\alpha+\beta)}{2(1+b)(\alpha+\beta+\alpha\beta)}+\cdots\,,
\end{align*}
so that all the higher order terms yield some finite contributions, since they become integrable as $b\uparrow\infty$. We have thus obtained that
\begin{align*}
    (\star)=\frac{2}{(4\pi)^3}
\int_0^{1/2r}\frac{\rmd b}{1+b}\int_1^\infty\int_1^\infty\frac{\rmd\alpha\rmd\beta}{(\alpha+\beta+\alpha\beta)^{3/2}}+\mathcal{O}(1)\,.
\end{align*}
We can compute
\begin{align*}
    \int_1^\infty
 \frac{ \rmd \beta }{(\alpha+\beta+\alpha\beta)^{3/2}}=\frac{2}{(1+\alpha)\sqrt{1+2\alpha}}\,.
\end{align*}
Then, setting $\gamma=\sqrt{1+2\alpha}$, we have
\begin{align*}
    \int_1^\infty\frac{2\rmd \alpha}{(1+\alpha)\sqrt{1+2\alpha}}=\int_{\sqrt{3}}^\infty \frac{4\rmd\gamma}{1+\gamma^2}=4[\arctan]^\infty_{\sqrt{3}}=\frac{2\pi}{3}\,.
\end{align*}
Overall, we thus end up with 
\begin{align*}
    (\star)=\frac13\times\frac{1}{16\pi^2}\int_0^{1/2r}\frac{\rmd b}{1+b}+\mathcal{O}(1)=\frac13\times\frac{1}{16\pi^2}|\log r|+\mathcal{O}(1)\,.
\end{align*}
\noindent This finally tells us that the divergent part of $\tau_{20}$ is exactly given by ${\varphi_i}\frac{b_r}{3}$ for every $ i\in I$. To conclude, note that what is left of $\tau_{20}$ after the subtraction of the divergent part defines a smooth function of $y_*$.

\subsubsection{Bounds on $\overline\tau_3$.}\label{subsec:tau3}

We first briefly discuss the counterterm for $\overline\tau_3$ and we will do the stochastic etimates in a second step.
Note that for any test function $\varphi$, it follows from the integration by parts formula (in fact the integration by parts formula can be taken as definition of the Laplace--Beltrami operator $\Delta_g$)  $$\int_M u \Delta_g v \rmd \mathrm{v}_g=-\int_M  \langle\nabla u,\nabla v \rangle_g \rmd \mathrm{v}_g,$$  that 
\begin{eqnarray*}
\int_{M} {\varphi_i} \vert \nabla \begin{tikzpicture}[scale=0.3,baseline=0cm]
\node at (0,0) [dot] (0) {};
\node at (0,0.5) [dot] (1) {};
\node at (-0.4,1)  [noise] (noise1) {};
\node at (0.4,1)  [noise] (noise2) {};
\draw[K] (0) to (1);
\draw[K] (1) to (noise1);
\draw[K] (1) to (noise2);
\end{tikzpicture}_r\vert^2 \, \varphi = -\int_M\left( \Delta \begin{tikzpicture}[scale=0.3,baseline=0cm]
\node at (0,0) [dot] (0) {};
\node at (0,0.5) [dot] (1) {};
\node at (-0.4,1)  [noise] (noise1) {};
\node at (0.4,1)  [noise] (noise2) {};
\draw[K] (0) to (1);
\draw[K] (1) to (noise1);
\draw[K] (1) to (noise2);
\end{tikzpicture}_r \right) \begin{tikzpicture}[scale=0.3,baseline=0cm]
\node at (0,0) [dot] (0) {};
\node at (0,0.5) [dot] (1) {};
\node at (-0.4,1)  [noise] (noise1) {};
\node at (0.4,1)  [noise] (noise2) {};
\draw[K] (0) to (1);
\draw[K] (1) to (noise1);
\draw[K] (1) to (noise2);
\end{tikzpicture}_r \, {\varphi_i}\varphi  - 
\int_M \big\langle \nabla  \begin{tikzpicture}[scale=0.3,baseline=0cm]
\node at (0,0) [dot] (0) {};
\node at (0,0.5) [dot] (1) {};
\node at (-0.4,1)  [noise] (noise1) {};
\node at (0.4,1)  [noise] (noise2) {};
\draw[K] (0) to (1);
\draw[K] (1) to (noise1);
\draw[K] (1) to (noise2);
\end{tikzpicture}_r, \nabla ({\varphi_i}\varphi) \big\rangle  \begin{tikzpicture}[scale=0.3,baseline=0cm]
\node at (0,0) [dot] (0) {};
\node at (0,0.5) [dot] (1) {};
\node at (-0.4,1)  [noise] (noise1) {};
\node at (0.4,1)  [noise] (noise2) {};
\draw[K] (0) to (1);
\draw[K] (1) to (noise1);
\draw[K] (1) to (noise2);
\end{tikzpicture}_r.
\end{eqnarray*}
The singular part of $\vert \nabla \begin{tikzpicture}[scale=0.3,baseline=0cm]
\node at (0,0) [dot] (0) {};
\node at (0,0.5) [dot] (1) {};
\node at (-0.4,1)  [noise] (noise1) {};
\node at (0.4,1)  [noise] (noise2) {};
\draw[K] (0) to (1);
\draw[K] (1) to (noise1);
\draw[K] (1) to (noise2);
\end{tikzpicture}_r\vert^2 $ as a random distribution is thus the same as the singular part of $\left( \Delta \begin{tikzpicture}[scale=0.3,baseline=0cm]
\node at (0,0) [dot] (0) {};
\node at (0,0.5) [dot] (1) {};
\node at (-0.4,1)  [noise] (noise1) {};
\node at (0.4,1)  [noise] (noise2) {};
\draw[K] (0) to (1);
\draw[K] (1) to (noise1);
\draw[K] (1) to (noise2);
\end{tikzpicture}_r \right) \begin{tikzpicture}[scale=0.3,baseline=0cm]
\node at (0,0) [dot] (0) {};
\node at (0,0.5) [dot] (1) {};
\node at (-0.4,1)  [noise] (noise1) {};
\node at (0.4,1)  [noise] (noise2) {};
\draw[K] (0) to (1);
\draw[K] (1) to (noise1);
\draw[K] (1) to (noise2);
\end{tikzpicture}_r $. The latter is also the singular part of

\makebox[\textwidth][c]{
\begin{minipage}{\dimexpr\textwidth+10cm}
\begin{align*}
\mathbb{E}\left[ P\IXtwo_r (x) \IXtwo_r(x) \right]
&= {2} \int_{(-\infty,t]^2} \hspace{-0.1cm}  \left( P e^{-(t-s_1)P} \circ G_r^{(2)}(s_1-s_2) \circ e^{-(t-s_2)P}\right) (x,x)\,\rmd s_1\rmd s_2   \\
&={2}\int_{(-\infty,t]^2}\left( G_r^{(2)}(s_1-s_2) \circ P e^{-(2t-s_1-s_2)P}\right) (x,x)\, \rmd s_1\rmd s_2 + \text{commutator}\,,
\end{align*}
\end{minipage}
}
\noindent
where the commutator of $G^{(2)}_r$ and the heat kernel gives a better contribution, which is not singular.

Next, changing variables for $u = t-s_1$ and $v=s_1-s_2$ and integrating by part in $u$ gives
\begin{align*}
2 &\int_{u>0,v>-u} \Big( G_r^{(2)}(v) \circ P e^{-(2u+v)P}\Big)(x,x) \, \rmd u\rmd v   \\
&= -\int_{\bbR^2}\boldsymbol{1}_{\geqslant0}\{u\}\boldsymbol{1}_{\geqslant0}\{u+v\} \Big( G_r^{(2)}(v) \circ \frac{\rmd}{\rmd u} e^{-(2u+v)P}\Big)(x,x) \, \rmd u\rmd v \\
&=\int_0^\infty G^{(2)}_r(v) \circ e^{-vP}(x,x)\,\rmd v+\int_{-\infty}^{0} G_r^{(2)}(v)\circ e^{vP}(x,x)\rmd v\\
&=2\int_0^\infty G^{(2)}_r(v) \circ e^{-vP}(x,x)\,\rmd v\,, 
\end{align*}
where in the last line we used the fact that $G_r^{(2)}$ is an even function of $v$.
The divergent part of $\tau_3$ thus coincides with that of the expression in \eqref{eq:equdontjaibesoin} which, in view of the calculation of the previous section, is the divergent part of $\tau_{20}$.

We next discuss the regularity of $\overline\tau_3$. First, we need to isolate the resonant part in the scalar product. Since we work in the manifold setting, note that we need to define carefully the resonant scalar product of two vector fields in $C^\infty(TM)$. For $s_1,s_2\in C^\infty(TM)^2$, using the notations and conventions in Appendix~\ref{SectionLPProjectors}, we define
$\left\langle s_1\odot_i s_2\right\rangle_{TM} $ as:
$$  \left\langle s_1\odot_i s_2\right\rangle_{TM}\defeq  \kappa_i^* \left( (\kappa_{i*}g)^{\mu\nu}\kappa_{i*}\psi_i \left( \kappa_{i*}({\varphi_i}s_1)_\mu \odot \kappa_{i*}(\widetilde{{\varphi}}_i s_2)_\nu\right)\right) $$
where $\kappa_i,\psi_i,{\varphi_i},\widetilde{{\varphi}}_i$ come from our definition of resonant product, $(\kappa_{i*}g)$ is the metric $g$ induced by the charts $\kappa_i:U_i\mapsto \kappa_i(U_i)\subset \bbR^d$.
Similarly we have $$\left\langle s_1\prec_i s_2\right\rangle_{TM}\defeq  \kappa_i^* \left( (\kappa_{i*}g)^{\mu\nu}\kappa_{i*}\psi_i \left( \kappa_{i*}({\varphi_i}s_1)_\mu \prec \kappa_{i*}(\widetilde{{\varphi}}_i s_2)_\nu\right)\right)$$
and we recover the usual decomposition 
$$
\left\langle s_1,s_2\right\rangle_{TM}=\sum_i \left\langle s_1\odot_i s_2\right\rangle_{TM}+\left\langle s_1\prec_i s_2\right\rangle_{TM}+\left\langle s_1 \succ_i s_2\right\rangle_{TM} 
$$ 
for the scalar product on sections of $TM$. We need to prove that such resonant scalar product satisfies some approximate integration by parts identity and we are careful since the Laplacian is no longer translation invariant. For every $Y\in C^\infty(M)$ a calculation relying on the definitions of both resonant product and resonant scalar product:
$$ 
\left\langle \nabla Y \odot_i \nabla Y \right\rangle_{TM}= {g}_1(Y\odot_1 Y)+(PY) \odot_2 Y+ Y \odot_i\Delta Y
$$ 
where ${g}_1$ is a smooth function, $P$ is a differential operator of order $1$ on $M$ with smooth coefficients, $\odot_1,\odot_2$ are localized resonant type products which might differ from the original $\odot_i$ in the choice of smooth cut--off functions involved in the definition but have the exact same analytical properties from Proposition~\ref{PropContinuityParaproductResonance}, and the last term involves the localized resonant product of $Y$ with $\Delta Y$. From the point of view of regularities, for $Y=\begin{tikzpicture}[scale=0.3,baseline=0cm]
\node at (0,0) [dot] (0) {};
\node at (0,0.5) [dot] (1) {};
\node at (-0.4,1)  [noise] (noise1) {};
\node at (0.4,1)  [noise] (noise2) {};
\draw[K] (0) to (1);
\draw[K] (1) to (noise1);
\draw[K] (1) to (noise2);
\end{tikzpicture}_r\in \mcC^{1-2\varepsilon}([0,T]\times M)$,  ${g}_1(\begin{tikzpicture}[scale=0.3,baseline=0cm]
\node at (0,0) [dot] (0) {};
\node at (0,0.5) [dot] (1) {};
\node at (-0.4,1)  [noise] (noise1) {};
\node at (0.4,1)  [noise] (noise2) {};
\draw[K] (0) to (1);
\draw[K] (1) to (noise1);
\draw[K] (1) to (noise2);
\end{tikzpicture}_r\odot_1 \begin{tikzpicture}[scale=0.3,baseline=0cm]
\node at (0,0) [dot] (0) {};
\node at (0,0.5) [dot] (1) {};
\node at (-0.4,1)  [noise] (noise1) {};
\node at (0.4,1)  [noise] (noise2) {};
\draw[K] (0) to (1);
\draw[K] (1) to (noise1);
\draw[K] (1) to (noise2);
\end{tikzpicture}_r)\in C_T\mcC^{2-4\varepsilon}(M)$ and 
$ 
(P\begin{tikzpicture}[scale=0.3,baseline=0cm]
\node at (0,0) [dot] (0) {};
\node at (0,0.5) [dot] (1) {};
\node at (-0.4,1)  [noise] (noise1) {};
\node at (0.4,1)  [noise] (noise2) {};
\draw[K] (0) to (1);
\draw[K] (1) to (noise1);
\draw[K] (1) to (noise2);
\end{tikzpicture}_r) \odot_2 \begin{tikzpicture}[scale=0.3,baseline=0cm]
\node at (0,0) [dot] (0) {};
\node at (0,0.5) [dot] (1) {};
\node at (-0.4,1)  [noise] (noise1) {};
\node at (0.4,1)  [noise] (noise2) {};
\draw[K] (0) to (1);
\draw[K] (1) to (noise1);
\draw[K] (1) to (noise2);
\end{tikzpicture}_r
\in C_T\mcC^{1-4\varepsilon}(M)
$
and finally
we are reduced to the study of $\begin{tikzpicture}[scale=0.3,baseline=0cm]
\node at (0,0) [dot] (0) {};
\node at (0,0.5) [dot] (1) {};
\node at (-0.4,1)  [noise] (noise1) {};
\node at (0.4,1)  [noise] (noise2) {};
\draw[K] (0) to (1);
\draw[K] (1) to (noise1);
\draw[K] (1) to (noise2);
\end{tikzpicture}_r\odot_i \Delta \begin{tikzpicture}[scale=0.3,baseline=0cm]
\node at (0,0) [dot] (0) {};
\node at (0,0.5) [dot] (1) {};
\node at (-0.4,1)  [noise] (noise1) {};
\node at (0.4,1)  [noise] (noise2) {};
\draw[K] (0) to (1);
\draw[K] (1) to (noise1);
\draw[K] (1) to (noise2);
\end{tikzpicture}_r $. This is now really very similar to what we did for the graph $\tau_2$ except there is an extra propagator $\Delta\underline\mcL^{-1}$ in all the Feynman amplitudes. However microlocal estimates from our companion work~\cite[Theorem 1.1]{BDFTCompanion} actually show that
\begin{align}\label{eq:DeltaL-1}
    \Delta\underline\mcL^{-1}\in \mcS^{-5}_{N^*\left(\{t=s\}\times \mathbf{d}_2\subset \bbR^2\times M^2\right)}\left( \bbR^2\times M^2\right)\,.
\end{align}
Now we repeat the stochastic estimates on $\bbR^4\times M^{10}$ taking this new kernel and its weak homogeneity into account.
For instance, the amplitude controlling the homogeneous chaos of orders $2$ and $4$ in the chaos decomposition of $\begin{tikzpicture}[scale=0.3,baseline=0cm]
\node at (0,0) [dot] (0) {};
\node at (0,0.5) [dot] (1) {};
\node at (-0.4,1)  [noise] (noise1) {};
\node at (0.4,1)  [noise] (noise2) {};
\draw[K] (0) to (1);
\draw[K] (1) to (noise1);
\draw[K] (1) to (noise2);
\end{tikzpicture}_r\odot_i \Delta \begin{tikzpicture}[scale=0.3,baseline=0cm]
\node at (0,0) [dot] (0) {};
\node at (0,0.5) [dot] (1) {};
\node at (-0.4,1)  [noise] (noise1) {};
\node at (0.4,1)  [noise] (noise2) {};
\draw[K] (0) to (1);
\draw[K] (1) to (noise1);
\draw[K] (1) to (noise2);
\end{tikzpicture}_r $ now read
\begin{align}    \label{eq:amplitude34}
    \mathcal A_{34}= &[\odot_i](y_*,y_1,y_2)[\odot_i](z_*,z_1,z_2)      \mcQ_x^\gamma(y_*,z_*)  \underline{\mcL}^{-1}(t-s_a,y_2,x_a) \underline{\mcL}^{-1}(t-s_b,z_2,x_b)\\&G^{(2)}_r
(s_a-s_b,x_a,x_b)\nonumber    
\Delta\underline\mcL^{-1}(t-s_c,y_1,x_c)  
\Delta\underline\mcL^{-1}(t-s_d,z_1,x_d)
G^{(2)}_r
(s_c-s_d,x_c,x_d)
\,, \nonumber\\
\label{eq:amplitude32}
    \mathcal A_{32}= &[\odot_i](y_*,y_1,y_2)[\odot_i](z_*,z_1,z_2)      \mcQ_x^\gamma(y_*,z_*)  \underline{\mcL}^{-1}(t-s_a,y_2,x_a) \underline{\mcL}^{-1}(t-s_b,z_2,x_b)\\&G^{(1)}_r
(s_a-s_b,x_a,x_b)\nonumber    
\Delta\underline\mcL^{-1}(t-s_c,y_1,x_c)  
\Delta\underline\mcL^{-1}(t-s_d,z_1,x_d)
G^{(1)}_r
(s_c-s_d,x_c,x_d)\\
&G^{(1)}_r
(s_a-s_c,x_a,x_c)G^{(1)}_r
(s_b-s_d,x_b,x_d)
\,. \nonumber
\end{align}
These two amplitudes respectively have the subamplitudes $\widetilde\mcC^{(2)}$ and $\widetilde\mcC^{(1)}$, which are dealt with in Section~\ref{subsubsec:shielding} (see Lemma~\ref{LemC^k}). Here again, it remains to calculate the weak homogeneities of the amplitudes and all the other subamplitudes, by summing the weak homogeneity of all analytical objects, the kernels $G_r^{(2)}, [\odot_i],\underline{\mcL}^{-1},\Delta\underline{\mcL}^{-1},\mcQ^\gamma_x$. For $\mcA_{34}$, this yields
\begin{align*}
\sum \text{weak homogeneities} &= 2(-2)+2(-6)+2(-3)-6-2\gamma+2(-5) =-38-2\gamma   \\ 
&> - \text{codim}_w\big(\big\{s_a=s_b=s_c=s_d=t, y_*=\dots=x_d=x\big\}\big)   \\
&= -8 -30 = -38\,,
\end{align*} 
hence we obtain $\gamma<0$. Repeating this verification for all subgraphs and for all amplitudes controlling the term homogeneous of order $2$ in the chaos decomposition yields the result that 
$ \begin{tikzpicture}[scale=0.3,baseline=0cm]
\node at (0,0) [dot] (0) {};
\node at (0,0.5) [dot] (1) {};
\node at (-0.4,1)  [noise] (noise1) {};
\node at (0.4,1)  [noise] (noise2) {};
\draw[K] (0) to (1);
\draw[K] (1) to (noise1);
\draw[K] (1) to (noise2);
\end{tikzpicture}_r\odot_i \Delta \begin{tikzpicture}[scale=0.3,baseline=0cm]
\node at (0,0) [dot] (0) {};
\node at (0,0.5) [dot] (1) {};
\node at (-0.4,1)  [noise] (noise1) {};
\node at (0.4,1)  [noise] (noise2) {};
\draw[K] (0) to (1);
\draw[K] (1) to (noise1);
\draw[K] (1) to (noise2);
\end{tikzpicture}_r \in \mcC^{-4\varepsilon}([0,T]\times M)$ almost surely.

 The locally covariant renormalisation for $\overline{\tau_3}$ differs from the Wick one by $f_1\cdot\nabla  \begin{tikzpicture}[scale=0.3,baseline=0cm]
\node at (0,0) [dot] (0) {};
\node at (0,0.5) [dot] (1) {};
\node at (-0.4,1)  [noise] (noise1) {};
\node at (0.4,1)  [noise] (noise2) {};
\draw[K] (0) to (1);
\draw[K] (1) to (noise1);
\draw[K] (1) to (noise2);
\end{tikzpicture}  +f_2$ where $f_1$ is a smooth vector and $f_2$ a smooth function which has higher regularity than $\vert \nabla  \begin{tikzpicture}[scale=0.3,baseline=0cm]
\node at (0,0) [dot] (0) {};
\node at (0,0.5) [dot] (1) {};
\node at (-0.4,1)  [noise] (noise1) {};
\node at (0.4,1)  [noise] (noise2) {};
\draw[K] (0) to (1);
\draw[K] (1) to (noise1);
\draw[K] (1) to (noise2);
\end{tikzpicture}\vert^2 $.

\subsubsection{The quintic term.}
\label{SubsectionQuinticTerm} 
The first two graphs respectively have the subgraphs with amplitude $\mcC^{(2)}$ and $\mcC^{(1)}$, which are dealt with in Section~\ref{subsubsec:shielding} (see Lemma~\ref{LemC^k}). To complete the argument, one extracts all the other closed, connected irreducible graphs and finds the range of parameter $\gamma$ so that they satisfy the criterion of Theorem \ref{ThmBlackBoxFeynmanGraphs}. For the reader's convenience, recall that the amplitude $\mcA_{45}$ of $\mcG_{45}=\begin{tikzpicture}[scale=0.6,baseline=-0.1cm]
\node at (0,-0.3) [dot] (1) {};
\node at (0,-0.1) [dot] (2) {};
\node at (0,0.3) [dot] (3) {};
\node at (0,0.5) [dot] (4) {};
\node at (0,0.8) [dot] (5) {};
\node at (-0.4,0.6) [dot] (6) {};
\node at (0.4,0.6) [dot] (7) {};
\node at (-0.8,0) [blackdot] (8) {};
\node at (0.8,0) [blackdot] (9) {};
\draw[K] (1) to (8);
\draw[K] (1) to (9);
\draw[K] (2) to (8);
\draw[K] (2) to (9);
\draw[K] (3) to (6);
\draw[K] (3) to (7);
\draw[K] (4) to (6);
\draw[K] (4) to (7);
\draw[K] (5) to (6);
\draw[K] (5) to (7);
\draw[DK] (8) to (6);
\draw[DK] (9) to (7);
\draw[K,purple] (8) to[bend right=+80] (9);
\end{tikzpicture}$
is given by 
\begin{equation*} \begin{split}
[\odot_i](y_*,y_1,y_2) \, [\odot_i](z_*,z_1,z_2) \, &\mcQ_x^\gamma(y_*,z_*) \, G^{(2)}_r( 0,y_1,z_1)   \\
&\times \underline{\mcL}^{-1}(t-s_a,y_2,x_a) \, \underline{\mcL}^{-1}(t-s_b,z_2,x_b) \, G^{(3)}_r(s_a-s_b,x_a,x_b),
\end{split} \end{equation*}
and the amplitude $\mcA_{43}$ of $\mcG_{43}=\begin{tikzpicture}[scale=0.6,baseline=-0.1cm]
\node at (0,-0.4) [dot] (1) {};
\node at (0,0.35) [dot] (2) {};
\node at (0,0.85) [dot] (3) {};
\node at (-0.4,0.6) [dot] (4) {};
\node at (-0.4,0) [dot] (5) {};
\node at (0.4,0.6) [dot] (6) {};
\node at (0.4,0) [dot] (7) {};
\node at (-0.9,0) [blackdot] (8) {};
\node at (0.9,0) [blackdot] (9) {};
\draw[K] (1) to (8);
\draw[K] (1) to (9);
\draw[K] (2) to (4);
\draw[K] (2) to (6);
\draw[K] (3) to (4);
\draw[K] (3) to (6);
\draw[DK] (4) to (8);
\draw[K] (4) to (5);
\draw[K] (5) to (8);
\draw[K] (6) to (7);
\draw[DK] (6) to (9);
\draw[K] (7) to (9);
\draw[K,purple] (8) to[bend right=+80] (9);
\end{tikzpicture}$
is
\begin{align*}
    \mathcal A_{43}= &[\odot_i](y_*,y_1,y_2)   [\odot_i](z_*,z_1,z_2)      \mcQ_x^\gamma(y_*,z_*)  G^{(1)}_r(0 ,y_1,z_1)   G^{(1)}_r( t-s_a ,y_1,x_a)  G^{(1)}_r(t-s_b  ,z_1,x_b)\\
    &\times \underline{\mcL}^{-1}(t-s_a,y_2,x_a)    \underline{\mcL}^{-1}(t-s_b,z_2,x_b) G^{(2)}_r(s_a-s_b,x_a,x_b)  \,.
\end{align*}
The reader can check that the scaling degrees of the subgraphs yield the range $\gamma<-\frac{1}{2}$.

\ssk

It remains to handle the last contribution, which verifies
$$
\bbE\big[\mcQ_x^\gamma\big((\tau_{41}(t,\bigcdot)-{\varphi_i}b_r\X_r),(\tau_{41}(t,\bigcdot)-{\varphi_i}b_r\X_r)\big)\big] = \langle\mcA_{41},1\rangle,
$$
where $\mcA_{41}$ is given by \eqref{eq:defA41}. The amplitude $\mcA_{41}$ is indeed well-defined as a distribution of $M^8\times \bbR^2$ by Theorem~\ref{ThmBlackBoxFeynmanGraphs} for $\gamma<-1/2$, taking the renormalized subamplitude $\mcB$ (that is constructed in Lemma~\ref{LemB}) into account in the induction over all subgraphs of $\mcA_{41}$.
\ssk

We conclude our discussion by pointing out the difference between the locally covariant and Wick renormalisations. The locally covariant renormalisation of $\begin{tikzpicture}[scale=0.3,baseline=0cm]
\node at (0,0) [dot] (0) {};
\node at (0,0.5) [dot] (1) {};
\node at (-0.4,1)  [noise] (noise1) {};
\node at (0,1.2)  [noise] (noise2) {};
\node at (0.4,1)  [noise] (noise3) {};
\draw[K] (0) to (1);
\draw[K] (1) to (noise1);
\draw[K] (1) to (noise2);
\draw[K] (1) to (noise3);
\end{tikzpicture} \odot_i \begin{tikzpicture}[scale=0.3,baseline=0cm]
\node at (0,0) [dot] (0) {};
\node at (0.3,0.6)  [noise] (noise1) {};
\node at (-0.3,0.6)  [noise] (noise2) {};
\draw[K] (0) to (noise1);
\draw[K] (0) to (noise2);
\end{tikzpicture} $ differs from
the Wick renormalisation by $\left(\underline{\mcL}^{-1}\left(f
\begin{tikzpicture}[scale=0.3,baseline=0cm] \node at (0,0)  [dot] (1) {}; \node at (0,0.8)  [noise] (2) {}; \draw[K] (1) to (2); \end{tikzpicture}\right)
\right)\odot_i \begin{tikzpicture}[scale=0.3,baseline=0cm]
\node at (0,0) [dot] (0) {};
\node at (0.3,0.6)  [noise] (noise1) {};
\node at (-0.3,0.6)  [noise] (noise2) {};
\draw[K] (0) to (noise1);
\draw[K] (0) to (noise2);
\end{tikzpicture}   +P(\begin{tikzpicture}[scale=0.3,baseline=0cm]
\node at (0,0) [dot] (0) {};
\node at (0,0.5) [dot] (1) {};
\node at (-0.4,1)  [noise] (noise1) {};
\node at (0,1.2)  [noise] (noise2) {};
\node at (0.4,1)  [noise] (noise3) {};
\draw[K] (0) to (1);
\draw[K] (1) to (noise1);
\draw[K] (1) to (noise2);
\draw[K] (1) to (noise3);
\end{tikzpicture} ) $ where $f$ is a smooth function and $P$ a smoothing operator and therefore $\left(\underline{\mcL}^{-1}\left(f
\begin{tikzpicture}[scale=0.3,baseline=0cm] \node at (0,0)  [dot] (1) {}; \node at (0,0.8)  [noise] (2) {}; \draw[K] (1) to (2); \end{tikzpicture}\right)
\right)\odot_i \begin{tikzpicture}[scale=0.3,baseline=0cm]
\node at (0,0) [dot] (0) {};
\node at (0.3,0.6)  [noise] (noise1) {};
\node at (-0.3,0.6)  [noise] (noise2) {};
\draw[K] (0) to (noise1);
\draw[K] (0) to (noise2);
\end{tikzpicture} \in \mathcal{C}^{\frac{1}{2}-\varepsilon} $, $\forall \varepsilon>0$ which is absorbed in the $\mathcal{C}^{-\frac{1}{2}-}$ regularity of $\begin{tikzpicture}[scale=0.3,baseline=0cm]
\node at (0,0) [dot] (0) {};
\node at (0,0.5) [dot] (1) {};
\node at (-0.4,1)  [noise] (noise1) {};
\node at (0,1.2)  [noise] (noise2) {};
\node at (0.4,1)  [noise] (noise3) {};
\draw[K] (0) to (1);
\draw[K] (1) to (noise1);
\draw[K] (1) to (noise2);
\draw[K] (1) to (noise3);
\end{tikzpicture} \odot_i \begin{tikzpicture}[scale=0.3,baseline=0cm]
\node at (0,0) [dot] (0) {};
\node at (0.3,0.6)  [noise] (noise1) {};
\node at (-0.3,0.6)  [noise] (noise2) {};
\draw[K] (0) to (noise1);
\draw[K] (0) to (noise2);
\end{tikzpicture}  $.

\medskip

The results of this section justify the convergence in $L^2(\Omega)$, hence in any $L^p(\Omega)$ with $p<\infty$, of $\widehat\xi_r$ to a limit random variable $\widehat\xi$ in its natural space. The convergence {\it in probability} of $v_r \in \llparenthesis \alpha_0 , 1+\epsilon'\rrparenthesis$ to a limit $v$ in that space follows as a consequence of the pathwise continuity of $v_r$ with respect to $\widehat\xi_r$ obtained from the fixed point construction of $v_r$. Formula \eqref{EqFromUtoV} relating $v_r$ to $u_r$ shows the convergence  {\it in probability} of $u_r\in C_T\mcC^{-1/2-\epsilon}(M)$ to a limit $u$ in that space.

\bigskip

\section{Non-constant coupling function}
\label{SectionCouplingFunction}

Denote by $\mu_{\textrm{GFF}}(\rmd\phi)$ the Gaussian free field measure on $M$. We construct in this section a $\Phi^4_3$ measure on $M$ with formal description 
\begin{equation} \label{EqPhi43Lambda}
\nu(\rmd\phi)\propto e^{-\int_M\lambda(x)\phi^4(x)\rmd x} \mu_{\textrm{GFF}}(\rmd\phi)
\end{equation}
corresponding to a space dependent coupling constant $\lambda(x)$, say a {positive $C^\infty$ function} on $M$. We considered so far the case $\lambda = 1$. We show that the results proved to analyse this special case allow to deal with the general case. We construct the measure \eqref{EqPhi43Lambda} as an invariant measure of a Markovian evolution in $\mcC^{-1/2-\epsilon}(M)$. The goal of this section is to prove that the \textbf{counterterms are local functionals in the coupling function} $\lambda\in C^\infty(M,\mathbb{R}_{>0})$ which is a deep feature of renormalisation: {\em The divergent counterterms you need to subtract at some point $x$ depends only on some finite jets of the Lagrangian functional density at the same point $x$}. We refer to the work \cite{Abdesselam3d} of Abdesselam which also deals with space-dependent couplings for some hierarchical model in $3$d.

\medskip

Let us make a detailed calculation to determine the fine structure of the divergences that arise in the equation as well as to establish the locality of the counterterms. 
We use the same regularisation $\xi_r$ as above.
Starting from $\mcL u_r=\xi_r-\lambda u_r^3$, we set a first decomposition $u_r=\begin{tikzpicture}[scale=0.3,baseline=0cm] \node at (0,0)  [dot] (1) {}; \node at (0,0.8)  [noise] (2) {}; \draw[K] (1) to (2); \end{tikzpicture}_{r} +Z_r$, hence $\mcL ( \begin{tikzpicture}[scale=0.3,baseline=0cm] \node at (0,0)  [dot] (1) {}; \node at (0,0.8)  [noise] (2) {}; \draw[K] (1) to (2); \end{tikzpicture}_{r} +Z_r )=\xi_r-\left(\begin{tikzpicture}[scale=0.3,baseline=0cm] \node at (0,0)  [dot] (1) {}; \node at (0,0.8)  [noise] (2) {}; \draw[K] (1) to (2); \end{tikzpicture}_{r} +Z_r \right)^3$. Therefore
\begin{align*}
\mcL Z_r& = -\lambda\big(\begin{tikzpicture}[scale=0.3,baseline=0cm] \node at (0,0)  [dot] (1) {}; \node at (0,0.8)  [noise] (2) {}; \draw[K] (1) to (2); \end{tikzpicture}_{r}^3 + 3Z_r\begin{tikzpicture}[scale=0.3,baseline=0cm] \node at (0,0)  [dot] (1) {}; \node at (0,0.8)  [noise] (2) {}; \draw[K] (1) to (2); \end{tikzpicture}_{r}^2  +3Z_r^2 \begin{tikzpicture}[scale=0.3,baseline=0cm] \node at (0,0)  [dot] (1) {}; \node at (0,0.8)  [noise] (2) {}; \draw[K] (1) to (2); \end{tikzpicture}_{r}+Z_r^3 \big) 
= -\lambda\left(\begin{tikzpicture}[scale=0.3,baseline=0cm]
\node at (0,0) [dot] (0) {};
\node at (-0.4,0.5)  [noise] (noise1) {};
\node at (0,0.7)  [noise] (noise2) {};
\node at (0.4,0.5)  [noise] (noise3) {};
\draw[K] (0) to (noise1);
\draw[K] (0) to (noise2);
\draw[K] (0) to (noise3);
\end{tikzpicture}_r+ {\color{red} 3 a_r  \begin{tikzpicture}[scale=0.3,baseline=0cm] \node at (0,0)  [dot] (1) {}; \node at (0,0.8)  [noise] (2) {}; \draw[K] (1) to (2); \end{tikzpicture}_{r}} +3Z_r\begin{tikzpicture}[scale=0.3,baseline=0cm]
\node at (0,0) [dot] (0) {};
\node at (0.3,0.6)  [noise] (noise1) {};
\node at (-0.3,0.6)  [noise] (noise2) {};
\draw[K] (0) to (noise1);
\draw[K] (0) to (noise2);
\end{tikzpicture}_r +3Z_r {\color{red} a_r}  +3Z_r^2 \begin{tikzpicture}[scale=0.3,baseline=0cm] \node at (0,0)  [dot] (1) {}; \node at (0,0.8)  [noise] (2) {}; \draw[K] (1) to (2); \end{tikzpicture}_{r}+Z_r^3 \right)\\
&=
-\lambda\left(\begin{tikzpicture}[scale=0.3,baseline=0cm]
\node at (0,0) [dot] (0) {};
\node at (-0.4,0.5)  [noise] (noise1) {};
\node at (0,0.7)  [noise] (noise2) {};
\node at (0.4,0.5)  [noise] (noise3) {};
\draw[K] (0) to (noise1);
\draw[K] (0) to (noise2);
\draw[K] (0) to (noise3);
\end{tikzpicture}_r +3Z_r\begin{tikzpicture}[scale=0.3,baseline=0cm]
\node at (0,0) [dot] (0) {};
\node at (0.3,0.6)  [noise] (noise1) {};
\node at (-0.3,0.6)  [noise] (noise2) {};
\draw[K] (0) to (noise1);
\draw[K] (0) to (noise2);
\end{tikzpicture}_r  +3Z_r^2 \begin{tikzpicture}[scale=0.3,baseline=0cm] \node at (0,0)  [dot] (1) {}; \node at (0,0.8)  [noise] (2) {}; \draw[K] (1) to (2); \end{tikzpicture}_{r}+Z_r^3 \right)
-3\lambda  {\color{red}  a_r u_r}
\end{align*}
where we compare the covariant Wick renormalized powers $\begin{tikzpicture}[scale=0.3,baseline=0cm]
\node at (0,0) [dot] (0) {};
\node at (0.3,0.6)  [noise] (noise1) {};
\node at (-0.3,0.6)  [noise] (noise2) {};
\draw[K] (0) to (noise1);
\draw[K] (0) to (noise2);
\end{tikzpicture}_r$ with the non-renormalized powers and the counterterms are shown in {\color{red}red} to clearly see the difference.

This motivates defining a new regularized equation for $u_r$ as
\begin{eqnarray*}
\mcL u_r=\xi_r-\lambda u_r^3+{\color{red} 3\lambda a_r u_r}
\end{eqnarray*}
where adding the counterterm in {\color{red} red} has the effect of Wick renormalising the trees appearing on the r.h.s of the equation for $Z_r$. 
Therefore, in what follows, all trees that appear are covariantly Wick renormalized.
We also define the following new stochastic trees decorated by the subscript $\lambda$:
$$\begin{tikzpicture}[scale=0.3,baseline=0cm] \node at (0,0) [dot] (0) {}; \node at (0,0.5) [dot] (1) {}; \node at (-0.4,1)  [noise] (noise1) {}; \node at (0,1.2)  [noise] (noise2) {}; \node at (0.4,1)  [noise] (noise3) {}; \draw[K] (0) to (1); \draw[K] (1) to (noise1); \draw[K] (1) to (noise2); \draw[K] (1) to (noise3); \end{tikzpicture}_{r,\lambda}\defeq \underline{\mathcal{L}}^{-1}(\lambda\begin{tikzpicture}[scale=0.3,baseline=0cm]
\node at (0,0) [dot] (0) {};
\node at (-0.4,0.5)  [noise] (noise1) {};
\node at (0,0.7)  [noise] (noise2) {};
\node at (0.4,0.5)  [noise] (noise3) {};
\draw[K] (0) to (noise1);
\draw[K] (0) to (noise2);
\draw[K] (0) to (noise3);
\end{tikzpicture}_r)\quad\text{and}\quad 
\begin{tikzpicture}[scale=0.3,baseline=0cm] \node at (0,0) [dot] (0) {}; \node at (0,0.4) [dot] (1) {}; \node at (-0.3,0.8)  [noise] (noise1) {}; \node at (0.3,0.8)  [noise] (noise2) {}; \draw[K] (0) to (1); \draw[K] (1) to (noise1); \draw[K] (1) to (noise2); \end{tikzpicture}_{r,\lambda}\defeq \underline{\mathcal{L}}^{-1}(\lambda \begin{tikzpicture}[scale=0.3,baseline=0cm]
\node at (0,0) [dot] (0) {};
\node at (0.3,0.6)  [noise] (noise1) {};
\node at (-0.3,0.6)  [noise] (noise2) {};
\draw[K] (0) to (noise1);
\draw[K] (0) to (noise2);
\end{tikzpicture}_r)\,.$$ These objets have an extra coupling function $\lambda$ inserted at the vertex to avoid confusion with all previous trees introduced earlier where the coupling function was set to $\lambda=1$. 

Next, we introduce a further decomposition
rewriting $u_r$ as $u_r=\begin{tikzpicture}[scale=0.3,baseline=0cm] \node at (0,0)  [dot] (1) {}; \node at (0,0.8)  [noise] (2) {}; \draw[K] (1) to (2); \end{tikzpicture}_{r}   %
\underset{=Z_r}{\underbrace{- \begin{tikzpicture}[scale=0.3,baseline=0cm] \node at (0,0) [dot] (0) {}; \node at (0,0.5) [dot] (1) {}; \node at (-0.4,1)  [noise] (noise1) {}; \node at (0,1.2)  [noise] (noise2) {}; \node at (0.4,1)  [noise] (noise3) {}; \draw[K] (0) to (1); \draw[K] (1) to (noise1); \draw[K] (1) to (noise2); \draw[K] (1) to (noise3); \end{tikzpicture}_{r,\lambda}+R_r}}$,
where again we used a subscript $\lambda$ to recall the fact that the nonlinear term $u_r^3$ has become $\lambda u_r^3$, which affects the trees in the equation. Then the next remainder $R_r$ satisfies the new equation:
\begin{eqnarray*}
\mcL R_r=-3\lambda \begin{tikzpicture}[scale=0.3,baseline=0cm]
\node at (0,0) [dot] (0) {};
\node at (0.3,0.6)  [noise] (noise1) {};
\node at (-0.3,0.6)  [noise] (noise2) {};
\draw[K] (0) to (noise1);
\draw[K] (0) to (noise2);
\end{tikzpicture}_r
\big(R_r- \begin{tikzpicture}[scale=0.3,baseline=0cm] \node at (0,0) [dot] (0) {}; \node at (0,0.5) [dot] (1) {}; \node at (-0.4,1)  [noise] (noise1) {}; \node at (0,1.2)  [noise] (noise2) {}; \node at (0.4,1)  [noise] (noise3) {}; \draw[K] (0) to (1); \draw[K] (1) to (noise1); \draw[K] (1) to (noise2); \draw[K] (1) to (noise3); \end{tikzpicture}_{r,\lambda}\big)
-3\lambda \begin{tikzpicture}[scale=0.3,baseline=0cm] \node at (0,0)  [dot] (1) {}; \node at (0,0.8)  [noise] (2) {}; \draw[K] (1) to (2); \end{tikzpicture}_{r}
 \big(R_r- \begin{tikzpicture}[scale=0.3,baseline=0cm] \node at (0,0) [dot] (0) {}; \node at (0,0.5) [dot] (1) {}; \node at (-0.4,1)  [noise] (noise1) {}; \node at (0,1.2)  [noise] (noise2) {}; \node at (0.4,1)  [noise] (noise3) {}; \draw[K] (0) to (1); \draw[K] (1) to (noise1); \draw[K] (1) to (noise2); \draw[K] (1) to (noise3); \end{tikzpicture}_{r,\lambda}\big)^2
-\lambda  \big(R_r- \begin{tikzpicture}[scale=0.3,baseline=0cm] \node at (0,0) [dot] (0) {}; \node at (0,0.5) [dot] (1) {}; \node at (-0.4,1)  [noise] (noise1) {}; \node at (0,1.2)  [noise] (noise2) {}; \node at (0.4,1)  [noise] (noise3) {}; \draw[K] (0) to (1); \draw[K] (1) to (noise1); \draw[K] (1) to (noise2); \draw[K] (1) to (noise3); \end{tikzpicture}_{r,\lambda}\big)^3\,,
\end{eqnarray*}
whose right hand side  is now Wick renormalized. Let us anticipate a bit on what follows and try
to guess what problematic terms we will encounter next.
We spot two problematic terms in the equation for $R_r$, first we expect that $R_r$ has regularity $1-$ and $\begin{tikzpicture}[scale=0.3,baseline=0cm]
\node at (0,0) [dot] (0) {};
\node at (0.3,0.6)  [noise] (noise1) {};
\node at (-0.3,0.6)  [noise] (noise2) {};
\draw[K] (0) to (noise1);
\draw[K] (0) to (noise2);
\end{tikzpicture}_r$ has regularity $-1-$ hence the product $R_r\begin{tikzpicture}[scale=0.3,baseline=0cm]
\node at (0,0) [dot] (0) {};
\node at (0.3,0.6)  [noise] (noise1) {};
\node at (-0.3,0.6)  [noise] (noise2) {};
\draw[K] (0) to (noise1);
\draw[K] (0) to (noise2);
\end{tikzpicture}_r$ is ill--defined with the borderline regularity.
Furthermore the product 
$3\lambda \begin{tikzpicture}[scale=0.3,baseline=0cm]
\node at (0,0) [dot] (0) {};
\node at (0.3,0.6)  [noise] (noise1) {};
\node at (-0.3,0.6)  [noise] (noise2) {};
\draw[K] (0) to (noise1);
\draw[K] (0) to (noise2);
\end{tikzpicture}_r
 \begin{tikzpicture}[scale=0.3,baseline=0cm] \node at (0,0) [dot] (0) {}; \node at (0,0.5) [dot] (1) {}; \node at (-0.4,1)  [noise] (noise1) {}; \node at (0,1.2)  [noise] (noise2) {}; \node at (0.4,1)  [noise] (noise3) {}; \draw[K] (0) to (1); \draw[K] (1) to (noise1); \draw[K] (1) to (noise2); \draw[K] (1) to (noise3); \end{tikzpicture}_{r,\lambda}$ is clearly ill--defined and requires a renormalisation. We will discuss later how to deal with this term.
 
As in the work of Jagannath--Perkowski, we introduce a Cole--Hopf transform $v_r 
\defeq e^{3\begin{tikzpicture}[scale=0.3,baseline=0cm] \node at (0,0) [dot] (0) {}; \node at (0,0.4) [dot] (1) {}; \node at (-0.3,0.8)  [noise] (noise1) {}; \node at (0.3,0.8)  [noise] (noise2) {}; \draw[K] (0) to (1); \draw[K] (1) to (noise1); \draw[K] (1) to (noise2); \end{tikzpicture}_{r,\lambda}} R_r$ to kill the 
borderline ill--defined product $R_r\begin{tikzpicture}[scale=0.3,baseline=0cm]
\node at (0,0) [dot] (0) {};
\node at (0.3,0.6)  [noise] (noise1) {};
\node at (-0.3,0.6)  [noise] (noise2) {};
\draw[K] (0) to (noise1);
\draw[K] (0) to (noise2);
\end{tikzpicture}_r $.
A long and tedious calculation yields the following equation for $v_r$:
\begin{eqnarray*}
&\mcL v_r= 9\vert\nabla \begin{tikzpicture}[scale=0.3,baseline=0cm] \node at (0,0) [dot] (0) {}; \node at (0,0.4) [dot] (1) {}; \node at (-0.3,0.8)  [noise] (noise1) {}; \node at (0.3,0.8)  [noise] (noise2) {}; \draw[K] (0) to (1); \draw[K] (1) to (noise1); \draw[K] (1) to (noise2); \end{tikzpicture}_{r,\lambda}\vert^2v_r  -3v_r\begin{tikzpicture}[scale=0.3,baseline=0cm] \node at (0,0) [dot] (0) {}; \node at (0,0.4) [dot] (1) {}; \node at (-0.3,0.8)  [noise] (noise1) {}; \node at (0.3,0.8)  [noise] (noise2) {}; \draw[K] (0) to (1); \draw[K] (1) to (noise1); \draw[K] (1) to (noise2); \end{tikzpicture}_{r,\lambda} - 6 \nabla( \begin{tikzpicture}[scale=0.3,baseline=0cm] \node at (0,0) [dot] (0) {}; \node at (0,0.4) [dot] (1) {}; \node at (-0.3,0.8)  [noise] (noise1) {}; \node at (0.3,0.8)  [noise] (noise2) {}; \draw[K] (0) to (1); \draw[K] (1) to (noise1); \draw[K] (1) to (noise2); \end{tikzpicture}_{r,\lambda}) \nabla v_r
+
3\lambda e^{3\begin{tikzpicture}[scale=0.3,baseline=0cm] \node at (0,0) [dot] (0) {}; \node at (0,0.4) [dot] (1) {}; \node at (-0.3,0.8)  [noise] (noise1) {}; \node at (0.3,0.8)  [noise] (noise2) {}; \draw[K] (0) to (1); \draw[K] (1) to (noise1); \draw[K] (1) to (noise2); \end{tikzpicture}_{r,\lambda}} 
\begin{tikzpicture}[scale=0.3,baseline=0cm]
\node at (0,0) [dot] (0) {};
\node at (0.3,0.6)  [noise] (noise1) {};
\node at (-0.3,0.6)  [noise] (noise2) {};
\draw[K] (0) to (noise1);
\draw[K] (0) to (noise2);
\end{tikzpicture}_r
 \begin{tikzpicture}[scale=0.3,baseline=0cm] \node at (0,0) [dot] (0) {}; \node at (0,0.5) [dot] (1) {}; \node at (-0.4,1)  [noise] (noise1) {}; \node at (0,1.2)  [noise] (noise2) {}; \node at (0.4,1)  [noise] (noise3) {}; \draw[K] (0) to (1); \draw[K] (1) to (noise1); \draw[K] (1) to (noise2); \draw[K] (1) to (noise3); \end{tikzpicture}_{r,\lambda}\\
 &\qquad\qquad\qquad-3\lambda e^{3\begin{tikzpicture}[scale=0.3,baseline=0cm] \node at (0,0) [dot] (0) {}; \node at (0,0.4) [dot] (1) {}; \node at (-0.3,0.8)  [noise] (noise1) {}; \node at (0.3,0.8)  [noise] (noise2) {}; \draw[K] (0) to (1); \draw[K] (1) to (noise1); \draw[K] (1) to (noise2); \end{tikzpicture}_{r,\lambda}} 
\big(e^{-3\begin{tikzpicture}[scale=0.3,baseline=0cm] \node at (0,0) [dot] (0) {}; \node at (0,0.4) [dot] (1) {}; \node at (-0.3,0.8)  [noise] (noise1) {}; \node at (0.3,0.8)  [noise] (noise2) {}; \draw[K] (0) to (1); \draw[K] (1) to (noise1); \draw[K] (1) to (noise2); \end{tikzpicture}_{r,\lambda}} v_r- \begin{tikzpicture}[scale=0.3,baseline=0cm] \node at (0,0) [dot] (0) {}; \node at (0,0.5) [dot] (1) {}; \node at (-0.4,1)  [noise] (noise1) {}; \node at (0,1.2)  [noise] (noise2) {}; \node at (0.4,1)  [noise] (noise3) {}; \draw[K] (0) to (1); \draw[K] (1) to (noise1); \draw[K] (1) to (noise2); \draw[K] (1) to (noise3); \end{tikzpicture}_{r,\lambda}\big)^2 \begin{tikzpicture}[scale=0.3,baseline=0cm] \node at (0,0)  [dot] (1) {}; \node at (0,0.8)  [noise] (2) {}; \draw[K] (1) to (2); \end{tikzpicture}_{r} 
-\lambda e^{3\begin{tikzpicture}[scale=0.3,baseline=0cm] \node at (0,0) [dot] (0) {}; \node at (0,0.4) [dot] (1) {}; \node at (-0.3,0.8)  [noise] (noise1) {}; \node at (0.3,0.8)  [noise] (noise2) {}; \draw[K] (0) to (1); \draw[K] (1) to (noise1); \draw[K] (1) to (noise2); \end{tikzpicture}_{r,\lambda}} 
\big(e^{-3\begin{tikzpicture}[scale=0.3,baseline=0cm] \node at (0,0) [dot] (0) {}; \node at (0,0.4) [dot] (1) {}; \node at (-0.3,0.8)  [noise] (noise1) {}; \node at (0.3,0.8)  [noise] (noise2) {}; \draw[K] (0) to (1); \draw[K] (1) to (noise1); \draw[K] (1) to (noise2); \end{tikzpicture}_{r,\lambda}} v_r- \begin{tikzpicture}[scale=0.3,baseline=0cm] \node at (0,0) [dot] (0) {}; \node at (0,0.5) [dot] (1) {}; \node at (-0.4,1)  [noise] (noise1) {}; \node at (0,1.2)  [noise] (noise2) {}; \node at (0.4,1)  [noise] (noise3) {}; \draw[K] (0) to (1); \draw[K] (1) to (noise1); \draw[K] (1) to (noise2); \draw[K] (1) to (noise3); \end{tikzpicture}_{r,\lambda}\big)^3.
\end{eqnarray*}

Let us again try to identify the origin of new divergences in this new equation. It may arise from the product $ e^{3\begin{tikzpicture}[scale=0.3,baseline=0cm] \node at (0,0) [dot] (0) {}; \node at (0,0.4) [dot] (1) {}; \node at (-0.3,0.8)  [noise] (noise1) {}; \node at (0.3,0.8)  [noise] (noise2) {}; \draw[K] (0) to (1); \draw[K] (1) to (noise1); \draw[K] (1) to (noise2); \end{tikzpicture}_{r,\lambda}} 
\big(\lambda
\begin{tikzpicture}[scale=0.3,baseline=0cm]
\node at (0,0) [dot] (0) {};
\node at (0.3,0.6)  [noise] (noise1) {};
\node at (-0.3,0.6)  [noise] (noise2) {};
\draw[K] (0) to (noise1);
\draw[K] (0) to (noise2);
\end{tikzpicture}_r
 \begin{tikzpicture}[scale=0.3,baseline=0cm] \node at (0,0) [dot] (0) {}; \node at (0,0.5) [dot] (1) {}; \node at (-0.4,1)  [noise] (noise1) {}; \node at (0,1.2)  [noise] (noise2) {}; \node at (0.4,1)  [noise] (noise3) {}; \draw[K] (0) to (1); \draw[K] (1) to (noise1); \draw[K] (1) to (noise2); \draw[K] (1) to (noise3); \end{tikzpicture}_{r,\lambda}\big)$ where we need to renormalize the quintic term in parenthesis. Once it is renormalized it will be of regularity $-1-$ and we still need to discuss how we can make sense of the product with   $e^{3\begin{tikzpicture}[scale=0.3,baseline=0cm] \node at (0,0) [dot] (0) {}; \node at (0,0.4) [dot] (1) {}; \node at (-0.3,0.8)  [noise] (noise1) {}; \node at (0.3,0.8)  [noise] (noise2) {}; \draw[K] (0) to (1); \draw[K] (1) to (noise1); \draw[K] (1) to (noise2); \end{tikzpicture}_{r,\lambda}} $ which is of regularity $1-$.
As usual, by the paraproduct decomposition, it is the resonant term
$ 
\lambda \begin{tikzpicture}[scale=0.3,baseline=0cm]
\node at (0,0) [dot] (0) {};
\node at (0.3,0.6)  [noise] (noise1) {};
\node at (-0.3,0.6)  [noise] (noise2) {};
\draw[K] (0) to (noise1);
\draw[K] (0) to (noise2);
\end{tikzpicture}_r
\odot
 \begin{tikzpicture}[scale=0.3,baseline=0cm] \node at (0,0) [dot] (0) {}; \node at (0,0.5) [dot] (1) {}; \node at (-0.4,1)  [noise] (noise1) {}; \node at (0,1.2)  [noise] (noise2) {}; \node at (0.4,1)  [noise] (noise3) {}; \draw[K] (0) to (1); \draw[K] (1) to (noise1); \draw[K] (1) to (noise2); \draw[K] (1) to (noise3); \end{tikzpicture}_{r,\lambda}$ which requires a renormalisation which is very similar to what we did for the quintic term $\tau_4$ except for the insertion of the coupling function $\lambda$ at the vertices. Let us indicate what changes must be made on the treatment of $\tau_{41}$ in order to renormalize this term with the coupling function $\lambda$. The divergent subamplitude
$\mcA_r$ now reads
$$\mcA_r\big(t,s,x_*,x_1,x_2,x_a\big)  \defeq [\odot](x_*,x_1,x_2)
\underline\mcL^{-1}
\big((t,x_1),(s,x_a)\big)G^{(2)}_r\big((t,x_2),(s,x_a)\big){\color{red}\lambda(x_2)\lambda(x_a)}$$
where the function $\lambda$ appearing in the subamplitude $\mcA_r$ is taken at two different points $x_2$ and $x_a$. Now if one repeats the proof of Lemma~\ref{LemB}  with the function $\lambda$ inserted at the right places, one ends up with a counterterm of the form
\begin{eqnarray*}
c_r(t_1,x_*)
&=&\mathcal{S}\int_{-\infty}^{t_1} 
\sum_{0\leqslant\vert k-\ell\vert\leqslant 1,i}  P_k^i(x_*,.) \circ \widetilde{{\varphi}}{\color{red} \lambda}\circ\left( G_r^{(2)}(t-s)\circ e^{-(t-s)P}\right)\circ{\color{red} \lambda}\widetilde{{\varphi}} \circ \widetilde{P}_\ell^i(.,x_*)\rmd s
\end{eqnarray*}
where the $P_k^i,\widetilde{P}_\ell^i$ are the Littlewood-Paley-Stein projectors, the functions $\widetilde{{\varphi}}\in C^\infty_c(U)$ are arbitrary test functions such that $\widetilde{\varphi}=1$ near $x_*$ and where we used the explicit definition of $[ \odot ]$ in terms of the Littlewood-Paley-Stein projectors.
Observe that changing the position of $\lambda$ at two places in the above composition of operators inserts two commutator terms and since the commutators lower the pseudodifferential order $[\Psi^{m_1},\Psi^{m_2}]\in \Psi^{m_1+m_2-1}$, we get 
\begin{eqnarray*}
c_r(t,x_*)
&=&
\mathcal{S}\int_{-\infty}^{t} 
\sum_{0\leqslant\vert k-\ell\vert\leqslant 1,i} {\color{red} \lambda^2(x_*)} P_k^i(x_*,.) \circ \widetilde{{\varphi}}\circ\left( G_r^{(2)}(t-s)\circ e^{-(t-s)P}\right)\circ\widetilde{{\varphi}} \circ \widetilde{P}_\ell^i(.,x_*)\rmd s\\
&+&\text{Trace density of some element in } \Psi^{-4}(M)\\
&=&\mathcal{S}\int_{-\infty}^{t} 
\sum_{0\leqslant\vert k-\ell\vert\leqslant 1,i} {\color{red} \lambda^2(x_*)} P_k^i(x_*,.) \circ \widetilde{{\varphi}}\circ\left( G_r^{(2)}(t-s)\circ e^{-(t-s)P}\right)\circ\widetilde{{\varphi}} \circ \widetilde{P}_\ell^i(.,x_*)\rmd s
\end{eqnarray*}
since elements in $\Psi^{-4}(M)$ are trace class. Therefore as we saw in Section \ref{SubsectionQuinticTerm}, we find that the term 
$$\lambda \begin{tikzpicture}[scale=0.3,baseline=0cm]
\node at (0,0) [dot] (0) {};
\node at (0.3,0.6)  [noise] (noise1) {};
\node at (-0.3,0.6)  [noise] (noise2) {};
\draw[K] (0) to (noise1);
\draw[K] (0) to (noise2);
\end{tikzpicture}_r
\odot
 \begin{tikzpicture}[scale=0.3,baseline=0cm] \node at (0,0) [dot] (0) {}; \node at (0,0.5) [dot] (1) {}; \node at (-0.4,1)  [noise] (noise1) {}; \node at (0,1.2)  [noise] (noise2) {}; \node at (0.4,1)  [noise] (noise3) {}; \draw[K] (0) to (1); \draw[K] (1) to (noise1); \draw[K] (1) to (noise2); \draw[K] (1) to (noise3); \end{tikzpicture}_{r,\lambda}-\lambda^2b_r \begin{tikzpicture}[scale=0.3,baseline=0cm] \node at (0,0)  [dot] (1) {}; \node at (0,0.8)  [noise] (2) {}; \draw[K] (1) to (2); \end{tikzpicture}_{r} $$
converges as $r\downarrow 0$ in $\mcC^{-1/2-5\epsilon}([0,T]\times M)$ in $L^2(\Omega)$. There is no problem for multiplying it with 
$$
\exp\big(3\begin{tikzpicture}[scale=0.3,baseline=0cm] \node at (0,0) [dot] (0) {}; \node at (0,0.4) [dot] (1) {}; \node at (-0.3,0.8)  [noise] (noise1) {}; \node at (0.3,0.8)  [noise] (noise2) {}; \draw[K] (0) to (1); \draw[K] (1) to (noise1); \draw[K] (1) to (noise2); \end{tikzpicture}_{r,\lambda}\big)\in\mcC^{1-2\epsilon}([0,T]\times M)
$$ 
and send $r$ to $0$.
For this reason, in the ill-defined product
 $ {e^{3\begin{tikzpicture}[scale=0.3,baseline=0cm] \node at (0,0) [dot] (0) {}; \node at (0,0.4) [dot] (1) {}; \node at (-0.3,0.8)  [noise] (noise1) {}; \node at (0.3,0.8)  [noise] (noise2) {}; \draw[K] (0) to (1); \draw[K] (1) to (noise1); \draw[K] (1) to (noise2); \end{tikzpicture}_{r,\lambda}} }
 {
\left(\lambda
\begin{tikzpicture}[scale=0.3,baseline=0cm]
\node at (0,0) [dot] (0) {};
\node at (0.3,0.6)  [noise] (noise1) {};
\node at (-0.3,0.6)  [noise] (noise2) {};
\draw[K] (0) to (noise1);
\draw[K] (0) to (noise2);
\end{tikzpicture}_r
 \begin{tikzpicture}[scale=0.3,baseline=0cm] \node at (0,0) [dot] (0) {}; \node at (0,0.5) [dot] (1) {}; \node at (-0.4,1)  [noise] (noise1) {}; \node at (0,1.2)  [noise] (noise2) {}; \node at (0.4,1)  [noise] (noise3) {}; \draw[K] (0) to (1); \draw[K] (1) to (noise1); \draw[K] (1) to (noise2); \draw[K] (1) to (noise3); \end{tikzpicture}_{r,\lambda}-\lambda^2b_r \begin{tikzpicture}[scale=0.3,baseline=0cm] \node at (0,0)  [dot] (1) {}; \node at (0,0.8)  [noise] (2) {}; \draw[K] (1) to (2); \end{tikzpicture}_{r}\right)}$
where the term in parenthesis is well-defined at the limit $r\downarrow 0$ but the product of the two terms is ill-defined in the limit $r\downarrow 0$.

\ssk

In our companion paper~\cite[
Theorem 3.13]{BDFTCompanion}, we prove the existence of
$$ e^{3\begin{tikzpicture}[scale=0.3,baseline=0cm] \node at (0,0) [dot] (0) {}; \node at (0,0.4) [dot] (1) {}; \node at (-0.3,0.8)  [noise] (noise1) {}; \node at (0.3,0.8)  [noise] (noise2) {}; \draw[K] (0) to (1); \draw[K] (1) to (noise1); \draw[K] (1) to (noise2); \end{tikzpicture}_{r,\lambda}} 
\left(\lambda
\begin{tikzpicture}[scale=0.3,baseline=0cm]
\node at (0,0) [dot] (0) {};
\node at (0.3,0.6)  [noise] (noise1) {};
\node at (-0.3,0.6)  [noise] (noise2) {};
\draw[K] (0) to (noise1);
\draw[K] (0) to (noise2);
\end{tikzpicture}_r
 \begin{tikzpicture}[scale=0.3,baseline=0cm] \node at (0,0) [dot] (0) {}; \node at (0,0.5) [dot] (1) {}; \node at (-0.4,1)  [noise] (noise1) {}; \node at (0,1.2)  [noise] (noise2) {}; \node at (0.4,1)  [noise] (noise3) {}; \draw[K] (0) to (1); \draw[K] (1) to (noise1); \draw[K] (1) to (noise2); \draw[K] (1) to (noise3); \end{tikzpicture}_{r,\lambda}
 -\lambda^2b_r \begin{tikzpicture}[scale=0.3,baseline=0cm] \node at (0,0)  [dot] (1) {}; \node at (0,0.8)  [noise] (2) {}; \draw[K] (1) to (2); \end{tikzpicture}_{r}-\lambda^2 b_r \begin{tikzpicture}[scale=0.3,baseline=0cm] \node at (0,0) [dot] (0) {}; \node at (0,0.5) [dot] (1) {}; \node at (-0.4,1)  [noise] (noise1) {}; \node at (0,1.2)  [noise] (noise2) {}; \node at (0.4,1)  [noise] (noise3) {}; \draw[K] (0) to (1); \draw[K] (1) to (noise1); \draw[K] (1) to (noise2); \draw[K] (1) to (noise3); \end{tikzpicture}_{r,\lambda} \right)$$
as a random variable valued in $C_T\mcC^{-2\varepsilon}(M)$ when $r\downarrow 0$.
Now we define a new element $\phi_r$ by the equation:
\begin{eqnarray*}
v_r=v_{\mathrm{ref},r}+\phi_r= \underset{=:v_{\mathrm{ref},r}}{\underbrace{ \underline{\mcL}^{-1}\left(3
 e^{3\begin{tikzpicture}[scale=0.3,baseline=0cm] \node at (0,0) [dot] (0) {}; \node at (0,0.4) [dot] (1) {}; \node at (-0.3,0.8)  [noise] (noise1) {}; \node at (0.3,0.8)  [noise] (noise2) {}; \draw[K] (0) to (1); \draw[K] (1) to (noise1); \draw[K] (1) to (noise2); \end{tikzpicture}_{r,\lambda}} 
\left(\lambda
\begin{tikzpicture}[scale=0.3,baseline=0cm]
\node at (0,0) [dot] (0) {};
\node at (0.3,0.6)  [noise] (noise1) {};
\node at (-0.3,0.6)  [noise] (noise2) {};
\draw[K] (0) to (noise1);
\draw[K] (0) to (noise2);
\end{tikzpicture}_r
 \begin{tikzpicture}[scale=0.3,baseline=0cm] \node at (0,0) [dot] (0) {}; \node at (0,0.5) [dot] (1) {}; \node at (-0.4,1)  [noise] (noise1) {}; \node at (0,1.2)  [noise] (noise2) {}; \node at (0.4,1)  [noise] (noise3) {}; \draw[K] (0) to (1); \draw[K] (1) to (noise1); \draw[K] (1) to (noise2); \draw[K] (1) to (noise3); \end{tikzpicture}_{r,\lambda}
 -\lambda^2b_r \left( \begin{tikzpicture}[scale=0.3,baseline=0cm] \node at (0,0)  [dot] (1) {}; \node at (0,0.8)  [noise] (2) {}; \draw[K] (1) to (2); \end{tikzpicture}_{r}+ \begin{tikzpicture}[scale=0.3,baseline=0cm] \node at (0,0) [dot] (0) {}; \node at (0,0.5) [dot] (1) {}; \node at (-0.4,1)  [noise] (noise1) {}; \node at (0,1.2)  [noise] (noise2) {}; \node at (0.4,1)  [noise] (noise3) {}; \draw[K] (0) to (1); \draw[K] (1) to (noise1); \draw[K] (1) to (noise2); \draw[K] (1) to (noise3); \end{tikzpicture}_{r,\lambda} \right) \right)
\right) }}
+\phi_r
\end{eqnarray*} 
where $v_{\mathrm{ref},r}\in \mcC^{1-\varepsilon}([0,T]\times M)$, $\forall \varepsilon >0$
as $r\downarrow 0$.

We rewrite the equation for $v_r$ in terms of both $(v_{\mathrm{ref},r},\phi_r)$, the goal of introducing $v_{\mathrm{ref},r}$ is that it is well-defined thanks to the double renormalisations we just performed and this equation makes appear new divergent terms
\begin{align*}
&\mcL \phi_r   \\ 
&= 9\vert\nabla \begin{tikzpicture}[scale=0.3,baseline=0cm] \node at (0,0) [dot] (0) {}; \node at (0,0.4) [dot] (1) {}; \node at (-0.3,0.8)  [noise] (noise1) {}; \node at (0.3,0.8)  [noise] (noise2) {}; \draw[K] (0) to (1); \draw[K] (1) to (noise1); \draw[K] (1) to (noise2); \end{tikzpicture}_{r,\lambda}\vert^2 (v_{\mathrm{ref},r}+\phi_r)  
-3(v_{\mathrm{ref},r}+\phi_r)\begin{tikzpicture}[scale=0.3,baseline=0cm] \node at (0,0) [dot] (0) {}; \node at (0,0.4) [dot] (1) {}; \node at (-0.3,0.8)  [noise] (noise1) {}; \node at (0.3,0.8)  [noise] (noise2) {}; \draw[K] (0) to (1); \draw[K] (1) to (noise1); \draw[K] (1) to (noise2); \end{tikzpicture}_{r,\lambda} - 6 \nabla( \begin{tikzpicture}[scale=0.3,baseline=0cm] \node at (0,0) [dot] (0) {}; \node at (0,0.4) [dot] (1) {}; \node at (-0.3,0.8)  [noise] (noise1) {}; \node at (0.3,0.8)  [noise] (noise2) {}; \draw[K] (0) to (1); \draw[K] (1) to (noise1); \draw[K] (1) to (noise2); \end{tikzpicture}_{r,\lambda}) \nabla (v_{\mathrm{ref},r}+\phi_r)
+\textcolor{red}{3\lambda^2b_re^{3\begin{tikzpicture}[scale=0.3,baseline=0cm] \node at (0,0) [dot] (0) {}; \node at (0,0.4) [dot] (1) {}; \node at (-0.3,0.8)  [noise] (noise1) {}; \node at (0.3,0.8)  [noise] (noise2) {}; \draw[K] (0) to (1); \draw[K] (1) to (noise1); \draw[K] (1) to (noise2); \end{tikzpicture}_{r,\lambda}}  \big( \begin{tikzpicture}[scale=0.3,baseline=0cm] \node at (0,0)  [dot] (1) {}; \node at (0,0.8)  [noise] (2) {}; \draw[K] (1) to (2); \end{tikzpicture}_{r}+ \begin{tikzpicture}[scale=0.3,baseline=0cm] \node at (0,0) [dot] (0) {}; \node at (0,0.5) [dot] (1) {}; \node at (-0.4,1)  [noise] (noise1) {}; \node at (0,1.2)  [noise] (noise2) {}; \node at (0.4,1)  [noise] (noise3) {}; \draw[K] (0) to (1); \draw[K] (1) to (noise1); \draw[K] (1) to (noise2); \draw[K] (1) to (noise3); \end{tikzpicture}_{r,\lambda} \big) }
\\
 &\quad-3\lambda e^{3\begin{tikzpicture}[scale=0.3,baseline=0cm] \node at (0,0) [dot] (0) {}; \node at (0,0.4) [dot] (1) {}; \node at (-0.3,0.8)  [noise] (noise1) {}; \node at (0.3,0.8)  [noise] (noise2) {}; \draw[K] (0) to (1); \draw[K] (1) to (noise1); \draw[K] (1) to (noise2); \end{tikzpicture}_{r,\lambda}} 
\big(e^{-3\begin{tikzpicture}[scale=0.3,baseline=0cm] \node at (0,0) [dot] (0) {}; \node at (0,0.4) [dot] (1) {}; \node at (-0.3,0.8)  [noise] (noise1) {}; \node at (0.3,0.8)  [noise] (noise2) {}; \draw[K] (0) to (1); \draw[K] (1) to (noise1); \draw[K] (1) to (noise2); \end{tikzpicture}_{r,\lambda}} (v_{\mathrm{ref},r}+\phi_r)- \begin{tikzpicture}[scale=0.3,baseline=0cm] \node at (0,0) [dot] (0) {}; \node at (0,0.5) [dot] (1) {}; \node at (-0.4,1)  [noise] (noise1) {}; \node at (0,1.2)  [noise] (noise2) {}; \node at (0.4,1)  [noise] (noise3) {}; \draw[K] (0) to (1); \draw[K] (1) to (noise1); \draw[K] (1) to (noise2); \draw[K] (1) to (noise3); \end{tikzpicture}_{r,\lambda}\big)^2 \begin{tikzpicture}[scale=0.3,baseline=0cm] \node at (0,0)  [dot] (1) {}; \node at (0,0.8)  [noise] (2) {}; \draw[K] (1) to (2); \end{tikzpicture}_{r} 
-\lambda e^{3\begin{tikzpicture}[scale=0.3,baseline=0cm] \node at (0,0) [dot] (0) {}; \node at (0,0.4) [dot] (1) {}; \node at (-0.3,0.8)  [noise] (noise1) {}; \node at (0.3,0.8)  [noise] (noise2) {}; \draw[K] (0) to (1); \draw[K] (1) to (noise1); \draw[K] (1) to (noise2); \end{tikzpicture}_{r,\lambda}} 
\big(e^{-3\begin{tikzpicture}[scale=0.3,baseline=0cm] \node at (0,0) [dot] (0) {}; \node at (0,0.4) [dot] (1) {}; \node at (-0.3,0.8)  [noise] (noise1) {}; \node at (0.3,0.8)  [noise] (noise2) {}; \draw[K] (0) to (1); \draw[K] (1) to (noise1); \draw[K] (1) to (noise2); \end{tikzpicture}_{r,\lambda}} (v_{\mathrm{ref},r}+\phi_r)- \begin{tikzpicture}[scale=0.3,baseline=0cm] \node at (0,0) [dot] (0) {}; \node at (0,0.5) [dot] (1) {}; \node at (-0.4,1)  [noise] (noise1) {}; \node at (0,1.2)  [noise] (noise2) {}; \node at (0.4,1)  [noise] (noise3) {}; \draw[K] (0) to (1); \draw[K] (1) to (noise1); \draw[K] (1) to (noise2); \draw[K] (1) to (noise3); \end{tikzpicture}_{r,\lambda}\big)^3
\,.
\end{align*}
The first divergent term in purple on the r.h.s comes from the counterterm in the definition of $v_{\mathrm{ref},r}$.
The next terms which are ill-posed read $9\vert\nabla \begin{tikzpicture}[scale=0.3,baseline=0cm] \node at (0,0) [dot] (0) {}; \node at (0,0.4) [dot] (1) {}; \node at (-0.3,0.8)  [noise] (noise1) {}; \node at (0.3,0.8)  [noise] (noise2) {}; \draw[K] (0) to (1); \draw[K] (1) to (noise1); \draw[K] (1) to (noise2); \end{tikzpicture}_{r,\lambda}\vert^2   $ and
$ \nabla( \begin{tikzpicture}[scale=0.3,baseline=0cm] \node at (0,0) [dot] (0) {}; \node at (0,0.4) [dot] (1) {}; \node at (-0.3,0.8)  [noise] (noise1) {}; \node at (0.3,0.8)  [noise] (noise2) {}; \draw[K] (0) to (1); \draw[K] (1) to (noise1); \draw[K] (1) to (noise2); \end{tikzpicture}_{r,\lambda})\cdot \nabla v_{\mathrm{ref},r}$
since $v_{\mathrm{ref},r}$ is a priori in $C_T\mcC^{1-2\varepsilon}(M)$ and 
$\nabla( \begin{tikzpicture}[scale=0.3,baseline=0cm] \node at (0,0) [dot] (0) {}; \node at (0,0.4) [dot] (1) {}; \node at (-0.3,0.8)  [noise] (noise1) {}; \node at (0.3,0.8)  [noise] (noise2) {}; \draw[K] (0) to (1); \draw[K] (1) to (noise1); \draw[K] (1) to (noise2); \end{tikzpicture}_{r,\lambda}) \in C_T\mcC^{-2\varepsilon}(M) $ hence the scalar product will be ill--defined.

\ssk

In the companion work~\cite[Theorem 3.13]{BDFTCompanion}, we also show how to extract the singular term from this scalar product.
More precisely,
\begin{eqnarray*}
 \nabla( \begin{tikzpicture}[scale=0.3,baseline=0cm] \node at (0,0) [dot] (0) {}; \node at (0,0.4) [dot] (1) {}; \node at (-0.3,0.8)  [noise] (noise1) {}; \node at (0.3,0.8)  [noise] (noise2) {}; \draw[K] (0) to (1); \draw[K] (1) to (noise1); \draw[K] (1) to (noise2); \end{tikzpicture}_{r,\lambda})\cdot \nabla v_{\mathrm{ref},r}=
\vert \nabla( \begin{tikzpicture}[scale=0.3,baseline=0cm] \node at (0,0) [dot] (0) {}; \node at (0,0.4) [dot] (1) {}; \node at (-0.3,0.8)  [noise] (noise1) {}; \node at (0.3,0.8)  [noise] (noise2) {}; \draw[K] (0) to (1); \draw[K] (1) to (noise1); \draw[K] (1) to (noise2); \end{tikzpicture}_{r,\lambda}) \vert^2 3
 e^{3\begin{tikzpicture}[scale=0.3,baseline=0cm] \node at (0,0) [dot] (0) {}; \node at (0,0.4) [dot] (1) {}; \node at (-0.3,0.8)  [noise] (noise1) {}; \node at (0.3,0.8)  [noise] (noise2) {}; \draw[K] (0) to (1); \draw[K] (1) to (noise1); \draw[K] (1) to (noise2); \end{tikzpicture}_{r,\lambda}} 
 \begin{tikzpicture}[scale=0.3,baseline=0cm] \node at (0,0) [dot] (0) {}; \node at (0,0.5) [dot] (1) {}; \node at (-0.4,1)  [noise] (noise1) {}; \node at (0,1.2)  [noise] (noise2) {}; \node at (0.4,1)  [noise] (noise3) {}; \draw[K] (0) to (1); \draw[K] (1) to (noise1); \draw[K] (1) to (noise2); \draw[K] (1) to (noise3); \end{tikzpicture}_{r,\lambda} +\text{well-defined}. 
\end{eqnarray*}
We isolated the singular term at the limit $r\downarrow 0$ as $\vert \nabla( \begin{tikzpicture}[scale=0.3,baseline=0cm] \node at (0,0) [dot] (0) {}; \node at (0,0.4) [dot] (1) {}; \node at (-0.3,0.8)  [noise] (noise1) {}; \node at (0.3,0.8)  [noise] (noise2) {}; \draw[K] (0) to (1); \draw[K] (1) to (noise1); \draw[K] (1) to (noise2); \end{tikzpicture}_{r,\lambda}) \vert^2 $. 
It remains to explain how to extract the counterterm of $\vert \nabla( \begin{tikzpicture}[scale=0.3,baseline=0cm] \node at (0,0) [dot] (0) {}; \node at (0,0.4) [dot] (1) {}; \node at (-0.3,0.8)  [noise] (noise1) {}; \node at (0.3,0.8)  [noise] (noise2) {}; \draw[K] (0) to (1); \draw[K] (1) to (noise1); \draw[K] (1) to (noise2); \end{tikzpicture}_{r,\lambda}) \vert^2 $ while taking into account the presence of the coupling function $\lambda$.
This was done at the end of Section~\ref{subsec:tau3}. Similarly,
the singular part of $\vert \nabla( \begin{tikzpicture}[scale=0.3,baseline=0cm] \node at (0,0) [dot] (0) {}; \node at (0,0.4) [dot] (1) {}; \node at (-0.3,0.8)  [noise] (noise1) {}; \node at (0.3,0.8)  [noise] (noise2) {}; \draw[K] (0) to (1); \draw[K] (1) to (noise1); \draw[K] (1) to (noise2); \end{tikzpicture}_{r,\lambda}) \vert^2 $ is the same as the singular part of $( \Delta\begin{tikzpicture}[scale=0.3,baseline=0cm] \node at (0,0) [dot] (0) {}; \node at (0,0.4) [dot] (1) {}; \node at (-0.3,0.8)  [noise] (noise1) {}; \node at (0.3,0.8)  [noise] (noise2) {}; \draw[K] (0) to (1); \draw[K] (1) to (noise1); \draw[K] (1) to (noise2); \end{tikzpicture}_{r,\lambda})\begin{tikzpicture}[scale=0.3,baseline=0cm] \node at (0,0) [dot] (0) {}; \node at (0,0.4) [dot] (1) {}; \node at (-0.3,0.8)  [noise] (noise1) {}; \node at (0.3,0.8)  [noise] (noise2) {}; \draw[K] (0) to (1); \draw[K] (1) to (noise1); \draw[K] (1) to (noise2); \end{tikzpicture}_{r,\lambda} $
whose divergent part now reads
\begin{eqnarray*}
\mathcal{S}\left( \int_{-\infty}^t tr_{L^2}\left(\lambda \circ e^{-s'_2P}\circ \lambda \circ G_r^{(2)}(s'_2) \rmd s'_2 \right) \right)
\end{eqnarray*}
where the $\lambda$ in the above expression is viewed as multiplication operator. Note that the difference  
$$ \int_{-\infty}^t tr_{L^2}\left((\lambda \circ e^{-s'_2P}\circ \lambda-\lambda^2\circ e^{-s'_2P}) \circ G_r^{(2)}(s'_2) \rmd s'_2 \right) $$ 
is regular when $r\downarrow 0$ simply by the fact that $\lambda(x)\lambda(y)-\lambda^2(x)$ vanishes near the diagonal $x=y$ and by simple power counting argument as we did when we studied amplitudes. Finally, the singular
part 
\begin{eqnarray*}
\mathcal{S}\left( \int_{-\infty}^t tr_{L^2}\left(\lambda^2 \circ e^{-s'_2P}\circ G_r^{(2)}(s'_2) \rmd s'_2 \right) \right)=\lambda^2(x)\frac{b_r}{3}
\end{eqnarray*}
as we did already calculate for quartic graphs.
This means that 
$$\nabla( \begin{tikzpicture}[scale=0.3,baseline=0cm] \node at (0,0) [dot] (0) {}; \node at (0,0.4) [dot] (1) {}; \node at (-0.3,0.8)  [noise] (noise1) {}; \node at (0.3,0.8)  [noise] (noise2) {}; \draw[K] (0) to (1); \draw[K] (1) to (noise1); \draw[K] (1) to (noise2); \end{tikzpicture}_{r,\lambda})\cdot \nabla v_{\mathrm{ref},r}-
\lambda^2(x)b_r
 e^{3\begin{tikzpicture}[scale=0.3,baseline=0cm] \node at (0,0) [dot] (0) {}; \node at (0,0.4) [dot] (1) {}; \node at (-0.3,0.8)  [noise] (noise1) {}; \node at (0.3,0.8)  [noise] (noise2) {}; \draw[K] (0) to (1); \draw[K] (1) to (noise1); \draw[K] (1) to (noise2); \end{tikzpicture}_{r,\lambda}} 
 \begin{tikzpicture}[scale=0.3,baseline=0cm] \node at (0,0) [dot] (0) {}; \node at (0,0.5) [dot] (1) {}; \node at (-0.4,1)  [noise] (noise1) {}; \node at (0,1.2)  [noise] (noise2) {}; \node at (0.4,1)  [noise] (noise3) {}; \draw[K] (0) to (1); \draw[K] (1) to (noise1); \draw[K] (1) to (noise2); \draw[K] (1) to (noise3); \end{tikzpicture}_{r,\lambda} 
$$
has a well-defined limit when $r\downarrow 0$.

In the same way the limit $9\vert\nabla \begin{tikzpicture}[scale=0.3,baseline=0cm] \node at (0,0) [dot] (0) {}; \node at (0,0.4) [dot] (1) {}; \node at (-0.3,0.8)  [noise] (noise1) {}; \node at (0.3,0.8)  [noise] (noise2) {}; \draw[K] (0) to (1); \draw[K] (1) to (noise1); \draw[K] (1) to (noise2); \end{tikzpicture}_{r,\lambda}\vert^2-9\lambda^2\frac{b_r}{3}$ has a well-defined limit. Let us add and subtract the divergent terms in the equation for $\phi_r$ to single out the divergences which are again represented in \textcolor{red}{red}, this now reads
\begin{align*}
\mcL \phi_r&=\left( 9\vert\nabla \begin{tikzpicture}[scale=0.3,baseline=0cm] \node at (0,0) [dot] (0) {}; \node at (0,0.4) [dot] (1) {}; \node at (-0.3,0.8)  [noise] (noise1) {}; \node at (0.3,0.8)  [noise] (noise2) {}; \draw[K] (0) to (1); \draw[K] (1) to (noise1); \draw[K] (1) to (noise2); \end{tikzpicture}_{r,\lambda}\vert^2 - 3\lambda^2b_r \right) (v_{\mathrm{ref},r}+\phi_r)  
-3(v_{\mathrm{ref},r}+\phi_r)\begin{tikzpicture}[scale=0.3,baseline=0cm] \node at (0,0) [dot] (0) {}; \node at (0,0.4) [dot] (1) {}; \node at (-0.3,0.8)  [noise] (noise1) {}; \node at (0.3,0.8)  [noise] (noise2) {}; \draw[K] (0) to (1); \draw[K] (1) to (noise1); \draw[K] (1) to (noise2); \end{tikzpicture}_{r,\lambda} - 6 \nabla( \begin{tikzpicture}[scale=0.3,baseline=0cm] \node at (0,0) [dot] (0) {}; \node at (0,0.4) [dot] (1) {}; \node at (-0.3,0.8)  [noise] (noise1) {}; \node at (0.3,0.8)  [noise] (noise2) {}; \draw[K] (0) to (1); \draw[K] (1) to (noise1); \draw[K] (1) to (noise2); \end{tikzpicture}_{r,\lambda}) \nabla (v_{\mathrm{ref},r}+\phi_r)\\
&\quad+6
\lambda^2(x)b_r
 e^{3\begin{tikzpicture}[scale=0.3,baseline=0cm] \node at (0,0) [dot] (0) {}; \node at (0,0.4) [dot] (1) {}; \node at (-0.3,0.8)  [noise] (noise1) {}; \node at (0.3,0.8)  [noise] (noise2) {}; \draw[K] (0) to (1); \draw[K] (1) to (noise1); \draw[K] (1) to (noise2); \end{tikzpicture}_{r,\lambda}} \begin{tikzpicture}[scale=0.3,baseline=0cm] \node at (0,0) [dot] (0) {}; \node at (0,0.5) [dot] (1) {}; \node at (-0.4,1)  [noise] (noise1) {}; \node at (0,1.2)  [noise] (noise2) {}; \node at (0.4,1)  [noise] (noise3) {}; \draw[K] (0) to (1); \draw[K] (1) to (noise1); \draw[K] (1) to (noise2); \draw[K] (1) to (noise3); \end{tikzpicture}_{r,\lambda}
+
 \textcolor{red}{3\lambda^2b_r (v_{\mathrm{ref},r}+\phi_r) }
-
\textcolor{red}{6
\lambda^2b_r
 e^{3\begin{tikzpicture}[scale=0.3,baseline=0cm] \node at (0,0) [dot] (0) {}; \node at (0,0.4) [dot] (1) {}; \node at (-0.3,0.8)  [noise] (noise1) {}; \node at (0.3,0.8)  [noise] (noise2) {}; \draw[K] (0) to (1); \draw[K] (1) to (noise1); \draw[K] (1) to (noise2); \end{tikzpicture}_{r,\lambda}} 
 \begin{tikzpicture}[scale=0.3,baseline=0cm] \node at (0,0) [dot] (0) {}; \node at (0,0.5) [dot] (1) {}; \node at (-0.4,1)  [noise] (noise1) {}; \node at (0,1.2)  [noise] (noise2) {}; \node at (0.4,1)  [noise] (noise3) {}; \draw[K] (0) to (1); \draw[K] (1) to (noise1); \draw[K] (1) to (noise2); \draw[K] (1) to (noise3); \end{tikzpicture}_{r,\lambda}
}
+\textcolor{red}{3\lambda^2b_re^{3\begin{tikzpicture}[scale=0.3,baseline=0cm] \node at (0,0) [dot] (0) {}; \node at (0,0.4) [dot] (1) {}; \node at (-0.3,0.8)  [noise] (noise1) {}; \node at (0.3,0.8)  [noise] (noise2) {}; \draw[K] (0) to (1); \draw[K] (1) to (noise1); \draw[K] (1) to (noise2); \end{tikzpicture}_{r,\lambda}}  \big( \begin{tikzpicture}[scale=0.3,baseline=0cm] \node at (0,0)  [dot] (1) {}; \node at (0,0.8)  [noise] (2) {}; \draw[K] (1) to (2); \end{tikzpicture}_{r}+ \begin{tikzpicture}[scale=0.3,baseline=0cm] \node at (0,0) [dot] (0) {}; \node at (0,0.5) [dot] (1) {}; \node at (-0.4,1)  [noise] (noise1) {}; \node at (0,1.2)  [noise] (noise2) {}; \node at (0.4,1)  [noise] (noise3) {}; \draw[K] (0) to (1); \draw[K] (1) to (noise1); \draw[K] (1) to (noise2); \draw[K] (1) to (noise3); \end{tikzpicture}_{r,\lambda} \big) }\\
 &\quad-3\lambda e^{3\begin{tikzpicture}[scale=0.3,baseline=0cm] \node at (0,0) [dot] (0) {}; \node at (0,0.4) [dot] (1) {}; \node at (-0.3,0.8)  [noise] (noise1) {}; \node at (0.3,0.8)  [noise] (noise2) {}; \draw[K] (0) to (1); \draw[K] (1) to (noise1); \draw[K] (1) to (noise2); \end{tikzpicture}_{r,\lambda}} 
\big(e^{-3\begin{tikzpicture}[scale=0.3,baseline=0cm] \node at (0,0) [dot] (0) {}; \node at (0,0.4) [dot] (1) {}; \node at (-0.3,0.8)  [noise] (noise1) {}; \node at (0.3,0.8)  [noise] (noise2) {}; \draw[K] (0) to (1); \draw[K] (1) to (noise1); \draw[K] (1) to (noise2); \end{tikzpicture}_{r,\lambda}} (v_{\mathrm{ref},r}+\phi_r)- \begin{tikzpicture}[scale=0.3,baseline=0cm] \node at (0,0) [dot] (0) {}; \node at (0,0.5) [dot] (1) {}; \node at (-0.4,1)  [noise] (noise1) {}; \node at (0,1.2)  [noise] (noise2) {}; \node at (0.4,1)  [noise] (noise3) {}; \draw[K] (0) to (1); \draw[K] (1) to (noise1); \draw[K] (1) to (noise2); \draw[K] (1) to (noise3); \end{tikzpicture}_{r,\lambda}\big)^2 \begin{tikzpicture}[scale=0.3,baseline=0cm] \node at (0,0)  [dot] (1) {}; \node at (0,0.8)  [noise] (2) {}; \draw[K] (1) to (2); \end{tikzpicture}_{r} 
-\lambda e^{3\begin{tikzpicture}[scale=0.3,baseline=0cm] \node at (0,0) [dot] (0) {}; \node at (0,0.4) [dot] (1) {}; \node at (-0.3,0.8)  [noise] (noise1) {}; \node at (0.3,0.8)  [noise] (noise2) {}; \draw[K] (0) to (1); \draw[K] (1) to (noise1); \draw[K] (1) to (noise2); \end{tikzpicture}_{r,\lambda}} 
\big(e^{-3\begin{tikzpicture}[scale=0.3,baseline=0cm] \node at (0,0) [dot] (0) {}; \node at (0,0.4) [dot] (1) {}; \node at (-0.3,0.8)  [noise] (noise1) {}; \node at (0.3,0.8)  [noise] (noise2) {}; \draw[K] (0) to (1); \draw[K] (1) to (noise1); \draw[K] (1) to (noise2); \end{tikzpicture}_{r,\lambda}} (v_{\mathrm{ref},r}+\phi_r)- \begin{tikzpicture}[scale=0.3,baseline=0cm] \node at (0,0) [dot] (0) {}; \node at (0,0.5) [dot] (1) {}; \node at (-0.4,1)  [noise] (noise1) {}; \node at (0,1.2)  [noise] (noise2) {}; \node at (0.4,1)  [noise] (noise3) {}; \draw[K] (0) to (1); \draw[K] (1) to (noise1); \draw[K] (1) to (noise2); \draw[K] (1) to (noise3); \end{tikzpicture}_{r,\lambda}\big)^3.
\end{align*}
We thus see that $u_r$ needs to be the solution to the renormalized equation
\begin{eqnarray}\label{EqSPDELambda}
\mcL u_r=\xi_r-\lambda u_r^3+{\color{red}\left( 3\lambda    a_r-3\lambda^2b_r\right) u_r}\,,
\end{eqnarray}
which has a well-defined solution in the limit $r\downarrow 0$. This solution enjoys all the properties we proved in the case where $\lambda$ is constant. Indeed, as in the case $\lambda=1$, $v_r$ is a solution to an equation of the form
$$
\mcL v_r = b_{r,\lambda}\nabla v_r - a_{r,\lambda} v_r^3 + Z_{2,r,\lambda} v_r^2 + Z_{1,r,\lambda} v_r + Z_{0,r,\lambda}
$$
where $b_{r,\lambda}\in C_T\mcC^{-\epsilon}(M), a_{r,\lambda}\in C_T\mcC^{1-2\epsilon}(M)$ and $Z_{i,r,\lambda}\in C_T\mcC^{-1/2-\epsilon}(M)$ all converge in their spaces as $r>0$ goes to $0$ in $L^2(\Omega)$, for all $\epsilon>0$. This analysis of Equation \eqref{EqSPDELambda} puts us in a position to go in the present setting over all the different steps that we have done above to construct the $\Phi^4_3$ measure when $\lambda=1$; it provides a construction of the $\Phi^4_3$ measure in a setting where the coupling constant is space dependent.

\bigskip

\appendix

\section{Littlewood-Paley-Stein projectors}
\label{SectionLPProjectors}

There are several ways of defining some Littlewood-Paley type projectors on function spaces over a manifold -- see \cite{KR, BB1, BB2, PoyferreGuillarmou, Mouzard} for a sample. We choose here an intermediate road and use the classical Littlewood-Paley projectors over $\bbR^d$ to define a number of operators on functions spaces over $M$ using local charts. This allows to import at low cost some known regularity properties of the corresponding objects from the flat to the curved setting. We denote as usual by $B^\gamma_{p,q}(M)$ the Besov spaces over $M$ and by $C^\gamma$(M) the Besov-H\"older space $B^{\gamma}_{\infty,\infty}(M)$, with associated norm denoted by $\Vert\cdot\Vert_{C^\gamma}$.

\ssk

Let then denote by 
$$
a'\prec b' \defeq \sum_{-1\leqslant j<k-1} (\Delta_j a')(\Delta_k b')
$$
the paraproduct of some distributions $a'$ and $b'$ on $\bbR^d$, and write
$$
a' \odot b' \defeq \sum_{\vert j-k\vert \leqslant 1} (\Delta_j a')(\Delta_k b')
$$
for the resonance of $a'$ and $b'$ whenever the latter is defined. Let $(U_i,\kappa_i)_i$ denote a finite open cover of $M$ by some charts, with $\kappa_i$ a smooth diffeomorphism between $U_i\subset M$ and $\kappa_i(U_i)\subset  \bbR^d$. Let $({\varphi_i})_i$ be a partition of unity subordinated to 
$(U_i)_i$, so $\sum_i {\varphi_i}=1$, with ${\varphi_i}\in C^\infty_c(U_i)$.
 Choose also for every index $i$ a function $\widetilde{{\varphi}}_i\in  C_c^\infty(U_i)$ such that $\widetilde{{\varphi}}_i$ equals $1$ on the support of ${\varphi_i}$ and some function $\psi_i\in C^\infty_c(\kappa_i(U_i))$ which equals $1$ on the support of $\kappa_{i*}(\widetilde{{\varphi}}_i)$. Given some smooth functions $a,b$ on $M$ we have the decomposition

\makebox[\textwidth][c]{
\begin{minipage}{\dimexpr\textwidth+10cm}
\begin{align*}
ab &= \sum_{i\in I} (a{\varphi_i}) (b\widetilde{{\varphi}}_i) = \sum_{i\in I} \kappa_i^*\big[\kappa_{i*}(a{\varphi_i})\big]   \kappa_i^*\big[\kappa_{i*}(b\widetilde{{\varphi}}_i)\big]   \\
&= \sum_{i\in I} \kappa_i^*\big[(\kappa_{i*}(a{\varphi_i})  (\kappa_{i*}(b\widetilde{{\varphi}}_i)\big] = \sum_{i\in I} \kappa_i^*\big[ \psi_i \kappa_{i*}(a{\varphi_i})  \kappa_{i*}(b\widetilde{{\varphi}}_i)\big]  \\
     &= \sum_{i\in I} \kappa_i^*\big[ \psi_i\big( \kappa_{i*}(a{\varphi_i}) \prec \kappa_{i*}(b\widetilde{{\varphi}}_i)\big)\big] + \sum_{i\in I} \kappa_i^*\big[ \psi_i\big( \kappa_{i*}(a{\varphi_i}) \odot \kappa_{i*}(b\widetilde{{\varphi}}_i)\big)\big]   \\
     &\hspace{5.3cm}+ \sum_{i\in I} \kappa_i^*\big[\psi_i\big( \kappa_{i*}(a{\varphi_i}) \succ \kappa_{i*}(b\widetilde{{\varphi}}_i)\big)\big].
\end{align*}
\end{minipage}
}
\esp

Actually for arbitrary ${\varphi_i},\widetilde{{\varphi}}_i\in C^\infty_c(U_i)^2$ such that $\widetilde{{\varphi}}_i=1$ on the support of ${\varphi_i}$,  
we set the generalized Littlewood-Paley-Stein projectors 
\begin{align}\label{eq:defLPblocks}
P_{k}^i(a) &\defeq \kappa_i^*\big[\psi_i\Delta_k\big( \kappa_{i*} ({\varphi_i} a)\big)\big]\,,\;\text{and}\;
\widetilde{P}_{k}^i(a) \defeq \kappa_i^*\big[\widetilde\psi_i\Delta_k\big( \kappa_{i*} (\widetilde{\varphi_i} a)\big)\big]\,,
\end{align}
where $\widetilde{\psi}_i\in C^\infty_c(\kappa_i(U_i))$ equals $1$ on the support of $\psi_i$.
We do not necessarily require that $\sum_{i\in I}\widetilde{\varphi_i}=1$.

 On the manifold $M$, recall $i\in I$ denotes a chart index,
we define generalized chart localized operations as:
$$
a\prec_i b \defeq  \sum_{-1\leqslant j<k-1} P^i_j a \widetilde{P}_k^i b\,,\;
a\succ_i b \defeq \sum_{i\in I} \sum_{-1\leqslant k<j-1}  P_j^i a\widetilde{P}^i_k b\,,\;\text{and}\;
a\odot_i b \defeq \sum_{i\in I} \sum_{\vert j-k\vert\leqslant 1} P^i_j a \widetilde{P}_k^i b\,.
$$
In particular when $\sum_{i\in I}{\varphi_i}=1$, the above operations decompose the product $ab$ on $M$ as:
$$
ab = \sum_{i\in I} \left(a\prec_i b + a\odot_i b + a\succ_i b\right).
$$
Note the important fact that the definition of the resonant product and paraproducts are asymmetrical, therefore they are noncommutative meaning that $a\prec b\neq b\succ a$, however all the regularity properties are similar as in the flat case.
We collect in the next two statements some regularity properties of these operators and refer the reader to our companion work~\cite[Proposition 2.6]{BDFTCompanion} for their proofs.

\ssk

\begin{prop} \label{PropContinuityParaproductResonance}
One has the following continuity estimates. For every chart index $i$,
\begin{itemize}
	\item[--] For $p,p_1,p_2, q, q_1,q_2$ in $[1,+\infty]$ with $\frac{1}{p_1} + \frac{1}{p_2} = \frac{1}{p}$ and $\frac{1}{q_1} + \frac{1}{q_2} = \frac{1}{q}$
	\begin{enumerate}
		\item For $\gamma_2\in\bbR$
		\begin{equation*} \begin{split}
		\Vert a \prec_i b \Vert_{B^{\gamma_2}_{p,q_2}} \lesssim \Vert a\Vert_{L^{p_1}} \Vert b\Vert_{B^{\gamma_2}_{p_2,q_2}},
		\end{split} \end{equation*}
		
		\item For $\gamma_1<0$ and $\gamma_2\in\bbR$
		\begin{equation*} \begin{split}
		\Vert a \prec_i b \Vert_{B^{\gamma_1+\gamma_2}_{p,q}} \lesssim \Vert a\Vert_{B^{\gamma_1}_{p_1,q_1}} \Vert b\Vert_{B^{\gamma_2}_{p_2,q_2}}
		\end{split} \end{equation*}
		
		\item For any $\gamma_1,\gamma_2\in\bbR$ with $\gamma_1+\gamma_2>0$ one has
		$$
		\Vert a\odot_i b\Vert_{B^{\gamma_1+\gamma_2}_{p,q}} \lesssim \Vert a\Vert_{B^{\gamma_1}_{p_1,q_1}} \Vert b\Vert_{B^{\gamma_2}_{p_2,q_2}}.
		$$
	\end{enumerate}

\end{itemize}
\end{prop}

\ssk

We recall from Lemma 7.2 of Mourrat \& Weber's work \cite{MourratWeber} the following comparison test that we used in our proof of Theorem \ref{ThmLpComingDown}.

\ssk

\begin{prop} \label{PropComparisonTest}
Let a continuous function $F : [0,T]\rightarrow[0,+\infty)$ that satisfies the inequality
\begin{equation} \label{EqMourratWeberInequality}
\int_s^t F(s_1)^\lambda \, \rmd s_1 \leqslant c \, (F(s)+1)
\end{equation}
for all $0\leqslant s\leqslant t\leqslant T$, for some exponent $\lambda>1$ and some positive constant $c$. Then there is a sequence of times $t_0=0<t_1<\cdots<t_N=T$ such that one has
$$
F(t_n) \leqslant 1 + 2^{\frac{\lambda}{\lambda-1}} \, \Big(\frac{c}{1-2^{-(\lambda-1)}}\Big)^{\frac{1}{\lambda-1}} \, t_{n+1}^{-\frac{1}{\lambda-1}},
$$
for all $0\leqslant n\leqslant N-1$.
\end{prop}

\ssk

\begin{Dem}
We include a proof following closely Mourrat-Weber's comparison test in a slightly different setting compared to them since we have $F(s)+1$ rather than $F(s)$ on the right hand side of the inequality \eqref{EqMourratWeberInequality}. We first define $t_0=0$, then given some time $t_n$, consider $t_{n+1}^*= t_n+c2^\lambda (1+F(t_n))^{1-\lambda}$, if $t_{n+1}^*\geqslant T$ we stop the algorithm and set $N=n+1$, $t_{n+1}=T$ and verify that the conclusion of the statement holds. Otherwise, choosing $t_{n+1}$ such that $F(t_{n+1})=\inf_{t_n <s< t_{n+1}^*} F(s)$ yields a bound of the form $F(t_{n+1})\leqslant \frac{1+F(t_n)}{2} $. By iteration, this yields a bound of the form $F(t_{n+1})\leqslant \frac{F(t_0)-1}{2^{n+1}}+1$. Note that for $n$ large enough, since $\lambda>1$
$$
t_{n+1}^*-t_n=c2^\lambda (1+F(t_n))^{1-\lambda}\geqslant c2^\lambda  \Big(2+\frac{F(t_0)-1}{2^{n}}\Big)^{1-\lambda}\geqslant c2^{\lambda} (2+ 1/3)^{1-\lambda}\geqslant 2c,
$$ 
hence the algorithm must terminate for $n$ large enough after finite number of iterations. Now we need to check the conclusion $t_n =\sum_{k=0}^{n-1} (t_{k+1}-t_{k})\leqslant  c2^\lambda \sum_i (1+ F(t_i))^{1-\lambda}$. Note that  since $F(t_i)\geqslant (F(t_n)-1)2^{n-i}+1$ then $(1+ F(t_i))^{1-\lambda}\leqslant ((F(t_n)-1)2^{n-i}+2)^{1-\lambda}$, so we have
\begin{equation*} \begin{split}
t_{n+1}\leqslant  c2^\lambda \sum_{i=0}^n \Big((F(t_n)-1)2^{n-i}+2\Big)^{1-\lambda} &\leqslant  c2^\lambda  (F(t_n)-1)^{1-\lambda}\sum_{i=0}^n 2^{(n-i)(1-\lambda)}   \\
&\leqslant  c2^\lambda  (F(t_n)-1)^{1-\lambda} \, \frac{1}{1-2^{1-\lambda}},
\end{split} \end{equation*}
which yields the estimate from the statement.
\end{Dem}

\medskip

Last we recall the fractional Leibniz rule and an elementary interpolation result used in the proof of the coming down property in Section \ref{SubsectionComingDown}, we prove these results in~\cite[Propositions 2.1 and 2.2]{BDFTCompanion}.

\ssk

\begin{prop} \label{PropLeibniz}
Let $\alpha>0, r\in \mathbb{N}$ and $ p,p_1,p_2,q \in [1,\infty]$ such that
\[  \frac{1}{p} =\frac{1}{p_1} + \frac{1}{p_2}.
\]
Then \[\| u^{r+1} \|_{B^\alpha_{p,q}}\lesssim \| u^r \|_{L^{p_1}}\| u \|_{B^{\alpha}_{p_2,q}}.\]
\end{prop}

\ssk

\begin{prop} \label{PropInterpolation}
Let $\alpha_1, \alpha_2 \in \mathbb{R}$ and $ p_1,p_2,q_1,q_2 \in [1,\infty]$ and $\theta\in [0,1]$. Define $\alpha= \theta \alpha_1+(1-\theta)\alpha_2 $, and $p,q\in [1,\infty]$ by
\[  \frac{1}{p} =\frac{\theta}{p_1} + \frac{1-\theta}{p_2}\, \text{ and  } \, \frac{1}{q} =\frac{\theta}{q_1} + \frac{1-\theta}{q_2}.
\]
Then \[\| u \|_{B^\alpha_{p,q}}\lesssim \| u \|^\theta_{B^{\alpha_1}_{p_1,q_1}}\| u \|^{1-\theta}_{B^{\alpha_2}_{p_2,q_2}}.\]
\end{prop}

{
\begin{prop}\label{prop_append:boundedpsidosLP}\cite[Proposition 2.11]{BDFTCompanion}
Let $(P^i_k,\widetilde{P}^i_\ell)$ be a pair of generalized Littlewood-Paley-Stein projectors,  where $i$ is a chart index and $k,\ell$ represents the frequencies $2^k,2^\ell$. For every pseudodifferential operator $A\in \Psi^m_{1,0}(M)$ the series of commutators  
\begin{eqnarray*}
\sum_{\vert k-\ell\vert\leqslant  1} \big( P^i_kA\widetilde{P}^i_\ell-AP^i_k\widetilde{P}^i_\ell \big)
\end{eqnarray*}
converges absolutely in $\Psi^{m-1}_{1,0}(M)$.
\end{prop}

\ssk

\section{Recollection of some results from the companion paper \cite{BDFTCompanion}}
\label{Appendix_companion_paper}

Recall the Cole-Hopf transform introduced in  \cite{JagannathPerkowski} which we used in our paper:
\begin{eqnarray*} \label{eq:decomp1}
u_r = \X_r - \IXthree_r + e^{-3\IXtwo_r} \big(v_{\textrm{ref},r} + v_r \big)
\end{eqnarray*}
where $v_{\textrm{ref},r}$ solves the equation
$$
\mathcal{L} v_{\textrm{ref},r} = 3e^{3\IXtwo_r} \left(\IXthree_r \Xtwo_r - b_r(\X_r + \IXthree_r)\right), \quad v_{\textrm{ref},r}(0)= 0,
$$ 
the function $v_r$ is the solution of Equation \eqref{EqJPFormulation}.

\smallskip

\begin{thm} \label{main_thm_1_companion} \cite[Theorem 3.13]{BDFTCompanion}
Fix $\varepsilon>0$. As $r\searrow0$, we have that 
$v_{\textrm{ref},r}$ converges in probability in $C_T\mcC^{1-\varepsilon}(M)$, and that  
\begin{equation} \label{EqFormerThm1.1}
\nabla \IXtwo_r \cdot \nabla v_{\textrm{\emph{ref}},r} - b_r(e^{3\IXtwo_r}\IXthree_r)
\end{equation}
converges in probability in $C_T\mcC^{-2\varepsilon}(M)$.
\end{thm}

\ssk
We recall the following collection of distributional kernels
\begin{equation*} \begin{split}
&\underline{\mathcal{L}}^{-1}\big((t,x),(s,y)\big) \defeq {\bf 1}_{(-\infty,t]}(s) \, e^{(t-s)(\Delta-1)}(x,y) \in \mathcal{D}^\prime( \mathbb{R}^2\times M^2 ) \\ 
& \Delta \underline{\mathcal{L}}^{-1}\big((t,x),(s,y)\big) \defeq {\bf 1}_{(-\infty,t]}(s) \, \left(\Delta e^{(t-s)(\Delta-1)}\right)(x,y) \in \mathcal{D}^\prime( \mathbb{R}^2\times M^2 )  \\
&G^{(p)}\big((t,x),(s,y)\big) \defeq  \left(\Big\{e^{\vert t-s\vert (\Delta-1)}(1-\Delta)^{-1}\Big\}(x,y)\right)^p \in  \mathcal{D}^\prime( \mathbb{R}^2\times M^2 ) , \qquad (1\leqslant p\leqslant 3) 
 \\
&[\odot_i](x,y,z) \defeq \sum_{\vert k-\ell\vert\leqslant  1} P^i_k(x,y) \, \widetilde{P}^i_\ell(x,z)  .\\
\end{split} \end{equation*}
where $P^i_k$ and $\widetilde{P}^i_\ell$ stand for some generalized Littlewood-Paley-Stein projectors that we introduce in previous section. 

\ssk

\begin{thm} \label{mainthm2_companion}\cite[Theorem 1.1]{BDFTCompanion}
With the conventions introduced above, we have the following microlocal estimates:
\begin{itemize}
	\item[--] The kernel $\underline{\mathcal{L}}^{-1}$ has scaling exponent $-3$ and wavefront set 
	$$
	N^*\left(\{t=s\}\times {\bf d}_2\subset \mathbb{R}^2\times M^2 \right).
	$$
	{
	\item[--] The kernel $\Delta\underline{\mathcal{L}}^{-1}$ has scaling exponent $-5$ and wavefront set 
	$$
	N^*\left(\{t=s\}\times {\bf d}_2\subset \mathbb{R}^2\times M^2 \right).
	$$}
	\item[--] The kernel $G^{(p)}$ have scaling exponent $-p$ and wavefront set 
	$$
	N^*\left(\{t=s\}\subset \mathbb{R}^2\times M^2 \right)\cup N^*\left(\{t=s\}\times {\bf d}_2 \subset \mathbb{R}^2\times M^2 \right).
	$$
	\item[--] The kernel $[\odot_i]$ has scaling exponent $-6$ and wavefront set
	$$
	 N^*\left(\{x=y=z\}\subset M^3 \right)
	$$
\end{itemize}
Any kernel $K$ of the above list satisfies some local diagonal bounds of the form
$$
\vert\partial_{\sqrt{t},\sqrt{s},x,y}^\alpha K\vert\lesssim \left(\sqrt{\vert t-s \vert}+\vert x-y\vert \right)^{a-\vert\alpha\vert}
$$
for the corresponding scaling exponent $a$. 
\end{thm}

\ssk

We recall some definitions of the heat calculus $\Psi_H^\alpha(M)$  and the parabolic calculus  $\Psi_P^\alpha(M)$.

\ssk

\begin{defn} \label{def:heatcalculus}
Let $M$ be a compact manifold and  $\alpha\leqslant 0$.
The space $\Psi_H^\alpha(M)$ is the set of smooth functions $K$ on $(0,+\infty)\times M^2$ satisfying the following properties for every $\delta\in [0,1)$.
\begin{itemize}
	\item We have the off-diagonal quantitative bounds, for any differential operator $P_{\sqrt{t},x,y}$, for all $N>0$, we find
\begin{eqnarray*}
\big\Vert P_{\sqrt{t},x,y}K\big\Vert_{L^\infty(M\times M)}\leqslant  C_{N,\alpha}\left(1+\frac{d(x,y)}{\sqrt{t}}\right)^{-N} \,  e^{-t\delta}.
\end{eqnarray*}

	\item For any $p$, for any open set $U$ endowed with a coordinate system near $p$, there is an element $\widetilde{A}\in C^\infty\big( [0,+\infty)_{\frac{1}{2}}\times \mathbb{R}^d\times U\big)$ such that
\begin{eqnarray*}
K_t(x,y)=t^{-\frac{d}{2}-1-\alpha}\widetilde{A}\Big(t,\frac{x-y}{\sqrt{t}},x\Big)
\end{eqnarray*}
and $\widetilde{A}\in C^\infty \big( [0,+\infty)_{\frac{1}{2}}\times \mathbb{R}^d\times U\big)$ satisfies the estimate
\begin{eqnarray*}
\big\Vert D^{\alpha}_{\sqrt{t},X,x}\widetilde{A} \big\Vert_{L^\infty([0,a]\times \mathbb{R}^d \times U)} \leqslant  C_{N,\alpha,\kappa(U)}\big(1+\Vert X\Vert\big)^{-N} \, e^{-t\delta}.
\end{eqnarray*}
The above bound holds true only for some compact time interval of the form $[0,a]$ for $0<a<+\infty$.
\end{itemize}
\end{defn}

\ssk

We denote by 
$
\mathcal{M}_P\subset C^\infty\big(T(\mathbb{R}\times M^2)\big)
$  
the finite $C^\infty$ module of vector fields  
tangent to the submanifold $\{0\}\times \text{Diag} \subset \mathbb{R}\times M^2$. 

\begin{defn} \label{defi:flatparabolic}
\cite[Definition 4.3]{BDFTCompanion}
Pick a negative real number $a$. Elements of the parabolic calculus in $\Psi_P^a(\mathbb{R}^2\times M^2)$ are operators whose  kernels $K\in C^\infty(\mathbb{R}^2\times M^2\setminus \textrm{\emph{Space time diagonal}})$ for which there exists a function $A\in C^\infty\left( [0,+\infty)\times \left(M^2\setminus \textrm{\emph{Space diagonal}}\right)\right)$ such that one has either  
$$
K(t,s,x,y) = A\big(\vert t-s\vert,x,y\big)
$$ 
or 
$$
K(t,s,x,y)={\bf 1}_{[0,\infty)}(t-s) A( t-s,x,y)
$$
and the following property hold. There exists $R>0$ such that for all $0<t<R$ and all  $V_1, \dots, V_k\in \mathcal{M}_P$,  one has
\begin{eqnarray} \label{ineq:Aderivatives}
\big\vert \big(V_1\dots V_k A\big)(t,x,y) \big\vert \lesssim_{V_1, \dots, V_k} \left\{ \begin{array}{ll} 
\left( \vert t\vert+d(x,y)^2\right)^{-\frac{2+2a+d}{2}} & \textrm{ if } 2+d+2a > 0\,,   \\
\big\vert\hspace{-0.05cm}\log\left( \vert t\vert+d(x,y)^2\right)\hspace{-0.07cm}\big\vert & \textrm{ if } 2+d+2a = 0\,.
\end{array} \right.
\end{eqnarray}
\end{defn}

\ssk

\begin{thm} \label{thm:compose_heat_parabolic} 
\cite[Theorem 4.17]{BDFTCompanion} 
Let $M$ be a smooth closed manifold of dimension $d$. Pick $A\in\Psi^{a}_P(M)$ and $B\in \Psi^b_H(M)$ with
$$
\left\{\begin{array}{ll} a,b\leqslant -1 &  \\  d+2+2a+2b \geqslant 0 &   \end{array} \right.
$$
For any $L>0$, set
\begin{eqnarray*}
C(t_1,t_2,x,y)\defeq\int_{-L}^{\inf(t_1,t_2)} A(t_2-s)\circ B(t_1-s)\rmd s\,.
\end{eqnarray*}
One has, for all $\epsilon>0$,
$$
\left\{\begin{array}{ll} C\in \Psi_P^{a+b} & \textrm{ if } d+2+2a+2b > 0 \,,\\  C\in \Psi_P^{a+b+\epsilon} & \textrm{ if } d+2+2a+2b = 0 \,. \end{array} \right.
$$
Moreover, the composition is bilinear hypocontinuous for the respective topologies.
\end{thm}
}

\ssk

\vspace{0.8cm}

\noindent \textcolor{gray}{$\bullet$} {\sf I. Bailleul} -- Univ Brest, CNRS UMR 6205, Laboratoire de Mathematiques de Bretagne Atlantique, France. {\it E-mail}: ismael.bailleul@univ-brest.fr   

\medskip
\noindent \textcolor{gray}{$\bullet$} {\sf N.V. Dang} -- Sorbonne Université and Université Paris Cité, CNRS, IMJ-PRG, F-75005 Paris, France.    \\
Institut Universitaire de France, Paris, France.
{\it E-mail}: dang@imj-prg.fr

\medskip
\noindent \textcolor{gray}{$\bullet$} {\sf L. Ferdinand} -- 
Laboratoire de Physique des 2 infinis Irène Joliot-Curie, UMR 9012, Universit\'e Paris-Saclay, Orsay, France.
{\it E-mail}: lferdinand@ijclab.in2p3.fr

\medskip

\noindent \textcolor{gray}{$\bullet$} {\sf T.D. T\^o} -- Sorbonne Université and Université Paris Cité, CNRS, IMJ-PRG, F-75005 Paris, France. {\it E-mail}: tat-dat.to@imj-prg.fr

\end{document}